\documentclass[11pt,a4paper,english]{article}
\usepackage{graphicx}
\usepackage{amssymb,latexsym}
\usepackage{amsmath}
\usepackage{epsf}
\usepackage{pdfsync}
\usepackage{epsfig}
\usepackage[parfill]{parskip}
\usepackage[matrix,arrow,color]{xy}
\usepackage[usenames]{color}
\usepackage{hyperref}

\input{xy}
\xyoption{all}

\input{epsf}

\makeatletter

\catcode`@=11
\renewcommand{\section}
{\@startsection{section}{1}{0pt}{\medskipamount}{\medskipamount}{\large\bf}}
\makeatletter\renewcommand{\subsection}
{\@startsection{subsection}{2}{\z@}{-3.25ex plus -1ex minus -.2ex}
{1.5ex plus .2ex}{\it }}

\numberwithin{equation}{section}
\catcode`@=12

\newcommand{\ba}{\begin{eqnarray*}}
\newcommand{\ea}{\end{eqnarray*}}
\newcommand{\ban}{\begin{eqnarray}}
\newcommand{\ean}{\end{eqnarray}}

\newcommand{\Tr}{{\rm Tr\,}}

\newcommand{\IZ}{\mathbb{Z}}
\newcommand{\IC}{\mathbb{C}}
\newcommand{\IP}{\mathbb{P}}
\newcommand{\IN}{\mathbb{N}}
\newcommand{\IR}{\mathbb{R}}
\newcommand{\IQ}{\mathbb{Q}}

\newcommand{\cI}{{\cal I}}
\newcommand{\cW}{{\cal W}}
\newcommand{\cN}{{\cal N}}
\newcommand{\cM}{{\cal M}}
\newcommand{\cS}{{\cal S}}
\newcommand{\cB}{{\cal B}}

\newcommand{\cH}{{\cal H}}
\newcommand{\cA}{{\cal A}}
\newcommand{\cE}{{\cal E}}
\newcommand{\cO}{{\cal O}}

\newcommand{\cQ}{{\cal Q}}
\newcommand{\cR}{{\cal R}}
\newcommand{\cZ}{{\cal Z}}

\newcommand{\cF}{{\cal F}}

\newcommand{\cT}{{\cal T}}
\newcommand{\cK}{{\cal K}}

\newcommand{\cV}{{\cal V}}

\newcommand{\sfA}{{\mathsf{A}}}
\newcommand{\sfe}{{\mathsf{e}}}

\newcommand{\sfb}{{\mathsf{b}}}
\newcommand{\sfP}{{\mathsf{P}}}
\newcommand{\sfZ}{{\mathsf{Z}}}
\newcommand{\sfQ}{{\mathsf{Q}}}
\newcommand{\sfD}{{\mathsf{D}}}

\newcommand{\sfR}{{\mathsf{R}}}

\newcommand{\oX}{{\overline{X}}}

\newcommand{\NDT}{{\tt NC}}
\newcommand{\DT}{{\tt DT}}

\def \nn{\nonumber}

\newcommand{\mbf}[1]{{\boldsymbol {#1} }}
\newcommand{\complex}{{\mathbb C}} 
\newcommand{\zed}{{\mathbb Z}} 
\newcommand{\real}{{\mathbb R}} 
\newcommand{\torus}{{\mathbb T}}

\def\e{{\,\rm e}\,}

\newcommand{\ch}{{\rm ch}}
\newcommand{\Ch}{{\rm Char}}

\def\ii{{\,{\rm i}\,}}
\def\dd{{\rm d}}

\newcommand{\Hom}{\mathrm{Hom}}

\newcommand{\End}{\mathrm{End}}

\newcommand{\bdiv}{\wp_{\infty}}
\newcommand{\Ext}{\mathrm{Ext}}

\newcommand{\Gammaw}{{\widehat\Gamma}}

\def\beq{\begin{equation}}
\def\bee{\begin{equation}}
\def\eeq{\end{equation}}
\def\bea{\begin{eqnarray}}
\def\eea{\end{eqnarray}}
\def\bd{\begin{displaymath}}
\def\ed{\end{displaymath}}

\newcommand{\Cint}{\int\kern-10.5pt-\kern7pt}

\newcommand{\PP}{{\mathbb{P}}}

\newcommand{\be}{\begin{equation}}
\newcommand{\ee}{\end{equation}}

\newcommand\fverbit{\egroup\item[\fbox{\unhbox\pippobox}]}
\newbox\pippobox

\def\a{\alpha}

\def\b{\beta}

{

\def\pa{\partial}

\def\eps{\epsilon}
\def\g{\gamma}

\def\bC{{\mathbb C}}

\def\th{\theta}

\def\w{\wedge}

\def\be{\begin{equation}}
\def\ee{\end{equation}}
\def\bea{\begin{eqnarray}}
\def\eea{\end{eqnarray}}
\def\nn{\nonumber\\}

\newtheorem{lemma}[equation]{Lemma}

\newcommand{\Proof}[1]{\noindent\underline{\textsf{Proof}}: #1 \hfill
  $\blacksquare$\\}

\begin{document}

\begin{titlepage}
\setcounter{page}{0}
\begin{flushright}
DAMTP--2010--124\\
HWM--10--35 \\ EMPG--10--25
\end{flushright}

\vskip 1.8cm

\begin{center}

{\Large\bf Instantons, Quivers \\ [3mm] and Noncommutative
  Donaldson--Thomas Theory}

\vspace{15mm}

{\large\bf Michele Cirafici$^{(a)}$, Annamaria Sinkovics$^{(b)}$ and
 Richard~J.~Szabo$^{(c)}$}
\\[6mm]
\noindent{\em $^a$ Centro de An\'{a}lise Matem\'{a}tica, Geometria e Sistemas
Din\^{a}micos\\ Departamento de Matem\'atica\\ Instituto Superior T\'ecnico\\
Av. Rovisco Pais, 1049-001 Lisboa, Portugal}\\ Email: \ {\tt cirafici@math.ist.utl.pt}\\[3mm]
\noindent{\em $^{(b)}$ Department of Applied Mathematics and
  Theoretical Physics
\\ Centre for Mathematical Sciences, University of
Cambridge \\ Wilberforce Road, Cambridge CB3 0WA, UK}\\ Email: {\tt
  A.Sinkovics@damtp.cam.ac.uk}\\[3mm]
\noindent{\em $^{(c)}$ Department of Mathematics\\ Heriot--Watt
  University\\
Colin Maclaurin Building, Riccarton, Edinburgh EH14 4AS, UK}\\ and\\ 
{\em Maxwell Institute for Mathematical Sciences, Edinburgh, UK}\\ Email: {\tt
  R.J.Szabo@ma.hw.ac.uk}

\vspace{15mm}

\begin{abstract}
\noindent

We construct noncommutative Donaldson--Thomas invariants associated with
abelian orbifold singularities by analysing the instanton
contributions to a six-dimensional topological gauge theory. The
noncommutative deformation of this gauge theory localizes on
noncommutative instantons which can be classified in terms of
three-dimensional Young diagrams with a colouring of
boxes according to the orbifold group. We
construct a moduli space for these gauge field configurations which
allows us to compute its virtual numbers via the counting of representations of a quiver with
relations. The quiver encodes the instanton dynamics of the noncommutative gauge theory, and is associated to the geometry of the
singularity via the generalized McKay correspondence. The index of BPS
states which compute the noncommutative Donaldson--Thomas invariants
is realized via topological quantum mechanics based on the quiver
data. We illustrate these constructions with several explicit
examples, involving also higher rank Coulomb branch invariants and
geometries with compact divisors, and connect our approach with other ones in the literature.
 
\end{abstract}

\end{center}
\end{titlepage}

\newpage

\tableofcontents

\section{Introduction}

Topological string theory on a smooth, six-dimensional toric
Calabi--Yau manifold is dual to a classical statistical mechanics
which describes the melting process of a three-dimensional
crystal. This duality was originally exhibited in a few examples
in~\cite{Okounkov:2003sp,qfoam} and subsequently extended to more
general (non-compact) toric Calabi--Yau threefolds
in e.g.~\cite{Saulina:2004da,Dijkgraaf:2004vp,Okuda:2004mb,Halmagyi:2005vk,Sulkowski:2006jp,Heckman:2006sk}. As the temperature is increased
the crystal melts and certain atomic configurations are removed. The
atomic configurations correspond in the dual picture to BPS states
that are geometrically enumerated by Donaldson--Thomas invariants,
which are invariant under deformations of the background. In turn,
these configurations are identified in the physical Type~IIA string
theory with stable bound states that a single D6 brane filling the
whole Calabi--Yau manifold can form with a gas of D0 and D2
branes.

As the physical moduli are continuously varied this picture gets modified. Stable states may become unstable and decay into more elementary constituents or new physical states can appear in the spectrum. This type of behaviour is at the core of the solution of $\cN=2$ supersymmetric Yang--Mills theory in four dimensions which was proposed in \cite{Seiberg:1994rs} and adapted to the supergravity setting in \cite{Denef:2007vg}. In Calabi--Yau compactifications it is only for a special region of the moduli space that the stable objects are enumerated via the Donaldson--Thomas invariants computed by topological string theory. 

As one moves around the moduli space, certain states can become lighter and different configurations become energetically favoured over others. The moduli space can be divided into chambers, each one with a physically distinct spectrum of stable BPS states. As the physical moduli are moved from one chamber to another, crossing a so-called wall of marginal stability, the index counting  BPS states jumps according to a wall-crossing formula. There is surmounting evidence that this wall-crossing formula is precisely the one found recently by Kontsevich and Soibelman \cite{Kontsevich:2009xt} in developing their theory of generalized Donaldson--Thomas invariants. This issue was extensively investigated in the context of gauge theory in \cite{Gaiotto:2010be,Gaiotto:2009hg,Gaiotto:2008cd} and further in the context of refined/motivic invariants in \cite{Dimofte:2009tm,Dimofte:2009bv}.

The usual Donaldson--Thomas invariants, at least as they are commonly
encountered in the context of topological string theory, are virtual
numbers of the moduli space of ideal sheaves with trivial
determinant. Since ideal sheaves are trivially stable, a generalized
theory of Donaldson--Thomas invariants is needed to fully account for
wall-crossing phenomena. This theory is naturally rooted in the
formalism of derived categories with the appropriate stability
conditions, which is widely believed to be the correct framework for
addressing questions concerning D~branes on Calabi--Yau manifolds
\cite{Aspinwall:2004jr}. The same sort of constructions have been pursued also by Joyce and Song in the less general but sometimes more concrete  framework of abelian categories \cite{joycesong}.

In many cases these constructions have been used to solve for the
physical spectrum of BPS states. This is the case for the class of
examples of local threefolds without compact four-cycles where the
chamber structure of the moduli space has been explicitly constructed
in \cite{nakanagao,nagao} and has found a clear physical
interpretation in \cite{Aganagic:2009kf} via a lift to M-theory. Here
the partition function of BPS states at a generic point of the moduli
space is seen as receiving competing contributions from both M2 and
anti-M2 branes. In a certain region of the moduli space the anti-M2
brane states are all unstable and the partition function of BPS states
is purely holomorphic. This is the region around the large radius
point described by the topological string partition function $\cZ_{\rm top}(q,Q)$, with the parameter $q$ weighting D0~branes and the parameters $Q$ weighting D2~branes. All the other regions can be reached by crossing walls of marginal stability and using the Kontsevich--Soibelman wall-crossing formula.

In another region of the moduli space the BPS state partition function has the form
\begin{equation}
\cZ_{\rm BPS} (q, Q) = \cZ_{\rm top} (q, Q) \ \cZ_{\rm top} (q, Q^{-1}) \ .
\label{ZBPSqQ}\end{equation}
This region corresponds to the noncommutative crepant resolution of a toric singularity where the BPS states are computed by noncommutative Donaldson--Thomas invariants. Ooguri and Yamazaki showed in \cite{Ooguri:2008yb} that these invariants count cyclic modules of a certain quiver which arises in a low-energy approximation of the theory governing a gas of D0 and D2 branes near the singularity in the sense of Douglas and Moore~\cite{Douglas:1996sw}. 

The quiver diagram is obtained from the toric diagram via a T-duality
transformation along the $\torus^2$ fibers of the toric
threefold. After the duality transformation the D0--D2 system becomes
an intricate configuration of D2 and NS5 branes. This configuration
has a low-energy description in terms of a quiver with a
superpotential. Adding the D6 brane modifies the quiver through the
addition of a new vertex and a single arrow from the new vertex to an
arbitrary reference vertex of the old quiver. This quiver construction
identifies a new kind of melting crystal \cite{Ooguri:2008yb}. The zero temperature configuration is obtained starting from the reference vertex  and consists of layers of coloured atoms, with each colour associated with a different node of the original quiver (the new vertex only labels the colour of the atom which sits at the top of the pyramid). Each layer represents a module of the path algebra of the quiver. Equivalently, in the first layer one draws a number of atoms corresponding to the nodes of the quiver that can be reached from the reference node in precisely one step. In the second layer one consider paths of the quiver consisting of two arrows, and so on. The general picture is obtained similarly, though some further combinatorial complications arise from the relations of the quiver, or equivalently the F-term constraints derived from the superpotential.

BPS states are counted by removing atomic configurations according to
a certain rule which roughly states that the crystal melts starting
from its peak; equivalently, if an atom is removed then so are all the
atoms above it. This implies that the complement of the atomic
configuration removed is algebraically an ideal in a certain
algebra; typically the ideals are generated by monomials in edge
variables associated to the pertinent quiver (before adding the
D6~brane node). In this way one computes the index of BPS
states in the region of the moduli space corresponding to a
noncommutative deformation of the toric variety, the so-called
noncommutative crepant resolution. It was proven by Van den Bergh
\cite{vandenbergh1} that the path algebra of a certain quiver with
relations associated with the toric singularity is a crepant
resolution of that singularity. The counting of BPS states in this
region was introduced by Szendr\H{o}i for the conifold
\cite{szendroi}, and by Mozgovoy and Reineke \cite{reineke} for
(essentially) generic singularities. 

This picture was further enriched in \cite{Aganagic:2010qr} where Aganagic and Schaeffer study generic toric Calabi--Yau threefolds, possibly with four-cycles. Their picture is general enough to include walls of the second kind which are elegantly described via mutations of the low-energy quiver. In particular, they offer a clear picture of the relation between melting crystal configurations and D~brane charges, thus resolving the apparent mismatch between the number of natural parameters associated to atomic colourings and the number of parameters in the topological string amplitude.

In this paper we take a rather different approach, which is somewhat
less ambitious. The central idea is to use a D~brane worldvolume
perspective and try to understand how much of these properties can be
captured by a study of the worldvolume gauge theory, and modifications
thereof. This approach was successful in the case of smooth toric
threefolds in the topological string chamber~\cite{qfoam}. The gauge
theory in question is the topological twist of six-dimensional $\cN=2$
supersymmetric Yang--Mills theory studied
in~\cite{Acharya:1997gp,Dijkgraaf:1996tz,Blau:1997pp,Hofman:2000yx,Baulieu:1997jx}
and the relevant BPS configurations are identified with generalized
instantons, solutions of the Donaldson--Uhlenbeck--Yau equations. A
 noncommutative deformation of the worldvolume gauge theory provides a natural compactification of the instanton moduli space and its virtual numbers can be evaluated via equivariant localization, reproducing the partition function for Donaldson--Thomas invariants~\cite{MNOP,MNOP2,MOOP}. Associated with this noncommutative gauge theory is a quantum mechanics which describes the dynamics of the collective coordinates on the instanton moduli space \cite{Jafferis:2007sg,Cirafici:2008sn}. We utilise and adapt to our problem the techniques of equivariant localization pioneered by Nekrasov in the context of Seiberg--Witten theory~\cite{Nekrasov:2002qd,Nekrasov:2003rj,Nekrasov:2003vi}; see e.g.~\cite{Szabo:2009vw} for a review geared at the context of the present paper.

We study this gauge theory on orbifolds of the form $\complex^3 /
\Gamma$, which we interpret as quotient stacks $[ \complex^3 / \Gamma
]$, where $\Gamma$ is a finite subgroup of $SL(3,\IC)$. The
topological gauge theory localizes by construction on
$\Gamma$-equivariant instanton configurations and thus poses a novel
enumerative problem. This enumerative problem, which reduces to
counting the virtual numbers of the moduli space of
$\Gamma$-equivariant ideal sheaves, is precisely equivalent to the
study of noncommutative Donaldson--Thomas invariants via a quiver
gauge theory. Indeed, the local structure of the instanton moduli
space on quotient stacks can be encoded in a quiver, which is a
modification of the McKay quiver associated to the singularity and
appears to be the same as the quiver used by Ooguri and Yamazaki
in~\cite{Ooguri:2008yb}.

Geometrically this problem reduces to the counting of $\Gamma$-equivariant closed subschemes of $\complex^3$, a problem which can be greatly simplified by using equivariant localization techniques with respect to the natural toric action on $\complex^3$. These techniques are only available when the toric action is compatible with the orbifold action, i.e. when the orbifold group is a subgroup of the torus group. In particular this is true for abelian orbifolds that respect the Calabi--Yau conditions, which is the case we will focus on in this paper.

We use this formalism to compute noncommutative Donaldson--Thomas invariants and assemble them into partition functions where the formal variables have a specific form which is derived from the instanton action. Much in the same way as in \cite{Aganagic:2010qr}, the counting variables are not all independent but related to geometrical quantities, via their relation to the D~brane charges in their work, and via their relation with the instanton action in ours. Our construction of the instanton moduli
space computes the instanton partition function in a clear and
self-consistent way with as many parameters as are present in the
formalism based on topological string theory. We also construct
Coulomb branch invariants associated to arbitrary numbers of
D6--D4--D2--D0 branes on these noncommutative crepant resolutions. We elucidate our formalism with plenty of examples.

This paper is written in a expository way, surveying various known mathematical results, and comparing them with our gauge theory
calculations; it is organised as follows. In Section~\ref{sec:BPS} we
describe the pertinent gauge theory, and the enumerative geometry problem it is supposed to address. In Section~\ref{sec:NCGT} we construct its noncommutative
instanton contributions, and describe how the worldvolume gauge theory
partition function naturally organises itself into a generating
function for coloured three-dimensional Young diagrams associated to
the $\Gamma$-invariant closed subschemes of $\complex^3$. In
Section~\ref{sec:Instmodsp} we propose our construction of the
instanton moduli space for $\complex^3 / \Gamma$ orbifolds; the
construction is inspired by the old construction of Kronheimer and
Nakajima~\cite{kronheimer} of the moduli variety of instantons on ALE
spaces. In Section~\ref{sec:InstQM} we reformulate this construction
in terms of a topological quiver quantum mechanics; we explain how the
instanton moduli space is characterized by the quiver, and how the
torus fixed points and local characters are
computed. Sections~\ref{sec:C3Z2}--\ref{sec:C3Z6} analyse explicit
examples with many detailed calculations, comparing our results with
the existing literature. Section~\ref{sec:Disc} summarises and
discusses some open technical aspects of our analysis. We have
included three appendices at the end of the paper containing some of
the more technical concepts and computations which are used in the
main text. 

\section{BPS states and gauge theory variables\label{sec:BPS}}

In the large radius limit the existence of bound states of
lower-dimensional branes with $N$ D6 branes wrapping a Calabi--Yau
threefold $X$ can be addressed directly from the study of the
Dirac--Born--Infeld theory defined on the D6 worldvolume. This gauge
theory is automatically twisted since on a Calabi--Yau manifold one
can identify spinor bundles with bundles of differential forms. In the case of a local threefold this is approximated in the low-energy limit by an ordinary supersymmetric gauge theory. This gauge theory is the topologically twisted version of six-dimensional $\cN=2$ Yang--Mills theory with gauge group $U(N)$. On an arbitrary K\"ahler threefold $X$ the bosonic
part of the action has the form
\begin{eqnarray}
S  &=& \frac{1}{2}\, \int_X\, \Tr \left( \dd_A \Phi \wedge * \dd_A
\overline{\Phi} + \big[\Phi \,,\, \overline{\Phi}~\big]^2 +
\big|F_A^{2,0} + \overline{\pa}\,_A^{\dagger} \rho\big|^2 +
\big|F_A^{1,1}\big|^2  \right) \nonumber\\  &&+\,
\frac{1}{2}\, \int_X \,\Tr \Big(   F_A \w F_A \w \omega +
\mbox{$\frac\vartheta3$}\, F_A
\w F_A \w F_A \Big) \ ,
\label{bosactionTGT}\end{eqnarray}
where $\Tr$ denotes the trace in the fundamental representation of $U(N)$, $\dd_A=\dd+\ii[A,-]$ is the gauge-covariant derivative, $\Phi$ is a complex adjoint scalar field, and
$F_A=\dd A+A\wedge A$ is the gauge field strength. Here $\rho$ is a $(3,0)$-form, $*$ is
the Hodge duality operator with respect to the K\"ahler metric of $X$, $\omega$ is the
background K{\"a}hler two-form of $X$, and $\vartheta$ is the
six-dimensional theta-angle which is identified with the
topological string coupling~$g_s$.

Since the gauge theory is cohomological, its quantum partition
function and supersymmetric observables localize onto the moduli space of solutions of the generalized instanton equations
\begin{eqnarray} \label{inste} F_A^{2,0} &=&
  \overline{\pa}\,_A^{\dagger} \rho
  \ , \nonumber\\[4pt]
F_A^{1,1} \w \omega \w \omega + \big[\rho\,\stackrel{\wedge}{,} \, \overline{\rho}\,\big] &=&
l~\omega \w \omega \w \omega \ , \nonumber\\[4pt] \dd_A \Phi &=& 0 \ ,
\end{eqnarray}
where the constant $l$ is related to the magnetic charge of the gauge
bundle. For a Calabi--Yau background we can consider minima where $\rho=0$.
On a smooth toric threefold the partition function of this gauge
theory can be evaluated via equivariant localization techniques. The
moduli space of solutions of the first-order equations (\ref{inste})
is desingularized by adding appropriate point-like
configurations. Since the gauge theory is cohomological every physical
observable can be expressed in terms of intersection integrals over
the moduli space. These integrals can be accordingly computed via the
localization formula.

For $N=1$ the problem is mathematically
well-formulated, and the resulting virtual numbers are the
Donaldson--Thomas invariants which count stable BPS bound states of D2
and D0 branes with a single D6 brane. For $N>1$ the problem is
similarly well-posed for an arbitrary collection of D6 branes in the
Coulomb branch of the gauge theory~\cite{Cirafici:2008sn} (see
also~\cite{Awata:2009dd,Liu:2010xd}). In this case the punctual
invariants computed in~\cite{Cirafici:2008sn} coincide with the
degenerate central charge limit of the non-abelian Donaldson--Thomas
invariants constructed in~\cite{stoppa} for D6--D0 bound states; Coulomb branch invariants are
also defined in~\cite{Diaconescu} corresponding to local D6--D2--D0
configurations. The non-abelian gauge theory in this branch does not seem to be naturally dual to topological string theory or even to
enumerate holomorphic curves.

This picture as it stands is literally true only in a certain
chamber. Generically on the whole Calabi--Yau moduli space stable BPS
states should be understood as stable objects in the bounded derived category of coherent sheaves on $X$. It is an interesting and ambitious project to understand how much of this picture can be captured in terms of gauge theory variables or modifications thereof. Such modifications can include turning on a noncommutative deformation of $X$ via a nontrivial $B$-field, or including nonlinear higher-derivative corrections to the gauge theory action and hence to the equations (\ref{inste}). It is likely that a mixture of these ingredients and string theory effects should capture the enumerative problem of stable BPS states at least in some chambers.

In this paper we will study this gauge theory on orbifolds of the form $\complex^3 / \Gamma$ and we shall propose that working equivariantly on $\complex^3$ with respect to the linear action of the finite group $\Gamma\subset SL(3,\IC)$, or equivalently working on the quotient stack $[\complex^3 / \Gamma]$, captures the enumerative problem corresponding to the noncommutative Donaldson--Thomas invariants.
The mathematical intuition behind this perspective comes from the work
of Bryan and Young~\cite{young} which studies deformation invariants
counting ideal sheaves of zero-dimensional $\Gamma$-equivariant
subschemes on $\complex^3$, or equivalently properly supported
substacks of $[\complex^3 / \Gamma]$. These invariants correspond precisely to the noncommutative Donaldson--Thomas invariants~\cite{joycesong}.

In the following we will interpret the ``gauge theory'' living in the
orbifold phase as a theory of $\Gamma$-equivariant sheaves on
$\complex^3$. Some technical aspects of the description of the gauge
theory in this sense are briefly discussed in
Appendix~\ref{App:Stacks}. This will allow us to define a  quiver which describes the dynamics of the instanton collective coordinates. The study of the representation theory of this quiver will then yield the noncommutative Donaldson--Thomas invariants.

\section{Noncommutative gauge theory\label{sec:NCGT}}

In this section we will study the noncommutative deformation of the
gauge theory introduced in Section~\ref{sec:BPS} for $U(1)$ gauge group; this corresponds to
subjecting the D6--D2--D0 system to a large Neveu--Schwarz
$B$-field. The idea is that string theory effects will resolve the
orbifold singularity $\IC^3/\Gamma$ and should make the gauge theory
well-behaved; see Appendix~\ref{App:Stacks} for some details. We are interested in the region of the K\"ahler
moduli space where the resolution is still small, for example when the
classical volume of the cycles is still vanishing, while the quantum
volume as measured by the $B$-field is non-zero but small. Even if the
$B$-field is vanishingly small, since the classical volume of the
cycles are zero, the gauge theory sits in the deep noncommutative
regime of the K\"ahler moduli space. We address this issue in more
detail in Section~\ref{subsec:Stability}. The BPS state counting in terms of D6--D0 bound states involves
fractional branes which can carry both D0 and D2 charge, where the
D2 charge originates in the large radius limit from D2~branes wrapped on two-cycles which vanish
at the orbifold point. In the more general BPS state counting problem that we consider later on, fractional branes can also come from wrapped D4~branes or bound states of these D2 and D4~branes.

\subsection{Noncommutative instantons on $\mathbb{C}^3$\label{subsec:InstC3}}

We begin by reviewing the construction of contributions from noncommutative instantons
to the partition function of
the six-dimensional $U(1)$ gauge theory on $X=\mathbb{C}^3$, following~\cite{qfoam,Cirafici:2008sn}. In this case there is a single patch in the geometry and only six-dimensional point-like instantons contribute. In particular, there are no contributions from four-dimensional instantons
stretched over two-spheres, since there are no non-trivial two-cycles in the geometry.

To compute the partition function for this gauge
theory we need to use localization and understand the moduli space
of solutions to the equations (\ref{inste}). To resolve short-distance
singularities of the moduli space, and
to find explicit instanton solutions, we use a
noncommutative deformation of the gauge theory~\cite{qfoam}. The coordinates $(x^i)$ of $\IC^3\cong \IR^6$ thus satisfy the Heisenberg algebra
\beq
[x^i, x^j] = \ii \theta^{ij} \ , \qquad i,j=1, \ldots ,6 \ ,
\eeq
where
\beq
\theta =(\theta^{ij}) = \left(
\begin{tabular}{cccccc}
0 &$\th_1$ &&&& \\
$-\th_1$ & 0 &&&& \\
&&0& $\th_2$&&\\
&&$-\th_2$&0&&\\
&&&&0& $\th_3$\\
&&&&$-\th_3$&0\\
\end{tabular} \right)
\eeq
is a constant matrix with $\theta_\alpha>0$ for $\alpha=1,2,3$. We change gauge theory variables to the covariant coordinates
\beq
X^i = x^i + \ii \theta^{ij}\, A_j \ ,
\eeq
and introduce complex combinations $Z^{\a} = {1 \over \sqrt{2
\theta_\a}}\, (X^{2\a-1} + \ii X^{2\a})$ for $\a=1,2,3$.

Then the instanton equations (\ref{inste})
can be rewritten in the ADHM form 
\bea \big[Z^{\a}\,,\, Z^{\b}\big] + \sum_{\g=1}^3\, \eps^{\a
\b \g}\, \big[Z^{\dagger}_{\gamma}\,,\, \rho\big] &=& 0 \ ,
 \nn[4pt] \relax
\sum_{\a=1}^3\, \big[Z^{\a}\,,\, Z^{\dagger}_{\a}\, \big] + \big[\rho\,,\,
\rho^{\dagger}\,\big] &=& 3 \ , \nn[4pt]
\relax \big[Z^{\a}\,, \, \Phi\big] &=& 0
\label{instenc}\eea 
for $\a,\b=1,2,3$. For the remainder of this section we set the $(3,0)$-form field $\rho$ to zero, as we work on a
Calabi--Yau geometry. We now introduce another deformation which
regulates the infrared singularities of the instanton moduli space, by
turning
on the $\Omega$-background with equivariant parameters
$\eps_{1},\eps_{2},\eps_{3}$ which parametrize the natural scaling action of
the three-torus $\torus^3$ on $\IC^3$. This deformation
changes the last equation of (\ref{instenc}) to
\beq
\big[Z^{\a} \,,\, \Phi\big] = \eps_{\a} \, Z^{\a} \ .
\eeq

The set of equations (\ref{instenc}) can be solved by harmonic oscillator
algebra. We represent the fields as operators on a three-particle
quantum mechanical Fock space $\cH$, which is the unique irreducible
module over the Heisenberg algebra, with the usual creation and
annihilation operators $a_\a^\dag= (x^{2\a-1} - \ii x^{2\a})/\sqrt{2
\theta_\a}$, $a_\a= (x^{2\a-1} + \ii x^{2\a}) /{\sqrt{2
\theta_\a}}$ for $\a=1,2,3$ and number
basis $|n_1,n_2,n_3\rangle= \prod_\a\,
{(a_\a^\dag)^{n_\a}}/{\sqrt{n_\a!}}\, |0,0,0\rangle$ with $n_\a\in\IN_0$. The vacuum solution is then given
by 
\bea 
Z^{\a} = a_{\a} \qquad \mbox{and} \qquad \Phi
= \sum_{\a=1}^{3}\, \eps_\a\, a_\a^{\dagger}\, a_\a  \ .
\eea 
Other solutions
are found with the solution generating technique~\cite{qfoam,Cirafici:2008sn}. The idea is that one can use
the partial isometry $U_n$ on $\cH$ obeying
\beq
U_n^{\dagger}\, U_n = 1 - \sum_{n_1+n_2+n_3<n}\, |n_1,n_2,n_3 \rangle \langle
n_1,n_2,n_3| \qquad \mbox{and} \qquad U_n\, U_n^{\dagger} =1
\eeq
to build a solution of the form 
\bea Z^{\alpha} &=& U_n \, a_{\alpha} \,
f(N) \, U_n^{\dagger} \qquad \mbox{with} \quad N= \sum_{\a=1}^3\,
a_\a^{\dagger}\, a_\a \ , \nn[4pt]
\Phi &=& U_n\,  \sum_{\a=1}^{3}\, \eps_\a\, a_\a^{\dagger}\, a_\a\,
U_n^{\dagger} \ .
\eea 
The function $f(N)$ of the number operator $N$ is then found by
substituting into the instanton equations to get
\beq
f(N) = \Big( 1 - {n\, (n+1)\, (n+2) \over N\, (N+1)\, (N+2)} \, 
\Big)^{1/2} \ .
\eeq

The Hilbert space $\cH=\IC[a_1^\dag,a_2^\dag,a_3^\dag]|0,0,0\rangle$ builds up from the states of the three-dimensional
harmonic oscillator, and the partial isometry $U_n$ maps
to its subspace generated by ideals
\beq
{\cal H}_{I} =  I(a_1^{\dagger}, a_2^{\dagger}, a_3^{\dagger}) | 0,0,0
\rangle \ .
\eeq
Such ideals are generated by monomials and are in one-to-one
correspondence with plane partitions (three-dimensional Young
diagrams). For an ideal $I$
corresponding to the plane partition $\pi$ one obtains the character
\bea
\Ch_\pi(t) &:=& \Tr_{{\cal H}_I}( \e^{t \, \Phi}) \label{ene}\\[4pt] &=& {1 \over (1 - \e^{t\, \eps_1} )\, (1 -
  \e^{t\, \eps_2} )\, (1 - \e^{t \, \eps_3} )}  -
\sum_{(n_1,n_2,n_3) \in \pi}\, \e^{t\, ( \eps_1\, (n_1-1) + \eps_2 \,
  (n_2 -1) +
  \eps_3\, (n_3-1))} \ , \nonumber
\eea
where the first term in the second line of (\ref{ene}) is the vacuum contribution and the
sum runs over box locations of
$\pi\subset\IN_0^3$. Proceeding with the localization calculation the
contribution of an instanton comes from the weight factor
\beq
\exp\Big({ - {\vartheta \over 48 \pi^3} \, \int_X\, F_A \w F_A \w F_A}
\Big) = \exp\Big(\, {  {\ii \vartheta\, {\cal{E}}_I^{(3)} \over \eps_1
    \,\eps_2 \, \eps_3}} \,\Big) \ ,
\label{descenteq}\eeq
where ${\cal{E}}_I^{(3)}$ is the coefficient of $t^3$ in the expansion
of the function
\be {\cal{E}}_I (t) = (1 - \e^{t \, \eps_1} )\, (1 - \e^{t\, \eps_2} )\,
(1 - \e^{t\, \eps_3} ) \, \Tr_{{\cal H}_I} (\e^{t\, \Phi} )
\ee
around $t=0$.
In the following we will choose the equivariant parameters such that
\be \eps_1 + \eps_2 + \eps_3 =0\ , \ee
which enforces $\torus^3$-invariance of the holomorphic three-form.
Then one finds
\beq
{\cal{E}}_I^{(3)} =  \eps_1 \, \eps_2 \, \eps_3 \, \sum_{(n_1,n_2, n_3) \in
  \pi} \, 1 =  \eps_1 \, \eps_2 \, \eps_3\, |\pi| \ ,
\eeq
and hence the weight of an instanton is given in terms of the total
number of boxes $|\pi|$ in the corresponding plane partition $\pi$ as
\be \e^{\ii \vartheta\, |\pi|} \ . \ee

In addition to the weight there is also a contribution $\cZ_\pi$ from
the fluctuation determinants to the measure on the instanton moduli space. They can be written as $\cZ_\pi=\cZ_{\rm
  vac} \, \chi_{\torus^3}(\cN_\pi)$, where $\cZ_{\rm vac}$ is the perturbative vacuum contribution from the
empty Young diagram $\pi=\emptyset$, which will be dropped throughout
in the following, and
\beq
\chi_{\torus^3}(\cN)=\int_\frak{M}\, \sfe_{\torus^3}(\cN)
\label{chicalN}\eeq
is the $\torus^3$-equivariant Euler characteristic of the obstruction bundle $\cN$ over the
instanton moduli space $\frak{M}$~\cite{qfoam,Hofman:2000yx,Cirafici:2008sn}; we will give a
precise and rigorous definition of the integral class (\ref{chicalN})
later on. 
This quantity was computed explicitly
in~\cite{qfoam,Cirafici:2008sn}, where it was shown that the instanton
measure at a fixed point $\pi$ is given by
\be
\chi_{\torus^3}(\cN_\pi) =  (-1)^{|\pi|} \ .
\ee
Putting everything together, the instanton part of
the full gauge theory partition function on $X=\IC^3$ is
\be
\cZ_{\IC^3} = \sum_{\pi}\, \big(-\e^{\ii \vartheta} \, \big)^{|\pi|} = M(q) \ ,
\ee
where $q= - \e^{\ii \vartheta} = \e^{-g_s}$. The quantity
\beq
M(q)=\prod_{n=1}^\infty\, \big(1-q^n\big)^{-n}
\label{MacMahon}\eeq
is the MacMahon function which is the generating function for plane
partitions. Each noncommutative instanton corresponds to a molten
crystal configuration $\IN_0^3\setminus \pi$.

\subsection{Noncommutative instantons on $\mathbb{C}^3/\mathbb{Z}_3$\label{subsec:InstC3Z3}}

We shall now discuss how to extend this calculation to the orbifolds of
interest in this paper. We begin with the
$\IC^3/\IZ_3$ orbifold in order to explicitly illustrate the generic features
which arise. We take a generator $g$ of the diagonal subgroup
$\IZ_3\subset\torus^3$ whose action on $\IC^3$ is given by
\beq
g\cdot(z_1,z_2,z_3)= ( \zeta \, z_1,\zeta \, z_2,\zeta \, z_3) \ ,
\label{Z3C3action}\eeq
where $\zeta=\e^{{2 \pi \ii / 3} }$. In the noncommutative gauge
theory the operators $a_\a,a_\a^\dag$ transform in the same way under
the action of $\Gamma=\IZ_3$. The orbifold quotient is defined by
taking the crossed product of the noncommutative algebra of fields
with the orbifold group $\Gamma$.

The Fock space 
\beq
\cH=\bigoplus_{r=0}^2\, \cH^{(r)}
\eeq
thus splits into three twisted sectors $\cH^{(r)}={\rm
  span}_{\IC}\big\{|n_1,n_2,n_3\rangle~|~ n_1+n_2+n_3\equiv r \ {\rm mod} \ 3\big\}$ of the orbifold according to the
irreducible representations of the group $\IZ_3$. The vacuum solution can now be written as
\beq
Z^\a = a_\a =  \left(
\begin{tabular}{ccc}
0& $ a_\a^{(1)}$ & 0 \\
0 & 0 & $ a_\a^{(2)}$ \\
$ a_\a^{(0)}$ &0 &0 \\
\end{tabular} \right) \ .
\eeq
The first instanton equation in (\ref{instenc}) then yields
\bea
a_\a^{(0)}\, a_\b^{(1)}&=& a_\b^{(0)}\, a_\a^{(1)} \ , \nonumber\\[4pt]
a_\a^{(1)}\, a_\b^{(2)} &=& a_\b^{(1)}\, a_\a^{(2)} \ , \nonumber\\[4pt]
a_\a^{(2)}\, a_\b^{(0)} &=& a_\b^{(2)}\, a_\a^{(0)}
\eea
for $\a,\b=1,2,3$.
 From the second instanton equation in (\ref{instenc}) we get the
 relations 
\bea
 a_\a^{(1)} \, a_\beta^{(1) \dagger} &=&  a_\beta^{ (0) \dagger} \,
 a_\a^{(0)} \ ,
\nn[4pt]  a_\a^{(2)}\,  a_\beta^{(2) \dagger} &=&  a_\beta^{ (1)
  \dagger} \,
 a_\a^{(1)} \ , \nn[4pt]  a_\a^{(0)} \, a_\beta^{(0) \dagger} &=&  a_\beta^{ (2)
\dagger} \, a_\a^{(2)} \eea 
for $\a \neq \b$, while for $\a=\b$ we have 
\bea
 a_\a^{(1)} \, a_\a^{(1) \dagger} -  a_\a^{(0) \dagger}\,  a_\a^{(0)} &=&
P^{(0)} \ , \nn[4pt]  a_\a^{(2)} \, a_\a^{(2) \dagger} -  a_\a^{(1)
  \dagger} \,
 a_\a^{(1)} &=& P^{(1)} \ , \nn[4pt]  a_\a^{(0)} \, a_\a^{(0) \dagger}
 -  a_\a^{(2) \dagger} \,
 a_\a^{(2)} &=& P^{(2)} \ ,
\eea 
where $P^{(r)}:\cH\to\cH^{(r)}$ for $r=0,1,2$ are the hermitian
projectors onto the twisted sectors of the Hilbert space.
The explicit expressions for the oscillator operators satisfying these
equations are 
\bea  a_1^{(r)} &=& \sum_{k=0}^{\infty}~
\sum_{n_1 + n_2 + n_3 = r +
3 k} \, \sqrt{n_1} \, |n_1 -1, n_2 , n_3 \rangle \langle n_1, n_2, n_3
| \ , \nn[4pt]  a_1^{(r) \dagger} &=& \sum_{k=0}^{\infty}~
\sum_{n_1 + n_2 + n_3 = r -1
+ 3 k}\, \sqrt{n_1 + 1} \, |n_1 +1 , n_2 , n_3 \rangle \langle n_1,
n_2, n_3 |
\eea 
for $r=0,1,2$, and analogously for the operators $ a_{2}^{(r)}$, $
a_{2}^{(r)\dag}$ and $ a_{3}^{(r)}$, $ a_{3}^{(r)\dag}$.
The vacuum solution for the scalar field $\Phi$ is now
\beq
\Phi= \sum\limits_{\a=1}^3\, \left(
\begin{tabular}{ccc}
$
\eps_\a\, N_\a^{(0)}$& 0& 0 \\
0& $\eps_\a \, N_\a^{(1)}$& 0 \\
0& 0& $\eps_\a \, N_\a^{(2)}$ \\
\end{tabular} \right)
\eeq
where the number operators $N_\a^{(r)}$ count states in twisted sectors
as
\beq
N_\a^{(r)} =  a_\a^{(r) \dagger}\,  a_\a^{(r)} = \sum_{k=0}^{\infty}~
\sum_{n_1 + n_2 + n_3 = r + 3 k}\, n_\a \, |n_1 , n_2 , n_3 \rangle \langle
n_1, n_2, n_3 | \ .
\eeq

We now look for the most general solution of the instanton equations
\bea
\big[Z^\a \,,\, Z^\b \big]
&=& 0 \ ,
 \nn[4pt] \relax
\sum_{\a=1}^3\, \big[Z^\a \,,\, Z^{\dagger}_\a \big]  &=& 3
 \left(
\begin{tabular}{ccc}
$P^{(0)}$& 0& 0 \\
0& $P^{(1)}$& 0 \\
0& 0& $P^{(2)}$ \\
\end{tabular} \right) \ ,
\nn[4pt]
\relax \big[Z^\a \,,\, \Phi\big] &=& \eps_\a\, Z^\a \ ,
\label{insteorb}\eea
where as before $\a,\b=1,2,3$ and $P^{(r)}$ are the projectors for the twisted sectors.
To construct these solutions we use the partial isometry operators $U_n$
from Section~\ref{subsec:InstC3} and split them into twisted sectors as
\bea
U_n \,U_n^{\dagger} &=& \left(
\begin{tabular}{ccc}
$P^{(0)}$& 0& 0 \\
0& $P^{(1)}$& 0 \\
0& 0& $P^{(2)}$ \\
\end{tabular} \right) \ , \nn[4pt] \relax
U_n^{\dagger}\, U_n &=&  \left(
\begin{tabular}{ccc}
$P^{(0)} - P_n^{(0)}$& 0& 0 \\
0& $P^{(1)}-P_n^{(1)}$& 0 \\
0& 0& $P^{(2)}-P_n^{(2)}$ \\
\end{tabular} \right) \ ,
\eea
where $P_n^{(r)}$ projects onto states with particle number $N<n$ in
the sector $\cH^{(r)}$ with
\beq
P_n^{(r)} = \sum_{k=0}^{\infty}~ \sum_{n_1 + n_2 + n_3 = r
+ 3 k < n}\, |n_1, n_2, n_3 \rangle \langle n_1, n_2, n_3 | \ .
\eeq
The general solution of the first two equations in (\ref{insteorb}) can then be written as
\be
Z^\a = U_n \, a_\a \ \underline{f}(N) \ U_n^{\dagger} \ .
\ee
Here $\underline{f}(N)$ is an operator-valued $3\times3$ matrix whose form is
determined by the instanton equations (\ref{insteorb}). It is easy to
show that any matrix of the form
\beq
\underline{f}(N) =  \left(
\begin{tabular}{ccc}
$f(N^{(0)})$& 0& 0 \\
0& $f(N^{(1)})$& 0 \\
0& 0& $f(N^{(2)})$ \\
\end{tabular} \right)
\qquad \mbox{with} \quad f\big(N^{(r)}\big)\big|_{N^{(r)}<n}=0
\eeq
is a solution, where $N^{(r)}=N_1^{(r)}+ N_2^{(r)}+ N_3^{(r)}$. The second equation in (\ref{insteorb}) then tells us
that the function $f$ satisfies the same recursion relation as in the
case of $\bC^3$~\cite{qfoam,Cirafici:2008sn}. There is also a more
general solution which satisfies the first two instanton equations in the same
way; it is of the form
\bea
\underline{f}(N) = 
\begin{pmatrix}
f(N^{(0)}) + f_{11} (N^{(1)}, N^{(2)})& f_{12}(N^{(2)})&  f_{13}(N^{(1)})\\
f_{21}(N^{(2)})& f(N^{(1)})+ f_{22} (N^{(0)}, N^{(2)})&  f_{23}(N^{(0)})\\
f_{31}(N^{(1)})& f_{32}(N^{(0)})& f(N^{(2)})+ f_{33}(N^{(0)}, N^{(1)})
\end{pmatrix} \nonumber \\
\eea
with no constraints on the extra functions.

The third equation in (\ref{insteorb}) is solved by
\be
\Phi=U_n \, \sum\limits_{\a=1}^3\, \left(
\begin{tabular}{ccc}
$
\eps_\a \, N_\a^{(0)}$& 0& 0 \\
0& $\eps_\a \, N_\a^{(1)}$& 0 \\
0& 0& $\eps_\a \, N_\a^{(2)}$ \\
\end{tabular} \right) \, U_n^{\dagger} \ .
\ee
A more general solution is of the form
\beq
\Phi=U_n \, \left(
\begin{tabular}{ccc}
$\sum\limits_{\a=1}^3\,
\eps_\a \, N_\a^{(0)}$& $\Phi_{12}(N^{(2)})$& $\Phi_{13}(N^{(1)})$ \\
$\Phi_{21}(N^{(2)})$& $\sum\limits_{\a=1}^3\, \eps_\a \, N_\a^{(1)}$& $\Phi_{23}(N^{(0)})$ \\
$\Phi_{31}(N^{(1)})$& $\Phi_{32}(N^{(0)})$& $\sum\limits_{\a=1}^3\,
\eps_\a \, N_\a^{(2)}$
\end{tabular} \right) \, U_n^{\dagger} \ .
\eeq
However, the extra functions here will play no role in the computation
of the gauge theory partition function, because
\beq
\Tr_\cH(\Phi) = \sum_{r=0}^2~
\sum_{\a=1}^3\,
\eps_\a \, \Tr_\cH\Big(N_\a^{(r)} \,\big(P^{(r)} - P_n^{(r)}\big) \Big) \
.
\eeq

The $\Gamma$-equivariant character for the vacuum solution of the noncommutative
gauge theory is given by
\be \Ch^\Gamma_{\emptyset}(t) = \Tr_{{\cal H}}(
\e^{t\, \Phi}) = \sum_{k=0}^{\infty}~ \sum_{r=0}^2 ~ \sum_{n_1 + n_2 +
  n_3 = r + 3 k}\, \e^{t \, \sum_\a\, \eps_\a\, n_\a} =:
\sum_{r=0}^2\, \Ch^\Gamma_r(t) \ ,
\ee
which splits into the twisted sectors. Projecting onto the
$\Gamma$-invariant sector corresponding to the trivial orbifold group representation gives
\bea
\Ch^\Gamma_{0}(t) &=&  \sum_{k=0}^{\infty}~ \sum_{n_1 + n_2 + n_3 = 3 k}\,
\e^{t \, \sum_\a\, \eps_\a \, n_\a} \nonumber\\[4pt]
&=&  {3 + 2\cosh(\eps_1 - \eps_2)\, t
+ 2\cosh(\eps_1 - \eps_3)\, t + 2\cosh(\eps_2 - \eps_3)\, t
\over (1 - \e^{3 \eps_1\, t})\, (1 - \e^{3 \eps_2\, t})\, (1 - \e^{3
  \eps_3\, t})}
\label{orbvacZ}\eea
where we have used $\eps_1 + \eps_2 + \eps_3 =0$. Taking the general
solution including the partial isometry $U_n$, the $\Gamma$-invariant character
is given by
\be
\Ch^\Gamma_\pi(t) = \Ch^\Gamma_0(t)  - \sum_{k=0}^{\infty}~ \sum_{\scriptstyle
  (n_1,n_2,n_3) \in \pi \atop \scriptstyle n_1 + n_2 + n_3 =  3k} \, 
\e^{t \, \eps_1 \, (n_1-1) + t \, \eps_2\, (n_2-1) + t \, \eps_3\,
  (n_3-1)} \ ,
\label{orbpiZ}\ee
where $\Ch^\Gamma_0(t)$ is the orbifold vacuum contribution (\ref{orbvacZ})
and $\pi$ is a plane partition. The sum in (\ref{orbpiZ}) corresponds
to $\IZ_3$-invariant zero-dimensional subschemes $Y\subset\IC^3$ for
which $\IZ_3$ acts trivially on $H^0(\cO_Y)$. In the following we
denote
them by ${\pi_0}$. Thus $(n_1,n_2,n_3) \in \pi_0$
if and only if $(n_1,n_2,n_3) \in \pi$ and $n_1 + n_2 + n_3 \equiv 0 \ {\rm
  mod}\ 3$.

The instanton weight can be obtained from localization as before through the descent equation (\ref{descenteq}), where now ${\cal{E}}_I^{(3)}$ is the coefficient of $t^3$ in the expansion of the function
\be 
{\cal{E}}^\Gamma_I (t) = {1 \over \Ch^\Gamma_0(t)}\, \Tr_{{\cal H}_I^{(0)}}( \e^{t\, \Phi})
\label{calEGammaIt}\ee
for the orbifold gauge theory around $t=0$. Expanding we obtain
\beq
 {\cal{E}}_I^{(3)} =  3 \eps_1\, \eps_2\, \eps_3\, \sum_{k=0}^{\infty}~ \sum_{\scriptstyle (n_1,n_2,n_3) \in \pi \atop \scriptstyle n_1 + n_2 + n_3 =  3k} \, 1 = 3 \eps_1\, \eps_2\, \eps_3 \,  \sum_{(n_1,n_2,n_3) \in {\pi_0}}\, 1 =
3 \eps_1\, \eps_2\, \eps_3 \, |{\pi_0}| \ .
\eeq
Hence the weight of an instanton in the sector corresponding to the trivial representation of $\zed_3$ is
\be \e^{3 \ii \vartheta \, |{\pi_0}|} \ . \ee
By including the other two twisted orbifold sectors, the
instanton contributions are characterized by plane partitions $\pi$ together with a
3-colouring $\pi=\pi_0\sqcup\pi_1 \sqcup\pi_2$, where $(n_1,n_2,n_3)\in\pi_r$
if and only if $n_1+n_2+n_3 \equiv r \ {\rm mod}\ 3$. The colours of the
boxes are in bijection with the set of irreducible representations of
the orbifold group $\Gamma=\IZ_3$.

\subsection{Coloured instanton partition functions\label{subsec:Colinst}}

More general abelian orbifolds $\IC^3/\Gamma$ can be
treated in exactly the same way and yield the same qualitative
behaviours. Let $\Gamma\subset\torus^3$ be a finite abelian group
acting linearly on $\IC^3$ with weights $r_1,r_2,r_3$ and with trivial
determinant, 
\beq
r_1+r_2+r_3\equiv0 \ .
\eeq
The set of irreducible
representations of $\Gamma$ forms a group $\widehat\Gamma\cong\Gamma$
under tensor product; we use an additive notation for the group
operation on weights $r$, a multiplicative notation for
corresponding characters $\chi_r:\Gamma\to\IC$, and tensor
product for representations $\rho_r$.

The Fock space of the noncommutative gauge
theory is a $\Gamma$-module which admits an isotopical decomposition into irreducible
representations as
\beq
\cH=\bigoplus_{r\in\widehat\Gamma}\, \cH^{(r)} \ ,
\label{hilsplitgen}\eeq
where
\bea
\cH^{(r)} &=& \Big(\, \frac1{|\Gamma|}\, \sum_{g\in\Gamma}\, \chi_r(g)\,
g^{-1}\, \Big) \cdot \IC\big[a_1^\dag\,,\,a_2^\dag \,,\,a_3^\dag \big]|0,0,0\rangle
\nonumber \\[4pt]
&=& {\rm span}_{\IC}\big\{|n_1,n_2,n_3\rangle~|~ n_1\, r_1+n_2\,r_2+n_3\,
r_3\equiv r\big\} \ .
\label{Hisotop}\eea
The covariant coordinates correspondingly decompose into operators
\beq
Z_\a=\bigoplus_{r\in\Gammaw}\, Z_\a^{(r)} \qquad \mbox{with} \quad
Z_\a^{(r)} \, :\, \cH^{(r)} \ \longrightarrow \ \cH^{(r+r_\a)}
\label{orbcovcoord}\eeq
for $\a=1,2,3$. The instanton equations (\ref{instenc}) then yield
\beq
Z_\a^{(r+r_\b)}\, Z_\b^{(r)}= Z_\b^{(r+r_\a)}\, Z_\a^{(r)}
\label{orbcomm}\eeq
for $\a, \b=1,2,3$ and $r\in\Gammaw$, and
\beq
\sum_{\a=1}^3\, \big( Z_\a^{(r-r_\a)}\, Z_\a^{(r-r_\a)}\,^\dag - Z_\a^{(r)}\,^\dag\, Z_\a^{(r)} \big) = 3P^{(r)}
\label{orbHeis}\eeq
where $P^{(r)}$ is the projection onto the isotopical component
$\cH^{(r)}$.
The partial isometries $U_n$ from Section~\ref{subsec:InstC3} are
accordingly decomposed as
\beq
U_n\, U_n^\dag=1 \qquad \mbox{and} \qquad U_n^\dag\, U_n =
1-\bigoplus_{r\in\widehat\Gamma}\, P^{(r)}\, {\mbf P}_{k_r(n)}\,
P^{(r)} \ ,
\label{Uncol}\eeq
where ${\mbf P}_{k_r(n)}$ is a projection operator of
finite rank $k_r(n)$, the number of states of $\cH^{(r)}$ with $N<n$. The
corresponding noncommutative instantons are labelled by
$\widehat\Gamma$-coloured plane partitions
$\pi=\bigsqcup_{r\in\widehat\Gamma}\, \pi_r$, where
$(n_1,n_2,n_3)\in\pi_r$ if and only if $n_1\, r_1+n_2\,r_2+n_3\,
r_3\equiv r$.

The orbifold field theory by construction naturally only keeps
contributions from $\Gamma$-invariant instanton configurations,
obtained by projection onto the trivial representation $r=0$ of the
orbifold group as in the example of Section~\ref{subsec:InstC3Z3}.
However, in what follows we would like to weigh the coloured instantons by a
set of variables $(p_r)_{r\in \widehat\Gamma}$ indexed by the
irreducible representations of the orbifold group $\Gamma$. For this,
rather than using the descent formula (\ref{descenteq}) from
localization, we will \emph{define} the instanton action of the D6 brane gauge
theory on $\IC^3/\Gamma$ via the Wess--Zumino coupling of constant
Ramond--Ramond fields $C$ dual to fractional D0 branes
(instantons); this enables the proper incorporation and weighting of
the twisted sectors $r\neq0$ in (\ref{hilsplitgen}) to match with
string theory expectations. Such fields decompose into twisted sectors of the closed
string
orbifold
as~\cite{Douglas:1996sw,Douglas:1997de,Diaconescu:1999dt,Szabo:2007wp}
\beq
C=\bigoplus_{r\in\widehat\Gamma}\, C^{(r)} \ .
\label{RRorb}\eeq
In Appendix~\ref{App:Stacks} we justify somewhat the formulation of
the gauge theory in this way.

Let $\rho$ denote the representation of the orbifold group on
(\ref{hilsplitgen}). For the example $\Gamma=\IZ_3$ considered in
Section~\ref{subsec:InstC3Z3}, the three-dimensional regular
representation $\rho(g)_{\a\b}=\zeta^\a\, \delta_{\a\b}$ naturally
corresponds to a superposition of fractional
instantons~\cite{Diaconescu:1999dt}. The corresponding
$\Gamma$-equivariant Chern character is given by~\cite{Szabo:2007wp}
\beq
\ch^\Gamma(F_A)=\bigoplus_{r\in\widehat\Gamma}\,
\Tr_{\cH^{(r)}}\big(\rho(r)\, \exp(-F_A^{(r)}/2\pi\ii)\big) \ ,
\label{eqChernchar}\eeq
where $F_A^{(r)}=[X,X]\big|_{\cH^{(r)}}\in\End_\complex\big(\cH^{(r)}\big)$ are the
diagonal components in the decomposition of the field strength $F_A$ on
the orbifold Hilbert space (\ref{hilsplitgen}) of the noncommutative
gauge theory.

Then the instanton action is defined by the anomalous coupling to the
D6~brane as~\cite{Douglas:1996sw,Douglas:1997de,Szabo:2007wp}
\beq
-\frac{\vartheta}{48 \pi^3}\, \int_X \, F_A \w F_A \w F_A
:= \frac{\vartheta}{6}\, \int_{\IC^3}\, \frac1{|\Gamma|}~
\sum_{r\in\widehat\Gamma}\, C^{(r)}\, \Tr_{\cH^{(r)}}\big(\rho(r)\,
\exp(-F_A^{(r)}/2\pi\ii)\big) \ ,
\label{WZorbdef}\eeq
which for the linear orbifold group actions on $\IC^3$ that we
consider in this paper can be expressed as
\beq
-\frac{\vartheta}{48 \pi^3}\, \int_X \, F_A \w F_A \w F_A
= \frac{\vartheta}{6}\,\frac\ii{8\pi^3}\, \frac1{|\Gamma|}\,
\sum_{r\in\widehat\Gamma}\, C^{(r)}\, \chi_\rho(r)\,
\Tr_{\cH^{(r)}}\big(F_A^{(r)}\w F_A^{(r)}\w F_A^{(r)}\big) \ ,
\label{WZorbtriv}\eeq
where $\chi_\rho:\Gamma\to\IC$ is the character of the
representation $\rho$. The number of instantons of colour
$r\in\widehat\Gamma$ is $\frac1{48\pi^3}\, \Tr_{\cH_I^{(r)}}\big(F_A^{(r)}\w F_A^{(r)}\w
F_A^{(r)}\big) = |\pi_r|$, and by defining $\xi_r:=C^{(r)}\,
\chi_\rho(r)/|\Gamma|$ the fractional instanton action becomes
\beq
-\frac{\vartheta}{6}\, \int_X \, F_A \w F_A \w F_A
= \ii\vartheta\, \sum_{r\in\widehat\Gamma}\, \xi_r \, |\pi_r| \ .
\label{WZorbfinal}\eeq
The weighting variables are thus related to D0~brane charges through $p_r=\e^{\ii\vartheta\, \xi_r}$,
and the instanton part of the orbifold gauge theory partition function
on $X=\IC^3/\Gamma$ with this definition takes the form
\beq
K^{\rm DT}_{\IC^3/\Gamma}=\sum_\pi\,
\chi_{\torus^3}(\cN_\pi)~\prod_{r\in\widehat\Gamma}\, p_r^{|\pi_r|}
\label{ZC3Gamma}\eeq
where the sum runs through $\widehat\Gamma$-coloured plane partitions
$\pi=\bigsqcup_{r\in\widehat\Gamma}\, \pi_r$. By rescaling $p_r\to p\,p_r$ we may also include a factor $p^{|\pi|}$ in (\ref{ZC3Gamma}) weighing the total charge $|\pi|=\sum_{r\in\Gammaw}\,|\pi_r|$ of the collection of fractional branes.

One of the goals of
subsequent sections will be to properly define and explicitly compute
the gauge theory fluctuation determinants $\chi_{\torus^3}(\cN_\pi)$ appearing in
(\ref{ZC3Gamma}) which determine a non-trivial measure on the moduli space of noncommutative instantons; we shall find that these are exactly the noncommutative
Donaldson--Thomas invariants. For this, we will properly define the
instanton moduli space (at toric fixed points) in terms of a
particular quiver with relations. The idea is that the effective
dynamics of the noncommutative gauge theory (at fixed points) can be
encoded in a quiver diagram. Each irreducible representation in
$\Gammaw$ is associated to a vertex of the quiver; in the context of
the noncommutative gauge theory, these vertices label isotopical
components of the Hilbert space
(\ref{hilsplitgen})--(\ref{Hisotop}). The links of the quiver diagram
specify the bifundamental matter field content, which arises from the
representation of the covariant coordinates (\ref{orbcovcoord}) on the orbifold
Hilbert space. The holomorphic conditions (\ref{orbcomm}) yield a set of relations for the quiver.

A general representation of this
quiver corresponds to $k_r= |\pi_r|$ noncommutative instantons of colour
$r\in\Gammaw$. A link joining representations $r$ and $r'$ has
multiplicity $|\pi_r|\, |\pi_{r'}|$ in the noncommutative gauge
theory. There are no moduli associated to a single instanton of
colour $r$ ($k_r=1$, $k_{r'}=0$ for $r'\neq r$). But for
instantons in the regular representation of the orbifold group
$\Gamma$, the $\Gamma$-orbit is the regular representation on the
coordinates $(z_1,z_2,z_3)\in \IC^3$ and so such instantons have a
non-trivial positional moduli space. This will generically require us
to impose appropriate stability conditions on the instanton moduli space, and to thereby restrict to fractional D0~branes whose orbits under the action of the complexified gauge group are closed.

\section{Instanton moduli spaces\label{sec:Instmodsp}}

In \cite{Cirafici:2008sn} we proposed an ADHM-type formalism to deal
with Donaldson--Thomas invariants on $\complex^3$ which is directly
derived from an analysis of the moduli space of torsion free
sheaves. The generalised ADHM equations arise from internal
consistency conditions when parametrizing the moduli space of framed
torsion free sheaves on $\PP^3$ via the Beilinson spectral
sequence. Under favourable conditions the spectral sequence
degenerates to a four-term complex whose cohomology is a generic
torsion free sheaf $\cE$. The purpose of this section is to extend
this construction to geometries which have the form of a crepant
resolution of a toric orbifold singularity.\footnote{Actually, as
  in~\cite{kronheimer}, the construction of the instanton moduli space
  does not need the assumption that the geometry is toric; our
  construction should hold whenever the tautological bundles generate
  the whole derived category (or even just the K-theory for some
  partial results). This
  assumption is however widely applied in the following sections, for
  example in order to use the localization formula.} We will do so via a
generalisation of the Kronheimer--Nakajima construction
\cite{kronheimer}. The result will be a construction of the instanton moduli
space via the representation theory of a  quiver with relations
which governs
the generalised ADHM data, much in the spirit of
\cite{Cirafici:2008sn}. In Section~\ref{sec:InstQM} we will argue that for certain choices of the
stability parameter of the quiver, one is working on the
noncommutative crepant resolution of the singularity. In the framework
of~\cite{Douglas:1997de}, the vacua of the worldvolume
quiver gauge theory of D0~branes on $\IC^3/\Gamma$ in the different
phases corresponding to different choices of Fayet--Iliopoulos (i.e. noncommutativity)
parameters all lead to moduli spaces that are simply the geometric phases of the
resolutions $X$ of $\IC^3/\Gamma$.

\subsection{Affine space}

Before considering orbifolds let us recall briefly the situation for the affine
Calabi--Yau space $\complex^3$. Here one would like to construct the instanton moduli space on
$\complex^3$, or better a moduli space of  sheaves on the
compactification $\PP^3$. In \cite{Cirafici:2008sn} we constructed a
model for the moduli space of framed sheaves
\begin{equation} \label{sheavesonP3}
{\frak M}_{N, k} \big(\mathbb{P}^3\big) = \left\{ \begin{array}{c}
\mathcal{E} =\text{torsion
free sheaf on}~ \mathbb{P}^3 \\ {\rm rank}(\mathcal{E})= N \ , \quad
c_1(\mathcal{E}) = 0 \\
\mathrm{ch}_2 (\mathcal{E}) = 0 \ , \quad \mathrm{ch}_3 (\mathcal{E})
= - k \\ \mathcal{E}\big|_{\wp_{\infty}} \cong\mathcal{O}^{\oplus N}_{\wp_{\infty}}
\end{array} \right\} ~\Big/~ {\rm isomorphisms}
\end{equation}
where $\wp_{\infty}$ is the plane at infinity. Since $\complex^3 \cong
\PP^3 / \PP^2$, in projective coordinates one has explicitly
$\wp_{\infty} =  [0 : z_1 :
z_2 : z_3] \cong \PP^2$. The strategy to construct this moduli space
is to adapt Beilinson's construction of the moduli space of torsion
free sheaves on $\PP^2$, via the Beilinson spectral sequence.

The construction starts from a sheaf $\cE$ on $\PP^3$ and considers the canonical projections 
\begin{equation}
\xymatrix@=15mm{
  &{\mathbb P}^3 \times {\mathbb P}^3 \ar[ld]_{p_1}\ar[rd]^{p_2}& \\
  {\mathbb P}^3 & & {\mathbb P}^3 \ .
}
\end{equation}
The spectral sequence descends from the Fourier--Mukai transform
\begin{equation}
{\mathbf R}^{\bullet} p_{1 \, *} \big( p_2^* \mathcal{E} \otimes
C^{\bullet} \big) \ ,
\label{FMtransf}\end{equation}
where
\begin{equation}
C^p := \mbox{$\bigwedge^{-p}$} \big( \mathcal{O}_{{\mathbb P}^3} (-1)
\boxtimes \mathcal{Q}^\vee \,\big)
\end{equation}
are the terms in the Koszul complex
\begin{eqnarray}
&0 ~\longrightarrow~ \bigwedge^3 \big( \mathcal{O}_{{\mathbb P}^3}
  (-1) \boxtimes \mathcal{Q}^\vee \,\big)~\longrightarrow~
    \bigwedge^2 \big( \mathcal{O}_{{\mathbb P}^3} (-1) \boxtimes
      \mathcal{Q}^\vee \,\big)~\longrightarrow~
    \mathcal{O}_{{\mathbb P}^3} (-1) \boxtimes \mathcal{Q}^\vee~
    \longrightarrow~
\nonumber\\ & \qquad\qquad~\longrightarrow~
   \mathcal{O}_{{\mathbb P}^3 \times {\mathbb P}^3}~
   \longrightarrow~\mathcal{O}_{\Delta} ~\longrightarrow~ 0 &
\label{Koszul}\end{eqnarray}
and we have used the notation $\mathcal{Q}^\vee = \Omega^1_{\mathbb{P}^3}
\otimes \mathcal{O}_{{\mathbb P}^3}(1)$; here $\Delta\cong\IP^3$ is
the diagonal in $\IP^3\times \IP^3$. The spectral sequence then has first term
\begin{equation}
E_1^{p,q} = \mathcal{F}_1^p \otimes H^q \left( \mathbb{P}^3 \, , \,
\mathcal{E} \otimes \mathcal{F}_2^p \right)
\label{spectralfirst}\end{equation}    
where $C^p = \mathcal{F}^p_1 \boxtimes \mathcal{F}^p_2$, and it converges to
\begin{equation}
{E}_\infty^{p,q} = \begin{cases}
  ~\mathcal{E}(-r) \ , \qquad \mathrm{if} ~ p+q=0  \ , \\
  ~0 \ , \qquad \mathrm{otherwise} \ . \end{cases}
\label{spectralconv}\end{equation}

To obtain a concrete model we were forced in \cite{Cirafici:2008sn} to impose an additional condition on the class of sheaves considered, namely $
H^1 \left( \PP^3 \,,\, \mathcal{E}(-2) \right) = 0 $. This condition
is of course restrictive and was introduced only as a matter of
convenience. However, it excludes certain configurations
of sheaves that we are interested in, namely the ideal sheaves of
points.\footnote{We are grateful to B.~Szendr\H{o}i for pointing this
  out to us.} Fortunately, this condition can be
traded for a different one which still has the virtue of collapsing
the spectral sequence to a four-term complex and includes the relevant
ideal sheaves of points in the parametrization of the moduli space. By imposing $
H^2 \left( \PP^3 \,,\, \mathcal{E}(-2) \right) = 0$ instead, the spectral sequence degenerates to the complex
\begin{equation}
\xymatrix{
  0 \ \ar[r] & \ V \otimes \mathcal{O}_{\PP^3} (-2) \ \ar[r]^a &
  \ {V}^{\oplus 3} \otimes \mathcal{O}_{\PP^3} (-1) \ \ar[r]^{\hspace{0.5 in}b} &
  \\ \ar[r]^{\hspace{-0.5 in}b} &
  \ (V^{\oplus 3} \oplus W) \otimes \mathcal{O}_{\PP^3} \ \ar[r]^c
  & 
  \ {V} \otimes \mathcal{O}_{\PP^3} (1) \ \ar[r] & \ 0 \ ,
} \label{finalcomplexmain}
\end{equation}
where $V$ and $W$ are complex vector spaces with $\dim_\IC(V)=k$ and
$\dim_\IC(W)=N$, while
\begin{equation}
\underline{\rm im}(a) = \underline{\ker}(b) \qquad \mbox{and} \qquad \cE =
\underline{\ker}(c)\, \big/ \, \underline{\rm im}(b) \ .
\label{C3sheafcoh}\end{equation}
This complex is just the dual of the complex considered
in~\cite{Cirafici:2008sn}. The precise form of the morphisms is not
important for the purposes of this section; they can essentially
be read off from~\cite[Section~4.6]{Cirafici:2008sn}. Indeed the choice of complex or
its dual is somewhat irrelevant for the construction that follows. In
particular, the two moduli spaces are isomorphic, so that we can
freely borrow results from~\cite{Cirafici:2008sn}. We
refer the reader to Appendix~\ref{App-Cohresults} and Appendix~\ref{App-Beilinson} for explicit proofs of these claims. In the following we will simply use the complex (\ref{finalcomplexmain}).

\subsection{Intersection theory\label{subsec:Inttheory}}

We begin our study of the instanton moduli space on $\complex^3 /
\Gamma$ by reviewing some facts from the work of Ito and
Nakajima~\cite{ItoNaka} that will be useful in what follows. Their
paper is an attempt to extend the McKay correspondence to
three-dimensional orbifolds of the form $\complex^3 / \Gamma$, where
$\Gamma \subset SL(3,\complex)$, and their natural smooth crepant\footnote{$X$ is \emph{crepant} if the canonical bundles are isomorphic, $K_X\cong\pi^*(K_{\IC^3/\Gamma})$; this constraint is required to obtain a Calabi--Yau structure on $X$ from that of $\IC^3/\Gamma$.} Calabi--Yau
resolutions given by the Hilbert--Chow morphism
\beq
\pi \,:\, X \ \longrightarrow \ \complex^3 \big/ \Gamma
\label{crepantres}\eeq
for the $\Gamma$-Hilbert scheme
$X=\mathrm{Hilb}^\Gamma(\complex^3)$ consisting of $\Gamma$-invariant
zero-dimensional subschemes $Y$ of $\complex^3$ of length $|\Gamma|$
such that $H^0(\cO_Y)$ is the regular representation of $\Gamma$. Roughly speaking, the McKay
correspondence in this setting is the statement that any well-posed
question about the geometry of the resolution $X$ should have a
$\Gamma$-equivariant answer on $\complex^3$. We will mostly use
geometrical notions in this section; in Section~\ref{sec:InstQM} we
will comment on the description of the McKay correspondence in terms
of derived categories. This correspondence has been studied from a
physical point of view analogous to ours in
e.g.~\cite{Tomasiello:2000ym,Mayr:2000as,Ezhuthachan:2006gu}; see
also~\cite{Degeratu} for a related description.

Consider the universal scheme $Z \subset X\times
\complex^3$ with correspondence diagram
\begin{equation}
\xymatrix@=15mm{
  & Z \ar[ld]_{q_1}\ar[rd]^{q_2}& \\
  X & & \complex^3
}
\end{equation}
and define the tautological bundle
\begin{equation}
\cR := q_{1 *} \cO_{Z} \ .
\label{tautdef}\end{equation}
Under the action of $\Gamma$ on $Z$, the bundle $\cR$ transforms in the
regular representation. Its fibres are the $|\Gamma|$-dimensional
vector spaces $\IC[z_1,z_2,z_3]/I\cong H^0(\cO_Y)$ for the regular
representation of $\Gamma$, where $I\subset \IC[z_1,z_2,z_3]$ is a
$\Gamma$-invariant ideal corresponding to a zero-dimensional subscheme
$Y$ of $\IC^3$ of length $|\Gamma|$. Multiplication on the fibres
$\IC[z_1,z_2,z_3]/I$ by the coordinates $z_\a$ of $\IC^3$ induces a
$\Gamma$-equivariant homomorphism $B:\cR\to Q\otimes\cR$, with
$B\wedge B=0$ as an element of
$\Hom_\Gamma(\cR,\bigwedge^2Q\otimes\cR)$. 
Here $Q$ is the fundamental three-dimensional representation of
$\Gamma\subset SL(3,\complex)$. If we denote the orbifold action
by $(z_1,z_2,z_3) \mapsto (r_1\cdot z_1, r_2\cdot z_2 , r_3\cdot z_3)$, with
$\rho_{r_\a}$ the irreducible one-dimensional representation of
$\Gamma$ with weight $r_\a$, then $Q = \rho_{r_1} \oplus
\rho_{r_2} \oplus \rho_{r_3}$. This defines a $\Gammaw$-colouring
$\IN_0^3\to \Gamma$ through the identification $\Gammaw\cong \Gamma$
by
\beq
(n_1,n_2,n_3) \ \longmapsto \ \rho_{r_1}^{\otimes n_1}\otimes
\rho_{r_2}^{\otimes n_2}\otimes \rho_{r_3}^{\otimes n_3} \ ,
\label{colouring}\eeq
which coincides with the colourings considered in
Section~\ref{sec:NCGT}. We will represent the element $B\in
Q\otimes\End_\Gamma(\cR)$ by a triple of endomorphisms
$B=(B_1,B_2,B_3)$, with $B\wedge B=\sum_{\a<\b}\, [B_\a,B_\b]$.

The decomposition of the regular representation induces a
decomposition of the tautological bundle into
irreducible representations
\begin{equation}
\cR = \bigoplus_{r\in\widehat\Gamma}\, \cR_r \otimes \rho_r \ ,
\label{cRdecomp}\end{equation}
where $\{ {\rho_r} \}_{r\in\widehat\Gamma}$ is the set of irreducible
representations; we denote the trivial representation by $\rho_0$. The
tautological bundles $\cR_r=\Hom_\Gamma(\rho_r,\cR)$ form an integral basis for the Grothendieck group
$K(X)$ of vector bundles on $X$, where the bundle corresponding to the trivial representation
is the trivial bundle $\cR_0 \cong \cO_X$. Note in particular that
they are not \textit{line} bundles in general, since their rank
depends on the dimension of the irreducible representations of
$\Gamma$; however, for $\Gamma$ abelian, the case considered in this
paper, they are always line bundles.\footnote{In the case of kleininan
  singularities the ranks are the entries of the vector which
  generates the kernel of the affine Cartan matrix. Geometrically this
  is related to the annihilator of the quadratic form which gives the
  intersection pairing.}

Similarly, one can introduce a dual basis $\cS_r$ of $K^c (X)$, the Grothendieck
group of coherent sheaves on $\pi^{-1} (0)$, or equivalently the
Grothendieck group of bounded complexes of vector bundles over $X$ which are
exact outside the exceptional locus $\pi^{-1} (0)$. The map
between the two descriptions is obtained by sending a coherent sheaf
on $\pi^{-1} (0)$ to the complex of vector bundles which is
a locally free resolution of it. The dual basis of $K^c (X)$ is 
\begin{equation}
\cS_r \ : \ \xymatrix{
 \cR^{\vee}_r \ \ar[r] & \
 \displaystyle{\bigoplus_{s\in\widehat\Gamma}\, a^{(2)}_{rs}\,
   \cR_s^{\vee} } \ \ar[r] & \ \displaystyle{
   \bigoplus_{s\in\widehat\Gamma}\, a^{(1)}_{rs}\, \cR_s^{\vee} } \
 \ar[r] & \ 
 \cR_r^{\vee}
}
\end{equation}
where the arrows arise from the decomposition, according to (\ref{cRdecomp}), of the maps
$\bigwedge^i Q\otimes\cR\xrightarrow{B\wedge}\bigwedge^{i+1} Q
\otimes\cR$, and
\begin{equation} \label{tensordecomp}
\mbox{$\bigwedge^i$}\, Q \otimes \rho_r =
\bigoplus_{s\in\widehat\Gamma}\, a^{(i)}_{sr}\, \rho_s \qquad
\mbox{with} \quad a^{(i)}_{sr}=\dim_\IC\Hom_\Gamma\big(\rho_s
\,,\,\mbox{$\bigwedge^i$}\, Q \otimes \rho_r \big) \ .
\end{equation}
Note that the determinant representation $\bigwedge^3 Q$ is trivial as
a $\Gamma$-module since we assume that $\Gamma$ is a subgroup of $SL(3,\complex)$, hence $\bigwedge^3 Q\otimes\cR\cong \cR$ and $a_{rs}^{(3)}=\delta_{rs}$. This also implies that $\bigwedge^2 Q \cong Q^{\vee}$ and therefore $a^{(2)}_{sr} = a^{(1)}_{rs}$. These multiplicities can be computed explicitly from the decompositions
\beq
Q\otimes\rho_r=(\rho_{r_1} \oplus
\rho_{r_2} \oplus \rho_{r_3})\otimes\rho_r=\rho_{r_1+r} \oplus
\rho_{r_2+r} \oplus \rho_{r_3+r} \ ,
\eeq 
which comparing with (\ref{tensordecomp}) gives
\beq
a_{rs}^{(1)}= \delta_{r,s+r_1}+\delta_{r,s+r_2}+\delta_{r,s+r_3} \qquad \mbox{and} \qquad a_{rs}^{(2)}= \delta_{r,s-r_1}+\delta_{r,s-r_2}+\delta_{r,s-r_3} \ .
\label{ars12expl}\eeq 

The purpose of this definition is to relate the representation theory
of $\Gamma$ with geometry. In the case of the familiar McKay
correspondence with ADE singularities, this is just the statement that
the representation theory of the discrete orbifold group contains all
the information about the intersection matrix of the ADE
singularity. In higher dimensions there is no such simple and direct
statement, and the tensor product decomposition (\ref{tensordecomp}) into
irreducible representations of the discrete group $\Gamma$ enters
rather more indirectly in the basis of $K^c(X)$.

To see this in more detail, define the collection of dual complexes $\{ \cS_r^{\vee} \}_{r\in\Gammaw}$ by
\begin{equation}
\cS_r^{\vee} \ : \  - \Big[ \xymatrix{
 \cR_r \ \ar[r] & \ \displaystyle{\bigoplus_{s\in\Gammaw}\,
   a^{(1)}_{rs}\, \cR_s} \ \ar[r] & \
   \displaystyle{\bigoplus_{s\in\Gammaw}\, a^{(2)}_{rs}\, \cR_s} \
     \ar[r] & \ 
 \cR_r
} \Big] \ .
\end{equation}
On $K^c (X)$ we can define a perfect pairing
\begin{equation}
(\cS , \cT  ) = \langle \Theta \cS , \cT \rangle= \int_X\, \ch(\Theta
\cS \otimes \cT)\w {\rm Todd}(X) \ ,
\end{equation}
where $\langle-,-\rangle$ is the dual pairing between $K(X)$ and
$K^c(X)$, and $\Theta :  K^c (X) \rightarrow K(X)$ is the map which sends a complex of vector bundles to the corresponding element in $K(X)$ (the alternating sum of the elements of the complex). For example
\begin{equation}
\Theta \cS_r^{\vee} = \sum_{s\in\Gammaw}\,  \big( -  \delta_{rs} +
a^{(2)}_{rs} - a^{(1)}_{rs} + \delta_{sr} \big)\, \cR_s =
\sum_{s\in\Gammaw}\, \big( a^{(2)}_{rs} - a^{(1)}_{rs} \big)\, \cR_s \ .
\end{equation}
It follows that
\begin{equation}
\big(\cS_r^{\vee} \, , \, \cS_s \big) = \langle \Theta \cS_r^{\vee} ,
\cS_s \rangle = \sum_{q\in\Gammaw} \, \big( a^{(2)}_{rq} -
a^{(1)}_{rq} \big) \, \langle \cR_q , \cS_s \rangle = a^{(2)}_{rs} -
a^{(1)}_{rs} \ ,
\end{equation}
where we have used the fact that $\{ \cR_s \}_{s\in\Gammaw}$ and $\{ \cS_r \}_{r\in\Gammaw}$ are dual bases of $K(X)$ and $K^c (X)$. This result underlies the relation between the tensor
product decomposition (\ref{tensordecomp}) and the intersection theory of~$X$.\footnote{For
  example, in complex dimension two the pairing $(\cS_r^{\vee} \, , \, \cS_s) $ would give the extended Cartan matrix of the ADE singularity.} 

These two bases correspond, via the McKay correspondence, with two
bases of $\Gamma$-equivariant coherent sheaves on $\complex^3$, as
shown in \cite{ItoNaka}. The Grothendieck groups of
$\Gamma$-equivariant sheaves on $\complex^3$, $K_{\Gamma}
(\complex^3)$ and $K^c_{\Gamma}(\complex^3)$ (with coherent sheaves of compact
support), have respective bases $\{ \rho_r \otimes \cO_{\complex^3}
\}_{r\in\Gammaw}$ and $\{ \rho_r \otimes \cO_0 \}_{r\in\Gammaw}$ where
$\cO_0$ is the skyscraper sheaf at the origin; the latter basis is
naturally identified as the set of fractional $0$~branes. All of these groups are
isomorphic to the representation ring $R(\Gamma)$ of the orbifold
group $\Gamma$. This correspondence will be used in the stability
analysis of Section~\ref{subsec:Stability}.

\subsection{Resolution of the diagonal}

The main ingredient in the construction of the instanton moduli space
is Beilinson's theorem which can be used to parametrize the
moduli space via a  spectral sequence. A requisite technical ingredient is the resolution of the diagonal sheaf $\cO_{\Delta}$ of $X \times X$, or better of its compactification. We will collectively use the notation $\Delta$ for the diagonal of a generic variety to simplify our notation; it will be clear from the context which variety we are considering. Recall that to construct the Beilinson spectral sequence on $\complex^3$ one needs the identity
\begin{equation} 
p_{1\,*} \big(p_2^* \mathcal{E} \otimes \mathcal{O}_{\Delta}\big) =
p_{1\,*} \big(p_2^* \mathcal{E}\big|_{\Delta}\big) = \mathcal{E}
\end{equation}
where $\Delta\cong\PP^3 \subset {\mathbb P}^3 \times {\mathbb P}^3$ is
the diagonal. This fact was extensively used in \cite{Cirafici:2008sn}
to construct explicitly the instanton moduli space on $\complex^3$. To
proceed with the same construction for the crepant resolution $X$ of
an orbifold singularity $\IC^3/\Gamma$, we need two ingredients: a
resolution of the diagonal sheaf $\cO_{\Delta}$ and a compactification
of $X$ (in the same way as $\PP^3$ is a compactification of
$\complex^3$). We will proceed in two steps: first we find a resolution of the diagonal sheaf of $X \times X$, and then we extend it to its compactification.

A resolution of the diagonal sheaf on $X \times X$ is obtained by
generalising the argument of \cite{nakaALE} to the higher-dimensional
case to get\footnote{In this exact sequence we explicitly use the
  Calabi--Yau condition to write the isomorphisms
  $\bigwedge^3Q\cong\IC$ and $\bigwedge^2Q\cong Q^\vee$.}
\begin{equation} \label{resX}
\xymatrix{
  0 \ \ar[r] & \ \big( \cR \boxtimes \cR^{\vee} \otimes \bigwedge^3
  Q^{\vee} \big)^{\Gamma} \ \ar[r]^{B\wedge} & \ 
  \big( \cR \boxtimes \cR^{\vee} \otimes \bigwedge^2 Q^{\vee}
  \big)^{\Gamma} \ \ar[r]^{\ \ \ \ \ \ \ \ \ \ \ \ \ \ \ \ B\wedge} & \\ \ar[r]^{\!\!\!\!\!\!\!\!\!\!\!\!\!\!\!\!\!\!\!\!\!\!\!\!\!B\wedge} & \ \big( \cR \boxtimes \cR^{\vee}
  \otimes Q^{\vee} \big)^{\Gamma} \ \ar[r]^{B\wedge} & \ \big( \cR \boxtimes
  \cR^{\vee} \big)^{\Gamma} \ \ar[r]^{ \ \ \Tr} & \ 
\cO_{\Delta} \ .}
\end{equation}
To unpack this complex a bit, one has to write explicitly the
decomposition of the tautological bundles from (\ref{cRdecomp}), use
the tensor product decomposition of the representations
(\ref{tensordecomp}), and then take the $\Gamma$-invariant part. This yields
\begin{eqnarray}
\Big( \cR \boxtimes \cR^{\vee} \otimes\mbox{$\bigwedge^i$} Q^\vee
\Big)^{\Gamma} &=& \Big(\, \bigoplus_{r,s,q\in\Gammaw}\, \cR_r \otimes
\rho_r \boxtimes a_{qs}^{(i)} \, \cR_s^{\vee} \otimes \rho_q^{\vee}\, \Big)^{\Gamma} 
\nonumber \\[4pt] &=&  \bigoplus_{r,s,q\in\Gammaw}\, \cR_r \boxtimes a_{sq}^{(i)}\, \cR_s^{\vee} \otimes \left( \rho_r \otimes\rho_s^{\vee} \right)^{\Gamma} 
\ = \ \bigoplus_{r,s\in\Gammaw}\, \cR_r \boxtimes a_{rs}^{(i)}\,
\cR_s^{\vee} \ ,
\end{eqnarray}
where in the last step we used Schur's lemma.
Therefore every term of the resolution (\ref{resX}) can be written as $C^{i} = \bigoplus_{r,s\in\Gammaw}\, \cR_r \boxtimes a_{rs}^{(i)}\, \cR_s^{\vee}$.

Now let us proceed to the compactification. The idea is that one can
mimic directly the compactification of $\complex^3$ into $\PP^3$ by
adding a boundary divisor. For $\PP^3$ one uses the Koszul resolution
(\ref{Koszul}) of the diagonal. We would like to think of the
compactification of $X$ to $\overline{X}$ as obtained by compactifying
$\complex^3 / \Gamma$ to $\PP^3 / \Gamma$ and then resolving the
singularity at the origin, leaving untouched the divisor at
infinity. This procedure is a generalisation of
\cite{kronheimer,nakaALE}. It corresponds to an orbifold
compactification $\overline{X} = X \sqcup \bdiv$ and we can regard
$\bdiv \cong \PP^2 / \Gamma$. More precisely, we glue objects in $X$
with $\Gamma$-invariant objects on $\bdiv$ to obtain something defined
globally. Then, in a neighbourhood of the boundary divisor $\bdiv$, $\overline{X}$ looks
like $\PP^3 / \Gamma$ and the gluing is compatible with the orbifold
action of $\Gamma$ on $\bdiv$. We can then glue the resolution of the diagonal
of $X \times X$ with the Koszul resolution of the diagonal in
$\PP^3/\Gamma \times\PP^3 / \Gamma$ (i.e., the Koszul resolution of
the diagonal of $\PP^3$ where the sheaves carry an appropriate
$\Gamma$-module structure) to get a globally defined resolution on $\overline{X} \times \overline{X}$. Indeed, given the nature of the tautological bundles, the complex (\ref{resX}) on $X\times X$ has a natural projection to $\complex^3 / \Gamma \times \complex^3 / \Gamma$.

Consider the Koszul resolution on $\PP^3 \times \PP^3 $ given in
(\ref{Koszul}). The sheaf $\cQ$ (which we would like to glue to $Q$,
the trivial bundle on $X$ which carries the fundamental representation
of $\Gamma$) is defined with a suitable $\Gamma$-module structure
through the dual Euler sequence
\begin{equation} \label{defQ}
\xymatrix{
   0 \ \ar[r] & \ \mathcal{O}_{{\mathbb P}^3} (-1) \ \ar[r] &
   \ \mathcal{O}_{{\mathbb P}^3}^{\oplus 4} \ \ar[r] & \ \mathcal{Q} \
   \ar[r]
   & \ 0 \ ,
}
\end{equation}
where we regard $\cO_{\PP^3}^{\oplus 4} \cong (Q \oplus \rho_0)
\otimes \cO_{\PP^3}$, i.e. we consider $\cO_{\PP^3}^{\oplus 4}$ as a
trivial bundle where each factor carries an action of $\Gamma$. The
first three factors are collectively taken care of by the fundamental
representation $Q$ (which acts on $\complex^3$), while the action of
$\Gamma$ on the fourth factor (which corresponds to the projective
coordinate $z_0$ centred at the plane at infinity) is
trivial. Equivalently, we projectivize the action of $\Gamma$ by
letting it act trivially on the fourth coordinate $z_0$ of $\PP^3 [z_0
: z_1 : z_2 : z_3 ]$ corresponding to the patch at infinity. This
induces a natural $\Gamma$-module structure on the sheaf $\cQ$ via the
last morphism in (\ref{defQ}). We can finally glue together the
resolution of the diagonal on $X$ and on $\PP^3$ to get a resolution
of the diagonal of $\overline{X} \times \overline{X}$ given by
\begin{eqnarray} \label{compX}
\xymatrix{
  0 \ \ar[r] & \ \big( \cR (-3 \bdiv) \boxtimes \cR^{\vee} \otimes
  \bigwedge^3 \cQ^{\vee} \big)^{\Gamma} \ \ar[r] &
  \ \big( \cR (-2 \bdiv) \boxtimes \cR^{\vee} \otimes \bigwedge^2
  \cQ^{\vee} \big)^{\Gamma} \ \ar[r] & \cr \ar[r] & \ \big( \cR
  (-\bdiv) \boxtimes \cR^{\vee} \otimes \cQ^{\vee} \big)^{\Gamma} \
  \ar[r] & \ \big( \cR \boxtimes \cR^{\vee} \big)^{\Gamma} \ \ar[r] &
\ \cO_{\Delta} \ . }
\end{eqnarray}

\subsection{Beilinson's theorem and the instanton moduli space\label{subsec:Beilinson}}

Consider the Fourier--Mukai transform (\ref{FMtransf}) of a torsion free coherent sheaf $\cE$
on $\overline{X}$, where now $C^{\bullet}$ denotes the resolution (\ref{compX}). Then by Beilinson's theorem a sheaf $\cE (-l) := \cE \otimes \cO_{\overline{X}} (-l\,\bdiv)$ on $\overline{X}$ is parametrized by a spectral sequence whose first term is
\begin{equation} E_1^{p,q} = \Big( \cR(p) \otimes H^q
\big(\, \overline{X} \,,\, \mathcal{E}(-l) \otimes \cR^{\vee} \otimes
 \mbox{$\bigwedge^{-p}$} \cQ^{\vee} \big) \Big)^{\Gamma} \ ,
\end{equation}
where our conventions are $p \le 0$. We will argue that all homological algebra based on
this spectral sequence reduce to the familiar case of $\PP^3$.

The
reason for this is that thanks to the tensor product decomposition
(\ref{tensordecomp}) all the relevant cohomology groups are of the
form $H^{\bullet} \left( \, \overline{X} , \cE (-l) \otimes \cR_r
\right)$, where $\cR_r$ is a line bundle (which restricts to a trivial
bundle at infinity, up to the $\Gamma$-module structure). Therefore
the relevant complex vector spaces are of the form
\begin{equation}
V = H^2 \left( \, \overline{X} \,,\, \cE (-l) \otimes \cR^{\vee}
\right) = H^2 \Big( \, \overline{X} \,,\, \cE (-l) \otimes
\mbox{$\bigoplus\limits_{r\in\Gammaw}$}\, \cR_r^{\vee} \otimes
\rho_r^{\vee} \Big) = \bigoplus_{r\in\Gammaw}\, V_r \otimes
\rho_r^{\vee} \ ,
\label{Vdecomp}\end{equation}
where $V_r$ are finite-dimensional vector spaces with trivial $\Gamma$-action.
Recall that the sheaves $\bigwedge^{-p} \cQ^{\vee}$, $p\leq0$ are
defined in such a way that, if one neglects the $\Gamma$-module
structure, they look like sheaves of differential forms $
\Omega_{\PP^3}^{-p} (-p)$ on $\PP^3$ near
infinity. Therefore, in a neighbourhood of $\bdiv$ one has
\begin{eqnarray}
\mathcal{E}(-l) \otimes \cR^{\vee} \otimes
 \mbox{$\bigwedge^{-p}$} \cQ^{\vee} &\cong& \mathcal{E}(-l) \otimes
 \bigoplus_{r\in\Gammaw} \, \cR_r^{\vee} \otimes \rho_r^{\vee} \otimes
 \mbox{$\bigwedge^{-p}$} \cQ^{\vee} \nonumber\\[4pt] &\cong&
 \bigoplus_{r,s\in\Gammaw}\, a^{(-p)}_{sr}\, \rho_s^{\vee} \otimes
 \big( \mathcal{E}(-l) \otimes  \Omega_{\PP^3}^{-p} (-p) \otimes
 \cR_r^{\vee} \big) \ .
\end{eqnarray}

Next we have to impose boundary conditions on $\cE$. This condition is
the same as the framing condition on $\PP^3$. In particular, the
tautological bundles $\cR_r$ only play a passive role since they are
trivial at infinity and their only function is to label a
representation $\rho_r$. Therefore we will again impose that
$\cE\big|_{ \bdiv}$ is trivial on a line $\ell_\infty\subset \bdiv$
with $\ell_\infty\cong \PP^1$, which
means that the associated gauge connection is flat. A choice of
different boundary conditions corresponds to using particular
configurations of D~branes as boundary conditions; although this might
be related to the noncommutative topological vertex formalism of
\cite{Nagao:2009rq}, or the orbifold topological vertex formalism
of~\cite{orbvertex}, we will leave such investigations for future work.

The sheaf cohomology groups are now the same as in the $\PP^3$ case
(see Appendix~\ref{App-Cohresults} and Appendix~\ref{App-Beilinson}) up
to multiplicity factors and the $\Gamma$-module structure. We can
therefore jump directly to the conclusion that we can represent the
sheaf $\cE$ as the single non-trivial cohomology of the complex
\begin{equation}
\xymatrix{
  0 \ \ar[r] & \ \big( V \otimes \mathcal{R} (-2) \big)^{\Gamma} \ \ar[r]^{\!\!\!\!\!\!\!\!\!\!a} &
\ \big(  V \otimes \bigwedge^2 Q^{\vee} \otimes \mathcal{R} (-1)
\big)^{\Gamma} \ \ar[r]^{\ \ \ \ \ \ \ \ \ \ \ \ \ \ \ \ b} & \\
  \ar[r]^{\!\!\!\!\!\!\!\!\!\!\!\!\!\!\!\!\!\!\!\!\!\!\!\!\!\!\!\!\!\!\!\!\!\!\!\!\!\!\!\!b} & \ 
 \big( (  V \otimes Q^{\vee} \oplus W ) \otimes \mathcal{R}
 \big)^{\Gamma} \ \ar[r]^{~~~~~c} & \ 
\big(  {V} \otimes \mathcal{R} (1) \big)^{\Gamma} \ \ar[r] & \ 0 \ ,
}
\label{ADHMcomplexfinal}\end{equation}
where the complex vector spaces appearing here are finite-dimensional. All the differentials involved are exactly
the same as in the $\PP^3$ case, but decomposed equivariantly
according to the $\Gamma$-module structure. The original sheaf $\cE$
is recovered as in (\ref{C3sheafcoh}), while the vector space $V$ is
given in (\ref{Vdecomp}) with $\dim_\complex V=k$. The virtual bundle defined by the cohomology of the complex (\ref{ADHMcomplexfinal}) is a representative (in equivariant K-theory) of the isomorphism class of the universal sheaf associated to the (fine) instanton moduli space.

The vector space
\begin{eqnarray}
W = H^0 \big( \PP^1 \,,\, \underline{\ker} (b)\big|_{ \ell_\infty} \big) = \bigoplus_{r\in\Gammaw}\, W_r \otimes \rho_r^{\vee}
\label{Wdecomp}\end{eqnarray}
parametrizes trivializations of the sheaf $\cE$, with $\dim_\complex
W=N$. Asymptotically the associated gauge connection is flat. The
spaces $X$ that we are considering all have the form of a crepant
resolution of an orbifold singularity $\complex^3 /
\Gamma$. By~\cite[Theorem 8.2.3]{joyce}, these resolutions have a
unique Ricci-flat ALE metric asymptotic to the flat geometry $\IC^3/\Gamma$. As a
smooth manifold, $X$ looks like the Lens space $S^5 / \,
\Gamma$ at infinity. As such, the flat connections are labelled by
representations of the orbifold group; since $\pi_1 (S^5) = 0$ we have
$\pi_1 (S^5/ \, \Gamma) = \pi_0 (\Gamma) = \Gamma$ where in our case
$\Gamma$ is a finite abelian group. Conjugacy classes of homomorphisms
from this fundamental group to the gauge group $U(N)$ are therefore in
correspondence with gauge equivalence classes of flat
connections. This classification is manifest in the decomposition into
irreducible $\Gamma$-modules of the fibre at infinity
(\ref{Wdecomp}). At infinity, the gauge sheaf is asymptotically in a
 representation $\rho$ of the orbifold group $\Gamma$ and the
dimensions $\dim_\complex W_r = N_r$ label the multiplicities of
the decomposition of $\rho$ into irreducible representations, with the
constraint
\beq
N=\sum_{r\in\Gammaw}\, N_r \ .
\label{Nsumwr}\eeq

From a string theory perspective the dimension $k_r$ of the vector space $V_r$ corresponds to the number of fractional D0 branes which transform in the representation $\rho_r$ of $\Gamma$. The vector space $W$ represents the D6 branes, to which the D0 branes are bound. Let us make a heuristic check. Since
\begin{equation}
V \otimes \mbox{$\bigwedge^i$} Q^{\vee} = \bigoplus_{r\in\Gammaw}\, V_r \otimes \rho_r^{\vee} \otimes \mbox{$\bigwedge^i$} Q^{\vee}  = \bigoplus_{r,s\in\Gammaw}\, a_{rs}^{(i)}\, V_s \otimes \rho_r^{\vee}
\end{equation}
we can write the complex (\ref{ADHMcomplexfinal}) as
\begin{equation}
\xymatrix{
  0 \ \ar[r] & \ \displaystyle{\bigoplus_{r\in\Gammaw}\, V_r \otimes \mathcal{R}_r (-2) \ \ar[r]^{ a}} & \ \displaystyle{ \bigoplus_{r,s\in\Gammaw}\,
  { a_{rs}^{(2)} \, V_s \otimes \cR_r (1) }} \ \ar[r]^{\ \ \ \ \ \ \ \ \ \ b} & \cr   \ar[r]^{\hspace{-1.3in}b} & \ \displaystyle{ \bigoplus_{r,s\in\Gammaw}\, \big( a_{rs}^{(1)}\, V_s \otimes \mathcal{R}_r \big)  \ \oplus \ \bigoplus_{r\in\Gammaw}\, W_r \otimes \mathcal{R}_r} \ \ar[r]^{~~~~~ \ \ \ \ \ c} & \ \displaystyle{
\bigoplus_{r,s\in\Gammaw}\, {V_r} \otimes \mathcal{R}_s (1)} \ \ar[r] & \ 0  \ .
}
\end{equation}
Let us consider the case of a single D0 brane, $k_r=1$ for some $r\in\Gammaw$ while $k_{r'}=0$ for all $r'\neq r$. Then the complex
\begin{equation} \label{singleD0}
\xymatrix{
  0 \ \ar[r] & \ \mathcal{R}_r (- 2) \ \ar[r]^{ \hspace{-0.3in}a} & \ \displaystyle{ \bigoplus_{s\in\Gammaw}\,
 { a_{sr}^{(2)} \, \cR_s (-1) }} \ \ar[r]^{\hspace{0.5in}b} & \\
\ar[r]^{\hspace{-1in}b} & \ \displaystyle{ \bigoplus_{s\in\Gammaw}\, \big(  a_{sr}^{(1)} \, \mathcal{R}_s \big)
\ \oplus \ \bigoplus_{r\in\Gammaw}\, W_r \otimes \mathcal{R}_r} \ \ar[r]^{~~~~~~~ \ \ \ \ \ \ c} & \
\mathcal{R}_r (1) \ \ar[r] & \ 0
}
\end{equation}
represents a single D0 brane bound to $N= \dim_{\complex}(W)$ D6 branes. Assume now that there are no D6 branes, $N=0$, and ignore the boundary divisor $\bdiv$ (as here we are only interested in a local argument, that can be set up in a neighbourhood of the exceptional locus). Then (\ref{singleD0}) becomes
\begin{equation}
\xymatrix{
  0 \ \ar[r] & \ \mathcal{R}_r \ \ar[r] & \ \displaystyle{ \bigoplus_{s\in\Gammaw}\, 
 { a_{sr}^{(2)} \, \cR_s }} \ \ar[r] & \ \displaystyle{ \bigoplus_{s\in\Gammaw}\, a_{sr}^{(1)} \, \mathcal{R}_s} \ \ar[r] & \ 
\mathcal{R}_r \ \ar[r] & \ 0 \ ,
}
\end{equation}
which is precisely the object that we called $\cS_r$ in Section~\ref{subsec:Inttheory} that denotes an element of a basis of $K^c (X)$. In other words, this is a sheaf supported on a cycle in the exceptional locus, precisely the behaviour we would expect from a fractional brane.\footnote{This is not quite true, as fractional branes in the derived category also have a shift in degree.}

It is natural now to interpret (\ref{singleD0}) as a bound state of
the D0 brane, wrapping a vanishing cycle, with the D6~branes. And in
full generality, we regard (\ref{ADHMcomplexfinal}) as a bound state of a
number of D0 and D6 branes, which is precisely what we wanted to
obtain. We can identify $N=\dim_\complex W$ with the number of D6
branes, which we will mostly assume to be just one in order to have a
$U(1)$ gauge theory. Note, however, that this is not necessary at this stage. Moreover, we will identify $k_r=\dim_\complex V_r$ with the fractional instanton charge associated to the representation $\rho_r^{\vee}$, which we will see later on represents the number of boxes in a three-dimensional Young diagram of a given colour as in Section~\ref{sec:NCGT}.

\subsection{Matrix equations\label{subsec:Matrixeq}}

In Section~\ref{subsec:Beilinson} we have explicitly constructed the moduli space of framed
instantons on $\overline{X}$ with fixed topological charges. This moduli space is parametrized via the Beilinson
spectral sequence by a sequence of linear maps between vector
spaces. The maps are implicitly determined by the homological
algebra and are nothing else than a particular equivariant
decomposition of the maps already derived in full generality in
\cite{Cirafici:2008sn}. While it is in principle possible to derive
them in a closed form, we will refrain from doing so and consider a
simpler case, which is the only one where the analysis can be
practically carried on. This is the case where the gauge theory is
abelian and $N=\dim_\complex W =1$, or the gauge theory is non-abelian but in its
Coulomb phase with the $U(N)$ gauge symmetry broken down to the maximal torus $U(1)^N$. In both instances the intersection indices of the moduli space can be computed directly via equivariant localization. For the more general non-abelian case we have in principle an ADHM-like parametrization of the moduli space, but the analysis is complicated by the lack of suitable techniques and a poor understanding of stability issues for generic torsion free coherent sheaves. 

In the $U(1)$ case most of the fields that enter in the generalized ADHM
parametrization derived in~\cite{Cirafici:2008sn} can be set to zero
or traded for stability conditions, and only the ``center of mass''
coordinates remain with
\begin{equation}
[  B_1 , B_2 ] = 0 \ , \qquad [  B_1 , B_3 ] = 0 \qquad \mbox{and} \qquad [  B_2 , B_3 ] = 0 \ .
\label{ADHMC3}\end{equation} 
The decomposition of these maps according to the $\Gamma$-action can
be neatly summarized in a  quiver diagram as we will explain in
the ensuing sections. For the moment we will just remark that the main
difference between the orbifold geometries and the flat case studied
in \cite{Cirafici:2008sn} is that the choice of a trivialization at
infinity carries also the information about inequivalent boundary
conditions. These are reflected in the form of the universal sheaf on
the instanton moduli space via the dependence on the framing vector $\mbf N =
(N_r)_{r\in\Gammaw}$. If one restricts to the $U(1)$ gauge
theory then only one of the dimensions $N_r$ can be non-vanishing and
precisely equal to one by (\ref{Nsumwr}). The non-abelian gauge theory, already in its Coulomb branch, offers a combinatorially non-trivial host of possibilities.

We therefore decompose the linear maps $B \in \Hom_{\Gamma} (V , Q \otimes V)$ with $V = \bigoplus_{r\in\Gammaw}\, V_r \otimes \rho^\vee_r$ as
\begin{equation} 
B = \bigoplus_{r\in\Gammaw}\, \left( B_1^{r} , B_2^{r} , B_3^{r} \right)
\label{Bdecomp}\end{equation}
where $B_\a^{r} : V_{r} \rightarrow V_{r+ r_\alpha}$. Then the orbifold
generalization of the ADHM equations (\ref{ADHMC3}) is
\begin{equation}
B_\b^{r + r_\a} \ B_\a^{r} = B_\a^{r + r_\b} \ B_\b^{r} \ , \qquad
r\in\Gammaw \ ,
\label{ADHMorb}\end{equation}
where $\a,\b=1,2,3$ and $Q = \rho_{r_1} \oplus
\rho_{r_2} \oplus \rho_{r_3}$. 
In the $U(1)$ gauge theory there are in principle several
moduli spaces, characterising instanton configurations with different
asymptotics.

We will argue that although these configurations are
physically distinct, the relevant moduli spaces are
isomorphic. Indeed, given a gauge field configuration one can
always change its asymptotic behaviour by tensoring its gauge bundle with a
tautological bundle. Tautological bundles can be thought of as line
bundles whose gauge field is the ``elementary'' configuration
asymptotic to a particular irreducible representation of the orbifold
group $\Gamma$. This heuristic picture is literally true since they
form a basis of the topological K-theory group of $X$. Tensoring with
line bundles thus establishes an isomorphism between different moduli
spaces of $U(1)$ instantons with fixed boundary conditions. In
particular the local structure of the moduli space is unchanged and so
is the contribution of an instanton configuration to the index of BPS
states. What changes is only the form of the instanton action (see
Section~\ref{subsec:instaction} for details). Because of this, when discussing partition functions of BPS states in the $U(1)$ gauge theory we will only consider a certain boundary condition, namely $N_0=1$ corresponding to trivial representations at infinity, with the understanding that the analysis for different boundary conditions is qualitatively similar.

\subsection{Cohomology of the $\Gamma$-Hilbert scheme\label{subsec:CohHilb}}

Let us now turn to the cohomology of $X$. We will review the
construction of~\cite{crawreid,craw} which gives two bases of $H^4 (X , \zed)$ and $H^2 (X , \zed)$ dual to the bases of exceptional surfaces and curves in the resolution $X$. The algorithm is combinatorial and allows a direct construction of these bases starting from the basis of tautological bundles, labelled by the irreducible representations of the orbifold group $\Gamma$. This will enable us to evaluate the integrals in the instanton action which involve the K\"ahler form $\omega$ of $X$. It is accomplished by expanding $\omega$ in the basis of $H^2 (X , \zed)$ and $\omega \wedge\omega$ in the basis of $H^4 (X , \zed)$ given by the tautological bundles.

One starts from the triangulation $\Sigma$ of the toric resolution $X = \mathrm{Hilb}^\Gamma (\complex^3)$ and associates to each line in the diagram the $\Gamma$-invariant ratio of monomials which parametrizes that curve. This naturally associates to each line a character of the orbifold group; here and in the following we will use the same symbol for a representation and its character. Then one can associate a character to each vertex of (the interior of) $\Sigma$. Only the following cases are possible:
\begin{itemize}
\item A vertex $v$ of valency 3. This vertex defines an exceptional
  projective plane $\PP^2$. All three lines meeting at $v$ are marked by the same character $\rho_r$. Then mark the vertex $v$ with the character $\rho_m = \rho_r \otimes \rho_r$.
\item A vertex $v$ of valency 4. This vertex defines an exceptional Hirzebruch surface $\mathbb{F}_r$. Two lines are marked with $\rho_r$ and two with $\rho_s$. Then mark the vertex $v$ with the character $\rho_m = \rho_r \otimes \rho_s$.
\item A vertex $v$ of valency 5 or 6 (excluding the case where three
  straight lines meet at a point). This vertex defines an exceptional Hirzebruch
  surface $\mathbb{F}_r$ blown up in one or two points respectively. There are two uniquely determined characters $\rho_r$ and  $\rho_s$ which each mark \textit{two} lines. The remaining line or two lines are marked with distinct characters. Then mark the vertex $v$ with the character $\rho_m = \rho_r \otimes \rho_s$.
\item A vertex $v$ of valency 6 at the intersection of three straight lines. This vertex defines an exceptional del Pezzo surface $\dd \PP_6$ of degree six. The straight lines are marked by three characters $\rho_r$, $\rho_s$ and $\rho_q$. Then the monomials defining the pair of morphisms $\dd \PP_6 \rightarrow \PP^2$ lie in uniquely determined character spaces $\rho_l$ and $\rho_m$ obeying the relation $\rho_l \otimes \rho_m = \rho_r \otimes \rho_s \otimes \rho_q$. Then mark the vertex $v$ with both characters $\rho_l$ and $\rho_m$. We will not need this case explicitly in this paper.
\end{itemize}

Once $\Sigma$ is ``decorated" by the characters of $\Gamma$ in this way, several geometrical properties are determined combinatorially. As a start one can show that every non-trivial character of $\Gamma$ appears in the toric fan $\Sigma$ precisely once as either:
\begin{itemize}
\item[(i)] marking a line;\footnote{This case also allows for a line,
    not necessarily straight, passing through several vertices, and
    thus stretches a bit the meaning of ``precisely once''. See
    \cite{crawreid,craw} for a discussion of this point.}
\item[(ii)] marking a vertex; or
\item[(iii)] the second character $\rho_l$ marking the intersection of three straight lines (we will not require this case).
\end{itemize}
To anticipate where this is all going, characters along lines will correspond to curves while characters on vertices will correspond to surfaces, in a precise sense. Indeed one can prove \cite{craw} that the first Chern classes $c_1 (\cR_r)$ associated with a character of the form (i) and (iii) form a basis of~$H^2 (X , \zed)$.

The above decoration encodes the following relations between tautological bundles in the Picard group $\mathrm{Pic} (X)$:
\begin{itemize}
\item If $\rho_m = \rho_r \otimes \rho_r$ marks a vertex of valency 3, then $\cR_m = \cR_r \otimes \cR_r$.
\item If $\rho_m = \rho_r \otimes \rho_s$ marks a vertex of valency 4, then $\cR_m = \cR_r \otimes \cR_s$.
\item If $\rho_m = \rho_r \otimes \rho_s$ marks a vertex of valency 5 or 6, then $\cR_m = \cR_r \otimes \cR_s$.
\item If $\rho_l$ and $\rho_m$ obeying $\rho_l \otimes \rho_m = \rho_r \otimes \rho_s \otimes \rho_q$ mark the intersection point $v$ of three straight lines, then $\cR_l \otimes \cR_m = \cR_r \otimes \cR_s \otimes \cR_q$.
\end{itemize}

Given all of the above ingredients, one can find virtual bundles whose second Chern classes form a basis of $H^4(X , \zed)$ dual to the basis of $H_4 (X , \zed)$ defined by the compact exceptional surfaces. The virtual bundles $\cV_m$ are defined as follows:
\begin{itemize}
\item[(a)] For each relation $\cR_m = \cR_r \otimes \cR_r$ arising from a vertex of valency 3, define $\cV_m = \left( \cR_r \oplus \cR_r \right) \ominus \left( \cR_m \oplus \cO_X  \right)$.
\item[(b)] For each relation $\cR_m = \cR_r \otimes \cR_s$ arising from a vertex of valency 4, define $\cV_m = \left( \cR_r \oplus \cR_s \right) \ominus \left( \cR_m \oplus \cO_X  \right)$.
\item[(c)] For each relation $\cR_m = \cR_r \otimes \cR_s$ arising from a vertex of valency 5 or 6, define $\cV_m = \left( \cR_r \oplus \cR_s \right) \ominus \left( \cR_m \oplus \cO_X  \right)$.
\item[(d)] For each relation $\cR_l \otimes \cR_m = \cR_r \otimes \cR_s \otimes \cR_q$ arising from a vertex where three straight lines intersect, define $\cV_m = \left( \cR_r \oplus \cR_s \oplus \cR_q \right) \ominus \left( \cR_l \oplus \cR_m \oplus \cO_X  \right)$.
\end{itemize}
This completes the characterization of the cohomology of the $\Gamma$-Hilbert scheme in terms of the tautological bundles.

\subsection{General form of the instanton action\label{subsec:instaction}}

We can now put together the results we have obtained so far to compute the instanton action in the general case. Recall that for the $U(1)$ gauge theory this action has the generic form
\begin{equation}
S_{\rm inst} = \frac{g_{s}}6 \, \int_{\overline{X}}\, F_A \wedge F_A \wedge F_A + \frac12\, \int_{\overline{X}}\, \omega \wedge F_A \wedge F_A + \int_{\overline{X}}\, \omega \wedge \omega \wedge F_A \ .
\label{Sinstgeneric}\end{equation}
Here $\omega$ is the K\"ahler form supported in
$X=\overline{X}\setminus\bdiv$, and in the following we will understand
(\ref{Sinstgeneric}) in terms of intersection indices of both compact
and non-compact divisors of $X$ and $\overline{X}$. 

Every term in (\ref{Sinstgeneric}) can be computed from the
knowledge of the Chern character $\mathrm{ch} (\cE)$ of the sheaf $\cE$ given as the only non-vanishing cohomology of the complex (\ref{ADHMcomplexfinal}). This yields
\begin{eqnarray}
\mathrm{ch} (\cE) &=& - \mathrm{ch} \Big( \big( V \otimes \mathcal{R} (-2) \big)^{\Gamma} \Big) +
 \mathrm{ch} \Big(
 \big(\mbox{$V \otimes \bigwedge^2 Q^{\vee}$} \otimes \mathcal{R} (-1) \big)^{\Gamma} \Big)
 \cr & & -\, \mathrm{ch} \Big(
\big(( {V \otimes Q^{\vee} \oplus W}) \otimes \mathcal{R} \big)^{\Gamma}
 \Big) + \mathrm{ch} \Big( \big(  {V} \otimes \mathcal{R} (1) \big)^{\Gamma} \Big) \ .
\end{eqnarray}
From this equation we get the instanton numbers
\begin{eqnarray} 
c_1 (\cE) &=& - \sum_{r,s\in\Gammaw}\, \bigg( \Big(N_s\,\delta_{rs} - \big( a^{(2)}_{rs} - a^{(1)}_{rs} \big)\, k_s \Big)\, c_1 ({\cR_r}) \nonumber \\ & & \qquad \qquad +\, \Big( \delta_{rs}\,  c_1 \big(\cO_{\overline{X}}(-2)\big) - a_{rs}^{(2)} \, c_1 \big(\cO_{\overline{X}}(-1)) - \delta_{rs} \, c_1 \big(\cO_{\overline{X}}(2)\big) \Big) \, k_s\bigg) \ , \nonumber \\[4pt]
\mathrm{ch}_2 (\cE) &=& - \sum_{r,s\in\Gammaw}\, \bigg( \Big( N_s\,\delta_{rs} - \big(  a^{(2)}_{rs} - a^{(1)}_{rs} \big) \, k_s \Big)\, \mathrm{ch}_2 ({\cR_r})
\label{genaction} \\ & & \qquad \qquad +\, c_1(\cR_r) \wedge \Big( \delta_{rs} \, c_1 \big(\cO_{\overline{X}}(-2)\big) - a_{rs}^{(2)} \, c_1 \big(\cO_{\overline{X}}(-1)) - \delta_{rs} \, c_1 \big(\cO_{\overline{X}}(1)\big) \Big) \, k_s 
 \cr & & \qquad \qquad + \, \Big( \delta_{rs} \, \mathrm{ch}_2 \big(\cO_{\overline{X}}(- 2)\big) - a_{rs}^{(2)} \, \mathrm{ch}_2 \big(\cO_{\overline{X}}(-1)\big) - \delta_{rs} \, \mathrm{ch}_2 \big(\cO_{\overline{X}}(1)\big) \Big) \, k_s \bigg) \ ,
\nonumber \\[4pt] \mathrm{ch}_3 (\cE) &=& - \sum_{r,s\in\Gammaw}\, \bigg( \Big( N_s\, \delta_{rs} - \big( a^{(2)}_{rs} - a^{(1)}_{rs} \big)\, k_s \Big)\, \mathrm{ch}_3 ({\cR_r})
 \cr & & \qquad \qquad + \, {c}_1 (\cR_r)\wedge \Big( \delta_{rs} \, \mathrm{ch}_2 \big(\cO_{\overline{X}}(-2)\big) - a_{rs}^{(2)} \, \mathrm{ch}_2 \big(\cO_{\overline{X}}(-1)\big) - \delta_{rs} \, \mathrm{ch}_2 \big(\cO_{\overline{X}}(1)\big) \Big) \, k_s 
  \cr & & \qquad \qquad + \, \mathrm{ch}_2 (\cR_r) \wedge \Big( \delta_{rs} \,{c}_1 \big(\cO_{\overline{X}}(-2)\big) - a_{rs}^{(2)} \, {c}_1 \big(\cO_{\overline{X}}(-1)\big) - \delta_{rs} \, {c}_1 \big(\cO_{\overline{X}}(1)\big) \Big) \, k_s 
   \cr & & \qquad \qquad + \, \Big( \delta_{rs} \, \mathrm{ch}_3 \big(\cO_{\overline{X}}(- 2)\big) - a_{rs}^{(2)} \, \mathrm{ch}_3 \big(\cO_{\overline{X}}(-1)\big) - \delta_{rs} \, \mathrm{ch}_3 \big(\cO_{\overline{X}}(1)\big) \Big) \, k_s\bigg) \nonumber
  \end{eqnarray}
which give an unambiguous instanton action upon integration. Note that $c_1 (\cR_0) = c_1 (\cO_{X}) = 0$.
From the behaviour of the Chern characters under tensor product we deduce
\begin{eqnarray}
c_1 \big(\cO_{\overline{X}} (- 1)\big) &=& - c_1 \big(\cO_{\overline{X}} (1)\big) \ , \nonumber\\[4pt] c_1 \big(\cO_{\overline{X}} (2)\big) &=& 2 c_1 \big(\cO_{\overline{X}} (1) \big) \ , \nonumber\\[4pt]
\mathrm{ch}_2 \big(\cO_{\overline{X}} (-1)\big) &=&  \mathrm{ch}_2 \big(\cO_{\overline{X}} (1)\big) \ , \nonumber\\[4pt] \mathrm{ch}_2 \big(\cO_{\overline{X}} (2)\big) &=& 4\, \mathrm{ch}_2 \big(\cO_{\overline{X}} (1)\big) \ , \nonumber\\[4pt] \mathrm{ch}_3 \big(\cO_{\overline{X}} (-1)\big) &=& - \mathrm{ch}_3 \big(\cO_{\overline{X}} (1)\big) \ , \nonumber\\[4pt] \mathrm{ch}_3 \big(\cO_{\overline{X}} (2)\big) &=& 8\, \mathrm{ch}_3 \big(\cO_{\overline{X}} (1)\big) \ .
\end{eqnarray}
Together with (\ref{ars12expl}) this allows us to simplify (\ref{genaction}) as
\begin{eqnarray}
c_1 (\cE) &=& - \sum_{r,s\in\Gammaw}\, \Big( N_s\,\delta_{rs} - \big(  a^{(2)}_{rs} - a^{(1)}_{rs} \big) \, k_s  \Big)\, c_1 ({\cR_r}) \ , \nonumber
\\[4pt]
\mathrm{ch}_2 (\cE) &=& - \sum_{r,s\in\Gammaw}\, \bigg( \Big(N_s\,
\delta_{rs} - \big( a^{(2)}_{rs} - a^{(1)}_{rs} \big)\, k_s \Big)\,
\mathrm{ch}_2 ({\cR_r}) \cr && \qquad \qquad \qquad +\, c_1 \big(\cO_{\overline{X}}(1)\big) \wedge c_1(\cR_r)\, \big(  a_{rs}^{(2)} - 3 \delta_{rs} \big) \, k_s \bigg) \ , 
\\[4pt] 
\mathrm{ch}_3 (\cE) &=& - \sum_{r,s\in\Gammaw}\, \bigg( \Big( N_s\, \delta_{rs} - \big( a^{(2)}_{rs} - a^{(1)}_{rs} \big)\, k_s \Big)\, \mathrm{ch}_3 ({\cR_r})
  + c_1 \big(\cO_{\overline{X}}(1)\big) \wedge \ch_2 (\cR_r)\, \big( a_{rs}^{(2)} - 3 \delta_{rs} \big) \, k_s 
  \cr & & \qquad \qquad \qquad -\, c_1 (\cR_r) \wedge \ch_2 \big(\cO_{\overline{X}} (1)\big)\, \big( a^{(2)}_{rs} - 3 \delta_{rs} \big) \, k_s
   - 6k_s\, \delta_{rs}\, \mathrm{ch}_3 \big(\cO_{\overline{X}}(1)\big) \bigg) \ . \nonumber
\end{eqnarray}

Given these Chern characters, to compute the instanton action we use the cohomology of the $\Gamma$-Hilbert scheme described in Section~\ref{subsec:CohHilb} which allows us to expand the integrals involving the K\"ahler form in the basis of cohomology derived from the tautological line bundles. Having identified such bases $c_1 (\cR_n) \in H^2 (X , \zed)$ and $c_2 (\cV_m) \in H^4(X ,\zed)$, we may expand
\begin{eqnarray}
\omega = \sum_{n\in\Gammaw}\, \varphi_n \ c_1 (\cR_n) \qquad \mbox{and} \qquad
\omega\wedge\omega = \sum_{m\in\Gammaw}\, \varsigma_m \ c_2 (\cV_m) \ ,
\end{eqnarray}
where the parameters $\varphi_n$ and $\varsigma_m$ play the role of
chemical potentials for the D4 and D2 branes. Strictly speaking the
K\"ahler class is not an integral class, but there is evidence that it
is quantised in topological string theory~\cite{qfoam}. It would be
interesting to make this identification more precise by computing the
D~brane charges in our formalism and make a connection with
\cite{Aganagic:2010qr}. 

In any case, the resulting integrals over $\oX$ compute
intersection indices among the various compact and non-compact
divisors in the geometry, which in general depend on the details of the
particular orbifold in question. In particular, the integral of
$\mathrm{ch}_3 \big(\cO_{\overline{X}}(1)\big)$ computes the triple
intersection of the divisor $\bdiv\cong\PP^2/\Gamma$ at
infinity. Since three planes $\PP^2$ intersect at a point in $\PP^3$
and we evaluate orbifold integrals by pullback (see Appendix~\ref{App:Stacks}), it is given by
\begin{equation}
\int_{\overline{X}}\, \mathrm{ch}_3 \big(\cO_{\overline{X}}(1)\big) =
\frac 1 6 \, \int_{\overline{X}}\, c_1 \big(\cO_{\overline{X}}
(1)\big) \wedge c_1 \big(\cO_{\overline{X}} (1)\big) \wedge c_1
\big(\cO_{\overline{X}} (1)\big)  = \frac{1}{6 | \Gamma |} \ .
\end{equation}
We have now arrived to the final form of the instanton action given by
\begin{eqnarray}
\int_{\overline{X}}\,\omega \wedge \omega \wedge c_1 (\cE) &=& -
\sum_{m,r,s\in\Gammaw}\, \varsigma_m \, \Big( N_s\, \delta_{rs}
- \big(  a^{(2)}_{rs} - a^{(1)}_{rs} \big)\, k_s \Big)\,
\int_{\overline{X}} \, c_2 (\cV_m) \wedge c_1 ({\cR_r})  \ , \label{c1action}
\\[4pt]
\int_{\overline{X}}\, \omega \wedge \mathrm{ch}_2 (\cE) &=& -
\sum_{n,r,s \in\Gammaw}\, \varphi_n\, \bigg( \Big( N_s\,\delta_{rs} -
\big( a^{(2)}_{rs} - a^{(1)}_{rs} \big)\, k_s \Big)\,
\int_{\overline{X}}\, c_1 (\cR_n) \wedge \mathrm{ch}_2 ({\cR_r})
 \label{ch2action} \\ && \qquad \qquad \qquad +\, \big(  a_{rs}^{(2)} - 3 \delta_{rs} \big) \, k_s \, \int_{\overline{X}}\, c_1 (\cR_n) \wedge c_1 \big(\cO_{\overline{X}}(1)\big) \wedge c_1(\cR_r) \bigg) \ , 
\nonumber
\\[4pt]
 \int_{\overline{X}}\, \mathrm{ch}_3 (\cE) &=& -
 \sum_{r,s\in\Gammaw}\, \bigg(\Big( N_s\,\delta_{rs} - \big(
 a^{(2)}_{rs} - a^{(1)}_{rs} \big)\, k_s \Big)\, \int_{\overline{X}}\,
 \mathrm{ch}_3 ({\cR_r}) - \frac{k_s}{|\Gamma|}\, \delta_{rs}
\cr && \qquad \qquad \qquad + \, \big( a_{rs}^{(2)} - 3 \delta_{rs} \big) \, k_s \, \int_{\overline{X}}\, c_1 \big(\cO_{\overline{X}}(1)\big) \wedge \ch_2 (\cR_r)
  \cr & & \qquad \qquad \qquad -\, \big( a^{(2)}_{rs} - 3 \delta_{rs} \big) \, k_s \, \int_{\overline{X}}\, c_1 (\cR_r) \wedge \ch_2 \big(\cO_{\overline{X}} \big(1)\big) \bigg) \ .
\label{ch3action}
\end{eqnarray}

Note that the ``regular'' instanton configurations are of the
form $k_r=k$ for all $r\in\Gammaw$ and can be naturally associated
with the regular representation $\bigoplus_{r\in\Gammaw}\, \rho_r$ of
the orbifold group $\Gamma$, or equivalently with the tautological
bundle (\ref{tautdef}). If we consider those instantons which
asymptote to the trivial representation at infinity, so that only
the framing integer $N_0$ is non-zero, then by (\ref{ars12expl}) and $c_1(\cR_0)=0$ the
integrals involving $c_1 (\cE)$ in (\ref{c1action}) and $\ch_2(\cE)$ in
(\ref{ch2action}) vanish identically, while the instanton charge
$\int_\oX\, \ch_3(\cE)=k$ is integer-valued.

\section{Instanton quantum mechanics and quiver moduli spaces\label{sec:InstQM}}

The instanton moduli space is characterized by a set of equations
which generalize the usual ADHM formalism and arise from the
degeneration of the Beilinson spectral sequence. But the spectral
sequence actually comes with more information. The parametrization of
the moduli space of torsion free sheaves can be realized via an
appropriate quiver, whose vertices correspond to the $\Gamma$-module
decomposition of the vector spaces $V$ and $W$ representing certain
cohomology groups in (\ref{Vdecomp}) and (\ref{Wdecomp}). The arrows
of the quiver are the elements in the decomposition of the
differentials which enter in the spectral sequence, according to their
$\Gamma$-module structure. We will argue that this quiver is precisely
the framed McKay quiver associated with the orbifold group
$\Gamma$. We will see that the effective action studied in
\cite{Ooguri:2008yb} as a low-energy limit of the theory of D0 and D2
branes in the background of a single D6 brane is recovered geometrically from the matrix quantum mechanics which governs the measure of the instanton moduli space.

\subsection{Quivers and their representations\label{subsec:Quivers}}

We will start with a quick review of some facts concerning quivers and
their representations, refering the reader to the
reviews~\cite{ginzburg,crawrev,reinekerev} for further details. A
quiver $\sf Q$ is a directed graph constructed from a set of vertices
${\sf Q}_0$ and a set of arrows ${\sf Q}_1$ connecting the
vertices. This information is encoded in maps ${\sf t} \, ,
\, {\sf h}: {\sf Q}_1 \rightrightarrows {\sf Q}_0$ that identify the
tail and head vertices of each arrow, respectively. A path $\sf p$ in the
quiver from a vertex $v$ to the vertex $w$ is a composition of arrows
${\sf p}= a_1 \cdots a_k$ such that ${\sf h} (a_m) = {\sf t} (a_{m+1})$ for $1 \le m < k$, and
${\sf t}({\sf p}):= {\sf t}(a_1)=v$ while ${\sf h}({\sf p}):= {\sf h}(a_k)=w$; in this case we say that the
path $\sf p$ has length $k$. In particular each vertex $v$ has associated to
it a {\it trivial} path ${\sf e}_v$ of length zero, which starts and
ends at the same vertex ${\sf t}({\sf e}_v)= {\sf h} ({\sf e}_v)=v$. This should
not be confused with a loop, which is a non-trivial path from a vertex
to itself of length one.

The collection of paths form an associative
noncommutative algebra, the path algebra $\complex \sf Q$ of the
quiver $\sf Q$, with the product of two paths defined by concatenation
if the paths compose and zero otherwise. It is graded by path length. The elements ${\sf e}_v$ for $v\in {\sf Q}_0$ are
orthogonal idempotents in this algebra, i.e. ${\sf e}_v \, {\sf e}_w =
\delta_{v,w}\, {\sf e}_v$, such that $\sum_{v \in {\sf Q}_0 }\, {\sf
  e}_v$ is the identity element of $\complex \sf Q$. A relation $\sf
r$ of the quiver is a $\complex$-linear combination of paths in $\complex \sf Q$ with the
same head and tail vertices, and length at least two. A bounded quiver
$({\sf Q} , {\sf R})$ is a quiver together with a finite set $\sf R$ of relations; they determine an ideal $\langle{\sf R}\rangle$ in the path algebra $\complex \sf Q$.

We are interested in representations of the quiver $\sf Q$. They form
a category which is equivalent to $\complex {\sf Q}$--mod, the category
of finitely-generated left $\complex \sf Q$-modules (or equivalently right $\complex \sf
Q$-modules). For any left $\complex \sf Q$-module $V$, we can form the
complex vector spaces $V_v =\mathsf{ e}_v V$ for $v\in{\sf Q}_0$ of dimension~$k_v$. We think of each vector space as living on a vertex of the
quiver. The arrows $v\to w$ of the quiver induce linear transformations between
the vector spaces $V_v\to V_w$. If the quiver has relations $\sf R$, we furthermore
require that the linear maps be compatible with the
relations, i.e. the sum of compositions of linear maps corresponding
to the relations gives the zero map, and similarly for the factor path
algebra ${\sf A} = \complex {\sf Q} / \langle{\sf R}\rangle$. The
category of representations of a quiver with relations,
$\mathrm{Rep}_{\complex} ({\sf Q} , {\sf R})$, is equivalent to the
category $\sf A$--mod of left $\sf A$-modules.

To each vertex $v\in{\sf Q}_0$ we can associate a one-dimensional
simple module ${\sf D}_v$ as the representation where
$V_v = \complex$ and $V_w = 0$ for all $w\neq v$. In the string theory
setting these modules correspond to fractional branes. Furthermore we
can define ${\sf P}_v = {\sf e}_v{\sf A}$, the subspace of the path
algebra generated by all paths that begin at the vertex $v$. The
usefulness of the modules ${\sf P}_v$ is that they are projective
objects in the category $\sf A$--mod which can be used to construct projective resolutions of the simple modules ${\sf D}_v$ through
\begin{equation}
\begin{xy}
\xymatrix{ \dots \ \ar[r] & \ \displaystyle{\bigoplus_{w \in {\sf
        Q_0}} \, d^{p}_{w,v}\, {\sf P}_w } \ \ar[r] & \ \dots \ \ar[r]
  & \ \displaystyle{ \bigoplus_{w\in {\sf Q_0}}\,d^{1}_{w,v}\, {\sf
      P}_w } \ \ar[r] & \ {\sf P}_v \ \ar[r]& \ {\sf
    D}_v \ \ar[r]  & \ 0
}
\end{xy}
\label{projres}\end{equation}
where
\begin{equation}
d^p_{w,v} = \dim_{\complex} \Ext^p_{\sf{A}} \left( {\sf D}_v ,
  {\sf D}_w \right) \ .
\end{equation}
When a quiver has an underlying geometrical interpretation, perhaps
via an equivalence of derived categories, it is often easier to
rephrase the geometrical computations in this algebraic
fashion.\footnote{When a bounded derived category of quiver representations is
  identified with a bounded derived category of coherent sheaves, the simple modules ${\sf D}_v$ correspond to the basis of compactly supported sheaves $\cS_v$ while the representations ${\sf P}_v$ which enter in the projective resolutions are related to the basis of tautological bundles $\cR_v$, which indeed give locally free resolutions of the sheaves $\cS_v$.}

On each node $v\in\sfQ_0$ of the quiver there is a natural
$GL(k_v,\complex)$-action by basis change automorphisms. We are thus
naturally led to consider the moduli space of isomorphism classes of
quiver representations, by factoring the action of the group
\beq
G_{\mbf k} = \prod_{v \in \sfQ_0}\, GL (k_v,\complex) \ , 
\eeq
where ${\mbf
  k}:=(k_v)_{v\in\sfQ_0}$ is the dimension vector characterising the quiver
representation. However the direct quotient is rather badly
behaved. The usual strategy in algebraic geometry is to resort to geometric invariant
theory. This produces a smooth quotient at the price of having to
discard certain orbits of the complexified $G_{\mbf k}$-action. One
restricts the quotient to only stable representations that are
defined in a purely algebraic manner via the slope stability parameter~\cite{king}, which
is given for any representation $V=\bigoplus_{v\in\sfQ_0}\, V_v$ with dimension vector $\mbf k$ as
\begin{equation}
\theta (V) = \theta(\mbf k)= \frac{\mbf\theta \cdot \mbf k}{\dim_{\complex} V}
\label{thetaV}\end{equation}
where $\mbf\theta \in \real^{\sfQ_0}$ and $\dim_{\complex}
V=\sum_{v\in\sfQ_0}\, k_v$. A representation $V$ is $\theta$-stable
(resp. $\theta$-semistable) if for any proper subrepresentation $V'\subset V$
one has $\theta(V'\,) < \theta(V)$ (resp. $\theta(V'\,) \le
\theta(V)$). Then the moduli space of $\theta$-stable representations
is well-behaved and fine (and for generic values of the stability
parameters $\mbf\theta$ we do not have to distinguish between stable
and semistable representations).

The other notion we need is that of a \textit{framing} of a
quiver $\sfQ$. There are several (equivalent) notions of framing of quiver
representations; here we will follow the treatment of Joyce and
Song~\cite[Section 7.4]{joycesong}. This operation consists in
defining a new quiver $\sfQ^f$ whose vertex set is doubled compared to that of $\sfQ$, i.e.
$\sfQ_0^{f} = \sfQ_0 \sqcup \sfQ_0$. To each vertex $v\in\sfQ_0$ of
the original quiver, there corresponds a new vertex $v'$ of $\sfQ^f$
(the double of $v$) and an
additional arrow $a_I: v'  \rightarrow v$. Similarly the
representation theory of $\sfQ^f$ now involves two sets of vector
spaces $V_v$ and $W_v$ together with additional maps $I_v : W_v \rightarrow
V_v$, and we can introduce the notion of framed
representations. The stability of framed representations is
essentially the same as stability of the representations
\textit{before} the framing. This defines the
moduli spaces of representations of framed quivers ${\sf Rep}_{\theta}
\left( \sfQ^{f} , \mbf k , \mbf N\right)$ with fixed dimension vectors $\mbf k$ and~$\mbf N$.

\subsection{The McKay quiver\label{subsec:McKay}}

We will now rephrase the construction of the instanton moduli space in
terms of an auxilliary quiver, the McKay quiver, derived from the
representation theory data for the action of the orbifold group~$\Gamma$. Recall that the problem we are studying has a double life:
the representation theory of $\Gamma$-equivariant
$\cO_{\complex^3}$-modules on $\complex^3$ and the geometry of the
crepant resolution of $\complex^3 / \Gamma$ given by the
$\Gamma$-Hilbert scheme $\mathrm{Hilb}^\Gamma (\complex^3)$ which
parametrizes $\Gamma$-invariant configurations of D0~branes. We shall
present a quiver which encodes this construction and has the
generalized ADHM equations as relations. 

We proceed from the point of view of representation theory. To begin
with, we consider all the irreducible representations of $\Gamma
\subset SL(3,\complex)$. To each of these representations we associate
a tautological bundle. We construct the quiver $\sfQ$ by declaring that each
node represents a different representation/tautological bundle
(including the trivial representation which corresponds to the trivial
bundle), i.e. the vertex set is $\sfQ_0= \Gammaw$. Two nodes are connected by
a number $a_{sr}^{(1)}$ of arrows going from $s$ to $r$ determined by
the tensor product decomposition (\ref{tensordecomp}) for $i=1$. Note
that in general the matrix $a_{sr}^{(1)}$ does not have any particular
symmetry property (in contrast to the familiar case of instantons on ALE
spaces where it would be symmetric). The resulting quiver is known as
the bounded McKay quiver $(\sfQ ,\sfR)$ and it is associated with an
ideal of relations $\langle {\sf R} \rangle$ in the corresponding path algebra $\complex\sfQ$.

In concrete applications
one is interested in the representations of this quiver, which are
obtained by associating with every vertex $r$ a $k_r$-dimensional complex
vector space $V_r$ and a linear map, represented by a $k_r \times k_s$
matrix, to every arrow from $V_s$ to $V_r$. Then the relations between
the arrows of the quiver induce equivalence relations between the
morphisms of the
representations. They can be compactly rewritten in terms of the
$\Gamma$-equivariant linear map $B \in \Hom_{\Gamma} (V , Q \otimes
V)$ introduced in Section~\ref{subsec:Beilinson}, and assume the
simple form given in (\ref{ADHMorb}).

Associated with this quiver is its moduli space of representations
${\sf Rep}(\sfQ , \sfR)$. This is not quite the end of the story, as
there is a natural $GL(V_r,\complex)$-action on each vector space $V_r$
which lifts to the linear maps $B^r_\a$ for $\a=1,2,3$ and $r\in\Gammaw$ as
\begin{equation}
B_\a^r \ \longmapsto \ g_{r+ r_\a}\, B^r_\a\, g_{r} \qquad
\mbox{with} \quad g_r\in GL(V_r,\complex) \ .
\end{equation} 
Therefore the relevant moduli space is actually the quotient in
geometric invariant theory of ${\sf Rep} (\sfQ , \sfR)$ by this
group action of $\prod_{r\in\Gammaw}\, GL(V_r,\complex)$.

\subsection{Noncommutative crepant resolutions\label{subsec:NCcrepant}}

The representation theory of quivers with dimension vectors $\mbf k =
(1 , \dots , 1)$ and $\dim_\complex W = 1$ is intimately related to the
smooth geometry of toric varieties. Many toric varieties can be realized as moduli spaces of representations of a quiver~\cite{CS08}. In many cases, and in
particular for the quivers we will consider in this paper, this moduli
space of representations, constrained by an appropriate ideal of
relations, is a smooth crepant resolution of a toric singularity for
generic values of the stability parameters. In particular, a moduli
space of representations of the bounded McKay quiver with fixed dimension
vector $\mbf k = (1,\dots,1)$ is precisely isomorphic to the natural crepant
resolution of an abelian orbifold singularity $\complex^3 / \Gamma$
provided by the
$\Gamma$-Hilbert scheme $\mathrm{Hilb}^{\Gamma}
(\complex^3)$~\cite{ItoNaka}; these quivers correspond to regular instantons. This holds in the chamber of moduli space in which the stability parameter $\theta_\rho$ is positive (see e.g.~\cite[Remark~4.20]{crawrev} for the definition of this stability parameter).

Under certain circumstances the path algebra $\sfA$ of the quiver
itself can be regarded as another desingularization, the noncommutative
crepant resolution of the
singularity~\cite{vandenbergh1}. In this case $\sfA$ represents a
noncommutative deformation of a variety which
contains the coordinate algebra of the singularity as its
center. Moreover, the noncommutative space dual to $\sfA$ ``knows'' about
all the other projective crepant resolutions, in the sense that there exists a
derived equivalence between the corresponding bounded derived
categories of $\sfA$-modules and of coherent sheaves~\cite{CI04}. For example, if
the bounded derived category $\mathbf{D}(X)$ of coherent sheaves on a
crepant resolution $X$ of the singularity is generated by a tilting
object $\cT$, then setting $\sfA = \End_{\mathbf{D}(X)}(\cT)$
induces a derived equivalence $\mathbf{D} (X) \cong \mathbf{D} (\sfA)$
and $\sfA$ is a noncommutative crepant resolution of its center, with
$\sfA$--mod the category of coherent sheaves on the noncommutative scheme
${\rm Spec}(\sfA)$. We
explain this equivalence in more detail in Section~\ref{subsec:Stability}.

The path algebra of the McKay quiver
for abelian orbifold singularities $\complex^3/\Gamma$ gives a
noncommutative crepant resolution. The path algebra of the \emph{bounded} McKay quiver is isomorphic to the skew group algebra~\cite[Proposition~2.8]{CMT07}, which is the standard noncommutative crepant resolution of the $\Gamma$-invariant ring as in~\cite{vandenbergh1}. Moreover, there exists a derived equivalence between the corresponding bounded derived categories of $\sfA$-modules and of coherent sheaves; this is a special case of~\cite{BKR}.

Consider for example the orbifold $\complex^3 / \zed_3$ with the diagonal action~\cite{Aspinwall:2008jk}, as in Section~\ref{subsec:InstC3Z3}. The relevant quiver is
\begin{equation}
\vspace{4pt}
\begin{xy}
\xymatrix@C=20mm{
& \ v_0 \ \bullet \ \ar@/^/[ddl] \ar@/_0.5pc/[ddl] \ar@//[ddl]  & \\
& & \\
v_1 \ \bullet \ \ar@//[rr] \ar@/^/[rr]  \ar@/_/[rr]   & &  \ \bullet \ v_2  \ar@/^/[uul] \ar@/_0.5pc/[uul] \ar@//[uul] 
}
\end{xy}
\vspace{4pt}
\label{quiverC3Z3}\end{equation}
with weights $r_\a=1$ for $\a=1,2,3$, i.e. in this case $Q = \rho_1 \oplus \rho_1 \oplus \rho_1$. The maps $\sfb_\a^{r} : \sfD_{r}
\rightarrow \sfD_{r +r_\a\ {\rm mod}\ 3}$ corresponding to the arrows of the quiver are of the form
\begin{eqnarray}
\sfb_{\a}^{0} \ &:& \ \sfD_{0} \ \longrightarrow \ \sfD_{1} \ , \nonumber\\[4pt]
\sfb_{\a}^{1} \ &:& \ \sfD_{1} \ \longrightarrow \ \sfD_{2} \ , \nonumber\\[4pt]
\sfb_{\a}^{2} \ &:& \ \sfD_{2} \ \longrightarrow \ \sfD_{0}
\end{eqnarray}
for $\a=1,2,3$. The relations are obtained by unpacking the generic form $\sfb_\b^{r +r_\a} \ \sfb_\a^{r} = \sfb_\a^{r+r_\b} \ \sfb_\b^{r}$; explicitly we find
\begin{equation}
\begin{matrix} 
\sfb_2^{1} \, \sfb_1^{0} = \sfb_1^{1} \, \sfb_2^{0}  \ , & \qquad
\sfb_3^{1} \, \sfb_1^{0} = \sfb_1^{1} \, \sfb_3^{0}  \ , & \qquad
\sfb_3^{1} \, \sfb_2^{0} = \sfb_2^{1} \, \sfb_3^{0}  \ , \\[4pt]
\sfb_2^{2} \, \sfb_1^{1} = \sfb_1^{2} \, \sfb_2^{1}  \ , & \qquad
\sfb_3^{2} \, \sfb_1^{1} = \sfb_1^{2} \, \sfb_3^{1}  \ , & \qquad
\sfb_3^{2} \, \sfb_2^{1} = \sfb_2^{2} \, \sfb_3^{1}  \ , \\[4pt]
\sfb_2^{0} \, \sfb_1^{2} = \sfb_1^{2} \, \sfb_2^{0}  \ , & \qquad
\sfb_3^{0} \, \sfb_1^{2} = \sfb_1^{2} \, \sfb_3^{0}  \ , & \qquad
\sfb_3^{0} \, \sfb_2^{2} = \sfb_2^{2} \, \sfb_3^{0}  \ .
\end{matrix}
\label{C3Z3rels}\end{equation}

Consider now the associated path algebra $\sfA$ and its center $\sfZ (\sfA)$. As a ring, $\sfZ(\sfA)$ is generated by elements of the form
\begin{equation}
\textsf{x}_{\a\b\gamma} = \sfb_\a^{2} \ \sfb_\b^{1} \ \sfb_\gamma^{0} \ \qquad \text{with} \quad \a \le \b \le \gamma \ .
\end{equation}
If we choose coordinates $(z_1, z_2 , z_3)$ on $\complex^3$ on which the orbifold group $\Gamma=\zed_3$ acts as in (\ref{Z3C3action}), then we can explicitly map these generators to the $\Gamma$-invariant elements of the polynomial algebra $\complex [z_1, z_2 , z_3]$ as ${\sf x}_{\a\b\gamma} = z_\a\, z_\b \, z_\gamma$. This means that
\begin{equation}
\text{Spec} \, \sfZ (\sfA) = \complex^3 \big/ \zed_3 \ ,
\end{equation}
and hence the path algebra $\sfA$ is a noncommutative resolution of
the $\complex^3 / \zed_3$ singularity which is seen as its center. By
the McKay correspondence there is a bounded derived equivalence $
\mathbf{D} (\sfA)\cong \mathbf{D} (X)$, where the local del~Pezzo
surface $X=\cO_{\PP^2}(-3)$ of degree zero is the unique crepant Calabi--Yau
resolution of $\complex^3/\zed_3$ obtained by blowing up the singular
point at the origin of $\complex^3$ to a projective plane $\PP^2$~\cite{Ezhuthachan:2006gu}.

\subsection{Quiver quantum mechanics on $\complex^3$ \label{subsec:QuiverC3}}

In this section we describe the quantum mechanics which govern the
dynamics of the instanton collective coordinates. It arises as the
dimensional reduction of the noncommutative D6~brane gauge theory of
Section~\ref{sec:NCGT} to the D0~branes. This model is
topological and exactly solvable; the study of these types of quantum
theories was pioneered in \cite{Moore:1998et,Moore:1997dj}. This
theory is in its simplest form on the affine Calabi--Yau space
$\complex^3$; we begin by briefly reviewing this model following~\cite{Cirafici:2008sn}. In this case it is based on  two vector spaces $V$ and $W$, of complex
dimensions $\dim_\complex V = k$ and $\dim_\complex W = N$, introduced
in (\ref{Vdecomp}) and (\ref{Wdecomp}). In the
D~brane picture $V$ is spanned by the gas of $k$ D0~branes, while $W$
represents the Chan--Paton bundle on the $N$ (spectator) D6~branes. In this description we fix
the topological sector and restrict attention to instantons of charge
$k$. 

The fields of the quiver are given by
\begin{eqnarray}
X_i = (B_1 , B_2 , B_3 , \varphi , I) \qquad \mbox{and} \qquad
\Psi_i = (\psi_1 ,\psi_2 , \psi_3 , \xi , \varrho) \ .
\end{eqnarray}
The matrices $B_\a$ arise from 0--0 strings and represent the
position of the coincident D0~branes inside the D6~branes. The field $I$
describes open strings stretching from the D6~branes to the D0~branes;
it characterizes the
size and orientation of the D0~branes inside the D6~branes, and is
required to make the system supersymmetric. In the noncommutative gauge theory the field $\varphi$ is the
dimensional reduction of the $(3,0)$-form field $\rho$. Consistently with this open string interpretation we can regard these fields as linear maps
\begin{eqnarray}
(B_1 , B_2 , B_3 , \varphi) \in \Hom_\bC (V,V) \qquad \mbox{and} \qquad
I\in\Hom_\bC(W,V) \ .
\label{bosnaive}\end{eqnarray}

The fields $B_\a$ and $\varphi$ all lie in the adjoint
representation of $U(k)$ where $k$ is the number of D0~branes (or the
instanton number). On the other hand,
$I$ is a $U(k) \times U(N)$ bifundamental field where $N$ is the
number of D6~branes (or the rank of the six-dimensional gauge
theory). Under the full symmetry group $U(k)\times U(N)\times\torus^3$
the transformation rules are
\begin{eqnarray}
B_\a ~&\longmapsto~& \e^{- \ii \epsilon_\a}\, g_{U(k)} \, B_\a \, g_{U(k)}^{\dagger} \ , \nonumber\\[4pt]
\varphi~&\longmapsto~& g_{U(k)} \, \varphi \,
g_{U(k)}^{\dagger} \ , \nonumber\\[4pt]
I ~&\longmapsto~& g_{U(k)}\, I\, g_{U(N)}^{\dagger}\ ,
\end{eqnarray}
where in the transformation of $\varphi$ we have imposed the Calabi--Yau condition $\epsilon_1 +
\epsilon_2 + \epsilon_3=0$. 
The corresponding BRST transformations read
\be
\begin{array}{rllrl}
\cQ B_\a &=& \psi_\a & \quad \mbox{and} \qquad \cQ \psi_\a& = \ [\phi , B_\a] -
\epsilon_\a\,
B_\a \ , \\[4pt] \cQ \varphi& =& \xi & \quad \mbox{and} \qquad
\cQ \xi &= \ [\phi ,\varphi] \ , \\[4pt]
\cQ I &=& \varrho & \quad \mbox{and} \qquad \cQ \varrho& = \
\phi \,I - I \,\mbf a \ ,
\end{array}
\label{BRSTmatrices}\ee
where $\phi$ is the generator of $U(k)$ gauge transformations and $\mbf a = \mathrm{diag} (a_1 , \dots , a_N)$ is a background
field which parametrizes an element of the Cartan subalgebra
$\mathfrak{u}(1)^{\oplus N}$; in the noncommutative gauge theory $\mbf
a$ plays the role of the vev of the Higgs field $\Phi$.

The information about the bosonic field content can be summarized in the quiver diagram
\begin{equation} \label{ADHMquiver}
\vspace{4pt}
\xymatrix@C=20mm{
& \ V \ \bullet \ \ar@(ul,dl)_{B_2} \ar@(ur,ul)_{B_1} \ar@(dr,dl)^{B_3}
\ar@{.>}@(ur,dr)^< < < <{\varphi} & \ \bullet \ W \ar@//[l]_{ \ \ \ \ \ I} 
}
\vspace{4pt}
\end{equation}
The relationship between the collective coordinates and quivers will
also hold for more general geometries.
The quiver quantum mechanics on $\complex^3$ is characterized by the
bosonic field equations
\begin{eqnarray}
\mathcal{E}_{\a} \,&:&\, [B_\a , B_\b] + \sum_{\g=1}^3\, \epsilon_{\a\b\gamma} \,\big[B_\gamma^{\dagger}
\,,\, \varphi\big] = 0 \ , \nonumber\\[4pt]
\mathcal{E}_\lambda  \,&:&\, \sum_{\a=1}^{3}\, \big[B_\a \,,\,
B_\a^{\dagger}\,\big] + \big[\varphi \,,\, \varphi^{\dagger}\,\big] + I\,
I^{\dagger} = \lambda \ , \nonumber\\[4pt] \mathcal{E}_I \,&:&\,
I^{\dagger} \,\varphi = 0 \ ,
\label{quiverdefeqs}\end{eqnarray}
where $\lambda>0$ is a Fayet--Iliopoulos parameter which originates
through the noncommutative deformation.

As in the standard formalism for topological field theories we add the Fermi multiplets $(\vec{\chi} ,
\vec{H}\,)$, which contain the antighost and auxilliary fields
$\vec{\chi}= (\chi_1 , \chi_2 , \chi_3 , \chi_\lambda , \zeta)$ and $\vec{H} =
(H_1 , H_2 , H_3 , H_\lambda , h)$. Since the auxilliary
fields are determined by the equations $\vec{\mathcal{E}}$, they carry the same quantum numbers. This implies that the
antighost fields are defined as maps
\begin{eqnarray}
(\chi_1 , \chi_2 , \chi_3 , \chi_\lambda) \in \Hom_\bC (V,V) \qquad \mbox{and}
\qquad \zeta \in \Hom_\bC (V,W) \ ,
\label{fermnaive}\end{eqnarray}
and that the BRST transformations associated with the new fields are
\be
\begin{array}{rllrl}
\cQ \chi_\a &=& H_\a & \quad \mbox{and} \qquad \cQ H_\a &= \ [\phi , \chi_\a]+\epsilon_\a\, \chi_\a \ , \\[4pt]
\cQ \chi_\lambda &=& H_\lambda & \quad \mbox{and} \qquad \cQ H_\lambda &= \ [\phi , \chi_\lambda ] \ ,
\\[4pt] \cQ \zeta &=& h & \quad \mbox{and} \qquad \cQ h &= \ \mbf a \,
\zeta - \zeta\, \phi \ .
\end{array}
\ee
To these fields we add the gauge multiplet $(\phi,\overline{\phi},\eta)$ to close the algebra
\be
\cQ \phi=0 \ , \quad \cQ \overline{\phi}=\eta \quad \mbox{and} \quad
\cQ \eta=\big[\phi\,,\,\overline{\phi}~\big] \ .
\label{gaugeBRST}\ee

The action that corresponds to this system of
fields and equations is given by
\begin{eqnarray}
S &=& \cQ \, \Tr\Big(\, \sum_{\a=1}^3\, \chi_\a^{\dagger}\,(H_\a -
\mathcal{E}_\a) + \chi_\lambda \,(H_\lambda - \mathcal{E}_\lambda ) + \zeta^{\dagger}\, (h -
\mathcal{E}_I)\cr && \qquad\quad +\,\sum_{\a=1}^3\, \psi_\a \,\big[~\overline \phi \,,\,
B_\a^{\dagger}\big] +
\xi \,\big[~\overline{\phi} \,,\, \varphi^{\dagger}\big] + \varrho\,
 \overline{\phi}\,I^{\dagger} + \eta \,\big[\phi \,,\,
 \overline{\phi}~
\big] + \mbox{h.c.} \, \Big) \ .
\end{eqnarray}
This action is topological and the path integral localizes onto the fixed points of the BRST charge $\cQ$ given by the equations
\begin{eqnarray} \label{fixedpoint}
\left( B_\a \right)_{ab} \,\left( \phi_a - \phi_b - \epsilon_\a
\right)=  0 \qquad \mbox{and} \qquad
I_{a , l}\, \left( \phi_a - a_l \right) = 0 \ .
\end{eqnarray}
Solutions to these equations can be completely characterized by
$N$-vectors of plane partitions $\vec \pi = \left( \pi_1 , \dots , \pi_N
\right)$ with $|\vec\pi|=\sum_l\, |\pi_l|=k$ boxes~\cite{Cirafici:2008sn}.

The fluctuation determinants can be evaluated via standard
techniques~\cite{Moore:1998et,Moore:1997dj,Cirafici:2008sn}. However, 
for practical purposes it is more efficient to construct a local model
of the instanton moduli space around each fixed point $\vec\pi$ of the $\torus^3$-action via the instanton deformation complex~\cite{Cirafici:2008sn}
\begin{equation} \label{adhmdefcomplexC}
\xymatrix{
  \Hom_\complex (V_{\vec\pi} , V_{\vec\pi})
   \quad\ar[r]^{\!\!\!\!\!\!\!\!\!\!\!\!\!\!\!\!\sigma} &\quad
   {\begin{matrix} \Hom_\complex (V_{\vec\pi} , V_{\vec\pi} \otimes Q )
   \\ \oplus \\
   \Hom_\complex (W_{\vec\pi} , V_{\vec\pi}) \\ \oplus  \\ \Hom_\complex (V_{\vec\pi} ,
   V_{\vec\pi}  \otimes \bigwedge^3 
   Q) \end{matrix}}\quad \ar[r]^{\tau} & \quad
   {\begin{matrix} \Hom_\complex (V_{\vec\pi} , V_{\vec\pi}  \otimes \bigwedge^2
       Q) \\ \oplus \\ 
       \Hom_\complex (V_{\vec\pi},W_{\vec\pi} \otimes \bigwedge^3 Q)
   \end{matrix}}
}
\end{equation}
where the module $Q\cong\complex^3$ is a carrier space for the torus
action with weights $t_\a^{-1}=\e^{-\ii\epsilon_\a}$. The character of this complex at a fixed point $\vec\pi$ is given by
\begin{equation} \label{character}
\Ch_{\vec\pi}(t_1,t_2,t_3) =  W_{\vec\pi}^\vee \otimes V_{\vec\pi} -
{V}_{\vec\pi}^\vee \otimes W_{\vec\pi} + (1-t_1)\, (1-t_2)\,
(1-t_3)~ {V}^\vee_{\vec\pi} \otimes V_{\vec\pi} \ ,
\end{equation}
where we have used the Calabi--Yau condition to set $t_1\,
t_2\, t_3=1$. From this character one can compute the fluctuation
determinant around an instanton solution by the standard conversion formula
\begin{equation} \label{conversion} \sum_{i=1}^n\,
n_i~\e^{w_i} ~\longrightarrow ~\prod_{i=1}^n\,w_i^{n_i} \ ,
\end{equation}
where $w_i=w_i(\epsilon_1,\epsilon_2,\epsilon_3)$ are the weights of the toric action on the instanton
moduli space. As explained in~\cite{Cirafici:2008sn}, the equivariant
index (\ref{character}) in this way computes the ratio of fluctuation determinants
in the noncommutative gauge theory on $\complex^3$. Below we use this
rule to compute the instanton measure in the noncommutative
gauge theory on $\complex^3/\Gamma$.

\subsection{Quiver quantum mechanics on $\complex^3/\Gamma$\label{QuiverQMorb}}

Let us consider now the orbifold case. As we have seen from the study
of the Beilinson spectral sequence the structure of the moduli space
can be roughly speaking obtained from the instanton moduli space on
$\complex^3$ by decomposing each morphism equivariantly according to
the $\Gamma$-action. This perspective has an obvious extension to the
instanton quantum mechanics as now each of the fields involved in the
multiplets should be regarded as an equivariant morphism which can be
decomposed analogously to the linear maps (\ref{Bdecomp}). The
relevant bosonic fields and their equations of motions can be
conveniently rephrased in terms of an auxilliary quiver which is
essentially the McKay quiver introduced in Section~\ref{subsec:McKay},
up to some modifications which we will now explain. 

A first rather irrelevant modification is the addition of the fields
$\varphi_r$ for $r\in\Gammaw$ which play the role of extra arrows. However, as discussed
in~\cite{Cirafici:2008sn} we are only interested in those field
configurations on which $\varphi$ vanishes identically. This
corresponds to adding the new fields on the quiver with new relations
in the path algebra that sets the fields to zero.

A somewhat different role is played by the field $I$. It corresponds to
the addition of a vector space $W_r$ for every vector space $V_r$ and
a set of linear maps $I_r : W_r \rightarrow V_r $ for each
$r\in\Gammaw$. This operation corresponds precisely to the framing of
the quiver discussed in Section~\ref{subsec:Quivers}, and indeed in
our setting it corresponds to the framing of the moduli space of
torsion free sheaves. Typically the framing operation in the usual
ADHM formalism involves further sets of fields $ J_r,K_r : V_r \rightarrow
W_r$, as reviewed for example in~\cite{ginzburg}. However, for the
$U(1)$ gauge theory or the non-abelian gauge theory in its Coulomb
branch the fields $J,K$ can also be set to zero and substituted by
suitable stability conditions~\cite{Cirafici:2008sn}.

When the gauge theory is considered on orbifolds $\complex^3 / \Gamma$
the construction of Section~\ref{subsec:QuiverC3} requires
modification. The orbifold quantum mechanics is constructed to count
$\Gamma$-equivariant coherent sheaves of compact support on $\complex^3$. One could in principle extend our analysis to the whole system of
quiver quantum mechanics fields. The resulting topological field theory is defined by an action which localizes on the relations of the McKay quiver and is invariant under a set of $\Gamma$-equivariant BRST transformations. It is a fairly easy exercise to obtain the explicit formulas, but we refrain from writing them down.

The quantum mechanics is constructed in essentially the same way as the
one for $\complex^3$, but instead of starting from the generalized
ADHM quiver (\ref{ADHMquiver}) and the associated equations
(\ref{quiverdefeqs}), one begins with a modified McKay quiver associated with the singularity and a different set of equations. Now the bosonic field content is made up of equivariant matrices
\begin{eqnarray}
(B_1 , B_2 , B_3 , \varphi) \in\Hom_{\Gamma} (V,V) \qquad \mbox{and} \qquad
I\in\Hom_{\Gamma} (W,V) \ .
\end{eqnarray}
If we decompose the vector spaces $V$ and $W$ as $\Gamma$-modules into
irreducible representations $r\in \Gammaw$ as in (\ref{Vdecomp})
and (\ref{Wdecomp}), then the non-vanishing isotopical components of these fields
are maps
\begin{eqnarray}
B_\a^{r} \, &:& \, V_{r} \ \longrightarrow \ V_{r+r_\a} \ , \nonumber \\[4pt]
\varphi^{r} \, & : & \, V_{r} \ \longrightarrow \ V_{r+r_1+r_2+r_3}
\cong V_{r} \ , \nonumber \\[4pt]
I^{r} \, & : & \, V_{r} \ \longrightarrow \ V_{r} \ ,
\end{eqnarray}
where as before we have parametrized the fundamental representation of the
orbifold group as $Q = \rho_{r_1} \oplus \rho_{r_2} \oplus \rho_{r_3}$
and used the fact that the determinant representation is trivial due
to the Calabi--Yau condition. These maps uniquely determine the
generalized framed McKay quiver in terms of the decomposition of the
fundamental representation into irreducible $\Gamma$-modules.

The BRST
transformations respect the $\Gamma$-module structure and are given by
\begin{equation}
\begin{array}{rllrl}
\cQ B_\a^{r} &=& \psi_\a^r & \quad \mbox{and} \qquad \cQ \psi_\a^{r}& = \ [\phi , B_\a^{r}] -
\epsilon_\a\,
B_\a^{r} \ , \\[4pt] \cQ \varphi^{r}& =& \xi^{r} & \quad \mbox{and} \qquad
\cQ \xi^{r} &= \ [\phi ,\varphi^{r}] \ , \\[4pt]
\cQ I^{r} &=& \varrho^{r} & \quad \mbox{and} \qquad \cQ \varrho^{r} & = \
\phi \,I^{r} - I^{r} \,\mbf a^{r} \ ,
\end{array}
\end{equation}
where now the vector $\mbf a^{r}$ collects all the Higgs field
eigenvalues $a_l$ associated with the irreducible representation
$\rho_r$. We will discuss their role more thoroughly in Section~\ref{subsec:NCDTtype}. The
bosonic equations of motion change as well into a set of matrix
equations labelled by the irreducible representations
$r\in\Gammaw$ as
\begin{eqnarray}
\mathcal{E}_\a^{r} \,&:&\, B_\a^{r+r_\b} \, B_\b^{r} -
B_\b^{r+r_\a} \, B_\a^{r} + \sum_{\g=1}^3\, \epsilon_{\a\b\gamma} \,\Big( \big(
B_\gamma^{r-r_\gamma} \big)^{\dagger} \, \varphi^{r} - \varphi^{r-r_\g} \, \big(
B_\gamma^{r-r_\gamma} \big)^{\dagger} \Big) = 0 
\ , \nonumber\\[4pt]
\mathcal{E}_\lambda^{r} \,&:&\, \sum_{\a=1}^{3}\, \Big( B_\a^{r-r_\a} \,
\left( B^{r-r_\a}_\a \right)^{\dagger} - \big(B^r_\a\big)^\dag\,B_\a^r \Big) + \big[\varphi^{r} \,,\, \left( \varphi^{r} \right)^{\dagger}\,\big] + I^{r}\,
\left( I^{r} \right)^{\dagger} = \lambda^{r} \ , \nonumber\\[4pt] \mathcal{E}_I^{r} \,&:&\,
\left( I^{r} \right)^{\dagger} \,\varphi^{r} = 0 \ ,
\label{McKayeqs}\end{eqnarray}
where $\lambda^r>0$ and again only the set $\{ \mathcal{E}_\a^{r} \, , \,
\mathcal{E}_I^{r} \}$ arises as an ideal of relations in the path
algebra of the generalized quiver.

The multiplets of antighost and auxilliary fields can be added in a
similar way. The $\Gamma$-module structure of the auxilliary fields is
dictated by the equations (\ref{McKayeqs}). The resulting antighost
fields decompose as equivariant maps
\begin{eqnarray}
(\chi_1 , \chi_2 , \chi_3 , \chi_\lambda) \in\Hom_{\Gamma} (V,V) \qquad
\mbox{and} \qquad \zeta \in\Hom_{\Gamma} (V,W) \ ,
\end{eqnarray}
and the BRST transformations close upon adding
\begin{equation}
\begin{array}{rllrl}
\cQ \chi_\a^{r} &=& H_\a^{r} & \quad \mbox{and} \qquad \cQ H_\a^{r} &=
\ [\phi , \chi_\a^{r}] + \epsilon_\a \, \chi_\a^{r} \ , \\[4pt]
\cQ \chi_\lambda^{r} &=& H_\lambda^{r} & \quad \mbox{and} \qquad \cQ H_\lambda^{r} &= \ [\phi , \chi_\lambda^{r}] \ ,
\\[4pt] \cQ \zeta^{r} &=& h^{r} & \quad \mbox{and} \qquad \cQ h^{r} &= \ \mbf a^{r} \,
\zeta^{r} - \zeta^{r} \, \phi
\end{array}
\end{equation}
together with the gauge multiplet.

The partition function of the
topological quantum mechanics can be computed by considering the
equivariant version of the instanton deformation complex and using the
localization formula as explained in Section~\ref{subsec:QuiverC3}. In this case the complex is
\begin{equation} \label{equivdefcomplex}
\xymatrix{
  \Hom_{\Gamma}  (V_{\vec\pi} , V_{\vec\pi})
   \quad\ar[r]^{\!\!\!\!\!\!\!\!\!\!\!\!\!\!\!\!\sigma} &\quad
   {\begin{matrix} \Hom_{\Gamma} (V_{\vec\pi} , V_{\vec\pi} \otimes Q )
   \\ \oplus \\
   \Hom_{\Gamma} (W_{\vec\pi} , V_{\vec\pi}) \\ \oplus  \\ \Hom_{\Gamma} (V_{\vec\pi} ,
   V_{\vec\pi}  \otimes \bigwedge^3 
   Q) \end{matrix}}\quad \ar[r]^{\tau} & \quad
   {\begin{matrix} \Hom_{\Gamma} (V_{\vec\pi} , V_{\vec\pi}  \otimes \bigwedge^2
       Q) \\ \oplus \\ 
       \Hom_{\Gamma} (V_{\vec\pi},W_{\vec\pi} \otimes \bigwedge^3 Q)
   \end{matrix}}
}
\end{equation}
where we decompose the morphisms of (\ref{adhmdefcomplexC}) according
to the $\Gamma$-action as dictated by the Beilinson spectral
sequence. The character at a fixed point is now the $\Gamma$-invariant
part of the character for the complex on $\complex^3$, i.e.
\begin{equation} \label{orbcharacter}
\Ch_{\vec\pi}^\Gamma(t_1,t_2,t_3)= \big( W_{\vec\pi}^\vee \otimes V_{\vec\pi} -
{V}_{\vec\pi}^\vee \otimes W_{\vec\pi}+ (1-t_1)\, (1-t_2)\,
(1-t_3) ~ {V}^\vee_{\vec\pi} \otimes V_{\vec\pi} \big)^{\Gamma} \ .
\end{equation}

Let us clarify the meaning of the formula
(\ref{orbcharacter}). Consider the
rank one case $N=1$. Using
the Calabi--Yau condition $t_1\,t_2\,t_3=1$ to eliminate the toric weight
$t_3=(t_1\,t_2)^{-1}$, we regard the character as an element in the
virtual representation ring
$R(\torus^3)\cong\IZ[t_1^{\pm\,1},t_2^{\pm\,1}]$ of the torus group
$\torus^3$. The inclusion $\Gamma\hookrightarrow\torus^3$ defines a
restriction map $R(\torus^3)\to R(\Gamma)\cong
\IZ\big[\rho_{r_1}^{\pm\,1},\rho_{r_2}^{\pm\,1},\rho_{r_3}^{\pm\,1} \big]\big/\langle
\rho_{r_1}\,\rho_{r_2}\, \rho_{r_3}-1\rangle$ by $(t_1,t_2)\mapsto
(\rho_{r_1},\rho_{r_2})$. Hence by substituting $t_\a=\rho_{r_\a}$ we
regard the index as an element of the representation ring $R(\Gamma)$
of the orbifold group, and compute (\ref{orbcharacter}) by composing
with the projection $R(\Gamma)\to\IZ$ onto the trivial
representation.

In the process of computing the character we identify the dimensions
$k_r=\dim_\complex (V_r)_{\vec\pi}$ with the number of boxes in a plane partition
which transform in the irreducible representations of
the orbifold group~$\Gamma$; in particular for $N=1$ one has
\begin{equation}
k_r = | \pi_r|
\end{equation} 
as in Section~\ref{subsec:Colinst}. The integer $k_r$ can be
identified with the number of fractional branes associated to the
representation $\rho_r$, which in our formalism is identified with the
instanton number. At the fixed points the instanton configurations are
parametrized by $\Gammaw$-coloured plane partitions and
the character (\ref{orbcharacter}) is expressed entirely in terms of combinatorial data.

Modulo the issue of stability, which we will discuss
momentarily, our quiver seems to reproduce precisely the quiver used
in~\cite{Ooguri:2008yb} to compute noncommutative Donaldson--Thomas
invariants. This quiver was obtained through the low-energy effective
field theory of a system of D~branes compactified on a local Calabi--Yau
threefold. However, our perspective partly clarifies its origin. Since
in that case we are dealing with the $U(1)$ gauge theory, our framing only
involves a single vector space $W_r$ since $\dim_\complex W$ is always equal to
the rank of the gauge theory. Moreover the pertinent space $W_r$ is $W_0$,
the vector space attached to $V_0$, which in turn is labelled by the
trivial representation or the trivial line bundle $\cR_0 =
\cO_X$ on the D6~brane. This corresponds to a choice of boundary conditions on the
gauge fields of our D6~brane gauge theory. Incidentally this also
explains the physical meaning of choosing a different reference vertex
in the formalism of~\cite{Ooguri:2008yb}: it corresponds to BPS
configurations whose asymptotic profile at infinity sits in a
non-trivial representation $\rho_r$ of the orbifold group $\Gamma$.

\subsection{Pair invariants for quivers\label{subsec:NCDTquivers}}

A particular class of quivers with superpotentials is deeply related to
the geometry of Calabi--Yau manifolds. This relation is at the core of
the definition of noncommutative Donaldson--Thomas invariants given by
Szendr\H{o}i via the counting of cyclic modules of the conifold quiver
\cite{szendroi}. Based on his work, Joyce and Song gave a fully
general definition of pair invariants associated to quivers with
superpotentials \cite{joycesong}; they are essentially weighted Euler
characteristics of the moduli space of framed quiver
representations. Our instanton quivers also come associated
with an ideal of relations on the corresponding path algebra,
generated by the generalized ADHM equations, which can be
easily ascribed to cyclic derivatives of a superpotential. Representations of the
framed McKay quiver are precisely the data that define generalised
instantons on $\complex^3/\Gamma$; this is the main link between our construction of the instanton quantum mechanics and Joyce--Song pair invariants.  

First let us review some facts from \cite{joycesong} to see
explicitly how our construction fits into their more general
framework; afterwards we will freely borrow from their results. Given
a quiver $\sfQ$, define a stability condition on quiver
representations as follows. Consider two sequences $\mbf\theta\in\IR^{\sfQ_0}$ and $\mbf\mu\in (0 , \infty)^{\sfQ_0}$. Then slope stability
can be 
defined on the category of representations of the quiver, considering
only non-zero objects to have a well-defined stability condition. For
any non-trivial quiver representation $V$ of dimension vector $\mbf
k=(k_v)_{v\in\sfQ_0}$, the slope stability parameter $\mu$ is defined as
\begin{equation}
\mu (V)= \mu({\mbf k})= \frac{\mbf\theta\cdot \mbf k}{\mbf\mu\cdot\mbf k} \ .
\end{equation}
This definition generalizes the usual $\theta$-stability parameter
(\ref{thetaV}), to which it reduces for the particular sequence $\mu_v=1$ for all $v\in\sfQ_0$. 

Consider now the moduli space of representations ${\sf Rep}
(\sfQ, \mbf k)$ associated with a superpotential
${\sf W} \in\complex\sfQ\big/[\complex\sfQ, \complex\sfQ\, ]$, which
gives a two-sided ideal of relations $\langle\sfR\rangle$ in the path
algebra $\complex\sfQ$ by taking cyclic derivatives
$\partial_a{\sf W}$ for $a\in\sfQ_1$. Let ${\sf V}_{\mbf k}(\sfA)$ be the $G_{\mbf
  k}$-invariant closed subscheme of ${\sf Rep} (\sfQ, \mbf k)$
cut out by the equations $\partial_a{\sf W}=0$ (here $\sfA= \complex
\sfQ /\langle \sfR\rangle$ is the factor path algebra). This allows us to define the
framed quiver moduli space as the quotient stack
\begin{equation}
\frak{M} (\sfQ^f,{\sf W};\mbf k, \mbf N) = \Big[ \big( {\mathsf
  V}_{\mbf k}(\sfA) \times\Hom_\complex(W,V) \big) \, \Big/
\,G_{\mbf k} \Big] \ .
\end{equation}
In particular one can define the moduli space of \textit{$\mu$-stable
  framed quiver
  representations of type} $(\mbf k, \mbf N)$, which is a fine moduli
space and an open substack
$\frak{M}_\mu (\sfQ^f,{\sf W};\mbf k , \mbf N) \subset \frak{M}
(\sfQ^f,{\sf W};\mbf k, \mbf N)$.

Then following Behrend
\cite{behrend}, noncommutative Donaldson--Thomas invariants associated
with this moduli space are
defined as the weighted topological Euler characteristics
\begin{equation}
\NDT_\mu (\mbf k , \mbf N) = \chi \big( \frak{M}_\mu (\sfQ^f,{\sf
  W};\mbf k , \mbf N) \, , \, \nu \big) = \sum_{n\in\zed}\, n \
\chi\big(\nu^{-1}(n)\big) \ ,
\label{Behrend}\end{equation}
where $\nu: \frak{M}_\mu (\sfQ^f,{\sf
  W};\mbf k , \mbf N) \to\zed$ is a $G_{\mbf k}$-invariant
constructible function related to the Euler characteristic of the Milnor fibre of the
superpotential $\sf W$; at any smooth point $V\in{\sf V}_{\mbf
  k}(\sfA)$ one has $\nu(V)=(-1)^D$ where $D=\dim_\complex\frak{M}_\mu (\sfQ^f,{\sf W};\mbf k , \mbf N)$.
In particular, this definition makes sense at $\mu=0$ for which
$\mu$-stability coincides with $\theta$-stability at $\mbf\theta = (0
, \dots , 0)$, with every object of the category $\sfA$--mod
$0$-semistable. The new
invariants then reproduce precisely the ones introduced by Szendr\H{o}i for
the conifold. In the case of orbifold singularities they enumerate
$\Gamma$-equivariant sheaves on $\complex^3$ via the McKay
correspondence; for ideal sheaves they coincide with the orbifold
Donaldson--Thomas invariants defined in~\cite{young}.

In many cases these invariants can be related to the
quiver generalized Donaldson--Thomas invariants
$\overline{\DT}_\mu(\mbf k) \in \mathbb{Q}$ defined by Joyce and
Song in~\cite{joycesong} via a certain infinite-dimensional Lie algebra morphism acting on
the moduli stack of left $\sfA$-modules. From these invariants one defines the quiver BPS invariants $\widehat{\DT}_\mu(\mbf k) \in \mathbb{Q}$ as
\begin{equation}
\widehat{\DT}_\mu(\mbf k) = \sum_{ m \ge
    1 \, : \, m | \mbf k }\, \frac{\text{M\"o}(m)}{m^2} \
\overline{\DT}_\mu(\mbf k/m)
\end{equation}
where $\text{M\"o} : \mathbb{N} \rightarrow \mathbb{Q}$ is the
M\"obius function. In special cases these invariants count BPS states,
and they generalize the integer Gopakumar--Vafa invariants of
Calabi--Yau threefolds~\cite{Gopakumar:1998jq}. In the general case
they are conjectured to be integer-valued. By the M\"obius inversion formula, this expression
has an inverse given by
\begin{equation}
\overline{\DT}_\mu(\mbf k) = \sum_{ m \ge
    1 \,: \, m | \mbf k}\, \frac{1}{m^2} \,
\widehat{\DT}_\mu(\mbf k/m) \ .
\label{BPSformula}\end{equation}
Noncommutative Donaldson--Thomas invariants are related to the quiver
generalized Donaldson--Thomas invariants
by~\cite[Theorem~7.23]{joycesong}
\bea
\NDT_\mu(\mbf k, \mbf N) &=& \sum_{m=1}^\infty~
\sum_{\stackrel{\scriptstyle \mbf k_1,\dots, \mbf k_m\neq\mbf
    0}{\scriptstyle \mbf k_1+\cdots + \mbf k_m=\mbf k \ , \ \mu(\mbf
    k_i)=\mu(\mbf k)}} \, \frac{(-1)^m}{m!} \label{NCDTgenrel} \\ &&
    \times\, \prod_{i=1}^m\, \Big((-1)^{\mbf k_i\cdot \mbf
      N-\bar\chi(\mbf k_1+\cdots+\mbf k_{i-1},\mbf k_i)}\, \big(\mbf
    k_i\cdot \mbf N-\bar\chi(\mbf k_1+\cdots+\mbf k_{i-1},\mbf k_i)
    \big)\, \overline{\DT}_\mu(\mbf k_i)\Big)
\nonumber \eea
where
\beq
\bar\chi(\mbf k,\mbf k'\,) =
\chi(V,V'\,)-\chi(V',V)=\sum_{a\in\sfQ_1}\,\big(k_{h(a)}\, k_{t(a)}' -
k_{t(a)}\, k_{h(a)}'\big)
\label{chibar}\eeq
is the antisymmetrization of the Euler form
$\chi(V,V'\,)=\sum_{p\geq0}\, \dim_\IC\, \Ext_{\sfA}^p(V,V'\,)$ for $V,V'\in\sfA\text{--mod}$.
Because of the condition on the partitions of $\mbf k$, the sum over
$m$ in (\ref{NCDTgenrel}) contains only a finite number of
non-zero terms. In the case of semi-small crepant
resolutions, the Euler forms $\bar\chi$ vanish and this relation
yields a useful relationship between the corresponding partition
functions
\begin{equation} \label{quivergeneralized}
1 + \sum_{\mbf k \, : \, \mu(\mbf k)=\mu}\, \NDT_\mu(\mbf k,\mbf N) \,
\mbf p^{\mbf
  k} = \exp\Big( - \sum_{\mbf k \, : \, \mu(\mbf k)=\mu}\, (-1)^{\mbf k
  \cdot \mbf N}\, \left( \mbf k \cdot \mbf N \right) \,
\overline{\DT}_\mu(\mbf k) \, \mbf p^{\mbf k} \Big)
\end{equation}
where $\mbf p^{\mbf k}:= \prod_{v\in\sfQ_0}\, p_v^{k_v}$.

\subsection{Quiver moduli spaces for Donaldson--Thomas data}

To see how the Joyce--Song construction is connected with our
perspective, let us start by reviewing the definition of the instanton moduli space on $\complex^3$ put forward in~\cite{Cirafici:2008sn}. Consider the two vector spaces $V$ and $W$, of dimensions $\dim_\complex V
= k$ and $\dim_\complex W =N$, in the quantum mechanics
of the gas of D0~branes and the D6~branes on which the
gauge theory lives. The moduli space of representations of the quiver (\ref{ADHMquiver}) is
given by
\begin{equation}
\cM (k  , N) =  \Hom_\complex (V , Q \otimes  V) \ \oplus \
 \Hom_\complex (V ,  \mbox{$\bigwedge^3$} Q \otimes  V) \ \oplus \ \Hom_\complex (W ,
V) \ ,
\end{equation}
on top of which we have the natural action of the complexified gauge group $GL(k,\complex)$. We call an
element $(B,\varphi,I)$ of $\cM(k,N)$ a Donaldson--Thomas
datum. We define a complex ``moment map''
$\mu_{\complex} = (\mathcal{E}_\a , \mathcal{E}_I)$ given
collectively by the ADHM type equations of the matrix quantum
mechanics in~(\ref{quiverdefeqs}).

The instanton moduli space is defined via $\theta$-stability as the geometric invariant theory quotient
\begin{equation}
\frak{M}_\theta(k,N)= \mu_{\complex}^{-1} (0) \ \big/\!\!\big/\!^{~}_{\theta} \ GL(k,\complex) \ .
\label{Mthetak1}\end{equation}
In the case of a single D6~brane $N=1$, we can now proceed to define a Donaldson--Thomas invariant
in the gauge theory formalism as the Euler characteristic (\ref{chicalN}) of the
obstruction bundle over the moduli space; we regard this number as the gauge theory realization of Behrend's local weighted Euler characteristic (\ref{Behrend}). Each invariant can be evaluated by using the localization formula with respect to the natural lift of the toric action on $\complex^3$ to the moduli space. Fixed points of the toric action are in natural correspondence with certain chains of maps which are classified by plane partitions. The local structure of the instanton moduli space around each fixed point is completely characterized by the equivariant index of the complex (\ref{adhmdefcomplexC}) generated by the derivative $\tau$ of the
moment map modulo linearized complex gauge transformations $\sigma$.

This construction can be generalized to $\complex^3/\Gamma$ orbifolds
by considering $\Gamma$-equivariant morphisms, as dictated by the
instanton deformation complex. The Donaldson--Thomas data now
decompose accordingly as
\begin{equation}
\cM^{\Gamma} (\mbf k , \mbf N) =  \Hom_{\Gamma} (V , Q \otimes V) \ 
 \oplus \ \Hom_{\Gamma} (V , \mbox{$\bigwedge^3$} Q \otimes V) \ 
\oplus \ \Hom_{\Gamma} (W , V) \ .
\end{equation}
We use the $\Gamma$-equivariant decomposition of the ADHM equations
(\ref{McKayeqs}) to define ``moment maps'' $\mu_{\complex}^{\Gamma} = (
\cE_\a^r , \cE_I^r )$ which correspond to the ideal of relations in
the instanton quiver path algebra. These equations define a subvariety
$(\mu_\complex^\Gamma)^{-1}(0) \subset \Hom_{\Gamma} (V , Q \otimes V) \oplus
\Hom_{\Gamma} (V , \bigwedge^3 Q \otimes V)$. This allows us to define
the Donaldson--Thomas quiver moduli space as the quotient stack
\begin{equation}
\frak{M}^\Gamma (\mbf k ,\mbf N) = \Big[ \big((\mu_\complex^{\Gamma})^{-1}(0) \times
\Hom_{\Gamma} (W , V) \big) \, \Big/ \, G_{\mbf k} \Big] \ .
\end{equation}
We regard this stack as a moduli space of stable framed
representations in the sense
of~\cite[Section~7.4]{joycesong}. Geometrically this moduli scheme is
an Artin stack over $\complex$ and has certain nice properties which allow
us to define enumerative invariants. We postpone a discussion of this
to Section~\ref{subsec:Stability}.

From our analysis of the
noncommutative gauge theory in Section~\ref{sec:NCGT} it follows that
the instanton moduli space is constructed as the moduli space of
$\complex^3$ but with a decomposition of the instanton equations
according to the equivariant structure lifted from the orbifold
action. Hence
the $\Gamma$-invariant fixed point set of the instanton moduli space
(\ref{Mthetak1}) for $N=1$ admits a decomposition
\begin{equation}
\mathfrak{M}_\theta(k,1)^{\Gamma} = \bigsqcup_{\mbf{k} \, : \, 
|\mbf{k}| = k}\, \mathfrak{M}^\Gamma(\mbf{k},1) \ ,
\label{frakMC3decomp}\end{equation}
where $|\mbf k|:=\dim_{\complex}(V)=\sum_{v\in\sfQ_0}\, k_v$. It is
straightforward to generalize this to the Coulomb branch of the
non-abelian gauge
theory where the vector space $W$ is no longer one-dimensional and decomposes according
to the $\Gamma$-action as in (\ref{Wdecomp}); we discuss this in
Section~\ref{subsec:NCDTtype}. 

Since the natural action of the torus group $\torus^3$ commutes with the 
action of the orbifold group $\Gamma$ on $\complex^3$ it lifts
naturally to the moduli space $\frak{M}^\Gamma (\mbf k,\mbf N)$. The transformations
of the fields are the usual ones and we can work equivariantly with
respect to the toric action. This allows us to immediately classify the torus fixed points that enter
into the localization formula which will be used to compute generating functions of BPS states, since they are precisely the instanton
configurations on $\complex^3$ which are fixed by the toric action and
are invariant under the $\Gamma$-action. Again we can express fixed
points of the toric action via plane partitions, where now each box
carries an additional colour associated with the $\Gamma$-action. By
(\ref{frakMC3decomp}) and since the orbifold action commutes with the
torus action on $\complex^3$, the $\torus^3$ fixed points on
$\frak{M}^\Gamma (\mbf k,\mbf N)$ coincide with the $\torus^3$ fixed points
of $\frak{M}_\theta (k,N)$ which are also invariant under the orbifold
action. We will substantiate these arguments further in
Section~\ref{subsec:Stability}.

\subsection{Noncommutative Donaldson--Thomas invariants of type $\mathbf
  N$\label{subsec:NCDTtype}}

In the following we will propose a physical interpretation of
the framed pair invariants $\NDT_\mu (\mbf k, \mbf N) $ for $\mu = 0$
as noncommutative Coulomb branch invariants of Donaldson--Thomas
type in $U(N)$ gauge theory. In the instanton quiver formalism the framing operation has a
clear physical meaning. It represents the choice of boundary
conditions for the gauge field living on the worldvolume of the stack
of D6 branes. Asymptotically the gauge connection is flat, and flat
connections on resolved geometries of abelian orbifolds $\complex^3/ \Gamma$ are classified by the irreducible representations of the orbifold group $\Gamma$.

Based on these definitions and on the identification of the relevant
quiver as the instanton quiver introduced in
Section~\ref{QuiverQMorb}, we can now construct a partition function
for the noncommutative invariants. It follows from the local character
of the instanton moduli space given by (\ref{orbcharacter}).
Neglecting the $\Gamma$-action, the two vector spaces $V$ and $W$ can
be decomposed at a
fixed point $\vec\pi=(\pi_1,\dots,\pi_N)$ of the $U(1)^N \times \torus^3$ action on the
instanton moduli space as~\cite{Cirafici:2008sn}
\begin{eqnarray}
V_{\vec\pi} = \sum_{l=1}^N \,e_l~ \sum_{(n_1,n_2,n_3)\in \pi_l}\,
t_1^{n_1-1} \,t_2^{n_2-1}\,t_3^{n_3-1} \qquad \mbox{and} \qquad W_{\vec\pi} =
\sum_{l=1}^N\,e_l \ .
\label{decompos}
\end{eqnarray}
Here we view the spaces as $U(1)^N \times \torus^3$ representations
regarded as polynomials in $t_\a=\e^{\ii\epsilon_\a}$, $\a=1,2,3$ and $e_l = \e^{\ii
  a_l}$, $l=1,\dots,N$,
with the sum over boxes of $\pi_l$ for each $l$ corresponding to the
$\torus^3$-character on $\complex[B_1,B_2,B_3]/I_l$, i.e. the
decomposition of $H^0(\cO_{Z_l})$ as a $\torus^3$-representation,
where $Z_l$ is the $\torus^3$-fixed subscheme of $\complex^3$
corresponding to the three-dimensional Young diagram~$\pi_l$.

Let us now consider the analogous decompositions for the resolved
geometry. We can further decompose the vector spaces according to the
$\Gamma$-action as in (\ref{Vdecomp}) and (\ref{Wdecomp}). 
Recall that the decomposition of $W$ corresponds to imposing boundary
conditions at infinity, which are classified by irreducible
representations of the orbifold group $\Gamma$. In this context each
$U(1)$ factor in the Coulomb phase is associated with a vacuum expectation
value of the Higgs field $a_l$ which corresponds to a certain
irreducible representation of $\Gamma$. Even if the maximal symmetry breaking
pattern $U(N) \rightarrow U(1)^N$ is fixed, one still has to
specify in which superselection sector one is working. This sector is
characterized by choosing which of the eigenvalues $a_l$ are in a particular
irreducible representation of $\Gamma$. The number of eigenvalues of
the Higgs field in the representation $\rho_r^{\vee}$ is precisely
$N_r = \dim_\complex W_r$.

Similarly the dimensions $k_r=\dim_\complex V_r$ give the instanton number of a multi-instanton configuration transforming in the representation $\rho_r^{\vee}$. However their $\torus^3$-module decompositions are somewhat complicated to compute directly since they do not correspond in a simple way to plane partitions. It follows from (\ref{decompos}) that each partition carries an action of $\Gamma$ on its own, but this action is ``offset'' by the prefactor $e_l$. On the other hand we can also write the decomposition
\begin{equation}
V_{\vec\pi} =\bigoplus_{l=1}^N ~ \bigoplus_{r\in\Gammaw} \, \big( E_l \otimes \rho_{b(l)}^{\vee} \big) \otimes \left( P_{l,r} \otimes \rho_r^{\vee} \right) =\bigoplus_{l=1}^N~ \bigoplus_{r\in\Gammaw} \, \big( E_l \otimes P_{l,r} \big) \otimes \rho_{r+b(l)}^{\vee}
\label{VpiGammadecomp}\end{equation}
where $E_l$ is the module generated by $e_l$, and we have introduced
the boundary function $b(l)$ which to each sector $l$ corresponding to
a module $E_l$ associates the weight of the corresponding
representation of $\Gamma$; if the vacuum expectation value $e_l$
transforms in the irreducible representation $\rho_s$, then
$b(l)=s$. Here $P_{l,r}$ is a module which corresponds to the
$\Gamma$-module decomposition of the sum $H^0(\cO_{Z_l})= \sum_{(n_1,n_2,n_3)\in \pi_l}\, t_1^{n_1-1} \,t_2^{n_2-1}\,t_3^{n_3-1} $. Recall that each fixed point is characterized by a vector of partitions $\vec \pi$. Each entry in this vector can be decomposed according to the $\Gamma$-action, taking further into account the transformation properties of the Higgs field vacuum expectation values $e_l$. In our decomposition (\ref{VpiGammadecomp}) we have factorized this contribution explicitly so that now $\dim_\complex P_{l,r}$ is the number of boxes in the plane partition at position $l$ of the fixed point vector $\vec \pi = (\pi_1 , \dots , \pi_N)$ which transform in the representation $\rho_r^{\vee}$, a number which we will call $|\pi_{l,r}|$. This should not be confused with the physical instanton configuration which transforms in the representation $\rho_r^{\vee} \otimes \rho_{b(l)}^{\vee}$. The two concepts only differ by the $\Gamma$-action.

On the other hand the module $V_r$ contains \textit{all} the instanton configurations transforming in the representation $r\in\Gammaw$ (for fixed topological charge $k_r=\dim_\complex V_r$). The two parametrizations are related by
\begin{equation}
(V_r)_{\vec\pi} = \bigoplus_{l=1}^N \,E_l \otimes P_{l,r-b(l)} \ .
\end{equation}
The instanton numbers are thus expressed via
\begin{equation} \label{NAinst}
k_r = \sum_{l=1}^N \, |\pi_{l,r-b(l)}| \ .
\end{equation}

This parametrization is useful for computing explicitly the local
contribution of an instanton. To compute the $\Gamma$-invariant
projection of the character (\ref{orbcharacter}), we write
(\ref{character}) as
\beq
\Ch_{\vec\pi}(t_1,t_2,t_3) = {\cal T}_{\vec\pi}^+ + {\cal
  T}_{\vec\pi}^-
\label{Chpisplit}\eeq
where
\bea
{\cal T}_{\vec\pi}^+ &=& W^\vee_{\vec\pi}\otimes V_{\vec\pi}
- V^\vee_{\vec\pi}\otimes  V_{\vec\pi}~ 
\frac{(1-t_1)\,(1-t_2)}{t_1\, t_2} \ , \nonumber\\[4pt] {\cal
  T}_{\vec\pi}^- &=& -V^\vee_{\vec\pi}\otimes
  W_{\vec\pi} + V^\vee_{\vec\pi}\otimes  V_{\vec\pi}~ 
(1-t_1)\,(1-t_2) \ ,
\label{calTpm}\eea
and we have used the Calabi--Yau condition $t_1\, t_2\, t_3=1$.
This splitting has the property
\beq
\big({\cal T}_{\vec\pi}^+\big)^\vee=- {\cal
  T}_{\vec\pi}^- \ ,
\eeq
where the dual involution acts on the weights as $(t_\a,e_l)^\vee=
(t_\a^{-1},e_l^{-1})$. As in~\cite{Cirafici:2008sn}, it follows that
one need only consider the partial character ${\cal T}_{\vec\pi}^+$,
and from (\ref{conversion}) the contribution of an instanton to the gauge theory fluctuation
determinant is given by
\beq
\chi_{\torus^3}(\cN_{\vec\pi})=(-1)^{{\cal K} (\vec\pi;\mbf N)} \qquad \mbox{with} \quad {\cal
  K}(\vec\pi;\mbf N)=\big( {\cal T}_{\vec\pi}^+\big)^\Gamma\, \Big|_{t_\a=e_l=1} \ .
\eeq
In particular, we need only compute the value of the equivariant index
$\big( {\cal T}_{\vec\pi}^+\big)^\Gamma$ modulo~$2$.

Let us first consider the term
\begin{eqnarray}
\big( W^\vee_{\vec\pi}\otimes V_{\vec\pi}\big)^{\Gamma} &=& \Big( \, \bigoplus_{l=1}^N ~\bigoplus_{r,s\in\Gammaw}\, E_l \otimes P_{l,r} \otimes \rho_{r+b(l)}^{\vee} \otimes {W}^\vee_s \otimes \rho_s \, \Big)^{\Gamma} \\[4pt] &=& \bigoplus_{l=1}^N ~ \bigoplus_{r,s\in\Gammaw}\, E_l \otimes P_{l,r} \otimes {W}^\vee_s \otimes \big( \rho_{r+b(l)}^{\vee} \otimes\rho_s \big)^\Gamma \ = \ \bigoplus_{l=1}^N ~ \bigoplus_{r\in\Gammaw}\, E_l \otimes P_{l,r} \otimes {W}^\vee_{r+b(l)} \ . \nonumber
\end{eqnarray}
The other terms involve the $\torus^3$ weights $t_1$ and $t_2$. As
explained in Section~\ref{QuiverQMorb}, the weights $t_\a$ should be
properly regarded as the $\Gamma$-modules $t_\a\mapsto \rho_{r_\a}$,
where $r_\a$ for $\a=1,2,3$ are the weights of the $\Gamma$-action on $\complex^3$. We thus find
\begin{eqnarray}
\Big(\, \frac{V^\vee_{\vec \pi} \otimes V_{\vec \pi}}{t_1 \, t_2} \, \Big)^{\Gamma} &=& \bigoplus_{l,l'=1}^N~ \bigoplus_{r,s\in\Gammaw}\, E^\vee_l \otimes P^\vee_{l,r} \otimes E_{l'} \otimes P_{l',s} \otimes \big( \rho_{r+b(l)} \otimes \rho_{s+b(l'\,)}^{\vee} \otimes \rho_{r_1}^{\vee} \otimes \rho_{r_2}^{\vee} \, \big)^{\Gamma} \nonumber \\[4pt]
&=& \bigoplus_{l,l'=1}^N~ \bigoplus_{r\in\Gammaw}\, E^\vee_l \otimes P^\vee_{l,r} \otimes E_{l'} \otimes P_{l',r+b(l)-b(l'\,)-r_1-r_2} \ ,
\end{eqnarray}
and similarly
\begin{eqnarray}
\Big(\, \frac{{V}^\vee_{\vec \pi} \otimes V_{\vec \pi}}{ t_\a} \, \Big)^{\Gamma} &=&  \bigoplus_{l,l'=1}^N~ \bigoplus_{r\in\Gammaw}\, E^\vee_l \otimes P^\vee_{l,r} \otimes E_{l'} \otimes P_{l',r+b(l)-b(l'\,)-r_\a} \ .
\end{eqnarray}

Therefore the projection of the partial character ${\cal T}_{\vec\pi}^+ $, evaluated at $(\epsilon_1 , \epsilon_2 , \epsilon_3 , \mbf a) = 0$, onto the trivial representation of the orbifold group $\Gamma$ gives
\begin{eqnarray}
\cK(\vec\pi;\mbf N) &=& \sum_{l=1}^N ~ \sum_{r\in\Gammaw}\, |\pi_{l,r}| \ N_{r+b(l)} - \sum_{l,l'=1}^N \ \sum_{r\in\Gammaw}\, |\pi_{l,r}|\, \Big( |\pi_{l',r+b(l)-b(l'\,)-r_1-r_2}| - |\pi_{l',r+b(l)-b(l'\,)-r_1}| \nonumber \\ && \hspace{6cm} -\, |\pi_{l',r+b(l)-b(l'\,)-r_2}| + |\pi_{l',r+b(l)-b(l'\,)}| \Big) \ .
\label{instmeasure}\end{eqnarray}
The fixed point values of the instanton action (\ref{Sinstgeneric}) in these new variables can be written as
\begin{eqnarray}
\int_{\overline{X}}\, \omega \wedge \omega \wedge c_1 (\cE_{\vec\pi}) &=& -
\sum_{m,r,s\in\Gammaw}\, \varsigma_m \, \Big( N_s\,\delta_{rs} -
\big(  a^{(2)}_{rs} - a^{(1)}_{rs} \big)\, \sum_{l=1}^N\,
|\pi_{l,s-b(l)}| \Big)\, \int_{\overline{X}}\, c_2 (\cV_m) \wedge c_1 ({\cR_r})
 \ , \cr && \label{c1Naction}
\\[4pt]
\int_{\overline{X}}\, \omega \wedge \mathrm{ch}_2 (\cE_{\vec\pi}) &=& - \sum_{n,r,s\in\Gammaw}\, \varphi_n\, \bigg( \Big( N_s\, \delta_{rs} - \big( a^{(2)}_{rs} - a^{(1)}_{rs} \big) \, \sum_{l=1}^N \, |\pi_{l,s-b(l)}| \Big)\, \int_{\overline{X}}\, c_1 (\cR_n) \wedge \mathrm{ch}_2 ({\cR_r}) \cr && \qquad +\,\big(  a_{rs}^{(2)}  - 3 \delta_{rs} \big) \,  \sum_{l=1}^N\, |\pi_{l,s-b(l)}| \, \int_{\overline{X}}\, c_1 (\cR_n) \wedge c_1 \big(\cO_{\overline{X}}(1)\big) \wedge c_1(\cR_r)
  \bigg) \ , \cr && \label{ch2Naction}
\\[4pt]
 \int_{\overline{X}}\, \mathrm{ch}_3 (\cE_{\vec\pi}) &=& - \sum_{r,s\in\Gammaw}\, \bigg(\Big( N_s\,\delta_{rs} - \big( a^{(2)}_{rs} - a^{(1)}_{rs} \big) \, \sum_{l=1}^N\, |\pi_{l,s-b(l)}| \Big)\, \int_{\overline{X}}\, \mathrm{ch}_3 ({\cR_r}) \cr && \qquad \qquad
  +\,\big( a_{rs}^{(2)} - 3 \delta_{rs} \big) \, \sum_{l=1}^N\, |\pi_{l,s-b(l)}| \, \int_{\overline{X}}\, c_1 \big(\cO_{\overline{X}}(1)\big) \wedge \ch_2 (\cR_r)
  \cr & & \qquad \qquad +\, \big(  a^{(2)}_{rs} - 3 \delta_{rs} \big)
  \, \sum_{l=1}^N \, |\pi_{l,s-b(l)}| \, \int_{\overline{X}}\,  c_1
  (\cR_r) \wedge \ch_2 \big(\cO_{\overline{X}} (1)\big) \bigg) \cr &&
  +\, \frac1{|\Gamma|}\, \sum_{s\in\Gammaw} \ \sum_{l=1}^N \, |\pi_{l,s-b(l)}|  \ .
\label{ch3Naction} \end{eqnarray}
Note that the choice of boundary condition enters not only explicitly in the dimensions $N_r$, but also implicitly in the plane partitions.
Finally, the partition function for noncommutative Donaldson--Thomas invariants of type $\mbf N$ is in full generality given by
\begin{equation}
\cZ_{\complex^3 / \Gamma}\left(\mbf N\right) = \sum_{\vec \pi} \, (-1)^{\cK(\vec\pi;\mbf N)}~ \e^{-S_{\rm inst} [\vec \pi ; \mbf N]} \ .
\end{equation}

\subsection{Stability conditions and BPS invariants\label{subsec:Stability}}

The problem we are facing now is that the concepts we have used
so far (e.g. coherent sheaves, tautological bundles, Beilinson's
theorem) are all \textit{large radius} concepts. We would like however
to use our construction to investigate Donaldson--Thomas invariants
near the singular orbifold point and argue that this can be done via the McKay correspondence by a suitable choice of the stability conditions. To do so we argue that the parameter which identifies the region of the moduli space of BPS states we are working in is the stability parameter which enters in the construction of the quiver varieties. While this resembles closely the rigorous mathematical construction of generalised Donaldson--Thomas invariants presented in e.g.~\cite{nakanagao,nagao,reineke}, our setting is quite different. We will argue that a particular choice of the stability parameter defines the noncommutative crepant resolution. Physically this corresponds to a highly non-geometrical limit in the noncommutative gauge theory, i.e. when the \textit{classical} volume of all the cycles goes to zero, while their \textit{quantum} volumes, as measured by the $B$-field, is still non-vanishing (though very small). 

Before exposing our arguments let us comment on the physical picture
that we expect. It is known that the string theory linear
sigma-model can be used to investigate the whole extended Calabi--Yau
moduli space, including in principle topology changing
transitions. These ideas can be made rather precise via the study of
the phase structure of the sigma-model
\cite{Witten:1993yc,Aspinwall:1993nu}. While generically the phases
have a geometric description, this is not at all necessary and certain
abstract ``non-geometries'' can be investigated as well. String theory
implies that the same point of view, while less studied in this
language, should apply also to the description in terms of weakly
coupled solitons~\cite{Greene:1997uf}. Indeed there is by now evidence
that the appropriate description for BPS solitons given by the derived
category behaves consistently with this picture.

Still it should be
possible to explore at least parts of the whole moduli space of stable
BPS solitons by studying directly the Dirac--Born--Infeld theory
at the quantum level. More precisely we expect to be able to describe
those chambers in the moduli space which can be accessed from a regime
where worldsheet instanton corrections are negligible (and of course
string theory loops as well). In this regime the theory describing a
system of branes is far from being a local quantum field theory: the full
Dirac--Born--Infeld action involves an infinite series of higher
derivative corrections, the presence of the $B$-field induces a
noncommutative deformation of the theory, and the other Ramond--Ramond
fields yield even more subtle effects which have not been completely
understood. Although the situation is unclear it is reasonable to
suggest \cite{Jafferis:2008uf} that the full non-linear instanton
equations derived in \cite{Marino:1999af} might be already able to
capture the chamber structure of the BPS soliton moduli space, at
least for local threefolds. These equations arise as BPS conditions on
the D~brane embedding from $\kappa$-symmetry of the full twisted
Dirac--Born--Infeld action.

Therefore we do not expect to be able to carry out a fully rigorous
derivation of the chamber structure by studying just the topological
Yang--Mills theory living on the D6~branes. We will need to make further assumptions. We know however that string theory defined on the backgrounds we are considering has a certain non-trivial behaviour that is believed to hold at the quantum level: the derived McKay correspondence.

The analysis of BPS states on a Calabi--Yau manifold should be
ultimately rephrased in terms of derived categories. Although most of
our construction uses abelian categories, it is
useful to keep in mind where we stand in the categorical
landscape. The category we wish to study is the bounded derived category
of coherent sheaves on $X$, $\mathbf{D} (X)$. This is generically a
very difficult problem. However for toric Calabi--Yau threefolds we
can have substantial simplification by using alternative
descriptions. If the Calabi--Yau space is a crepant resolution of an
orbifold singularity, then two alternative models are available. One can
characterize the derived category $\mathbf{D} (X)$ via an equivalent
category, the derived category of representations of a certain quiver;
or one can use the McKay correspondence at the derived level and deal
with the category $\mathbf{D}^{\Gamma}
(\complex^3)$, the derived category of $\Gamma$-equivariant sheaves on $\complex^3$. We will now review these equivalences.

The variety $X$ is the ``natural'' Calabi--Yau crepant resolution of the singular orbifold
$\complex^3 / \Gamma$ given by the $\Gamma$-Hilbert scheme $
\mathrm{Hilb}^\Gamma (\complex^3)$. The McKay
correspondence is established as a derived equivalence between
$\mathbf{D} (X)$ and $\mathbf{D}^{\Gamma} (\complex^3)$~\cite{BKR}. This
equivalence descends to K-theory as follows: there exists a natural
basis $\{ \rho_r \otimes \cO_{\complex^3} \}_{r\in\Gammaw}$ of the
Grothendieck group of $\Gamma$-equivariant sheaves on $\complex^3$,
$K_{\Gamma} (\complex^3)$, and a natural basis $\{ \rho_r \otimes
\cO_0 \}_{r\in\Gammaw}$ of its restriction to coherent sheaves with
compact support, $K_{\Gamma}^c (\complex^3)$, as discussed in Section~\ref{subsec:Inttheory}. These bases give an
explicit ring isomorphism of $K_{\Gamma} (\complex^3)$ and
$K_{\Gamma}^c (\complex^3)$ with the representation ring
$R(\Gamma)$. On the other hand the equivalence between the derived
categories give an explicit isomorphism between these groups and the
K-theory groups of $X$, $K(X)$ and $K^c (X)$. Under this isomorphism
the two basis sets are mapped respectively to the basis of
tautological bundles $\{ \cR_r \}_{r\in\Gammaw}$ and of fractional branes $\{ \cS_r \}_{r\in\Gammaw}$ introduced in Section~\ref{subsec:Inttheory}.

The $\Gamma$-Hilbert scheme $\mathrm{Hilb}^\Gamma (\complex^3)$ can be
equivalently realized as the fine moduli space of representations of the McKay quiver. This result underlies another
derived equivalence, between $\mathbf{D} (X)$ and the bounded derived category of finitely-generated left modules over a certain algebra. To properly define this category we need to introduce the notion of a \textit{tilting object}. Geometrically a coherent sheaf $\cT$ is said to be tilting if the following three conditions hold:
\begin{itemize}
\item $\cT$ can be decomposed as a sum of simple sheaves $\cT = \cF_1 \oplus \cdots \oplus \cF_{n}$;
\item $\Ext^p_{\cO_X} (\cF_i , \cF_j) = 0$ for all $p>0$ and for all
  $i,j=1,\dots,n$; and
\item $\cT$ generates $\mathbf{D} (X)$.
\end{itemize}
Physically the morphisms of the derived category which combine the
direct summands $\cF_i$ of $\cT$ to generate the whole of $\mathbf{D}
(X)$ correspond to combining D~branes via tachyon condensation.

Consider now the endomorphism algebra of $\cT$, $\sfA =
\mathrm{End}_{\mathbf{D} (X)} (\cT)$. This is a noncommutative algebra
with respect to the composition of morphisms. In practice its elements
can be seen as matrices whose entries are elements of the spaces of
morphisms $\Hom_{\cO_X} (\cF_i , \cF_j)$. The tilting object $\cT$
induces a derived equivalence through the adjoint functors
\begin{eqnarray}
\mathbf{R} \Hom_{\cO_X} (\cT , - ) &:& \mathbf{D} (X) \
\longrightarrow \ \mathbf{D} (\sfA^{\rm op}) \ , \nonumber\\[4pt]
\cT \stackrel{\mathbf L} \otimes_{\sfA} - &:& \mathbf{D} (\sfA^{\rm
  op}) \ \longrightarrow \ \mathbf{D} (X) 
\end{eqnarray}
where $\mathbf{D} (\sf{A})$ denotes the bounded derived category
of finitely-generated left $\sfA$-modules and $\sfA^{\rm op}$ is the
opposite algebra of $\sfA$. The algebra $\sfA$ can be identified with
the path algebra of a quiver with relations $\complex \sfQ /
\langle\sfR\rangle$; the idempotent elements ${\sf e}_i\in\sfA$ are
the projections $\cT\to\cF_i$, with $\Hom_{\cO_X}(\cF_i,\cF_j)$ the space of
paths from node $j$ to node $i$ and $\Ext^1_{\cO_X}(\cF_i,\cF_j)$ the
space of independent relations imposed on these paths. The moduli space of $\theta$-stable representations of this quiver with dimension vector $\mbf k= (1 , \dots , 1)$ is isomorphic to the crepant resolution $X = \mathrm{Hilb}^\Gamma (\complex^3)$.

In our case the quiver $\sfQ$ is the McKay quiver, whose vertices are
labelled by the irreducible representations of the orbifold group
$\Gamma$ and whose arrow set $\sfQ_1$ is dictated by the decomposition
into irreducible representations of the tensor products $Q\otimes\rho_r$. In this case the tilting bundle determining
the equivalence between the derived categories is the sum of the
tautological bundles over $X$~\cite{BKR,IU1,Aspinwall:2008jk} (see also~\cite[Remark~7.17]{crawrev}),
\begin{equation}
\cT = \bigoplus_{r\in\Gammaw}\, \cR_r \ ,
\label{tilting}\end{equation}
where only $\Hom_{\cO_X}(\cR_r,\cR_s)$ can be non-trivial.
Note that the sheaves in the tilting set
are the projective objects $\sfP_r=\cR_r$, not the fractional branes
$\sfD_r=\cS_r$. The relationship between the two sets of D~branes is
given by projective resolutions (\ref{projres}).

Our gauge theory construction naturally produces $0$-semistable
objects of the category $\sfA$--mod. It is vastly based on the McKay
correspondence for threefolds. We have used the $\Gamma$-equivariant
geometry of $\complex^3$ to describe our instanton moduli space. This
is in perfect harmony with the fact, discussed in
Section~\ref{subsec:NCDTquivers}, that the orbifold
Donaldson--Thomas invariants of~\cite{young} counting
$\Gamma$-equivariant ideal sheaves on $\complex^3$ coincide exactly
with the noncommutative invariants $\NDT_{\mu=0}(\mbf k)$~\cite[Section~7.4]{joycesong}.

Let us now turn to the construction of the partition function
for these invariants. We have explained how to use the geometry of
$\mathrm{Hilb}^\Gamma (\complex^3)$ to evaluate explicitly the
instanton action in terms of large radius data. By using the fact that
the set $\{ \cR_r\}_{r\in\Gammaw}$ generates the topological K-theory
group $K(X)$ and therefore has a direct relation with the homology
$H_{\bullet} (X)$, we found precise combinations of the bundles
$\cR_r$ which correspond to divisors and curves in the resolved
geometry. On the other hand, since the
object $\cT$ in (\ref{tilting}) is a tilting generator, it is by
definition constant and well-defined over the entire K\"ahler moduli space
(though the way in which the derived category is generated changes as
we move around the moduli space).

It is tempting to speculate that our
instanton action makes sense also at any point of the moduli
space. The only change in the partition function is eventually encoded
in the chemical potentials $\varphi_n$ and $\varsigma_m$ which specify
the strengths of the couplings between D~branes, and of course in the
fact that for given topological instanton charges $\mbf k$ the moduli
space might be empty. The latter condition is however automatically
taken care of by the measure on the instanton moduli space, at least
in those regions of the K\"ahler moduli space where we can compute it
explicitly. Therefore the problem is reduced to computing the
instanton measure for any value of the stability parameter and not
just for $\mu=0$ as we have done above. Note that by working with the
tilting set (\ref{tilting}) we bypass the question of what
are the stable fractional branes in each region of the moduli space,
or equivalently what happens to the basis of coherent sheaves $\cS_r$
supported on the exceptional set when we blow down the exceptional
cycles. This is somewhat in line with the proposal of
\cite{Aspinwall:2008jk} that the D~branes in the tilting set are
everywhere $\Pi$-stable over the whole K\"ahler moduli space.

\section{An example without compact four-cycles: \ $\complex^3 /  \zed_2  \times \zed_2$\label{sec:C3Z2}}

\subsection{Geometry and representation theory}

Our first example will be the resolution of $\complex^3 / \zed_2
\times \zed_2$, where the action of
$\IZ_2\times\IZ_2=\{1,g_1,g_2,g_3\}$ on $\IC^3$ is
given by
\bea
g_1\cdot(z_1,z_2,z_3)&=&(-z_1,-z_2,z_3) \ , \nonumber\\[4pt] g_2\cdot
(z_1,z_2,z_3)&=& (-z_1,z_2,-z_3) \ , \nonumber\\[4pt]
g_3\cdot(z_1,z_2,z_3) &=&
(z_1,-z_2,-z_3) \ .
\eea
This singular orbifold has a fan $\Sigma\subset\zed^3$
generated by the lattice vectors $D_1 = (0,2,1)$, $D_2 = (0,0,1)$ and $D_3 = (2,0,1)$.
These three vectors also correspond to the three non-compact divisors,
obtained by setting to zero the corresponding coordinate of $\IC^3$,
i.e. $D_\a = \{(z_1,z_2,z_3)\in\IC^3~|~ z_\a = 0 \}$.

\begin{figure}[h]
 \centering
  \includegraphics[width=5cm]{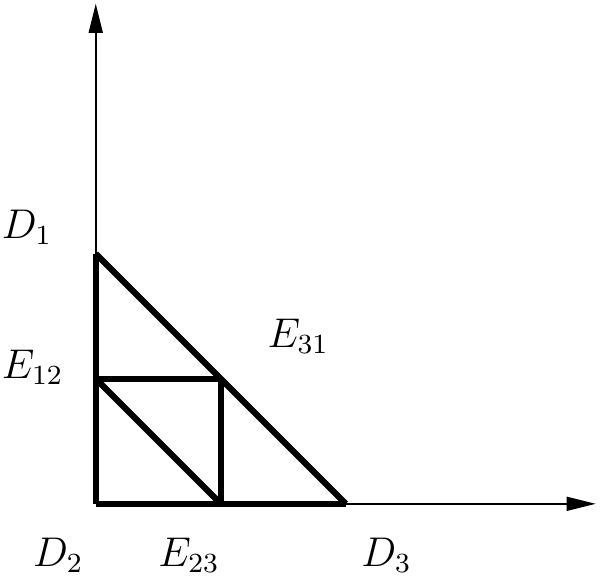}
 \caption{Toric fan for the closed topological vertex geometry.}
 \label{C3z2z2}
\end{figure}

One can resolve the singularity in several ways, corresponding to the distinct
possible triangulations of the toric diagram. Here we
only consider the symmetric resolution given by $X=\mathrm{Hilb}^{\zed_2 \times
  \zed_2} (\complex^3) $, which has the geometry of the closed
topological vertex~\cite{BryanKarp,KarpLiuMarino}, whose fan is depicted in
Figure~\ref{C3z2z2}. 
This resolution has three non-compact divisors; we will denote by
$E_{\a\b}$ the divisor whose vector lies between $D_\a$ and
$D_\b$. They all have the topology of $\complex \times \PP^1$. Finally
there are three compact curves given by the intersections $C_\a =
E_{\g\a} \cdot E_{\a\b}$. This geometry does not have any compact
divisors, as all the compact holomorphic submanifolds are curves,
which have
codimension two.

The linear equivalences between the non-compact divisors are
\begin{eqnarray}
2 D_1 + E_{31} + E_{12} &\sim& 0 \ , \nonumber \\[4pt]
2 D_2 + E_{23} + E_{12} &\sim& 0 \ , \nonumber \\[4pt]
2 D_3 + E_{23} + E_{31} &\sim& 0 \ .
\end{eqnarray}
The non-vanishing triple intersections are
\begin{eqnarray}
E_{12} \cdot D_2 \cdot E_{23} &=& 1 \ , \nonumber \\[4pt]
E_{12} \cdot E_{31} \cdot E_{23} &=& 1 \ , \nonumber \\[4pt]
E_{31} \cdot E_{23} \cdot D_3 &=& 1 \ , \nonumber \\[4pt]
E_{12} \cdot D_1 \cdot E_{31} &=& 1 \ .
\end{eqnarray}
In particular $D_1$, $D_2$ and $D_3$ generate the K\"ahler cone, while
$C_1$, $C_2$ and $C_3$ are the dual generators of the Mori cone with
respect to the intersection pairing.\footnote{Elements of the K\"ahler
cone are cohomology classes $\eta\in H^2(X,\IQ)$ such that
$\int_C\,\eta\geq0$ for every effective cycle $C\in H_2(X,\IQ)$ (and
similarly for $\eta^{\wedge r}$ on higher-dimensional
subvarieties). The Mori cone consists of linear combinations of
compact algebraic cycles with non-negative coefficients and is
generated by the exceptional curves.} This means that the tautological
bundles are $\cR_\a = \cO_X (D_\a)$, whose first Chern classes form a
basis of $H^2(X,\IZ)$ with
\begin{equation}
\int_{C_\a}\, c_1 (\cR_\b) = \delta_{\a\b}
\end{equation}
for $\a,\b=1,2,3$. Upon including the trivial bundle $\cR_0 = \cO_X$,
which generates $H^0(X,\IZ)$, 
these bundles form a canonical integral basis of $K(X)$.

Let us now turn to the representation theory data. The orbifold group
is $\Gamma=\zed_2 \times \zed_2$ and it acts on $\IC^3$ with weights
$r_1=(1,1,0 )$, $r_2=(1,0,1)$ and $r_3=r_1+r_2=(0,1,1)$. It has four
irreducible representations $\rho_r$ where $\rho_0$ is the trivial
representation, $\rho_1$ and $\rho_2$ correspond to the weights $r_1$
and $r_2$, and $\rho_3 = \rho_1 \otimes \rho_2$ corresponds to the
weight $r_3$. The tensor product decomposition of the defining representation
$Q$ gives a matrix
\begin{equation}
\big(a_{rs}^{(1)}\big) = \left(  \begin{matrix}  0 & 1 & 1 & 1 \\ 1 &
    0 & 1 & 1 \\ 1 & 1 & 0 & 1 \\ 1 & 1 & 1 & 0 \end{matrix}\right) \ .
\label{ars1Z2Z2}\end{equation}
Note that this matrix is symmetric. Since $\bigwedge^2 Q \cong
Q^{\vee}$, one has $a^{(2)}_{rs} = a^{(1)}_{sr}$ and in this
particular case the intersection product vanishes identically, $(\cS^{\vee}_r , \cS_s) = 0$. This reflects the fact that the resolved geometry has no compact divisors.  
The quiver constructed from representation theory is thus
\begin{equation}
\vspace{4pt}
\begin{xy}
\xymatrix@C=20mm{
& \ v_0 \ \bullet \ \ar@/^/[ddr] \ar@/_1pc/[ddl]  \ar@/^/[d]& \\
& \ v_3 \ \bullet \ \ar@/^/[u] \ar@/^/[dl] \ar@/^/[dr]& \\
v_1 \ \bullet \ \ar@/^/[uur] \ar@/^/[ur] \ar@//[rr] & & \ \bullet \ v_2  \ar@/_1pc/[uul] \ar@/^/[ul] \ar@/^/[ll]
}
\end{xy}
\vspace{4pt}
\end{equation}

\subsection{BPS partition functions\label{subsec:C3Z2Z2BPS}}

To evaluate the partition function we will choose the boundary
condition $\mbf N = (1,0,0,0)$ corresponding to $U(1)$ gauge field
configurations that are trivial at infinity. We begin by computing the
action given by (\ref{c1action})--(\ref{ch3action}). Since there are no
compact four-cycles, we cannot wrap compact D4~branes anywhere and the
integral $\int_{\overline{X}}\, \omega \wedge \omega \wedge c_1 (\cE)$
in (\ref{c1action}) must vanish
identically. Indeed this is the case, since the first Chern class $c_1
(\cE)$ itself is zero. Because of our boundary condition, there are no
terms proportional to $N_r$ since $c_1 (\cR_0) = 0$. Moreover the
intersection matrix $a_{rs}^{(2)} - a_{rs}^{(1)} $ vanishes, since
$a_{rs}^{(1)}$ is symmetric.

Now let us turn to the integral $\int_{\overline{X}}\, \omega \wedge \mathrm{ch}_2
(\cE)$ involving the second Chern class. The first line on
the right-hand side of (\ref{ch2action}) vanishes for the same reasons
as above. The remaining term from (\ref{ch2action}) is 
\beq
\int_{\overline{X}}\, \omega \wedge \mathrm{ch}_2 (\cE) = - \sum_{n,r,s=0}^3\,
\Big( a_{rs}^{(2)} - 3 \delta_{rs} \Big) \, k_s \,\varphi_n\, 
\int_{\overline{X}}\, c_1 (\cR_n) \wedge c_1(\cR_r)\wedge c_1 \big(\cO_{\overline{X}}(1)\big) \
,
\label{ch2Z2Z2rem}\eeq
where we recall that $c_1 (\cR_0) = 0$. This integral computes a
triple intersection number which we evaluate explicitly below.

The remaining integral is (\ref{ch3action}), which by the same
reasoning as above reduces to
\begin{eqnarray}
\int_{\overline{X}}\, \mathrm{ch}_3 (\cE) &=& - \sum_{r,s=0}^3 \, \bigg(\Big( a_{rs}^{(2)} - 3 \delta_{rs} \Big) \, k_s \, \int_{\overline{X}}\, c_1 \big(\cO_{\overline{X}}(1)\big) \wedge \ch_2 (\cR_r)
 - \frac{
 k_s}{ | \Gamma |}  \, \delta_{rs}  \nonumber \\ & & \qquad \qquad -\, \Big( a^{(2)}_{rs} - 3 \delta_{rs} \Big) \, k_s \, \int_{\overline{X}}\, c_1 (\cR_r) \wedge \ch_2 \big(\cO_{\overline{X}} \big(1)\big)\bigg) \ .
\label{ch3Z2Z2rem}\end{eqnarray}
Recall that in our
case $|\Gamma|=4$. To evaluate this integral, we note that the
integrals on the right-hand side of (\ref{ch3Z2Z2rem}) measure, in various
forms, all the triple intersections of the non-compact divisors involving
the divisor $\bdiv$ at infinity at least once. To evaluate these
integrals, we assume
that the divisor at infinity has no intersection with the compact
curves that resolve the singularity. Then we can evaluate the
intersection numbers as if they were effectively taken in $\PP^3$ (and
take care of the orbifold action only when evaluating the pullback by
dividing by the order of the orbifold group), i.e. upon
compactification the divisors $D_\a$ are topologically $\PP^2$. Therefore all the intersection products involve three divisors in $\PP^3$, which intersect at a point. By symmetry we can identify two independent integrals, those involving the triple intersection of two non-compact divisors with the divisor at infinity, say $D_1 \cdot D_2 \cdot \bdiv$, and those involving the self-intersection of a non-compact divisor, say $D_1 \cdot D_1 \cdot \bdiv$. We will parametrize these integrals with two integers, $b$ and $a$.

We can now evaluate the integral (\ref{ch2Z2Z2rem}) to be
\begin{eqnarray}
\int_{\overline{X}}\, \omega \wedge \mathrm{ch}_2 (\cE) = - \sum_{n,r,s=0}^3\,
\Big( a_{rs}^{(2)} - 3 \delta_{rs} \Big) \,\frac{ k_s
  \,\varphi_n}{|\Gamma|} \, I_{nr}
\end{eqnarray}
where we have introduced the intersection matrix
\begin{equation}
(I_{nr}) = \left(  \begin{matrix} 0 & 0 & 0 & 0 \\ 0 & a & b & b \\  0 & b & a & b \\  0 & b & b & a  \end{matrix} \right) \ .
\end{equation}
The zeroes come from the fact that $\cR_0$ is the trivial bundle with
vanishing Chern classes, while the remaining entries come from the
intersection products. We finally obtain
\begin{eqnarray}
\int_{\overline{X}}\, \omega \wedge \mathrm{ch}_2 (\cE) &=& - \frac{1}{4}\, \Big(
2 b\, \big( \varphi_1\, (k_0 + k_1 - k_2 - k_3) + \varphi_2\, (k_0 -
k_1 + k_2 - k_3)  \cr & & \qquad +\,  
    \varphi_3\, (k_0 - k_1 - k_2 + k_3) \big) + 
 a\, \big( \varphi_3\, (k_0 + k_1 + k_2 - 3 k_3)  \cr & & \qquad +\, \varphi_2\, (k_0 + k_1 - 3 k_2 + k_3) + 
    \varphi_1\, (k_0 - 3 k_1 + k_2 + k_3) \big) \Big) \ .
\end{eqnarray}

Now let us consider the last term in the instanton action
\begin{eqnarray}
\int_{\overline{X}}\, \ch_3 (\cE) = - \sum_{r,s=0}^3 \, \bigg(\Big(
a_{rs}^{(2)} - 3 \delta_{rs} \Big) \,\frac{ k_s \,
  \gamma_r}{2|\Gamma|} -\, \Big( a^{(2)}_{rs} - 3 \delta_{rs} \Big)
\,\frac{ k_s\, \alpha_r}{2|\Gamma|}
- \delta_{rs}\, \frac{
 k_s }{|\Gamma|}\, \bigg) \ ,
\end{eqnarray}
where we have introduced the vectors $(\gamma_r) = (0,c,c,c)$ and
$(\alpha_r) = (0,a,a,a)$. The integer $c$ is the triple intersection
product $D_\a \cdot \bdiv \cdot \bdiv$. The factors of $\frac12$ come
from the expansion of the Chern character $\ch_2$. We arrive finally at
\begin{eqnarray}
\int_{\overline{X}}\, \ch_3 (\cE) = - \mbox{$\frac{1}{8}$}\, \left( a - c \right)\, \left( 3 k_0 -k_1-k_2-k_3 \right) + \mbox{$\frac{1}{4}$}\, \left( k_0 + k_1 + k_2 + k_3 \right) \ ,
\end{eqnarray}
where again we have used $|\Gamma|=4$.

Let us now compute the values of the triple intersection numbers $a$,
$b$ and $c$ appearing above. We know the intersections between the
non-compact divisors in $X$ (when they make sense, i.e. when they involve at
least some compact curve), and we are modelling the behaviour
of $X$ at infinity as $\PP^3 / \Gamma$. We will momentarily ignore the
orbifold action. In the compactified geometry the divisors $D_\a$ look
like the compact divisor of $\PP^3$ at infinity. Therefore two of them
intersect with the divisor at infinity as three ordinary planes
$\PP^2$ inside $\PP^3$, i.e. at a point. Thus we conclude
$b=c=1$. However $a$ counts the self-intersection of a divisor with
$\bdiv$. Let us call this compactified divisor $\tilde{D}$. If our
space $X$ were an ordinary $\PP^3$ then we could conclude $\tilde{D}
\cdot \tilde{D} \cdot \bdiv = 1$; the usual argument would be
that one can consider a generic intersection with another divisor
$\tilde{D} \cdot \tilde{D}' \cdot \bdiv$ and then ``transport''
$\tilde{D}'$ back to $\tilde{D}$ to compute the intersection
product. However this ``transport'' is not permitted in our case since
our variety looks like $\PP^3$ only at infinity, and one cannot
``transport'' a divisor without intersecting the compact curves in the
exceptional locus. Therefore to compute the number $a$ the heuristic
``transporting'' argument is not sufficient. On the
other hand we know from the toric diagram (see Figure~\ref{C3z2z2}) that
the non-compact divisors $D_\a$ give always zero intersection whenever
their self-intersection appears in a triple intersection product, i.e. $D_\a \cdot D_\a \cdot E_{\a\b} = 0$ for every choice of $\a$ and $\b$. Therefore we take $a = D_\a \cdot D_\a \cdot \bdiv = 0$.

To compute the index of BPS states we need to compute the $\zed_2
\times \zed_2$-invariant part of the character~(\ref{character}). We
rewrite it as in (\ref{Chpisplit})--(\ref{calTpm}) and decompose the
vector space $V$ at a fixed point $\pi$ as $ V_\pi = 
\mathsf{V}_0 \oplus \mathsf{V}_1 \oplus \mathsf{V}_2 \oplus
\mathsf{V}_3$, where each subspace $\mathsf{V}_r$ is associated to the
group element represented by $\rho_r$ on $\complex^3$. Note that each
element is nilpotent. We can write the partial character $\mathcal{T}_{\pi}^+ $ as
\begin{eqnarray}
\mathcal{T}_{\pi}^+ &=& (\textsf{V}_0 \oplus \textsf{V}_1 \oplus
\textsf{V}_2 \oplus\textsf{V}_3) \nonumber\\ && - \, \Big(\, \frac{1}{t_1\, t_2} - \frac{1}{t_2} - \frac{1}{t_1} +1
\Big) \, \left( \textsf{V}_0 \oplus \textsf{V}_1 \oplus
\textsf{V}_2 \oplus\textsf{V}_3\right) \otimes \left( \textsf{V}^\vee_0 \oplus \textsf{V}^\vee_1 \oplus
\textsf{V}^\vee_2 \oplus\textsf{V}^\vee_3 \right) \ ,
\end{eqnarray}
which upon substituting $t_\a\mapsto \rho_{r_\a}$ gives
\begin{equation}
\big(\mathcal{T}_\pi^+\big)^{\zed_2 \times \zed_2} = |\pi_1| + 
|\pi_2| + |\pi_3| 
\end{equation}
and hence
\begin{equation}
\chi_{\torus^3}(\mathcal{N}_\pi) = (-1)^{|\pi_1| + 
|\pi_2| + |\pi_3|} \ .
\end{equation}
Combining all of these ingredients together, we can write the partition function of orbifold Donaldson--Thomas invariants as
\begin{eqnarray} 
\cZ_{\complex^3/\zed_2 \times \zed_2} &=& \sum_{\pi}\, \chi_{\torus^3}(\cN_\pi)~
\e^{ - g_s\, \int_{\overline{X}}\, \ch_3 (\cE_\pi)} ~ \e^{-
  \int_{\overline{X}}\, \omega \wedge \ch_2 (\cE_\pi)} \nonumber
\\[4pt]
&=&  \sum_{\pi} \, (-1)^{|\pi_1| + 
|\pi_2| + |\pi_3|} \, q^{  \frac{1}{8}\, \left( 3 |\pi_0|
  -|\pi_1|-|\pi_2|-|\pi_3| \right) + \frac{1}{4}\, \left( |\pi_0| +
  |\pi_1| + |\pi_2| + |\pi_3| \right) }  \cr && \qquad \times\, 
Q_1^{|\pi_0| +  |\pi_1| - |\pi_2| - |\pi_3|} \,
Q_2^{|\pi_0| -  |\pi_1| + |\pi_2| - |\pi_3|} \,
Q_3^{|\pi_0| -  |\pi_1| - |\pi_2| + |\pi_3|} 
\label{DTz2z2first}\end{eqnarray}
where we have introduced the weighting variables $q= \e^{-g_s}$ and
$Q_\a = \e^{-\varphi_\a}$ for $\a=1,2,3$.

\subsection{Coloured Young diagram partition functions}

Now we compare our construction with the available literature. In~\cite[Definition~1.3]{young}, adapted to our case, we learn of a combinatorial partition function 
\begin{equation}
K_{\zed_2 \times \zed_2} = \sum_{\pi}\, p_0^{|\pi_0|}\, p_1^{|\pi_1|}\, p_2^{|\pi_2|}\, p_3^{|\pi_3|}
\end{equation}
in formal variables $p_r$ which enumerates $\IZ_2\times\IZ_2$-coloured
three-dimensional Young diagrams. In~\cite[Theorem~A.3]{young} it is
proven that this partition function is related to the
Donaldson--Thomas partition function of the quotient stack $[\IC^3/\IZ_2\times\IZ_2]$ through
\begin{equation}
K_{\IC^3/\IZ_2\times\IZ_2}^{\rm DT} (p_0 , p_1 , p_2 , p_3) =
K_{\zed_2 \times \zed_2} (p_0 , - p_1 ,-  p_2 , - p_3) \ .
\end{equation}
This formula only depends on the four variables $p= p_0\, p_1\, p_2\, p_3$, $p_1$, $p_2$ and $p_3$, and can be written as
\begin{equation} \label{young}
K^{\rm DT}_{\IC^3/\zed_2 \times \zed_2} = \sum_{\pi} \, (-1)^{|\pi_1| + |\pi_2| + |\pi_3|} \, p^{|\pi|}\, p_1^{|\pi_1|}\, p_2^{|\pi_2|}\, p_3^{|\pi_3|} \ .
\end{equation}

A simple computation shows that after the change of variables
\begin{eqnarray}
p &=& q^{5/8}\, Q_1 \, Q_2 \, Q_3 \ , \nonumber\\[4pt]
p_1 &=& q^{-1/2}\,Q^{-2}_2 \, Q^{-2}_3 \ , \nonumber\\[4pt]
p_2 &=& q^{-1/2}\, Q^{-2}_1 \, Q^{-2}_3 \ , \nonumber\\[4pt]
p_3 &=& q^{-1/2}\,Q^{-2}_1 \, Q^{-2}_2 \ ,
\end{eqnarray}
our partition function (\ref{DTz2z2first}) coincides with (\ref{young}).
While our original variables seem somewhat apt to an interpretation in terms of D~brane charges,
the physical meaning of this redefinition is unclear. The D~brane
charge corresponding to each configuration represented by a plane
partition is however expected to be a rather non-trivial function of the D2
and D0 charges~\cite{Ooguri:2008yb}.

In this case the
partition function has an explicit description as a product of
generalized MacMahon functions, which generate weighted plane partitions, given by~\cite[Theorem~1.5]{young}
\beq
K_{\zed_2 \times \zed_2} = M(p)^4\, \frac{\widetilde{M}(p_1\,p_2,p)\,
  \widetilde{M}(p_1\,
  p_3,p)\,\widetilde{M}(p_2\,p_3,p)}{\widetilde{M}(-p_1,p)\,
  \widetilde{M}(-p_2,p)\, \widetilde{M}(-p_3,p)\,
  \widetilde{M}(-p_1\,p_2\, p_3,p)} \ ,
\eeq
where
\begin{eqnarray}
M (x , q) = \prod_{n=1}^{\infty} \, \big(  1 - x \, q^n \big)^{-n}
\qquad \mbox{and} \qquad \widetilde{M} (x , q) = M (x , q) \, M(x^{-1}
, q)
\end{eqnarray}
with $M(q)=M(1,q)$. Moreover, since in this case $X\to\IC^3/\Gamma$ is
a semi-small resolution, i.e. it contains no compact four-cycles, by~\cite[Proposition~A.7]{young} the Donaldson--Thomas
partition functions of $[\IC^3/\IZ_2\times \IZ_2]$ and its natural crepant resolution
$X={\rm Hilb}^{\IZ_2\times \IZ_2}(\complex^3)$ are related through
\beq
K_{\IC^3/\IZ_2\times\IZ_2}^{\rm DT} (p_0 , p_1 , p_2 , p_3) =
M(-p)^{-4}\, K_X^{\rm top}(p,p_1,p_2,p_3)\, K_X^{\rm
  top}(p,p^{-1}_1,p^{-1}_2,p^{-1}_3)
\eeq
where the topological string partition function
\beq
K_X^{\rm top}(p,p_1,p_2,p_3)= M(-p)^4\, \frac{M(p_1\,p_2,-p)\,
  M(p_1\,
  p_3,-p)\,M(p_2\,p_3,-p)}{M(p_1,-p)\,
  M(p_2,-p)\, M(p_3,-p)\,
  M(p_1\,p_2\, p_3,-p)}
\eeq
is computed via the topological vertex
formalism~\cite{KarpLiuMarino}. Here the variables $p_1$, $p_2$ and $p_3$ correspond to the basis
of curve classes (D2~branes) in $X$ and $p$ to the Euler number (D0~branes). In this way the gauge theory we have
constructed on $[\IC^3/\IZ_2\times \IZ_2]$ realizes the anticipated
wall-crossing behaviour of
the BPS partition function (\ref{ZBPSqQ}), connecting in this case
the orbifold point with the large radius point in the
K\"ahler moduli space. This partially justifies some of our arguments
from Section~\ref{subsec:Stability}.

\section{$A$-fibred threefolds\label{sec:Afib}}

\subsection{Geometry and representation theory}

Next we will consider another set of examples of semi-small crepant
resolutions, this time obtained as fibrations of hyper-K\"ahler ALE
spaces over the complex plane \cite{Cachazo:2001gh,szendroiADE}. These
ALE spaces are obtained by blowing up an abelian quotient singularity of the form $\complex^2 / \zed_n$. The resulting smooth geometry is then trivially fibred over the complex plane to obtain a threefold. The theory of Donaldson--Thomas invariants and their wall crossings on these geometries was studied in~\cite{gholampour}.

We are interested in the local surfaces which are semi-small crepant resolutions of the form $X \rightarrow \complex^2 / \Gamma \times \complex$, where the orbifold action is
\begin{equation}
g\cdot (z_1 , z_2 , z_3 ) = (\zeta \, z_1 , \zeta^{-1}\, z_2 , z_3)
\label{Anaction}\end{equation}
with $g$ a generator of $\Gamma=\IZ_n$ and $\zeta$ an $n$-th root of unity. The resolved Calabi--Yau
geometry is therefore a (trivial) fibration of a resolved $A_{n-1}$
singularity over the affine line. These $A$-singularities are
abelian. The resolved geometry is toric and is in particular a small
resolution of the singularity, i.e. a birational morphism such that
the exceptional locus consists of curves. The corresponding toric
diagram is obtained by subdividing the long edge of the toric diagram
for $\IC^3$ into $n$ parts of equal length via the insertion of $n-1$
additional vertices, and taking the unique triangulation
corresponding to the unique minimal resolution of the double point
singularity $\IC^2/\Gamma$. A choice of lattice vectors generating
the toric fan is given by
\beq
D_0=(1,0,0) \ , \qquad D_1=(0,1,0) \ , \qquad \dots \ , \qquad
D_{n}=(-n+1,n,0) \ ,
\eeq
with the linear equivalences among toric divisors
\beq
D_0-D_2-\cdots -(n-1)\,D_n\sim0 \qquad \mbox{and} \qquad
D_1+2D_2+\cdots+ n\, D_n\sim0 \ .
\eeq

The intersection matrix $C=(C_{rs})$ of
the exceptional curves $D_1,\dots,D_{n-1}$ is minus the Cartan matrix of the $A_{n-1}$ Lie
algebra. The set of generators of the cohomology groups is given in
terms of the set of tautological bundles, which in this case are
simply the line bundles corresponding to the divisors, plus the
trivial bundle. Each tautological bundle $\cR_r$ corresponds naturally
to an irreducible representation $\rho_r$ which labels the monodromy
of its canonical connection at infinity; equivalently it can be read off
from the orbifold action on the monomials which are dual to the
divisors. The K\"ahler cone generators are given by the first Chern
classes of the tautological bundles, $e_r=c_1(\cR_r)= \sum_s\,
(C^{-1})_{rs}\, D_s$, and they are dual to the Mori cone generators
with respect to the intersection pairing $D_r\cdot e_s=\delta_{rs}$. 

The whole geometric structure is encoded in the affine extension of
the Cartan matrix of $A_{n-1}$. All of the geometry can be rephrased in
terms of the representation theory of the $A_{n-1}$ Lie algebra and
its affine extension via the McKay correspondence. The representation
theory data can also be compactly encoded in the McKay quiver
associated with the singularity, whose arrow structure is dictated by
the decomposition into irreducible representations of the tensor product
$Q \otimes \rho_r$, where $Q$ is the fundamental representation of
$\Gamma\subset SU(2)\subset SU(3)$. This construction closely parallels the construction of the
McKay quiver for threefolds.

These sets of data can be used to study the $A$-fibred singularities. They can be realized via the representation theory of a certain quiver, which is a modification of the McKay quiver associated with the singularity. This quiver is obtained from the usual McKay quiver by adding a set of loops, arrows from each vertex to itself. Small crepant resolutions of these singularities have an alternative description as the moduli space of representations of the modified McKay quivers. The path algebras of these quivers are noncommutative crepant resolutions of the fibred singularities. 

\begin{figure}[h]
 \centering
  \includegraphics[width=5cm]{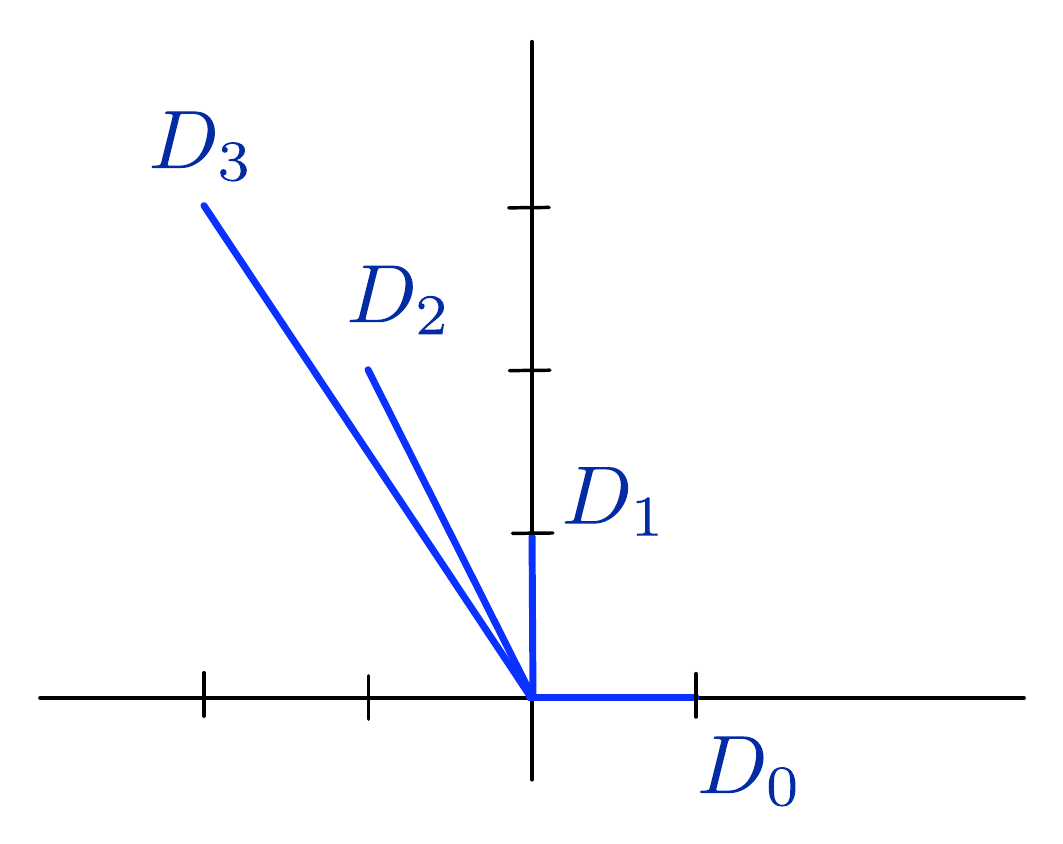}
 \caption{Toric fan for the $A_2$-fibration.}
 \label{A2fan}
\end{figure}

For example, consider the $\complex^2 / \zed_3\times \IC$ orbifold
with weights $r_1=1$, $r_2=2$, $r_3=0$. Its toric fan is depicted in
Figure~\ref{A2fan}.
The regular representation is
now $Q = \rho_1 \oplus \rho_2 \oplus \rho_0$ and the tensor product
decomposition gives symmetric matrices
\begin{equation}
\big(a_{rs}^{(1)}\big) = \big(a_{rs}^{(2)}\big) = \left(
  \begin{matrix} 1 & 1 & 1 \\ 1 & 1 & 1 \\ 1 & 1 & 1 \end{matrix}
\right) \ .
\end{equation}
The associated quiver is
\begin{equation}
\vspace{4pt}
\begin{xy}
\xymatrix@C=20mm{
& \ v_0 \ \bullet \ \ar@/^/[ddl] \ar@/_0.5pc/[ddr] \ar@(ur,ul) & \\
& & \\
 \ v_1 \ \bullet \ \ar@/_/[rr] \ar@/^/[uur] \ar@(lu,ld) & & \ \bullet \
v_2 \ \ar@/_/[uul] \ar@/_/[ll]  \ar@(rd,ru)
}
\end{xy}
\vspace{4pt}
\end{equation}

\subsection{BPS partition functions\label{subsec:C2Z3BPS}}

Let us start with the $U(1)$ gauge theory. The combinations
$a_{rs}^{(1)} - a_{rs}^{(2)}$ vanish identically, and the boundary
condition is imposed by choosing $\mbf N = (1,0,\dots,0)$. Again the
contribution (\ref{c1action}) vanishes, as it should, while the second
Chern character in the instanton action (\ref{ch2action}) has the same
form as in (\ref{ch2Z2Z2rem}) but with the sums now ranging over
$0,1,\dots,n-1$. Let us evaluate this integral. It measures the triple intersection
between two divisors, corresponding to two compact curves, with the
divisor $\bdiv$ at infinity. The geometry we are considering is a
trivial fibration of an $A_{n-1}$ singularity over an affine
line. Therefore all the intersection numbers are essentially given by
the intersections of curves in the exceptional locus of the blown up
singularity in $\IC^2$. However these intersections are still
``fibred'' over the affine line. For example, if two exceptional
curves intersect at a point in the ALE geometry, then their
intersection in the full Calabi--Yau threefold has the form
${\mathrm{pt}} \times \complex$. In the full Calabi--Yau geometry the
exceptional curves are actually non-compact divisors of the form
$\PP^1 \times \complex$. The intersection with the boundary divisor is
only due to the non-compact factor $\complex$.

We can therefore write the action as
\begin{equation}
\int_{\overline{X}}\, \omega \wedge \ch_2 (\cE) = - \sum_{m,r,s=0}^{n-1}\,
\varphi_m \, \Big( a^{(2)}_{rs} - 3 \, \delta_{rs} \Big) \ k_s\,
\frac{a}{|\Gamma|}\, \tilde{C}_{mr} \ ,
\end{equation}
where the constant $a$ parametrizes the intersections with $\bdiv$ and
\begin{equation}
\tilde{C} = \left( \begin{matrix} 0 & 0 \\ 0 & C^{-1} \end{matrix} \right)
\end{equation}
with $C$ the intersection matrix (minus the $A_{n-1}$
Cartan matrix). The rest of the instanton action (\ref{ch3action})
reads as in (\ref{ch3Z2Z2rem}). The integrals involved have the form
\begin{eqnarray}
\Big(\,\int_{\overline{X}}\, c_1 \big(\cO_{\overline{X}}(1)\big) \w \ch_2
(\cR_r)\,\Big)_{r=0,1,\dots,n-1} &=& \frac{a}{|\Gamma|} \, (0,-2 ,-2 ,\dots ,
-2) \ , \nonumber \\[4pt]
\Big(\, \int_{\overline{X}}\, c_1 (\cR_r) \w \ch_2
\big(\cO_{\overline{X}}(1)\big)\,\Big)_{r=0,1,\dots,n-1} &=& \frac{1}{2 |\Gamma|} \, (0 , b , b
, \dots , b) \ ,
\label{ch3Anints}\end{eqnarray}
where we have parametrized the intersection indices of non-compact divisors of the form ${\mathrm{pt}} \times \complex$ and $\PP^1 \times \complex$ with the divisor at infinity by two integers $a$ and $b$ (whose precise values are not important at the moment).

To compute the instanton measure we consider the
$\Gamma$-invariant part of the partial character
\begin{equation}
{\cal T}^{+}_{\pi} = V_{\pi}
- V_{\pi} \otimes  V^\vee_{\pi}~ 
\frac{(1-t_1)\,(1-t_2)}{t_1\, t_2} \ .
\label{Anpartialchar}\end{equation}
For the orbifold action (\ref{Anaction}) the terms proportional to
$V_{\pi} \otimes V^\vee_{\pi}$ cancel pairwise upon substituting
$t_1=\zeta$, $t_2=\zeta^{-1}$, and therefore
\begin{equation}
\big({\cal T}^{+}_{\pi}\big)^\Gamma = |\pi_0| \ .
\end{equation}
The instanton partition function for the rank one invariants is thus
\begin{equation}
\cZ_{\complex^2 / \zed_{n}\times\IC} = \sum_{\pi}\, (-1)^{|\pi_0|} \
\e^{-S_{\mathrm{inst}}[\pi]} \ .
\label{rank1partfn}\end{equation}

The orbifold group $\Gamma=\zed_{n}$ has $n$ irreducible representations. Therefore it is natural to parametrize the partition function of orbifold invariants as
\begin{equation}
K^{\rm DT}_{\complex^2 / \zed_{n}\times\IC} = \sum_{\pi}\, (-1)^{|\pi_0|} \
p_0^{|\pi_0|} \  p_1^{|\pi_1|} \ \cdots \ p_{n-1}^{|\pi_{n-1}|} \ ,
\end{equation}
which precisely coincides with the result of \cite{young}. This
partition function can also be written in a product form~\cite{young}
\beq
K^{\rm DT}_{\complex^2 / \zed_{n}\times\IC} = M(-p)^n\, \prod_{0<r\leq s<n}\,
\widetilde{M}(p_{[r,s]},-p) \ ,
\label{ZAnproduct}\eeq
where $p=p_0\,p_1\cdots p_{n-1}$ and $p_{[r,s]}=p_r\, p_{r+1}\cdots
p_s$. As before, the large radius partition function is that of closed
topological string theory on the 
crepant resolution $X={\rm Hilb}^{\IZ_n}(\IC^3)$ and can be computed
with the topological vertex formalism~\cite{young,BryanGholampour}; it can be obtained
from the orbifold partition function (\ref{ZAnproduct}) via the
substitution $\widetilde{M}(p_{[r,s]},-p)\to M(p_{[r,s]},-p)$ of
generalized MacMahon functions and suitable reparametrization. The
partition function at the large radius point is connected to the
reparametrized partition function at the orbifold point via the
wall-crossing factor $M(-p)^{-n}\, K_X^{\rm
  top}(p,p_1^{-1},\dots,p_{n-1}^{-1})$ as in (\ref{ZBPSqQ}).

For example, let us again consider the particular case
$\complex^2 / \zed_3\times \IC$, where
\begin{eqnarray} \label{youngformula}
K^{\rm DT}_{\complex^2 / \zed_{3}\times\IC} &=& \sum_{\pi}\, (-1)^{|\pi_0|} \  p_0^{|\pi_0|} \  p_1^{|\pi_1|} \ p_{2}^{|\pi_2|} \\[4pt]
&=& M(-p_0 \, p_1 \, p_2)^3 \, \widetilde{M} (p_1 , - p_0 \, p_1
\, p_2) \, \widetilde{M} (p_1 \, p_2 , - p_0 \, p_1 \, p_2) \,
\widetilde{M} (p_2 , - p_0 \, p_1 \, p_2) \ . \nonumber
\end{eqnarray}
The instanton action is given by
\bea
\int_{\overline{X}}\, \omega \wedge \ch_2 (\cE) &=&
\mbox{$\frac{a}{3}$}\, \big(\varphi_1\, (k_0-k_1)+\varphi_2 \,
   (k_0-k_2) \big) \ , \nonumber \\[4pt]
\int_{\overline{X}}\, \mathrm{ch}_3 (\cE) &=&
-\mbox{$\frac{a}{3}$}\, (-4 k_0+2 k_1+2 k_2)+\mbox{$\frac{1}{3}$}\, (
   k_0+ k_1+ k_2)-\mbox{$\frac{b}{6}$}\, (-2  k_0+
   k_1+ k_2) \ .
\eea
Identifying $k_r=|\pi_r|$ at the fixed points
$\pi=\bigsqcup_r\, \pi_r$, we can write the instanton partition function
(\ref{rank1partfn}) as
\begin{eqnarray}
\cZ_{\complex^2 / \zed_{3}\times\IC} = \sum_{\pi} \, (-1)^{|\pi_0|} \
q^{\frac{1}{6}\, ((8 a+2 b+2)\, |\pi_0|-(4 a+b-2)\, (|\pi_1| + |\pi_2|
  ))} \ Q_1^{-\frac{a}{3}\, (|\pi_0| - |\pi_1|)} \ Q_2^{-\frac{a}{3}\, (|\pi_0| - |\pi_2|)} 
\label{A2partfn}\end{eqnarray}
where as before we have introduced the weighting variables $q=\e^{-g_s}$ and $Q_r = \e^{-\varphi_r}$ for
$r=1,2$. The two partition functions (\ref{youngformula}) and (\ref{A2partfn}) are related by the simple change
of variables
\begin{eqnarray} 
p_0 &=& q^{\frac{1}{6}\, (8 a + 2 b +2)} \ Q_1^{-a/3} \ Q_2^{-a/3} \ , \nonumber \\[4pt]
p_1 &=& q^{-\frac{1}{6}\, (4 a + b -2)} \ Q_1^{a/3} \ , \nonumber \\[4pt]
p_2 &=& q^{-\frac{1}{6}\, (4 a + b -2)} \ Q_2^{a/3} \ ,
\label{changevar}\end{eqnarray}
with $p=p_0\, p_1\, p_2=q$.

\subsection{Coulomb branch invariants}

Let us now turn to the non-abelian gauge theory. We consider the
$U(N)$ gauge theory where the gauge symmetry is broken to $U(1)^N$
according to the pattern dictated by the framing vector $\mbf
N=(N_0,N_1,\dots,N_{n-1})$. From (\ref{c1Naction})--(\ref{ch3Naction}) the instanton action
now has the form
\begin{eqnarray}
\int_{\overline{X}}\, \omega\w \omega\w c_1(\cE) &=&0 \ , \nonumber \\[4pt]
\int_{\overline{X}}\, \omega \w \ch_2 (\cE)  &=& - \sum_{m,r,s=0}^{n-1}\,
\varphi_m \Big( a^{(2)}_{rs} - 3 \, \delta_{rs} \Big) \ \sum_{l=1}^N
\, |\pi_{l,s-b(l)}|\, \frac{a}{|\Gamma|}\, \tilde{C}_{mr} \ ,
\\[4pt]
\int_{\overline{X}}\, \mathrm{ch}_3 (\cE) &=&  - \sum_{r,s=0}^{n-1} \
\sum_{l=1}^N \, |\pi_{l,s-b(l)}| \, \bigg(
\left( a_{rs}^{(2)}  - 3 \delta_{rs} \right) \ \int_{\overline{X}}\,
c_1 \big(\cO_{\overline{X}} (1)\big) \w \ch_2 (\cR_r)
  \cr & & \qquad \qquad \qquad +\, \left( a^{(2)}_{rs} - 3 \delta_{rs}
  \right) \ \int_{\overline{X}}\, c_1 (\cR_r) \w \ch_2 \big(\cO_{\overline{X}}
  (1)\big) \bigg) -\frac{\delta_{rs}}{|\Gamma|} \ , \nonumber
\end{eqnarray}
up to constant terms
which are proportional to $N_r$ but independent of the instanton
numbers; such terms can be safely ignored and absorbed into the
normalization of the partition function.\footnote{These factors are
  irrelevant for the computation of the invariants. On the other hand,
  they would be crucial for establishing the modular properties of the
  partition function.} The integrals appearing here are given by
(\ref{ch3Anints}). 

Similarly we have to compute the instanton
measure from (\ref{instmeasure}). The second set of sums vanishes
identically because of the choice of orbifold action (\ref{Anaction}),
and we are left with
\begin{equation}
\cK(\vec\pi;\mbf N) = \sum_{l=1}^N \ \sum_{r=0}^{n-1}\, |\pi_{l,r}| \
N_{r+b(l)} \ .
\label{calKAn}\end{equation}
We can therefore write down the partition function for noncommutative
Donaldson--Thomas invariants of type $\mbf N$ in the concise form
\bea
\cZ_{\complex^2 / \zed_n\times \IC}(\mbf N) &=& \sum_{\vec \pi} \,
(-1)^{\sum_{l=1}^N \ \sum_{r=0}^{n-1} \, |\pi_{l,r}| \ N_{r+b(l)}}
\nonumber\\ && \qquad \qquad \times \,
q^{\int_{\overline{X}}\, \ch_3 (\cE_{\vec \pi})} \ \prod_{m=1}^n \,
Q_m^{\sum_{r,s=0}^{n-1}\, ( a^{(2)}_{rs} - 3 \delta_{rs}) \
  \sum_{l=1}^N \, |\pi_{l,s-b(l)}|\, \frac{a}{|\Gamma|}\, \tilde{C}_{mr}} 
\eea
where $Q_m = \e^{- \varphi_m}$.

To clarify the content of this formula, let us return to our
particular example $\complex^2 / \zed_3\times \IC$. For concreteness,
let us choose the rank $N=5$ with the boundary condition $\mbf N =
(2,2,1)$. For the boundary function $b(l)$, which to an index
$l=1,\dots,N$ associates the index of the irreducible representation
associated with the Higgs field vacuum expectation value $e_l$, this
implies the assignments
\begin{eqnarray}
b(1) &=& 0  \ \Longrightarrow \ e_1 \longleftrightarrow \rho_0 \ , \nonumber\\[4pt]
b(2) &=& 0  \ \Longrightarrow \ e_2 \longleftrightarrow \rho_0 \ , \nonumber\\[4pt]
b(3) &=& 1  \ \Longrightarrow \ e_3 \longleftrightarrow \rho_1 \ , \nonumber\\[4pt]
b(4) &=& 1  \ \Longrightarrow \ e_4 \longleftrightarrow \rho_1 \ , \nonumber\\[4pt]
b(5) &=& 2  \ \Longrightarrow \ e_5 \longleftrightarrow \rho_2 \ .
\end{eqnarray}
Since in this case one has
\begin{eqnarray}
\sum_{l=1}^5\, |\pi_{l,0-b(l)}| &=& |\pi_{1,0}| +  |\pi_{2,0}|  +
|\pi_{3,2}|  +  |\pi_{4,2}|  +  |\pi_{5,1}| \ , \nonumber \\[4pt]
\sum_{l=1}^5\, |\pi_{l,1-b(l)}| &=& |\pi_{1,1}| +  |\pi_{2,1}|  +
|\pi_{3,0}|  +  |\pi_{4,0}|  +  |\pi_{5,2}| \ , \nonumber \\[4pt]
\sum_{l=1}^5\, |\pi_{l,2-b(l)}| &=& |\pi_{1,2}| +  |\pi_{2,2}|  +
|\pi_{3,1}|  +  |\pi_{4,1}|  +  |\pi_{5,0}| \ ,
\end{eqnarray}
the instanton action can be written as (again up to
irrelevant constant terms)
\begin{eqnarray}
\int_{\overline{X}}\, \omega \wedge \ch_2 (\cE_{\vec\pi}) &=&
\frac{a}{3}\, \Big( \varphi_1\, \big( |\pi_{1,0}| - |\pi_{1,1}|
+|\pi_{2,0}| - |\pi_{2,1}| + |\pi_{3,2}| - |\pi_{3,0}| \nonumber \\
&& \qquad \qquad +\, |\pi_{4,2}| - |\pi_{4,0}| + |\pi_{5,1}| -
|\pi_{5,2}| \big) \nonumber \\ 
&& \qquad +\, \varphi_2\, \big( |\pi_{1,0}| - |\pi_{1,2}|+ |\pi_{2,0}|
- |\pi_{2,2}| + |\pi_{3,2}| - |\pi_{3,1}| \cr && \qquad\qquad \qquad +\,
|\pi_{4,2}| - |\pi_{4,1}| + |\pi_{5,1}| - |\pi_{5,0}| \big) \Big)
\end{eqnarray}
and
\begin{eqnarray} 
\int_{\overline{X}}\, \mathrm{ch}_3 (\cE_{\vec\pi}) &=& 
\frac{1}{6}\, \Big( (8 a+2 b+2 ) \, \big(|\pi_{1,0}| + |\pi_{2,0}| +
|\pi_{3,2}| + |\pi_{4,2}|+ |\pi_{5,1}| \big) \cr && \qquad - \, (4
a+b-2)\, \big( |\pi_{1,1}|+ |\pi_{1,2}|+ |\pi_{2,1}|+ |\pi_{2,2}| +
|\pi_{3,0}|+ |\pi_{3,1}| \cr && \qquad\qquad\qquad \qquad \qquad +\, |\pi_{4,0}|+
|\pi_{4,1}| + |\pi_{5,2}|+ |\pi_{5,0}| \big) \Big) \ .
\end{eqnarray}

Similarly the instanton measure (\ref{calKAn}) in this case becomes 
\begin{equation}
\cK\big(\vec\pi\,;\, \mbf N = (2,2,1)\big) = |\pi_{1,2}| + |\pi_{2,2}| + |\pi_{3,1}| + |\pi_{4,1}| +
|\pi_{5,0}| \ ,
\end{equation}
where we have dropped the even parity terms which do not affect the
alternating sign of the fluctuation determinant. By using the change
of variables (\ref{changevar}) of the $U(1)$ gauge theory we can write
the partition function as
\begin{eqnarray}
\cZ_{\complex^2 / \zed_{3}\times\IC}\big(\mbf N = (2,2,1)\big) &=& 
\sum_{\pi_1} \, (-1)^{|\pi_{1,2}|} \, p_0^{|\pi_{1,0}|} \,
p_1^{|\pi_{1,1}|} \, p_2^{|\pi_{1,2}|} \ 
\sum_{\pi_2} \, (-1)^{|\pi_{2,2}|} \, p_0^{|\pi_{2,0}|} \,
p_1^{|\pi_{2,1}|} \, p_2^{|\pi_{2,2}|} \cr && \times \
\sum_{\pi_3} \, (-1)^{|\pi_{3,1}|} \, p_0^{|\pi_{3,2}|} \,
p_1^{|\pi_{3,0}|} \, p_2^{|\pi_{3,1}|} \ 
\sum_{\pi_4} \, (-1)^{|\pi_{4,1}|} \, p_0^{|\pi_{4,2}|} \,
p_1^{|\pi_{4,0}|} \, p_2^{|\pi_{4,1}|} \cr && \times \
\sum_{\pi_5} \, (-1)^{|\pi_{5,0}|} \, p_0^{|\pi_{5,1}|} \,
p_1^{|\pi_{5,2}|} \, p_2^{|\pi_{5,0}|}  \ .
\end{eqnarray}
By repeatedly applying (\ref{youngformula}) this partition function
can be expressed in closed form as a product of generalized MacMahon
functions. For example the sum
\begin{equation}
\sum_{\pi_3} \, (-1)^{|\pi_{3,1}|} \, p_0^{|\pi_{3,2}|} \, p_1^{|\pi_{3,0}|} \, p_2^{|\pi_{3,1}|}
=
\sum_{\pi_3} \, (-1)^{|\pi_{3,0}|} \, p_0^{|\pi_{3,2}|} \, (- p_1)^{|\pi_{3,0}|} \, (- p_2)^{|\pi_{3,1}|}
\end{equation}
is equal to (\ref{youngformula}) upon redefining $p_0 \rightarrow p_2$,
$p_1 \rightarrow -p_0$ and $p_2 \rightarrow -p_1$. One then finds
\begin{eqnarray}
\cZ_{\complex^2 / \zed_{3}\times\IC}\big(\mbf N = (2,2,1)\big) &=&
M(-q)^{15} \, \widetilde{M} (p_0 , - q)^3 \, \widetilde{M} (p_1 , - q)^3
\, \widetilde{M} (- p_2 , - q)^4 \cr && \times \,\widetilde{M} (p_0 \, p_1
, - q) \,\widetilde{M} (- p_1 \, p_2  , - q)^2 \, \widetilde{M} ( -
p_0 \, p_2  , - q)^2 \ .
\label{NDT221}\end{eqnarray}

By expanding these functions one obtains expressions for the noncommutative Donaldson--Thomas
invariants $\NDT_0(\mbf k,\mbf N)$ of type $\mbf N= (2,2,1)$ for the orbifold geometry
$\complex^2 / \zed_3\times \IC$.
These invariants, although numerically different from those of
the $U(1)$ gauge theory, can also be derived from rank one quiver
generalized Donaldson--Thomas invariants via the formula
(\ref{quivergeneralized}). By taking the logarithm of the partition
function (\ref{NDT221}) we obtain the free energy
\begin{eqnarray}
\cF_{\complex^2 / \zed_{3}\times\IC}\big(\mbf N = (2,2,1)\big)&=&
\sum_{n,l=1}^{\infty}\, (-1)^{n\,l}\, \frac n l \, \Big( 15 \,
p_0^{n\, l} \, p_1^{n\, l} \, p_2^{n\, l} + 3 \, p_0^{n\, l} \,
p_1^{(n+1)\, l} \, p_2^{n\, l} + 3 \, p_0^{n\, l} \, p_1^{(n-1)\, l}
\, p_2^{n\, l} \nonumber \\ && \qquad
+\, 2 (-1)^l \, p_0^{n\,l} \, p_1^{(n+1)\, l} \, p_2^{(n+1)\, l} + 2
(-1)^l \, p_0^{n\, l} \, p_1^{(n-1)\, l} \, p_2^{(n-1)\, l} \nonumber\\ && \qquad +\, 4 (-1)^l
\, p_0^{n\, l} \, p_1^{n\, l} \, p_2^{(n+1)\, l}+ 4 (-1)^l \,
p_0^{n\, l} \, p_1^{n\, l} \, p_2^{(n-1)\, l} \\ && \qquad
+\, 2 (-1)^l \, p_0^{(n+1)\, l} \, p_1^{n\, l} \, p_2^{(n+1)\, l} + 2
(-1)^l \, p_0^{(n-1)\, l} \, p_1^{n\, l} \, p_2^{(n-1)\, l}  \cr && \qquad
+\, 
3 \, p_0^{(n+1)\, l} \, p_1^{n\,l} \, p_2^{n\, l}  + 3 \, p_0^{(n-1)\,
  l} \, p_1^{n\, l} \, p_2^{n\, l} \nonumber \\ && \qquad
+ \, p_0^{(n+1)\, l} \, p_1^{(n+1)\, l} \, p_2^{n\, l} + p_0^{(n-1)\, l}
\, p_1^{(n-1)\, l} \, p_2^{n\, l} \Big) \ . \nonumber
\end{eqnarray}
By combining terms one finds that this expression indeed fits the
pattern of (\ref{quivergeneralized}). For example
\begin{eqnarray}
&& \sum_{n,l=1}^{\infty} \, (-1)^{n\, l} \, 3 \, \frac n l \, p_0^{n\, l}
\, p_1^{(n-1)\, l} \, p_2^{n\, l} + \sum_{n,l=1}^{\infty} \,
(-1)^{n\, l+l} \, 2 \, \frac n l \, p_0^{(n+1)\, l} \, p_1^{n\, l} \,
p_2^{(n+1)\, l} \cr && \qquad \qquad \ = \ 
\sum_{n,l=1}^{\infty} \, (-1)^{n\, l}\, \frac{3n + 2n - 2}{l} \,
p_0^{n\, l} \, p_1^{(n-1)\, l} \, p_2^{n\, l} \\[4pt]
&& \qquad \qquad \ = \ \sum_{n,l=1}^{\infty}\, (-1)^{2 n\, l + 2(n-1)
  \, l + n\, l} \ \big(2 n\, l + 2 (n-1)\, l + n\, l\big) \ \left(
  \frac{1}{l^2} \right) \, p_0^{n\, l} \, p_1^{(n-1)\, l} \, p_2^{n\,
  l} \ , \nonumber
\end{eqnarray}
which implies
\begin{equation}
\overline{\DT}_0(\mbf k) = - \frac{1}{l^2} \qquad \text{for} \quad
\mbf k = (n\, l, n\, l-l, n\, l) \qquad \text{with} \quad n,l\ge 1 \ .
\end{equation}

Proceeding in a similar way for the remaining terms, we obtain the
non-vanishing rank one quiver invariants
\beq
\overline{\DT}_0(\mbf k) = - \frac{1}{l^2} \qquad \text{for} \quad
\mbf k = \left\{ \begin{matrix} (n\,l,n\,l+l,n\,l) \ , \quad n\geq0 \
    , \ l\geq1 \ , \\ (n\,l-l,n\,l,n\,l) \ , \quad n,l\geq1 \ , \\ (n\,l+l,n\,l,n\,l) \ , \quad n\geq0 \
    , \ l\geq1 \ , \\ (n\,l-l,n\,l-l,n\,l) \ , \quad n,l\geq1 \ , \\ (n\,l+l,n\,l+l,n\,l) \ , \quad n\geq0 \
    , \ l\geq1 \ , \end{matrix} \right.
\eeq
and
\beq
\overline{\DT}_0\big(\mbf k=(k,k,k)\big) = -3 \, \sum_{l\geq1 \,:\,
  l|k}\, \frac1{l^2} \qquad \mbox{for} \quad k\geq1 \ .
\eeq
Comparing with (\ref{BPSformula}) we obtain the non-vanishing integer
BPS invariants
\beq
\widehat{\DT}_0(\mbf k)= -1 \qquad \text{for} \quad
\mbf k = \left\{ \begin{matrix} (n, n-1, n) \ , \quad n\geq1 \ ,
    \\ (n,n+1,n) \ , \quad n\geq0 \
    , \\ (n-1,n,n) \ , \quad n\geq1 \ , \\ (n+1,n,n) \ , \quad n\geq0 \
    , \\ (n-1,n-1,n) \ , \quad n\geq1 \ , \\ (n+1,n+1,n) \ , \quad n\geq0 \
    , \end{matrix} \right.
\eeq
and
\beq
\widehat{\DT}_0\big(\mbf k=(k,k,k)\big) = -3 \qquad \mbox{for} \quad k\geq1 \ .
\eeq
These invariants agree with those obtained in~\cite{joycesong} from
the rank one partition function (\ref{youngformula}).

We can similarly generalize our arguments to get a compact solution
for the noncommutative Donaldson--Thomas invariants for any
$A$-fibred threefold of the form  $\complex^2 / \zed_n\times \IC$ and
any boundary condition fixed by a framing vector $\mbf N$. The
partition function then assumes the form
\begin{equation}
\cZ_{\complex^2 / \zed_{n}\times \IC}(\mbf N ) = \sum_{\vec \pi}\,
(-1)^{ \sum_{l=1}^N\ \sum_{r=0}^{n-1}\, |\pi_{l,r}| \ N_{r+b(l)}} \
p_0^{\sum_{l=1}^N\, |\pi_{l,0-b(l)}| } \ \cdots \
p_{n-1}^{\sum_{l=1}^N \, |\pi_{l,n-1-b(l)}| } \ .
\label{ZC2mbfN}\end{equation}
As we did above, for a given fixed framing vector $\mbf N$ it is possible to express
this formula in a closed form as a product of MacMahon
functions $M(q)$ and $\widetilde{M}(x,q)$. The corresponding quiver invariants, independent of $N$ and $\mbf N$, can again be computed
explicitly from the formulas (\ref{quivergeneralized}) and
(\ref{BPSformula}). The integer BPS invariants in this case are
computed in~\cite[Section~7.5.4]{joycesong} from the rank one
partition function~(\ref{ZAnproduct}).

\subsection{Higher rank wall-crossing formulas}

We will now describe the wall-crossing formula for Coulomb branch
invariants. Although the wall contributions are all contained in the
quiver BPS invariants $\widehat{\DT}_\mu(\mbf k)$, which are unchanged
by wall-crossing in this case~\cite{joycesong}, it is interesting
to examine if the noncommutative invariants $\NDT_\mu(\mbf k,\mbf N)$
have wall-crossings of their own. Here we consider only very
particular walls of stability. In general, there will be walls
corresponding to separated D6~branes colliding and forming a bound
state; these walls are not included in our analysis below, and to get
them one should use a non-primitive wall-crossing formula. On the
other hand, since the D6~branes are well-separated in the Coulomb
branch, it is reasonable to expect that the walls affecting D2--D0
bound states are reached before the walls corresponding to D6 bound states.

The large radius partition function
for Coulomb branch invariants can be computed from the $U(N)$
instanton contributions to the noncommutative gauge
theory on the ALE resolution $X$ of the
$A_{n-1}$-fibration~\cite{Cirafici:2008sn}. It is given by a simple
modification of the partition function for topological string theory
on $X$ as
\beq
K_X^{\rm DT}(q,p_1,\dots,p_{n-1};N)= M\big((-1)^N\, q\big)^{n\, N}\,
\prod_{0< r\leq s<n}\, M\big(p_{[r,s]}\,,\, (-1)^N\,q\big)^N \ .
\eeq
From the wall-crossing formula in the rank one case and the explicit
expression (\ref{ZC2mbfN}), it follows that this function is related to the partition
function for Coulomb branch invariants in the case that the asymptotic
Higgs fields $e_l$ all transform in the trivial representation
$\rho_0$ of the orbifold group $\Gamma=\IZ_n$, i.e. $b(l)=0$ for all $l=1,\dots,N$, or equivalently $N_0=N$
while $N_r=0$ for all $r=1,\dots,n-1$. One then finds the non-abelian
wall-crossing formula
\bea
&& \cZ_{\complex^2 / \zed_{n}\times \IC}\big(q,p_1,\dots,p_{n-1}\,;\,
\mbf N= (N,0,\dots,0) \big)  \label{nonabinfwallcross} \\ && \qquad \qquad \ = \ M\big((-1)^N\, q\big)^{-n\, N}\, K_X^{\rm
  DT}(q,p_1,\dots,p_{n-1};N)\, K_X^{\rm
  DT}(q,p_1^{-1},\dots,p^{-1}_{n-1};N) \ . \nonumber
\eea

The wall-crossing factor $M\big((-1)^N\, q\big)^{-n\, N}\, K_X^{\rm
  DT}(q,p_1^{-1},\dots,p^{-1}_{n-1};N)$ in (\ref{nonabinfwallcross})
describes the crossing of an infinite number of walls of marginal
stability, separating different chambers in the K\"ahler moduli space,
in going from the orbifold point to the large radius point. To
identify the individual walls, let us first recall the situation in
the $U(1)$ gauge theory as studied in~\cite{gholampour} (see also~\cite{Sulkowski:2009rw}). In this context the walls are determined by the affine Lie algebra structure associated with the McKay quiver. If we denote by $\mbf\theta$ a $\theta$-stability parameter for the (unframed) quiver, then the walls of marginal stability are determined by
\begin{equation}
\cW_{\hat{\mbf k}} = \big\{ \mbf\theta \in \real^{n} \ \big\vert \
\mbf\theta \cdot \hat{\mbf k} = 0 \ , \ \hat{\mbf k} \in \hat{\Delta}^+ \big\}
\end{equation}
where $\hat\Delta^{+}$ is the set of affine positive roots. These
walls connect different chambers. Of particular relevance among them
is the wall determined by the imaginary root $\hat{\mbf k}{}^{\rm
  im}$ of the affine $A_{n-1}$ Lie algebra, which corresponds to
the regular representation of $\Gamma=\IZ_n$ and separates the Donaldson--Thomas chamber from the Pandharipande--Thomas chamber. Our wall-crossing formulas do not include this wall.

Fix a real positive root $\mbf k$ and consider stability parameters on both
sides of the associated wall, $\mbf\theta_{\mbf k}^+$ and
$\mbf\theta_{\mbf k}^-$. Then our vector space $V$ can be
identified with the unique $\mbf\theta_{\mbf k}$-stable module
over the path algebra of dimension vector $\mbf k$ constructed
in~\cite{gholampour}. Therefore the wall-crossing formula
of~\cite[Theorem~4.15]{gholampour} is given by
\begin{equation}
\cZ_{\complex^2 / \zed_{n}\times \IC}^{\mbf\theta_{\mbf k}^-} (\mbf p) = \big( 1 - (-1)^{k_0}\, \mbf
p^{\mbf k} \big)^{- k_0} \, \cZ_{\complex^2 / \zed_{n}\times \IC}^{\mbf\theta_{\mbf k}^+} (\mbf p) \ ,
\end{equation}
where the instanton charge $k_0=|\pi_0|$ is singled out by the framing
condition. It follows from (\ref{nonabinfwallcross}) that the proof
of~\cite[Theorem~4.15]{gholampour} can be adapted to our more general
situation to give
\begin{equation}
\cZ_{\complex^2 / \zed_{n}\times \IC}^{\mbf\theta_{\mbf k}^-}\big(\mbf
p\,;\,
\mbf N=(N,0,\dots,0) \big) = \big( 1 - (-1)^{k_0\, N}\, \mbf p^{\mbf
  k}  \big)^{- k_0\, N}\, \cZ_{\complex^2 / \zed_{n}\times
  \IC}^{\mbf\theta_{\mbf k}^+}\big(\mbf p \,;\,
\mbf N=(N,0,\dots,0) \big) \ ,
\end{equation}
which establishes the wall-crossing formula for noncommutative Donaldson--Thomas
invariants of type $\mbf N=(N,0,\dots,0)$.

This formula is just a mild
generalization of the Kontsevich--Soibelman wall-crossing formula. In
this case the invariants all jump together; in each $U(1)$ sector
the wall-crossing formula is the same up to a redefinition of the
assignments of parameters to irreducible representations and some
signs. Thus each $U(1)$ factor jumps
separately while the walls are the same for every sector. We do not
know how to extend these considerations to generic framing vectors
$\mbf N$ corresponding to more general boundary conditions on the
Higgs fields.

\section{An example with compact four cycles: \ $\complex^3 / \zed_3$\label{sec:C3Z3}}

\subsection{Geometry and representation theory}

\begin{figure}[h]
 \centering
  \includegraphics[width=7cm]{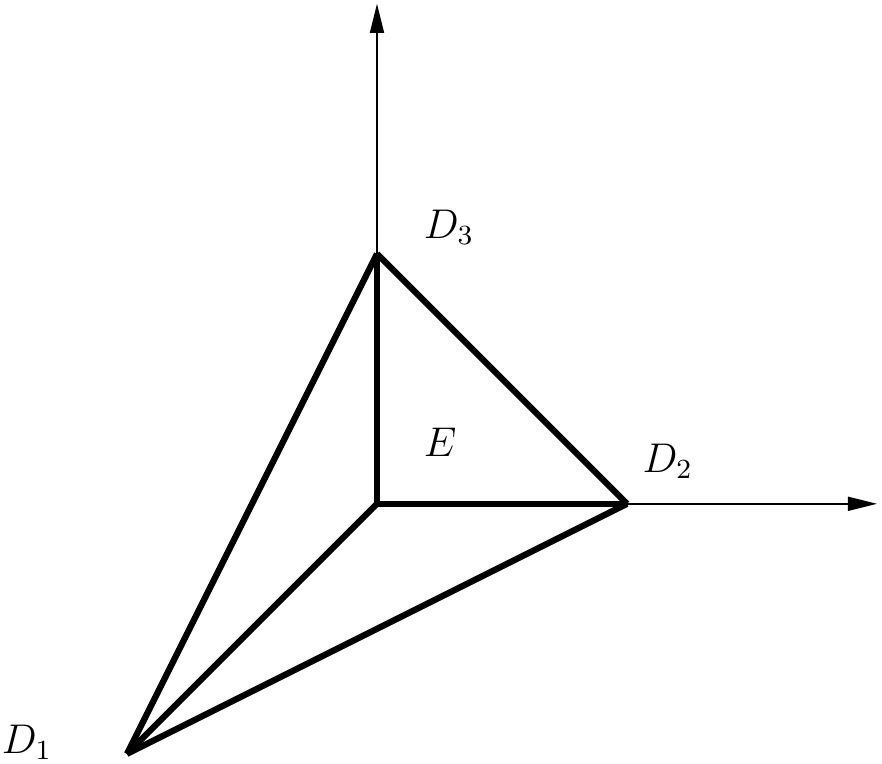}
 \caption{Toric fan for the local del~Pezzo surface of degree zero.}
 \label{C3z3}
\end{figure}

Our next example is the $\complex^3 / \zed_3$ orbifold with weights
$r_1=r_2=r_3=1$ that was studied in Section~\ref{subsec:InstC3Z3} and
Section~\ref{subsec:NCcrepant}; its unique Calabi--Yau crepant
resolution given by the
$\zed_3$-Hilbert scheme $X= \mathrm{Hilb}^{\zed_3} (\complex^3)$ is the
total space of the fibration $\cO_{\PP^2}(-3) \rightarrow \PP^2$, also
known as the local del~Pezzo surface of degree zero. This geometry has
one compact divisor, the base $E \cong \PP^2$ of the fibration, and
three rational curves which are homologous. Its fan
$\Sigma\subset\IZ^3$ is generated by the vectors
\begin{equation}
D_1 = (-1  , 1 , 1 ) \ , \qquad D_2 = (1,0,1) \qquad \mbox{and} \qquad D_3 = (0,1,1)
\ ,
\end{equation}
and the toric diagram for the resolved geometry is depicted in Figure~\ref{C3z3}.
The linear equivalences between the divisors are given by
\begin{equation}
D_\a \sim D_\b \qquad \mbox{and} \qquad 3 D_\a + E \sim 0 \qquad \mbox{for} \quad
\a,\b=1,2,3  \ .
\end{equation}
The Mori cone has a single generator
\begin{equation}
C = D_1 \cdot E = D_2 \cdot E = D_3 \cdot E
\end{equation}
dual to the generator of the K\"ahler cone which is given by any of the
linearly equivalent non-compact divisors $D_\a$. In particular $E\cdot
E\cdot E= 9$.

The basis of tautological bundles is indexed by the irreducible
representations $\rho_r$ with $r=0,1,2$. The defining representation,
which describes the action (\ref{Z3C3action}) of $\zed_3$ on the
coordinates of $\complex^3$, is $Q = \rho_1 \oplus \rho_1 \oplus
\rho_1$. The tensor product decompositions (\ref{tensordecomp}) for
$i=1,2$ 
yield the matrices
\begin{equation}
 \big(a_{rs}^{(1)}\big) = \left( \begin{matrix} 0 & 0 & 3 \\ 3 & 0 & 0 \\  0 & 3 & 0 \end{matrix} \right)
 \qquad \mbox{and} \qquad 
 \big(a_{rs}^{(2)}\big) = \left( \begin{matrix} 0 & 3 & 0 \\ 0 & 0 & 3 \\  3 & 0 & 0 \end{matrix} \right)
\label{C3Z3matrices}\end{equation}
with $a_{rs}^{(2)}= a_{sr}^{(1)}$. The associated quiver is given in (\ref{quiverC3Z3}).

If we write the original coordinates of $\complex^3$ as
$(z_1,z_2,z_3)$, then the rational curves are locally described by
invariant ratios of monomials of the form $z_1/z_2$ and cyclic
permutations thereof. Therefore they correspond to the character
$\rho_1$, and a generator of $H^2 (X , \zed)$ dual to the curve class $C$ is
given by $c_1 (\cR_1)$. In the fan $\Sigma$ there are now three toric
curves intersecting in a vertex of valence~$3$ and therefore, following the decoration procedure described in Section~\ref{subsec:CohHilb}, we
associate the character $\rho_2 = \rho_1 \otimes \rho_1$ to the
vertex. This gives one relation in the Picard group $\mathrm{Pic}
(X)$, namely $\cR_2 = \cR_1 \otimes \cR_1$. In particular the second
Chern class of $\cV = \left( \cR_1 \oplus \cR_1 \right) \ominus \left(
  \cR_2 \oplus \cO_X \right)$ generates $H^4 (X , \zed)$ and is dual
to the exceptional divisor with
\begin{equation}
\int_{\PP^2}\,  c_2 (\cV) = 1 \ .
\end{equation}
In this case we have $c_1 (\cR_2) = 2 c_1 (\cR_1)$ and $c_2 (\cV) = c_1 (\cR_1)\w c_1(\cR_1)$, with $\cR_\a=\cO_X(\a\,D)$ where $D$ is one of the linearly equivalent divisors $D_\a$ which is dual to the class $c_1(\cR_1)$ corresponding to~$\PP^2$.

\subsection{BPS partition functions}

We begin again with the $U(1)$ gauge theory. 
As explained in Section~\ref{subsec:Matrixeq} we need only consider the boundary condition given by the framing vector $\mbf N=(1,0,0)$. In the present geometry we can also consider a term involving the first Chern class, which measures the D4~brane charge. Firstly, having determined both a basis of $H^4 (X , \zed)$ and $H^2 (X , \zed)$ we can write
\begin{eqnarray}
\omega= \varphi \ c_1 (\cR_1) \qquad \mbox{and} \qquad
\omega \wedge \omega= \varsigma \ c_2 (\cV) = \varsigma \  c_1 (\cR_1)\w c_1(\cR_1) \ .
\end{eqnarray}
From the general form of the instanton action (\ref{c1action}) we have
\begin{eqnarray}
\int_{\overline{X}}\, \omega \wedge \omega \wedge c_1 (\cE) &=& - \varsigma\, \sum_{r,s=0}^2\, \left( a^{(1)}_{rs} - a^{(2)}_{rs} \right) \, k_s \, \int_\oX\, c_2 (\cV) \w c_1 (\cR_r) \nonumber\\[4pt] &=& \varsigma \, \left( 3 k_0 - 6 k_1 + 3 k_2 \right)\, \int_\oX\, c_1 (\cR_1)\w c_1 (\cR_1)\w c_1 (\cR_1) \ .
\end{eqnarray}
On the other hand since $E\cdot E\cdot E = 9$ and $E \sim - 3 D$, we see that
\begin{equation}
\int_\oX\, c_1 (\cR_1)\w c_1 (\cR_1)\w c_1 (\cR_1) = D \cdot D \cdot D =  - \frac 1 3 \ .
\end{equation}

Similarly, from (\ref{ch2action}) one has
\begin{eqnarray}
\int_\oX\, \omega \wedge \ch_2 (\cE) &=& - \varphi \, \sum_{r,s=0}^2\,\bigg( \left( a^{(1)}_{rs} - a^{(2)}_{rs} \right) \ k_s \ \int_\oX\, c_1 (\cR_1) \w \ch_2 (\cR_r) \cr & &\qquad\qquad\qquad + \, \left( a^{(2)}_{rs} - 3 \, \delta_{rs} \right) \ k_s \, \int_\oX\, c_1 (\cR_1) \w c_1 (\cR_r) \ c_1 \big(\cO_\oX (1)\big) \bigg) \nonumber\\[4pt] &= & -\mbox{$ \frac 1 6$}  \, \varphi \, \left( 9 k_0 - 12 k_1 + 3 k_2 \right) \cr & & -\, \varphi\, \left( 6 k_0 - 3 k_1 - 3 k_2 \right) \, \int_\oX\, c_1 (\cR_1)\w c_1(\cR_1) \w c_1 \big(\cO_\oX (1)\big) \ .
\end{eqnarray}
The last term of the instanton action (\ref{ch3action}) reads
\begin{eqnarray}
\int_\oX\, \ch_3 (\cE) &=& - \sum_{r,s=0}^2\, \bigg(\left(
  a^{(1)}_{rs} - a^{(2)}_{rs} \right)\, k_s\, \int_\oX\, \ch_3 (\cR_r)
- \left( a^{(2)}_{rs} - 3 \delta_{rs} \right) \, k_s \, \int_\oX\, c_1
(\cR_r) \w \ch_2 \big(\cO_\oX(1)\big) \cr && \qquad \qquad +\, \left(
  a^{(2)}_{rs} - 3 \delta_{rs} \right) \ k_s \ \int_\oX\, \ch_2
(\cR_r) \w c_1 \big(\cO_\oX(1)\big) -\delta_{rs}\, \frac{k_s}3 \bigg) \\[4pt] &=& - \mbox{$\frac 1 6$}\, \left( 7 k_0 - 8 k_1 + k_2 \right) + \mbox{$\frac 1 2$}\, \left( 6 k_0 - 3 k_1 - 3 k_2 \right) \ \int_\oX\, c_1 (\cR_1) \w c_1 \big(\cO_\oX(1)\big)\w c_1 \big(\cO_\oX(1)\big) \cr && -\, \mbox{$\frac 1 2$}\, \left( 12 k_0 - 3 k_1 - 9 k_2 \right) \ \int_\oX\, c_1 (\cR_1)\w c_1(\cR_1) \w c_1 \big(\cO_\oX(1)\big) + \mbox{$\frac 1 3$}\, \left( k_0 + k_1 + k_2 \right) \ . \nonumber
\end{eqnarray}
To evaluate the intersection indices arising here we work as follows. This geometry contains a compact divisor. By using linear equivalence, it is possible to rephrase the analysis of Section~\ref{subsec:C2Z3BPS} in such a way that only compact cycles enter, the divisor and rational curves. Therefore all the integrals involving $c_1 (\cO_\oX (1))$ and $c_1 (\cR_1)$ compute the intersection indices between elements of the exceptional locus with the divisor at infinity. We \textit{assume} that the compactification divisor $\bdiv$ is chosen so that these intersections vanish. Under this assumption, the instanton action finally becomes
\begin{eqnarray}
\int_\oX\, \ch_3 (\cE) &=& - \mbox{$\frac 1 6$}\, \left(5 k_0 - 10 k_1 - k_2 \right) \ , \nonumber \\[4pt]
\int_\oX\, \omega \wedge \ch_2 (\cE) &=& - \mbox{$\frac 1 2$}\, \varphi \, \left( 3 k_0 - 4 k_1 + k_2  \right) \ , \nonumber \\[4pt]
\int_\oX\, \omega \wedge \omega \wedge c_1 (\cE) &=& - \varsigma\, \left( k_0 - 2 k_1 + k_2 \right) \ .
\end{eqnarray}

We compute the instanton measure by taking the
$\zed_3$-invariant part of the character (\ref{character}).
For this, let us decompose
$V_\pi=\textsf{V}_0\oplus \textsf{V}_1\oplus \textsf{V}_2$
at a fixed point $\pi$ according to the $\zed_3$-action as before, where now each term is of order three. The explicit form of the partial character (\ref{Anpartialchar}) is now
\begin{eqnarray}
\mathcal{T}_{\pi}^+ &=& (\textsf{V}_0 \oplus \textsf{V}_1 \oplus \textsf{V}_2 ) \\ && 
-\, \Big(\, \frac{1}{t_1\, t_2} - \frac{1}{t_2} - \frac{1}{t_1} +1
\, \Big) \, \big( \textsf{V}_0
\otimes \textsf{V}^\vee_0 \ \oplus \ \textsf{V}_0 \otimes \textsf{V}^\vee_1 \ \oplus \ 
\textsf{V}_0 \otimes \textsf{V}^\vee_2 \ \oplus \ \textsf{V}_1
\otimes \textsf{V}^\vee_0 \nonumber \\ && \qquad \qquad \qquad \qquad \oplus\ \textsf{V}_1 \otimes \textsf{V}^\vee_1 \ \oplus \ 
\textsf{V}_1 \otimes \textsf{V}^\vee_2 \ \oplus \ \textsf{V}_2
\otimes \textsf{V}^\vee_0 \ \oplus \ \textsf{V}_2 \otimes \textsf{V}^\vee_1 \ \oplus \ 
\textsf{V}_2 \otimes \textsf{V}^\vee_2 \big) \ . \nonumber
\end{eqnarray}
It is easy to find the invariant part by substituting $t_\a = \zeta$ to get the virtual dimension (modulo~$2$)
\begin{eqnarray}
\big(\mathcal{T}_\pi^+\big)^{\zed_3} &=& {\rm vdim}_\IC\big(\textsf{V}_0 - ( \textsf{V}_0
\otimes \textsf{V}^\vee_0 \ \oplus \ \textsf{V}_1 \otimes \textsf{V}^\vee_1 \ \oplus \ 
\textsf{V}_2 \otimes \textsf{V}^\vee_2 ) \nonumber \\ && +\,2 (
\textsf{V}_0 \otimes \textsf{V}^\vee_2 \ \oplus \ \textsf{V}_1
\otimes \textsf{V}^\vee_0 \ \oplus \ \textsf{V}_2 \otimes \textsf{V}^\vee_1
) - ( \textsf{V}_0 \otimes \textsf{V}^\vee_1 \ \oplus \ \textsf{V}_1
\otimes \textsf{V}^\vee_2 \ \oplus \ \textsf{V}_2 \otimes \textsf{V}^\vee_0) \big) \nonumber \\[4pt] &=& |\pi_1| + 
|\pi_2| + |\pi_0| \, |\pi_1| + |\pi_0| \, |\pi_2| + |\pi_1|\, |\pi_2| \ ,
\label{C3Z3char}\end{eqnarray}
where we recall that the fixed points are classified by coloured partitions $\pi
=\pi_0 \sqcup \pi_1 \sqcup \pi_2$. This result agrees with that quoted in~\cite[Remark~A.5]{young}, and it shows that the
equivariant Euler characteristic of the obstruction bundle on the
quiver variety at a fixed point $\pi$ is given by
\begin{equation}
\chi_{\torus^3}(\mathcal{N}_\pi) = (-1)^{|\pi_1| + |\pi_2| + |\pi_0|\, |\pi_1|
+ |\pi_0|\, |\pi_2| + |\pi_1|\, |\pi_2|} \ .
\end{equation}

We can construct a partition function for these Euler
characteristics. As we did in Section~\ref{subsec:C3Z2Z2BPS}, we are led to define
\begin{equation}
K^{\rm DT}_{\complex^3 / \zed_3} = \sum_{\pi} \, (-1)^{|\pi_1| + |\pi_2| + |\pi_0|\, |\pi_1| + |\pi_0|\, |\pi_2| + |\pi_1|\,
|\pi_2|}\, p^{|\pi|}\, p_0^{|\pi_0|} \,
p_1^{|\pi_1|} \, p_2^{|\pi_2|} \ .
\label{KDTC3Z3}\end{equation}
Here $p$ is a formal parameter that weighs the number of boxes in a plane partition $\pi$, while the other formal variables $p_r$, $r=0,1,2$ keep track of the decomposition of the dimension vector
$\mbf{k} = (k_0 , k_1 , k_2)$. The first few terms of 
this partition function can be calculated explicitly to be
\begin{eqnarray}
K^{\rm DT}_{\complex^3 / \zed_3} &=& 1 + p\, p_0 + 3
p^2\, p_0\, p_1 + 3 p^3\, p_0\, p_1^2 - 3 p^3\,
p_0 \, p_1\, p_2 \cr && +\, 9 p^4\, p_0\,
p_1^2\, p_2 + p^4 \, p_0\, p_1^3 - 3 p^4\,
p_0^2\, p_1 \, p_2 \cr && -\, 6 p^5\, p_0 \,
p_1^3\, p_2 + 9 p^5\, p_0\, p_1^2 \,
p_2^2 - 9 p^5\, p_0^2\, p_1^2\, p_2 \cr && -\,
9 p^6\, p_0^2\, p_1^3\, p_2 + 15 p^6\, p_0\,
p_1^3\, p_2^2 + 21 p^6\, p_0^2\, p_1^2 \,
p_2^2 + 3 p^6\, p_0\, p_1^2\, p_2^3 + \cdots
\nonumber \\[4pt] &=& \Big( 1 + 3 (-p^3\, p_0\, p_1\, p_2) + 12
(-p^3\, p_0\, p_1\, p_2)^2 +  \cdots \Big) \cr &&
\times \, \Big( 1 + p\, p_0 + 3 p^2\, p_0\, p_1 + 3
p^3 \, p_0 \, p_1^2 + p^4\, (9 p_0\, p_1^2\,
p_2 + p_0\, p_1^3) \cr && \qquad +\, p^5 \, (9 p_0\,
p_1^2\, p_2^2 - 6 p_0 \, p_1^3\, p_2)
\cr && \qquad +\, p^6\, (15 p_0\, p_1^3\, p_2^2 + 9
p_0^2\, p_1^2\, p_2^2 + 3 p_0\, p_1^2\,
p_2^3) + \cdots \Big) \ .
\label{KDTC3Z3exp}\end{eqnarray}
In this expression we recognise a factor of the MacMahon function raised
to the power of the topological Euler characteristic $\chi(X)=\chi(\PP^2)=3$ of the target space.
Indeed the generating
function of $\zed_3$-invariant holomorphic polynomials decomposes
as
\begin{eqnarray}
&& \frac{1}{3}\, \sum_{r=0}^2\, \frac{1}{(1-\zeta^r\, z_1) \, (1 -
\zeta^r\, z_2) (1-\zeta^r\, z_3)} \\ && \qquad \ = \ \frac{1}{(1 - \frac{z_1}{z_3}) \,
(1 - \frac{z_2}{z_3})\, (1 - z_3^3)} + \frac{1}{(1 -
\frac{z_1}{z_2})\, (1 - \frac{z_3}{z_2}) \, (1 - z_2^3)}  + \frac{1}{(1
- \frac{z_2}{z_1})\, (1 - \frac{z_3}{z_1})\, (1 - z_1^3)} \nonumber 
\end{eqnarray}
into three invariant copies of the generating function for $\complex^3$. We thus expect the
topological string partition function to contain the factor
\begin{equation}
M (x)^3 = 1 + 3\,x + 12\,x^2 + 37\,x^3 + 111\,x^4 + 303\,x^5 +
804\,x^6 + \cdots
\end{equation}
with $x=-q= -p^3\, p_0\, p_1\, p_2$, corresponding to contributions
from ``regular'' instantons (see Section~\ref{subsec:instaction}).

We will now compare the combinatorial partition function (\ref{KDTC3Z3})--(\ref{KDTC3Z3exp}) with our BPS partition function
\begin{eqnarray}
\mathcal{Z}_{\complex^3 / \zed_3} &=& \sum_{\pi}\, (-1)^{
|\pi_1| + |\pi_2| + |\pi_0|\, |\pi_1| + |\pi_0|\, |\pi_2| + |\pi_1|\,
|\pi_2|} \ q^{\frac 1 3\, |\pi| - \frac1 6\, (7 |\pi_0| - 8 |\pi_1| +|\pi_2|)} \nonumber\\ && \qquad \qquad \qquad \times \ Q^{\frac 1 2\, (3 |\pi_0| - 4 |\pi_1| + |\pi_2|)} \ U^{|\pi_0|-2 |\pi_1|+ |\pi_2|} \ ,
\end{eqnarray}
where $q=\e^{-g_s}$,  $Q= \e^{-\varphi}$ and $U=\e^{-\varsigma}$. A quick computation shows that the two partition functions are related by the change of variables
\begin{eqnarray}
p &=& q^{1/ 3} \ , \nonumber \\[4pt]
p_0 &=& q^{- 7 /6} \ Q^{3/ 2} \ U \ , \nonumber \\[4pt]
p_1 &=& q^{ 4/ 3} \ Q^{- 2} \ U^{-2} \ , \nonumber \\[4pt]
p_2 &=& q^{-1/ 6} \ Q^{1/ 2} \ U \ .
\label{changelocalP2}\end{eqnarray}
This BPS partition function contains contributions from non-vanishing D4~brane charge and has an expansion
\begin{eqnarray}
\mathcal{Z}_{\complex^3 / \zed_3}&=& -
\frac{15 q^{9/2}}{U^3\, Q^{7/2}}-\frac{6 q^{13/3}}{U^4 \,Q^4}+\frac{q^{25/6}}{U^5\, Q^{9/2}}-\frac{9 q^{7/2}}{U^3\, Q^{5/2}}+\frac{3 q^3}{Q}+\frac{9 q^{17/6}}{U\,
   Q^{3/2}}+\frac{9 q^{8/3}}{U^2\, Q^2} \nonumber
\\ \nonumber
&& +\, \frac{3 q^{5/2}}{U^3 \,Q^{5/2}} +21 q^2-\frac{9 q^{11/6}}{U\,Q^{1/2}}-3 q+\frac{3 q^{5/6}}{U\,Q^{1/2}}-3 U\, Q^{3/2} q^{1/6}+1+\frac{U\, Q^{3/2}}{q^{5/6}} + \cdots \\[4pt] &=& M(-q)^3\, \Big( 1 + \frac{U\,Q^{3/2}}{q^{5/6}}+ \frac{3 q^{5 /6}}{U \,Q^{1/2}} + \frac{3
q^{5/ 2}}{U^3 \, Q^{5/2}} + \frac{9 q^{8/3}}{U^2 \,Q^{2}} + \frac{q^{25/6}}{U^5 \,Q^{9/2}} \nonumber \\ && \qquad \qquad \qquad +\, \frac{9 q^{17/6}}{U\,Q^{3/2}}
- \frac{6 q^{13/3}}{U^4\,Q^4} + \frac{15 q^{9/ 2}}{U^3\,Q^{7/2}}+ 9 q^2 + \frac{3 q^3}{Q} + \cdots \Big) \ .
\end{eqnarray}

\subsection{Coulomb branch invariants}
 
We will now describe the noncommutative Donaldson--Thomas
invariants $\NDT_0(\mbf k,\mbf N)$ of type $\mbf N$ in the present case where the Calabi--Yau threefold has compact four-cycles. Although explicit closed (product) formulas are no longer available, it is possible to compute the invariants order by order as above for a fixed boundary condition labelled by the framing vector $\mbf N$.
In this case the instanton measure (\ref{instmeasure}) is given by
\begin{eqnarray} \label{instmeasureC3Z3}
\cK(\vec\pi;\mbf N) = \sum_{l=1}^N \ \sum_{r=0}^2\, |\pi_{l,r}| \ N_{r+b(l)} + \sum_{l,l'=1}^N \ \sum_{r=0}^2\, |\pi_{l,r}| \,\Big(|\pi_{l',r+b(l)-b(l'\,)-2}|+ |\pi_{l',r+b(l)-b(l'\,)}| \Big)
\end{eqnarray}
where we have dropped irrelevant even parity terms. The rank $N$ action is obtained from the $U(1)$ action by writing the instanton charges as in (\ref{NAinst}). Therefore the partition function of noncommutative Donaldson--Thomas invariants of type $\mbf N$ is
\begin{eqnarray}
\mathcal{Z}_{\complex^3 / \zed_3}( \mbf N) &= & \sum_{\vec \pi} \ (-1)^{\sum_{l=1}^N\ \sum_{r=0}^2\,|\pi_{l,r}| \, ( N_{r+b(l)} + \sum_{l'=1}^N \,( |\pi_{l',r+b(l)-b(l'\,)-2}| + |\pi_{l',r+b(l)-b(l'\,)}|) )
} \cr && \qquad \times \ q^{\frac1 6 \, \sum_{l=1}^N \, (-5   |\pi_{l,0-b(l)}|
 + 10 |\pi_{l,1-b(l)}|
 + |\pi_{l,2-b(l)}|
)} \\ && \qquad \times\ Q^{\frac 1 2 \, \sum_{l=1}^N \, (3 |\pi_{l,0-b(l)}|
 - 4 |\pi_{l,1-b(l)}|
 +|\pi_{l,2-b(l)}|
)} \ U^{\sum_{l=1}^N \, ( |\pi_{l,0-b(l)}|
 - 2 |\pi_{l,1-b(l)}|
 +|\pi_{l,2-b(l)}|
)} \ . \nonumber
\end{eqnarray}

These invariants, although related to the noncommutative
Donaldson--Thomas invariants, appear to be new. They differ from the
definitions of~\cite{joycesong} by the parameters involved. In our
formulation the invariants of~\cite{joycesong} have the form
\begin{eqnarray}
K^{\rm DT}_{\complex^3 / \zed_3}(\mbf N) &= & \sum_{\vec \pi} \ (-1)^{\sum_{l=1}^N\ \sum_{r=0}^2\,|\pi_{l,r}| \, ( N_{r+b(l)} + \sum_{l'=1}^N \,( |\pi_{l',r+b(l)-b(l'\,)-2}| + |\pi_{l',r+b(l)-b(l'\,)}|) )
} \cr && \qquad \qquad \times \ \tilde p_0^{\,\sum_{l=1}^N\, |\pi_{l,0-b(l)}|} \ \tilde p_1^{\, \sum_{l=1}^N\, |\pi_{l,1-b(l)}|
} \ \tilde p_2^{\, \sum_{l=1}^N\, |\pi_{l,2-b(l)}|} \ ,
\end{eqnarray}
where $\tilde p_r=p\, p_r=q^{1/3}\, p_r$ for $r=0,1,2$.
Our invariants are numerically different but are related via the
change of variables (\ref{changelocalP2}) which allows one set of
invariants to be expressed uniquely via the other set. Our formulation
seems however to be more physically motivated. The Coulomb branch
invariants are also related to the quiver generalized
Donaldson--Thomas invariants, and hence to the quiver BPS invariants,
via the formula (\ref{NCDTgenrel}). However, in this case the relative
Euler form (\ref{chibar}) is non-zero and hence explicit infinite
product forms for the partition functions are not available, thus
making the explicit determination of these invariants somewhat more
involved.

\subsection{Large radius partition functions}

We will now discuss the relationship between the orbifold and large
radius phases of the local $\PP^2$ geometry. The large radius
partition function for rank one BPS states is that of topological string theory on
$X=K_{\PP^2}=\cO_{\PP^2}(-3)$. It can be computed from the
topological vertex~\cite{marino} as
\be 
K_{X}^{\rm top}= \sum_{\lambda_1, \lambda_2, \lambda_3} \,
(-Q)^{\sum_\a\, 
|\lambda_\a|} \ q^{-\sum_\a\, \kappa_{\lambda_\a}} \
C_{\emptyset\lambda_2'\lambda_3}(q) \, C_{\emptyset\lambda_1'
\lambda_2}(q)\, C_{\emptyset\lambda_3'\lambda_1}(q)
\ee 
where the sum runs over ordinary partitions $\lambda=(\lambda_i)$ (Young
tableaux) labelling the three internal lines of the toric diagram dual
to the fan of Figure~\ref{C3z3},
with conjugate partitions $\lambda'$ and $\kappa_\lambda := |\lambda| +
\sum_{i}\, \lambda_i \, (\lambda_i - 2i)$, $Q=\e^{-t}$ 
with $t=\int_{\PP^2}\,\omega$ the K\"ahler parameter corresponding to the hyperplane class
in ${\mathbb P}^2$, and $q=\e^{-g_s}$. In the melting crystal
formulation, the topological vertex can be expressed as a sum over plane
partitions $\pi$ which asymptote to boundary partitions
$(\lambda_1,\lambda_2,\lambda_3)$ along the three coordinate axes
as~\cite{Okounkov:2003sp}
\beq
C_{\lambda_1\lambda_2\lambda_3}(q)= M(q)^{-1}\ q^{\frac12\,
  (\|\lambda_1'\|^2 + \|\lambda_2'\|^2 + \|\lambda_3'\|^2)} \
\sum_{\pi\, :\, \partial\pi=(\lambda_1,\lambda_2,\lambda_3)}\,
q^{|\pi|} \ ,
\eeq
where $\|\lambda\|^2:= \sum_i\, \lambda_i^2$ and in this expression
$|\pi|$ denotes the renormalized volume of the infinite
three-dimensional Young diagram $\pi$. 
We can rewrite this expansion in terms of Schur functions as
\bea 
K_{X}^{\rm top} &=&
\sum_{\lambda_1, \lambda_2, \lambda_3} \, (-Q)^{\sum_\a\,
|\lambda_\a|} \ q^{-\sum_\a\, \kappa_{\lambda_\a} / 2} \nn && \qquad
\qquad \times \ 
s_{\lambda_1'}(q^{\rho})\, s_{\lambda_2'}(q^{\rho})\,
s_{\lambda_3'}(q^{\rho}) \,
s_{\lambda_1'}(q^{\rho + \lambda_3'})\, s_{\lambda_2'}(q^{\rho +
  \lambda_1'}) \, 
s_{\lambda_3'}(q^{\rho + \lambda_2'})
\eea 
where $q^{\lambda+ \rho}:= (q^{\lambda_i -i + 1/2})$. In the
non-abelian gauge theory, the contributions from noncommutative $U(N)$
instantons in the Coulomb branch can be computed
following~\cite{Cirafici:2008sn} and yield the rank~$N$ BPS partition
function
\beq
K_X^{\rm DT}(q,Q;N)=K_X^{\rm top}\big((-1)^{N+1}\,q \, ,\, Q\big)^N \ .
\eeq
The orbifold phase is recovered by blowing down the compact divisor
${\mathbb P}^2$, i.e. by formally setting $Q\to1$ in the B-model
topological string theory, as explained below.

The D~brane bound states enumerated by these partition functions have no D4 charges, i.e. we should set $U=1$ in the
orbifold partition functions.
In contrast to our previous examples, here there is no closed form for
the partition function either at large radius or in the noncommutative
crepant resolution chamber, and the locations of the walls of marginal
stability are not known. However, wall-crossing is always described by
the Kontsevich--Soibelman formula, and in the present case since the
D6~brane charge is unity one can use the semi-primitive wall-crossing
formula. Note that for Calabi--Yau threefolds with compact
four-cycles, a simple formula such as (\ref{ZBPSqQ}) connecting the
orbifold point to the large radius point is not anticipated. In the
present case, this is because the geometry contains a divisor
which lies over the singular point of $\IC^3/\IZ_3$, namely the base
$\PP^2$ of the fibration, i.e. in this case the crepant resolution
$\pi:X\to \IC^3/\IZ_3$ is not semi-small. Hence the conditions of the
crepant resolution conjecture of~\cite{young,orbvertex} are not met;
the essence of the problem is that the additional non-vanishing
homology group $H_4(X)$ introduces more variables into the counting problem
on $X$ than is dictated by the classical McKay correspondence. Moreover, since in this case $\bar\chi\neq0$, the quiver BPS invariants $\widehat{\DT}_\mu(\mbf k)$ vary by wall-crossing formulas under changes of stability condition in the derived category ${\bf D}(X)$. To illustrate the moduli space phase structure of marginally stable D~brane states
on the Calabi--Yau ALE space $X=\cO_{\PP^2}(-3)$, we will now briefly review
the approach of~\cite{Aganagic:2006wq} to relating the large
radius and orbifold points in the context of topological string
theory. This approach follows the spirit of the crepant resolution conjecture for partition functions of orbifold Gromov--Witten invariants~\cite{bryan2}.

The setup of~\cite{Aganagic:2006wq} is the topological B-model.
The partition function can be interpreted as a wavefunction and
consequently has definite transformation properties on the
complex structure moduli space. If one knows the
topological string amplitude in some region of the moduli space
(say the large radius phase) then these transformation properties are enough
to compute it at any other point in the moduli space. Similarly if
one knows the partition function as a function of some coordinates
(say in the real polarization where the coordinates are the period
integrals given by the special geometry) then one can compute it as a
function of other coordinates (say in the holomorphic polarization
where the coordinates are given by the Hodge decomposition of the
holomorphic three-form $\Omega$). Going from one region to another in the
moduli space is in fact a change of coordinates. 

In terms of the quantum mechanical
system in which the topological string amplitude is a wavefunction,
this change of coordinates in the moduli space is a
canonical transformation. One interprets this relation as
a path integral computed perturbatively in terms
of Feynman diagrams. The propagator $\Delta$ is given by the generator
of the canonical transformation. The vertices are the derivatives of the free
energy $\mathcal{F}_g$ in the coordinates computed on the
saddle point. This prescription can be summarized by the rule
\begin{equation} \label{feynman}
\widetilde{F}_g = \mathcal{F}_g + {\mit\Gamma}_g \left( \Delta ,
\partial_{i_1} \cdots \partial_{i_n} \mathcal{F}_{r<g}\right) \ ,
\end{equation}
where $\mit\Gamma_g$ is a functional
obtained through the Feynman rules (only genus $r<g$ diagrams
contribute at fixed genus $g$). Then one solves for
$\widetilde{F}_g$ genus by genus.

One can now apply this construction to compute Gromov--Witten invariants at the
orbifold point $\mathbb{C}^3 / \mathbb{Z}_3$. Starting from the A-model on local $\mathbb{P}^2$, the orbifold
point is the point in the K\"ahler moduli space where the K\"ahler parameter $t\to0$. To reach this point one considers the
A-model in the large radius phase where classical geometry is a
good concept. Then one looks at the B-model on the mirror
manifold (see~\cite[Section~6]{Diaconescu:1999dt} for a review\footnote{In~\cite{Diaconescu:1999dt} the orbifold point is incorrectly set at $t\neq0$.}). The
mirror is described by a family of elliptic curves, as prescribed
by the rules of local mirror symmetry. The moduli space of this
family is one-dimensional and can be regarded as a projective line $\mathbb{P}^1$ with
three punctures at $z=0,1,\infty$. The point $z=0$ is the large
volume point where the exceptional divisor in the mirror
A-model has infinite size, the point $z= \infty$ is the
orbifold point with no blow-up, and $z=1$ is the conifold
point where the underlying worldsheet conformal field theory is singular. The B-model is solved by a three-dimensional vector of periods that satisfies the Picard--Fuchs
equation, which can be solved
in each of the three neighbourhoods around $z=0,1,\infty$. Then in
each of these neighbourhoods we have a set of good coordinates that
solve the Picard--Fuchs equation. 

The orbifold mirror map sends linear combinations of these solutions to a basis for the orbifold cohomology of $[\IC^3/\IZ_3]$. Thus if we know the topological string free
energy in the large radius neighbourhood $\mathcal{F}_g^{\infty}$ then
we can recover it around the orbifold point
$\mathcal{F}_g^{\rm orb}$ by using (\ref{feynman}) as
\begin{equation} \label{feynmanorb}
\mathcal{F}_g^{\rm orb} = \mathcal{F}_g^{\infty} +
{\mit\Gamma}_g \left( \Delta ,
\partial_{i_1} \cdots \partial_{i_n} \mathcal{F}_{r<g}^{\infty} \right)
\end{equation}
up to normalization. The left-hand and right-hand sides are computed with
different sets of coordinates. The genus $g$ A-model free energy $\cF_g^\infty$ at large radius generates genus $g$ Gromov--Witten invariants in homology classes of $H_2(X,\IZ)$, while
\begin{equation}
\mathcal{F}_g^{\rm orb} (\sigma) = \sum_{n\in 3\IN_0}\, \big\langle
(O_{\sigma})^n \big\rangle\, \frac{\sigma^n}{n!}
\label{calFgorb}\end{equation}
is the free energy at the orbifold point, where the topological
observable $O_{\sigma}$ corresponds to the generator of the orbifold cohomology group
$H_{\rm orb}^{1,1}([\mathbb{C}^3 / \mathbb{Z}_3])$ (see Appendix~\ref{App:Stacks}). The
closed string Hilbert space splits into three twisted sectors. The
orbifold action is given by (\ref{Z3C3action}) and the supersymmetric ground states of
the $k$-th twisted sector are given by the generators of the cohomology $H^{\bullet}(\zeta^k)$ of the
subset of $\mathbb{C}^3 / \mathbb{Z}_3$ that is invariant under the action of
$\zeta^k$. As we review in Appendix~\ref{App:Stacks}, this is the
cohomology of the inertia stack and the orbifold
cohomology of $\mathbb{C}^3 / \mathbb{Z}_3$ splits into the three
twisted sectors as
\begin{equation}
H^{\bullet}_{\rm orb} \big([\mathbb{C}^3 / \mathbb{Z}_3]\big) = H^{\bullet}
(\zeta^0) \oplus H^{\bullet+(1,1)} (\zeta^1) \oplus
H^{\bullet+(2,2)} (\zeta^2) \ ,
\end{equation}
where the constant shifts of the
degree are called (twice) the age. Note that here the generator $\zeta$ ``behaves like'' a divisor, while $\zeta^2=\bar\zeta$ ``behaves like'' a cycle in codimension two.

Thus knowledge of the large radius
free energy $\mathcal{F}_g^{\infty}$ as a function of the large
radius coordinates determines the relevant coordinates
around the orbifold points in the B-model as the
solutions of the Picard--Fuchs equation which are mirror to the
generators of $H^{k,k}_{\rm orb} ([\mathbb{C}^3 / \mathbb{Z}_3])$. By
the properties of mirror symmetry one can find also the functional
relation between the two sets of coordinates, and plugging everything into
(\ref{feynmanorb}) yields the Gromov--Witten invariants at the orbifold
point in terms of those at large radius.
 Some further mathematical details of this construction are described in Appendix~\ref{App:Stacks}.

In the general case, we also wish to enumerate cycles in codimension two.
In the absence of D6~branes, D~brane charges are related at the large radius and
orbifold points in~\cite{Douglas:1997de,Diaconescu:1999dt,DFR,Aspinwall:2004vm}.
In the large radius limit, the spectrum of BPS branes contains D4~branes on
$\PP^2$, which are classified by stable holomorphic vector bundles on
$\PP^2$. In this case the Beilinson monad construction (of the moduli space of coherent
sheaves on $\PP^2$) associates the Beilinson quiver
\beq
\vspace{4pt}
\begin{xy}
\xymatrix@C=20mm{
& \ v_0 \ \bullet \ \ar@/^/[ddl] \ar@/_0.5pc/[ddl] \ar@//[ddl]  & \\
& & \\
v_1 \ \bullet \ \ & &  \ \bullet \ v_2 \ar@/^/[uul] \ar@/_0.5pc/[uul] \ar@//[uul] 
}
\end{xy}
\vspace{4pt}
\label{quiverBeilinson}\end{equation}
with relations given by the first column of (\ref{C3Z3rels}). The representations of this quiver
correspond to coherent sheaves on $\PP^2$; more generally the derived category of coherent sheaves ${\bf D}(\PP^2)$ on $\PP^2$ is equivalent to the derived category of left modules over the path algebra of the Beilinson quiver. Using local mirror symmetry techniques similar to those sketched above, these approaches establish a correspondence between fractional branes at the orbifold point and bound states of D0, D2 and D4~branes (and their antibranes) described by vector bundles on the exceptional $\PP^2$ cycle.

In our case, the addition of a D6~brane corresponds to extending the
D4--D2--D0 quiver by a node; the gauge bundle on the D6~brane is taken
to be one of the tilting line bundles of
Section~\ref{subsec:Stability}~\cite{Aganagic:2010qr}. The direct image map from coherent $\cO_{\PP^2}$-modules to coherent $\cO_X$-modules induced by the embedding $\PP^2\hookrightarrow X$ maps objects of
${\bf D}(\PP^2)$ injectively to objects in ${\bf D}(X)$. However, there can be
more morphisms in the derived category ${\bf D}(X)$. As explained
in~\cite{Aspinwall:2004vm}, in order to account for the extra open
strings induced by the embedding one needs to ``complete'' the
Beilinson quiver (\ref{quiverBeilinson}) with additional arrows. The
completed gauge quiver in this case is precisely the McKay quiver
(\ref{quiverC3Z3}) of $\IC^3/\IZ_3$. Conversely, since any vector
bundle retracts to its zero section, certain topological
characteristics on $X=\cO_{\PP^2}(-3)$ are determined entirely by
those of the exceptional divisor $E= \PP^2$. Indeed, in~\cite{DFR} it is demonstrated
that holomorphic objects near the orbifold point come from
representations of the Beilinson quiver, or equivalently from large
volume gauge sheaves. Using these facts and local mirror symmetry, it
should be possible to map orbifold and large radius phase objects into
one another, along the lines of~\cite{Aganagic:2010qr}.

In order to make sense of the (argument of the) central charge of the
non-compact D6~branes, one needs to consider the local $\PP^2$
geometry as a large radius limit of a compact Calabi--Yau
space. According to~\cite{Jafferis:2008uf}, the (conjecturally) proper
limit involves an extra parameter: a component of the $B$-field normal
to the base survives the local limit. This parameter plays a crucial
role in stability and wall-crossing analyses. This would also explain
why the large radius Donaldson--Thomas theory behaves in such a
complicated manner when trying to approach the orbifold phase, even
though slope stability is trivial for ideal sheaves: the new stability
condition crucially involves this extra parameter. Furthermore, there
are various terms in the Dirac--Born--Infeld action which correspond
to the possible deformations of the D~brane inside the Calabi--Yau
manifold. In the local limit, which corresponds to zooming in on a
neighbourhood of $\PP^2$ in $X$, many of these terms should be dropped
since the brane in the local geometry has much fewer allowed
deformations. However, by the above arguments, the result of the local
limit is \emph{not} the six-dimensional Yang--Mills theory we have
been considering.

\section{The $\complex^3 / \zed_6$ orbifold\label{sec:C3Z6}}

\subsection{Geometry and representation theory}

\begin{figure}[h]
 \centering
  \includegraphics[width=7cm]{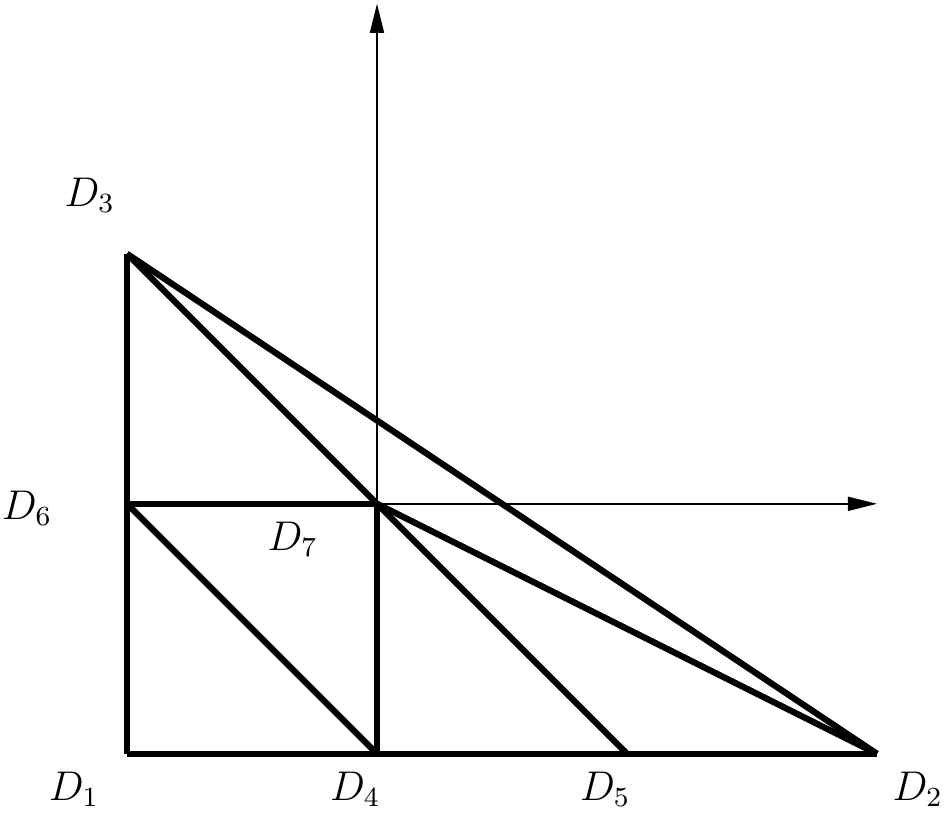}
 \caption{Toric fan for the resolution of the $\complex^3 / \zed_6$ singularity used in the main text.}
 \label{C3z6}
\end{figure}

Our final example is the orbifold $\complex^3 / \zed_6$ with weights
$r_1=1$, $r_2=2$, $r_3=3$. It has a toric diagram given by three lattice vectors
\begin{equation}
D_1 = (-1,-1,1) \ , \qquad D_2 = (2 , -1, 1) \qquad \mbox{and} \qquad D_3 = (-1 , 1 , 1)
\end{equation}
which represent an isolated quotient singularity. We consider the crepant resolution given by the Hilbert scheme $X=\mathrm{Hilb}^{\IZ_6} (\complex^3)$ which is obtained by adding the vectors
\begin{equation}
D_4 = (0,-1,1) \ , \qquad D_5 = (1,-2,1) \ , \qquad D_6 = (-1,0,1) \ , \qquad D_7 = (0,0,1)
\end{equation}
and the appropriate triangulation shown in Figure~\ref{C3z6}. 
This is only one of the five distinct possible crepant resolutions. 

The resolved geometry has six non-compact divisors $D_i$ with $i=1,\dots, 6$ and one compact divisor $D_7$ with the topology of a Hirzebruch surface blown up at one point, together with the linear equivalences
\begin{eqnarray}
6 D_1 + D_7 + 2 D_6 + 4 D_4 &\sim& 0 \ , \nonumber \\[4pt]
3 D_2 + D_7 + 2 D_5 + D_4 &\sim& 0 \ , \nonumber \\[4pt]
2 D_3 + D_7 + D_6 &\sim& 0 \ ,
\end{eqnarray}
which we use to solve for the divisors
\begin{eqnarray}
D_5 &\sim& -D_1 -D_2+D_3 - D_4 \ , \nonumber \\[4pt]
D_6 &\sim& -2 D_1 + D_2 - D_4 \ , \nonumber \\[4pt]
D_7 &\sim& 2 D_1 - D_2 - 2 D_3 + D_4 \ .
\end{eqnarray}
The non-vanishing intersection numbers between three \textit{distinct} divisors can be read off from the toric diagram by checking whether or not the three divisors lie at the corners of a basic triangle. This gives
\begin{eqnarray}
D_2 \cdot D_7 \cdot D_3 = 1 \ , \qquad D_3 \cdot D_7 \cdot D_6 = 1 \ , \qquad D_1 \cdot D_6 \cdot D_4 = 1 \ , \nonumber \\[4pt] D_4 \cdot D_7 \cdot D_6 = 1 \ , \qquad D_7 \cdot D_5 \cdot D_4 = 1 \ , \qquad D_7 \cdot D_5 \cdot D_2 = 1 \ ,
\end{eqnarray}
and the triple intersection numbers of each divisor can be found from these integers by using linear equivalence; for example
\begin{equation}
D_7 \cdot D_7 \cdot D_7 = D_7 \cdot (- 2 D_3 - D_6 ) \cdot (-3 D_2 - 2 D_5 - D_4) = 7 \ .
\end{equation}

\begin{figure}[h]
 \centering
  \includegraphics[width=7cm]{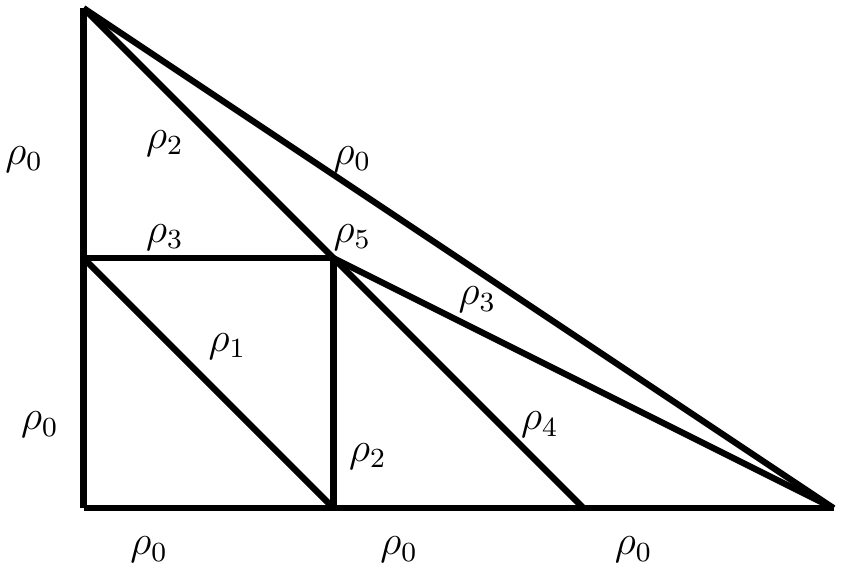}
 \caption{Decorated fan for the resolved $\complex^3 / \zed_6$ geometry.}
 \label{C3z6dec}
\end{figure}

The Mori cone is generated by the four compact curves $C_n$,
$n=1,2,3,4$ which are dual to the generators of the K\"ahler cone
given by $D_1$, $D_2$, $D_3$ and $-D_1+D_2+D_3-D_4$. Consequently the tautological bundles corresponding to the six one-dimesional irreducible representations $\rho_r$, $r=0,1,2,3,4,5$ of the orbifold group $\Gamma=\IZ_6$ are
\begin{eqnarray}
\cR_0 &=& \cO_X \ , \nonumber \\[4pt]
\cR_1 &=& \cO_X(D_1) \ , \nonumber \\[4pt]
\cR_2 &=& \cO_X (D_2) \ , \nonumber \\[4pt]
\cR_3 &=& \cO_X (D_3) \ , \nonumber \\[4pt]
\cR_4 &=& \cO_X (-D_1 + D_2 + D_3 - D_4) \ , \nonumber \\[4pt]
\cR_5 &=& \cR_2 \otimes \cR_3 \ = \ \cO_X (D_2 + D_3) \ .
\end{eqnarray}
The corresponding decoration of the toric fan is depicted in Figure~\ref{C3z6dec}. 
This decoration was also obtained in~\cite{Cacciatori:2008fq}, which
contains a study of the Gromov--Witten theory of the symmetric
resolution $X$ as well as a description of its mirror manifold.

From the decoration we immediately obtain the generators of the cohomology groups given by
\begin{eqnarray}
H^2 (X , \zed)=\IZ\big\langle c_1 (\cR_1 ) \ , \ c_1 (\cR_2) \ , \ c_1 (\cR_3) \ , \ c_1 (\cR_4) \big\rangle  \qquad \mbox{and} \qquad
H^4 (X , \zed)=\IZ\big\langle c_2 (\cV) \big\rangle
\end{eqnarray} 
where
\begin{equation}
\cV = \left( \cR_2 \oplus \cR_3 \right) \ominus \left( \cR_5 \oplus \cO_X \right) \ .
\end{equation}
Let us compute $c_2 (\cV)$ more explicitly. First of all the first Chern class of $\cV$ vanishes
\begin{equation}
c_1 (\cV) = c_1 (\cR_2) + c_1 (\cR_3) - c_1 (\cR_5) = 0 
\end{equation}
due to $\cR_5 = \cR_2 \otimes \cR_3$ and the additivity of the first Chern class under tensor product. Therefore $c_2 (\cV) = - \ch (\cV)$ which simplifies the computation. By the additivity of the Chern character
\begin{eqnarray}
\ch_2 (\cV) &=& \ch_2 (\cR_2) + \ch_2 (\cR_3) - \ch_2 (\cR_5) \nonumber \\[4pt]
&=& \mbox{$\frac 1 2$}\, \big( c_1 (\cR_2)\w c_1(\cR_2) + c_1 (\cR_3)\w c_1(\cR_3) - c_1 (\cR_5)\w c_1(\cR_5) \big) \nonumber \\[4pt]
&=& \mbox{$\frac 1 2$}\, \big( c_1 (\cR_2)\w c_1 (\cR_2) + c_1 (\cR_3)\w c_1 (\cR_3) - \left( c_1 (\cR_2) + c_1 (\cR_3)  \right)\w \left( c_1 (\cR_2) + c_1 (\cR_3)  \right)\big) \nonumber \\[4pt]
&=& - c_1 (\cR_2) \w c_1 (\cR_3) \ ,
\end{eqnarray}
which implies
\begin{equation}
c_2 (\cV) = c_1 (\cR_2) \w c_1 (\cR_3) \ .
\end{equation}

From the representation theory data we can construct the matrices
$a_{rs}^{(1)}$ and $a_{rs}^{(2)}=a_{sr}^{(1)}$. They are given by the
tensor product decompositions of $Q= \rho_1 \oplus \rho_2 \oplus \rho_3$; explicitly 
\begin{equation}
\big(a^{(1)}_{rs}\big) = \left( \begin{matrix} 
0 & 0 & 0 & 1 & 1 & 1 \\
1 & 0 & 0 & 0 & 1 & 1 \\
1 & 1 & 0 & 0 & 0 & 1 \\
1 & 1 & 1 & 0 & 0 & 0 \\
0 & 1 & 1 & 1 & 0 & 0 \\
0 & 0 & 1 & 1 & 1 & 0 \\
\end{matrix} \right)
\qquad \mbox{and} \qquad
\big(a^{(2)}_{rs}\big) = \left( \begin{matrix} 
0 & 1 & 1 & 1 & 0 & 0 \\
0 & 0 & 1 & 1 & 1 & 0 \\
0 & 0 & 0 & 1 & 1 & 1 \\
1 & 0 & 0 & 0 & 1 & 1 \\
1 & 1 & 0 & 0 & 0 & 1 \\
1 & 1 & 1 & 0 & 0 & 0 \\
\end{matrix} \right) \ .
\end{equation}
The associated quiver is
\begin{equation}
\vspace{4pt}
\begin{xy}
\xymatrix@C=20mm{
& \ v_0 \ \bullet \ \ar@/^/[dr] \ar@/^0.5pc/[ddr] \ar@/^/[ddd] & \\
v_5 \ \bullet \ \ar@/^/[ur]  \ar@/^/[rr] \ar@/^/[drr] & & \ v_1 \ \bullet \ar@/^/[d] \ar@/^/[ddl] \ar@/^/[dll] \\
v_4 \ \bullet \ \ar@/^/[u] \ar@/^/[uur] \ar@/^/[urr]   & &  \ v_2 \ \bullet \ar@/^/[dl] \ar@/^/[ll] \ar@/^/[ull] \\
& \ v_3 \ \bullet \ \ar@/^/[uuu] \ar@/^/[ul] \ar@/^/[uul] & 
}
\end{xy}
\vspace{4pt}
\end{equation}

\subsection{BPS partition functions}

We now compute the instanton action with boundary condition $\mbf N = (1 , 0, 0,0,0, 0)$. First of all (\ref{c1action}) becomes
\begin{equation}
\int_\oX\, \omega \wedge \omega \wedge c_1 (\cE) = - \varsigma\, \sum_{r,s=0}^5\, \left( a_{rs}^{(1)} - a_{rs}^{(2)} \right) \ k_s\, \int_\oX\, c_1 (\cR_2) \w c_1 (\cR_3) \w c_1 (\cR_r) \ .
\end{equation}
To evaluate this contribution we need the integrals
\begin{eqnarray}
\int_\oX\, c_1 (\cR_2) \w c_1 (\cR_3) \w c_1 (\cR_0) &=& 0  \ , \nonumber \\[4pt]
\int_\oX\, c_1 (\cR_2) \w c_1 (\cR_3) \w c_1 (\cR_1) &=& D_2 \cdot D_3 \cdot D_1 \ = \ 0 \ , \nonumber \\[4pt]
\int_\oX\, c_1 (\cR_2) \w c_1 (\cR_3) \w c_1 (\cR_2) &=& D_2 \cdot D_3 \cdot D_2 \ = \ D_2 \cdot D_3 \cdot \big(-\mbox{$ \frac 1 3$}\, (D_7 + 2 D_5 + D_4) \big) \ = \ - \mbox{$\frac 1 3$} \ , \nonumber \\[4pt]
\int_\oX\, c_1 (\cR_2) \w c_1 (\cR_3) \w c_1 (\cR_3) &=& D_2 \cdot D_3 \cdot D_3 \ = \ D_2 \cdot D_3 \cdot \big(- \mbox{$\frac 1 2$}\, (D_7 + D_6) \big) \ = \ - \mbox{$\frac 1 2$} \ , \nonumber \\[4pt]
\int_\oX\, c_1 (\cR_2) \w c_1 (\cR_3) \w c_1 (\cR_4) &=& D_2 \cdot D_3 \cdot (-D_1 + D_2 + D_3 - D_4) \ = \ \mbox{$- \frac 1 3 - \frac 1 2 \ = \ - \frac 5 6$} \ , \nonumber \\[4pt]
\int_\oX\, c_1 (\cR_2) \w c_1 (\cR_3) \w c_1 (\cR_5) &=& D_2 \cdot D_3\cdot (D_2 + D_3) \ = \ - \mbox{$\frac 5 6$} \ , 
\end{eqnarray}
and finally
\begin{equation}
\int_\oX\,\omega \wedge \omega \wedge c_1 (\cE) = \frac\varsigma6\, \Big( 
5 k_0-3k_1-8 k_2-5
   k_3+3k_4+8 k_5 \Big) \ .
\end{equation}

Consider now the second Chern character (\ref{ch2action}) given by
\begin{eqnarray}
\int_\oX\, \omega \wedge \mathrm{ch}_2 (\cE) &=& - \sum_{n=1}^4~ \sum_{r,s=0}^5\, \varphi_n \, k_s\, \bigg( \left(  a^{(1)}_{rs} - a^{(2)}_{rs} \right) \ \int_\oX\, c_1 (\cR_n) \w \mathrm{ch}_2 ({\cR_r}) \\ && \qquad \qquad\qquad +\, \left(  a_{rs}^{(2)}  - 3 \delta_{rs} \right) \, \int_\oX\, c_1 (\cR_n) \w c_1 \big(\cO_\oX(1)\big) \w c_1(\cR_r) \bigg)
  \ . \nonumber
\end{eqnarray}
To evaluate the first term one has to compute the triple intersection numbers between the divisors. For example the triple intersection
\begin{equation}
\int_\oX\, c_1 (\cR_3) \w \mathrm{ch}_2 ({\cR_3}) = \mbox{$\frac 1 2$}\, D_3 \cdot D_3 \cdot D_3 
\end{equation}
is linearly equivalent to
\begin{equation}
\mbox{$\frac 1 2$}\, D_3 \cdot (D_5 + D_1 + D_2 + D_4 ) \cdot \big( -\mbox{$ \frac 1 2$}\, (D_7 + D_6)\big) = - \mbox{$\frac 1 4$}\, D_3 \cdot D_7 \cdot D_2 =  - \mbox{$\frac 1 4$} \ .
\end{equation}
Laborious manipulations give altogether
\begin{equation}
\Big(\,\int_\oX\, c_1 (\cR_n) \w \mathrm{ch}_2 ({\cR_r})\, \Big)_{n,r=0}^{5} = \left( \begin{matrix}
0 & 0 & 0 & 0 & 0 & 0 \\[4pt]
0 & \frac 1 6 & 0 & 0 & - \frac 1 6 & 0 \\[4pt]
0 & 0 & 0 & - \frac 1 4 & - \frac{7}{12} & - \frac{7}{12} \\[4pt]
0 & 0 & - \frac 1 6 & - \frac 1 4 & - \frac{11}{12} & - \frac{11}{12} \\[4pt]
0 & 0 & - \frac 1 6 & -  \frac 1 2 & - \frac 7 6 & - \frac 3 2
 \end{matrix} \right) \ .
\end{equation}

To evaluate the intersections with the boundary divisor $\bdiv$ we proceed as follows. As a start by using linear equivalence we can write
\begin{equation}
D_4 \sim D_7 - 2 D_1 + D_2 + 2 D_3 \ ,
\end{equation}
so that all the intersection products involve either the original non-compact divisors $D_1$, $D_2$ and $D_3$ of $\complex^3$, or the compact divisor $D_7$ which by assumption has no intersection with $\bdiv$. Furthermore we can argue by symmetry that the intersection indices can be parametrized by two numbers
\begin{eqnarray}
\bdiv \cdot D_i \cdot D_i &=& a \qquad \mbox{for} \quad i=1,2,3 \ , \nonumber \\[4pt]
\bdiv \cdot D_i \cdot D_j &=& b \qquad \mbox{for} \quad i\neq j=1,2,3 \ .
\end{eqnarray}
The intersection matrix can be therefore parametrized as
\begin{equation}
\Big(\,\int_\oX\, c_1 (\cR_n) \w  c_1 \big(\cO_\oX (1)\big) \w c_1 ({\cR_r})\,\Big)_{n,r=0}^5 = \frac{1}{|\Gamma|}\, \left( \begin{matrix}
0 & 0 & 0 & 0 & 0 & 0 \\
0 & a & b & b & -2 a +3 b & 2 b \\
0 & b & a & b & a & a+b\\ 
0 & b & b & a & 2a-b & a+b \\
0 & -2a + 3 b & a & 2 a -b & 9a-8b & 3a-b \\
 \end{matrix} \right)
\end{equation}
where here $|\Gamma|=6$.

As in Section~\ref{subsec:C3Z2Z2BPS} we can set $b=1$ and $a=0$. We will however for the time being keep both parameters arbitrary. Under these conditions the second Chern character term in the instanton action (\ref{ch2action}) becomes
\begin{eqnarray}
\int_\oX\, \omega \wedge \mathrm{ch}_2 (\cE) &=& -
\frac 1 6 \, \Big( 2 \varphi_1 \,(k_0 - k_2 - k_3 + k_5) + 
   5 \varphi_4\, (3 k_0 + k_1 - 2 k_2 - 3 k_3 - k_4 + 2 k_5) \cr && \qquad\qquad +\, 
   \varphi_2 \,(7 k_0 + 2 k_1 - 5 k_2 - 7 k_3 - 2 k_4 + 5 k_5) \cr && \qquad\qquad +\, 
   \varphi_3 \, (10 k_0 + 3 k_1 - 7 k_2 - 10 k_3 - 3 k_4 + 7 k_5)\Big)
\cr && +\, \frac a6\,\Big(-\varphi_3\, (4 k_0+3 k_1+k_2-3 k_3-5 k_4) \cr && \qquad\qquad +\, \varphi_4\, (-14 k_0-18 k_1+2 k_2+7 k_3+26 k_4-3
   k_5) \cr && \qquad\qquad +\, \varphi_2\, (-2 k_0-2 k_1+2 k_2-k_3+2 k_4+k_5)\cr && \qquad \qquad +\, \varphi_1\, (2 k_0+5 k_1-k_2-k_3-7
   k_4+2 k_5)\Big) \cr && +\,\frac b6\, \Big(2 \varphi_2\, (-k_0+k_1-k_2+k_3-k_4+k_5)\cr && \qquad\qquad +\, \varphi_1\, (-6 k_0-5 k_1+k_2+2
   k_3+7 k_4+k_5) \cr && \qquad\qquad +\, 2 \varphi_4\, (5 k_0+9 k_1-k_2-3 k_3-13 k_4+3 k_5)\cr && \qquad\qquad +\,\varphi_3\, (3 k_1+k_2-2
   k_3-5 k_4+3 k_5)\Big) \ .
\end{eqnarray}

Finally we are left with the last part of the instanton action (\ref{ch3action}). We have to evaluate the integrals involving the Chern classes. Fortunately we have already computed most of the intersection products. In vector notation we have
\begin{eqnarray}
\Big(\, \int_\oX\, \ch_3 (\cR_r)\,\Big)_{r=0}^5 &=& \Big(\, \frac 1 6\, \int_\oX\, c_1 (\cR_r)\w c_1 (\cR_r)\w c_1 (\cR_r)\, \Big)_{r=0}^5 \nonumber\\[4pt] &=& \frac 1 6\, \Big( 0 \,,\, \frac 1 3 \,,\, 0 \,,\, -\frac 1 2 \,,\, - \frac 7 6 \,,\, -3 \Big) \ , \nonumber \\[4pt]
\Big(\,\int_\oX\, c_1 \big(\cO_\oX(1)\big) \w \ch_2 (\cR_r)\,
\Big)_{r=0}^5 &=& \Big(\, \frac 1 2\, \int_\oX\, c_1 \big(\cO_\oX
(1)\big) \w c_1 (\cR_r) \w c_1 (\cR_r) \, \Big)_{r=0}^5 \nonumber \\[4pt] &=& \frac{1}{2 |\Gamma|}\, \big( 0 , a , a , a , 9a-8b , 2 a + 2 b \big) \ .
\end{eqnarray}
The remaining integrals involve the double intersection of the divisor at infinity. Arguing again by symmetry we can parametrize these integrals with a single number
\bea
\Big(\,\int_\oX\, c_1 (\cR_r) \w \ch_2 \big(\cO_\oX(1)\big)\, \Big)_{r=0}^5 &=&\Big(\, \frac 1 2\, \int_\oX\, c_1 (\cR_r) \w c_1 \big(\cO_\oX (1)\big)\w c_1 \big(\cO_\oX (1)\big)\, \Big)_{r=0}^5 \nonumber\\[4pt] &=& \frac{1}{2 |\Gamma|}\, \big( 0,c,c,c,c,2c \big) \ ,
\eea
where we have expressed $D_4$ in terms of $D_1$, $D_2$ and $D_3$ by using linear equivalence, and taken the intersection with $D_7$ to be zero so that
\begin{equation}
D_4 \cdot \bdiv \cdot \bdiv = (D_7 - 2D_1 + D_2 + 2 D_3) \cdot \bdiv \cdot \bdiv = -2c+c+2c = c \ .
\end{equation}
Finally the last term in the instanton action is
\begin{eqnarray}
\int_\oX\,  \mathrm{ch}_3 (\cE) &=& \frac{1}{12}\Big(- (9 k_0 + 5 k_1 - 4 k_2 - 9 k_3 - 5 k_4 + 4 k_5)
\cr && \qquad +\,
a\, (-12 k_0 - 8 k_1 +  k_3 + 24 k_4 - 5 k_5) + 
 2 b\, ( 3 k_0 +  3 k_1 -  k_2  - 12 k_4 + 7 k_5)
\cr && \qquad +\,
 c\, (4 k_0 - k_3 - 3 k_5) +
2(k_0 + k_1 + k_2 + k_3 + k_4 + k_5)\Big) \ .
\end{eqnarray}

We now need to compute the $\zed_6$-invariant part of the character (\ref{character}). For this, we decompose the vector space $V$ at a fixed point $\pi$ according to the $\zed_6$-action as
\begin{equation}
V_\pi=\textsf{V}_0\oplus\textsf{V}_1\oplus \textsf{V}_2 \oplus \textsf{V}_3 \oplus \textsf{V}_4 \oplus \textsf{V}_5 \ .
\end{equation}
The explicit form of the partial character (\ref{Anpartialchar}) is now
\begin{eqnarray}
\mathcal{T}_{\pi}^+ &=& (\textsf{V}_0\oplus\textsf{V}_1\oplus \textsf{V}_2 \oplus \textsf{V}_3 \oplus \textsf{V}_4 \oplus \textsf{V}_5) \cr
&& -\, \Big(\, \frac{1}{t_1\, t_2} - \frac{1}{t_2} - \frac{1}{t_1} +1 \,
\Big) \, \big( \textsf{V}_0\oplus \textsf{V}_1\oplus \textsf{V}_2 \oplus \textsf{V}_3 \oplus \textsf{V}_4 \oplus \textsf{V}_5 \big) \cr &&\qquad \qquad \qquad \qquad \qquad \qquad \otimes \, \big( \textsf{V}^\vee_0\oplus \textsf{V}^\vee_1\oplus \textsf{V}^\vee_2 \oplus \textsf{V}^\vee_3 \oplus \textsf{V}^\vee_4 \oplus \textsf{V}^\vee_5 \big) \ .
\end{eqnarray}
The invariant part is given by substituting $t_\a=\zeta^\a$ with $\zeta^6=1$ to get
\begin{eqnarray}
\big(\mathcal{T}_{\pi}^+\big)^{\IZ_6} &=& {\rm vdim}_\IC \big( \textsf{V}_0 - ( \textsf{V}_0\otimes
\textsf{V}^\vee_0  \ \oplus \  \textsf{V}_1 \otimes\textsf{V}^\vee_1  \ \oplus \ 
\textsf{V}_2 \otimes\textsf{V}^\vee_2  
 \ \oplus \  \textsf{V}_3 \otimes\textsf{V}^\vee_3 
 \ \oplus \  \textsf{V}_4 \otimes\textsf{V}^\vee_4 
 \ \oplus \  \textsf{V}_5 \otimes\textsf{V}^\vee_5 
)
\cr && \qquad \qquad +\, (
\textsf{V}_0 \otimes\textsf{V}^\vee_5 
 \ \oplus \  \textsf{V}_1 \otimes\textsf{V}^\vee_0 
 \ \oplus \ \textsf{V}_2 \otimes\textsf{V}^\vee_1 
 \ \oplus \ \textsf{V}_3 \otimes\textsf{V}^\vee_2 
 \ \oplus \ \textsf{V}_4 \otimes\textsf{V}^\vee_3 
 \ \oplus \ \textsf{V}_5 \otimes\textsf{V}^\vee_4 
)
\cr 
&& \qquad \qquad +\, (
\textsf{V}_0 \otimes\textsf{V}^\vee_4 
 \ \oplus \  \textsf{V}_1 \otimes\textsf{V}^\vee_5 
 \ \oplus \ \textsf{V}_2 \otimes\textsf{V}^\vee_0 
 \ \oplus \ \textsf{V}_3 \otimes\textsf{V}^\vee_1 
 \ \oplus \ \textsf{V}_4 \otimes\textsf{V}^\vee_2 
 \ \oplus \ \textsf{V}_5 \otimes\textsf{V}^\vee_3 
)
 \cr && \qquad \qquad - \,
( 
\textsf{V}_0 \otimes\textsf{V}^\vee_3 
 \ \oplus \  \textsf{V}_1 \otimes\textsf{V}^\vee_4 
 \ \oplus \ \textsf{V}_2 \otimes\textsf{V}^\vee_5 
 \ \oplus \ \textsf{V}_3 \otimes\textsf{V}^\vee_0 
 \ \oplus \ \textsf{V}_4 \otimes\textsf{V}^\vee_1 
 \ \oplus \ \textsf{V}_5 \otimes\textsf{V}^\vee_2 
) \big) \nonumber \\[4pt]
&=& -|\pi_1| -
|\pi_2| - |\pi_3| -|\pi_4|-|\pi_5| + |\pi_0|\, \big(|\pi_1|+|\pi_2|+|\pi_3|+|\pi_4|\big)
\cr && +\, |\pi_1|\, \big(|\pi_2|+|\pi_3|+|\pi_4|\big) + |\pi_2|\, \big(|\pi_3|+|\pi_4|\big) + |\pi_3|\, \big(|\pi_4|+|\pi_5|\big) + |\pi_4|\, |\pi_5|
\cr && -\, 2 \big(
|\pi_0|\, |\pi_3| + |\pi_1|\, |\pi_4| + |\pi_2|\, |\pi_5| \big) \ ,
\end{eqnarray}
where we have introduced a $\IZ_6$-colouring of the partitions $\pi
= \pi_0 \sqcup \pi_1 \sqcup \pi_2 \sqcup \pi_3 \sqcup \pi_4 \sqcup \pi_5$. Therefore
the equivariant Euler characteristic of the obstruction bundle on the
quiver variety at a fixed point $\pi$ is
\begin{equation}
\chi_{\torus^3}(\mathcal{N}_\pi) = (-1)^{\mathcal{K}(\pi)} \ ,
\end{equation}
where the phase factor is 
\begin{eqnarray}
\mathcal{K}(\pi) &=& |\pi_1| +
|\pi_2| + |\pi_3| +|\pi_4|+|\pi_5| + |\pi_0|\, \big(|\pi_1|+|\pi_2|+|\pi_3|+|\pi_4| \big)
\cr && +\, |\pi_1| \,\big(|\pi_2|+|\pi_3|+|\pi_4| \big) + |\pi_2|\, \big(|\pi_3|+|\pi_4|\big) + |\pi_3| \, \big(|\pi_4|+|\pi_5|\big) + |\pi_4|\, |\pi_5|
\end{eqnarray}
since the even parity terms do not contribute.

We will now evaluate the partition function
\begin{eqnarray}
K^{\rm DT}_{\complex^3 / \zed_6} &=& \sum_{\pi} \, (-1)^{\mathcal{K}(\pi)} \ p^{|\pi|} \ p_0^{|\pi_0|} \ p_1^{|\pi_1|} \ p_2^{|\pi_2|} \ p_3^{|\pi_3|} \ p_4^{|\pi_4|} \ p_5^{|\pi_5|} \\[4pt]
&=& 1 - p \,p_1 +p^2\, \big( -p_1 \, p_2-p_1 \, p_3-p_1\, p_4 \big) \cr
&& +\,  p^3 \, \big(
-p_4\, p_1^2+2 p_2\, p_3\, p_1+p_2\, p_4\, p_1+p_3\, p_4\, p_1-p_3\, p_5\, p_1
\big) \cr
&& +\, p^4\, \big(
p_4^2\, p_1^2-p_2\, p_4\, p_1^2-p_3\, p_4\, p_1^2-p_3\, p_5\, p_1^2-p_2\, p_3^2\, p_1-p_0\,
   p_3\, p_4\, p_1+4 p_2\, p_3\, p_4\, p_1
   \cr 
 &&  \qquad +\,p_2\, p_3\, p_5\, p_1+p_2\, p_4\, p_5\,
   p_1-p_3\, p_4\, p_5\, p_1
\big) \cr
&& +\, p^5\, \big(
-p_4^2\, p_1^3-p_2\, p_4^2\, p_1^2-p_3\, p_4^2\, p_1^2-p_0\, p_3\, p_4\, p_1^2+2 p_2\, p_3\,
   p_4\, p_1^2-p_3^2\, p_5\, p_1^2-p_2\, p_3\, p_5\, p_1^2
   \cr && \qquad -\,p_2\, p_4\, p_5\, p_1^2+2 p_3\,
   p_4\, p_5\, p_1^2+2 p_2\, p_3\, p_4^2\, p_1+3 p_2\, p_3^2\, p_4\, p_1+p_0\, p_2\, p_3\, p_4\,
   p_1+p_2\, p_3^2\, p_5\, p_1
   \cr && \qquad +\,p_0\, p_3\, p_4\, p_5\, p_1-5 p_2\, p_3\, p_4\, p_5\, p_1
\big) \cr
&& +\, p^6\, \big(
-p_4^3\, p_1^3-p_2\, p_4^2\, p_1^3-p_3\, p_4^2\, p_1^3-p_3\, p_4\, p_5\, p_1^3+p_0\, p_3\,
   p_4^2\, p_1^2-4 p_2\, p_3\, p_4^2\, p_1^2+p_3^2\, p_5^2\, p_1^2
   \cr && \qquad +\, p_0\, p_3^2\, p_4\, p_1^2-p_2\,
   p_3^2\, p_4\, p_1^2 
   - p_0\, p_2\, p_3\, p_4\, p_1^2+2 p_2\, p_3^2\, p_5\, p_1^2+p_2\, p_4^2\,
   p_5\, p_1^2-p_3\, p_4^2\, p_5\, p_1^2
   \cr && \qquad -\,p_2^2\, p_4\, p_5\, p_1^2-p_3^2\, p_4\, p_5\, p_1^2+2
   p_0\, p_3\, p_4\, p_5\, p_1^2-5 p_2\, p_3\, p_4\, p_5\, p_1^2-3 p_2\, p_3^2\, p_4^2\,
   p_1 \cr && \qquad -\, p_0\, p_2\, p_3\, p_4^2\, p_1
   +p_2\, p_3\, p_4\, p_5^2\, p_1-p_0\, p_2\, p_3^2\, p_4\,
   p_1+4 p_2\, p_3\, p_4^2\, p_5\, p_1 \cr && \qquad +\, 6 p_2\, p_3^2\, p_4\, p_5\, p_1-6 p_0\, p_2\, p_3\, p_4\,
   p_5\, p_1
\big) + \cdots \ . \nonumber
\end{eqnarray}
Again one observes the factorization
\begin{eqnarray}
K^{\rm DT}_{\complex^3 / \zed_6} &=& \Big( 1 - 6 p^6\, p_0\, p_1\, p_2\, p_3\, p_4\, p_5 + 33 p^{12}\, p_0^2\, p_1^2\, p_2^2\, p_3^2\, p_4^2\, p_5^2 - 
 146 p^{18}\, p_0^3\, p_1^3\, p_2^3\, p_3^3\, p_4^3\, p_5^3 + \cdots \Big) \cr 
&& \times\, \Big(
1-p_1\, p+ \big(-p_1\, p_2-p_1\, p_3-p_1\, p_4 \big)\, p^2
\cr && \qquad
+\, \big(-p_4\, p_1^2+2 p_2\, p_3\, p_1+p_2\, p_4\,
   p_1+p_3\, p_4\, p_1-p_3\, p_5\, p_1 \big)\, p^3
 \\ && \qquad +\, \big(p_4^2\, p_1^2-p_2\, p_4\, p_1^2-p_3\,
   p_4\, p_1^2-p_3\, p_5\, p_1^2-p_2\, p_3^2\, p_1-p_0\, p_3\, p_4\, p_1+4 p_2\, p_3\, p_4\,
   p_1\cr && \qquad +\, p_2\, p_3\, p_5\, p_1
    +p_2\, p_4\, p_5\, p_1-p_3\, p_4\, p_5\, p_1\big)\, p^4 +
\cdots \Big) \ , \nonumber
\end{eqnarray}
where
\begin{eqnarray}
M (x)^6 = 1 + 6 \, x + 33\,x^2 +
 146\, x^3 + \cdots
\end{eqnarray}
with $x=-q=-p^6 \, p_0\, p_1\, p_2\, p_3\, p_4\, p_5$. This factor is
again expected to appear in the large radius regime as the
contribution from degree zero curve classes (``regular D0~branes'').

On the other hand we can use our computation of the instanton action
to write down the partition function in a more ``physical''
form. According to our discussion in Section~\ref{subsec:C3Z2Z2BPS},
we set $a=0$ and $b=c=1$. The instanton partition function has the form
\begin{equation}
\cZ_{\complex^3 / \zed_6} = \sum_{\pi} \ (-1)^{\mathcal{K}(\pi)} \
\e^{-g_s\, \int_\oX\, \ch_3 (\cE_\pi)} \ \e^{-\int_\oX\, \omega\wedge \ch_2
  (\cE_\pi) } \ \e^{-\int_\oX\, \omega \wedge \omega \wedge c_1 (\cE_\pi)}
\end{equation}
where the universal sheaf is evaluated on the fixed points $\pi$ (i.e. $k_r
= |\pi_r|$). Introducing the K\"ahler parameters $U= \e^{- \varsigma}$
and $Q_n = \e^{-\varphi_n}$ for $n=1,2,3,4$, we can then write
\begin{equation}
\cZ_{\complex^3 / \zed_6} = \sum_{\pi} \ (-1)^{\mathcal{K}(\pi)} \
q^{\cI(\pi)} \ Q_1^{\cB_1(\pi)} \ Q_2^{\cB_2(\pi)} \ Q_3^{\cB_3(\pi)}
\ Q_4^{\cB_4(\pi)} \ U^{\cA(\pi)}
\end{equation}
where from our computations above we have
\begin{eqnarray}
\cI(\pi) &=& \mbox{$\frac{1}{12}$}\, \big(3 |\pi_0| +3 |\pi_1| +4
|\pi_2| +10 |\pi_3| -17 |\pi_4| +9 |\pi_5| \big) \ , \nonumber \\[4pt]
\cB_1(\pi) &=& \mbox{$\frac{1}{6}$}\, \big(8 |\pi_0| +5 |\pi_1| -3
|\pi_2| -4 |\pi_3| -7 |\pi_4| +|\pi_5| \big) \ , \nonumber\\[4pt]
\cB_2(\pi) &=& \mbox{$\frac{1}{2}$}\, \big(3 |\pi_0| -|\pi_2| -3
|\pi_3| +|\pi_5| \big) \ , \nonumber \\[4pt]   
\cB_3(\pi) &=& \mbox{$\frac{1}{3}$}\, \big(5 |\pi_0| -4 |\pi_2| -4
|\pi_3| +|\pi_4| +2 |\pi_5| \big) \ , \nonumber \\[4pt]
\cB_4(\pi) &=&\mbox{$\frac{1}{6}$}\, \big(5 |\pi_0| -13 |\pi_1| -8
|\pi_2| -9 |\pi_3| +21 |\pi_4| +4 |\pi_5| \big) \ , \nonumber \\[4pt]
\cA(\pi) &=& \mbox{$\frac16$}\, \big(5 |\pi_0|-3|\pi_1|-8 |\pi_2|-5
   |\pi_3|+3|\pi_4|+8 |\pi_5| \big) \ .
\end{eqnarray}
In these new variables the partition function has an expansion
\begin{small}
\begin{eqnarray}
\cZ_{\complex^3 / \zed_6} &=& \cdots -\frac{Q_3\, Q_4^4}{q^{7/2}\, Q_1}-\frac{Q_4^{17/6}}{q^{7/6}\, U^{5/6} \,Q_1^{4/3} \,\sqrt{Q_2}\,
   \sqrt[3]{Q_3}}+\frac{Q_3^{2/3}\, Q_4^{8/3}}{q^{7/3}\, Q_1^{2/3}}+\frac{4 Q_4^{8/3}}{q^{2/3}\, \sqrt[3]{U}\, Q_1^{5/2}\,
   Q_2^{3/2}\, Q_3^{4/3}}\cr && +\, \frac{Q_3\, Q_4^2}{q^{5/4}}
   +\frac{Q_4^2}{q^{5/4}\, Q_1}+\frac{2 Q_4^2}{q^{17/12}\, U^{5/3}\,
   Q_1^{8/3}\, Q_2^2\, Q_3^2}-\frac{\sqrt{U}\, Q_4^{11/6}}{q^{3/4}\, Q_1^{7/6}\, Q_2}
   \cr && +\, q^{2/3}\, U^{4/3}\, \sqrt{Q_1}\, \sqrt{Q_2}\,
   Q_3^{4/3}\, Q_4^{4/3} 
   -\frac{\sqrt[3]{Q_3}\, Q_4^{4/3}}{q^{7/6}\, \sqrt[3]{Q_1}}-\frac{Q_4^{4/3}}{q^2\, U^{4/3}\, Q_1^{7/6}\,
   \sqrt{Q_2}\, Q_3^{2/3}}\cr && -\, \frac{Q_4^{7/6}}{q^{3/2}\, U^{5/6}\, Q_1^{4/3}\, Q_2^{3/2}\, Q_3^{2/3}}-\frac{\sqrt[3]{Q_1}\,
   Q_3^{2/3}\, Q_4^{2/3}}{\sqrt[12]{q}}
   +\frac{Q_4^{2/3}}{\sqrt[12]{q}\, Q_1^{2/3}\, \sqrt[3]{Q_3}}-\frac{\sqrt[6]{Q_1}\,
   Q_3^{2/3}\, \sqrt{Q_4}}{q^{25/12}\, \sqrt{U}} \cr && -\, \frac{q^{5/12}\, \sqrt{U}\, \sqrt{Q_4}}{Q_1^{5/6}\, Q_2 \sqrt[3]{Q_3}}-\frac{3
   \sqrt{Q_4}}{q^{7/12}\, U^{5/2}\, Q_1^{10/3}\, Q_2^{7/2}\, Q_3^{10/3}}-6 q
   +\frac{1}{q^{5/6}\, U^{4/3}\, Q_1^{5/6}\, \sqrt{Q_2}\,
   Q_3} \cr && +\, 1+\frac{q^{3/2}\, \sqrt{U}}{Q_1^{7/6}\, Q_2\, Q_3 \,\sqrt[6]{Q_4}}+\frac{1}{\sqrt[3]{q}\, U^{5/6}\, Q_1\, Q_2^{3/2}\, Q_3\,
   \sqrt[6]{Q_4}}
   \cr && -\,\frac{4}{q^{7/6}\, U^{13/6}\, Q_1^{11/6}\, Q_2^2 \, Q_3^2\, \sqrt[6]{Q_4}}+\frac{\sqrt[4]{q}}{U^{4/3}\, \sqrt[6]{Q_1}\,
   \sqrt{Q_2}\, Q_3^{2/3}\, Q_4^{2/3}} \cr && +\, \frac{2 q^{11/12}\, U^{5/6}\, Q_1^{4/3}\, \sqrt{Q_2}\,
   Q_3^{4/3}}{Q_4^{5/6}}
   -\frac{\sqrt{Q_1}\, \sqrt[3]{Q_3}}{q^{11/12}\, \sqrt{U}\, Q_4^{5/6}}-\frac{1}{q^{7/4}\, U^{11/6}\, \sqrt[3]{Q_1}\,
   \sqrt{Q_2}\, Q_3^{2/3}\, Q_4^{5/6}} \cr && -\, \frac{5 q^{3/4}}{U^{5/6}\, Q_1^{4/3}\, Q_2^{3/2}\, Q_3^{5/3}\, Q_4^{5/6}}
   -\frac{1}{q^{5/4}\,
   U^{4/3}\, \sqrt{Q_1}\, Q_2^{3/2}\, Q_3^{2/3}\, Q_4}-\frac{\sqrt[6]{q}\, Q_1^{7/6}\, Q_3^{2/3}}{\sqrt{U}
   \,Q_4^{3/2}} \cr && -\, \frac{\sqrt[6]{q}\, \sqrt[6]{Q_1}}{\sqrt{U}\, \sqrt[3]{Q_3}\, Q_4^{3/2}}+\frac{4}{U^{13/6}\, Q_1^{3/2}\, Q_2^2\,
   Q_3^{7/3}\, Q_4^{3/2}}
   +\frac{2 q^{2/3}}{Q_2\, \sqrt[3]{Q_3}\, Q_4^{5/3}}-\frac{\sqrt[4]{q}\, Q_1^{5/6}}{\sqrt{U}\,
   Q_4^{13/6}} \cr && -\, \frac{1}{q^{7/12}\, U^{11/6}\, \sqrt{Q_2}\, Q_3\, Q_4^{13/6}}-\frac{q^{13/12}}{U^{13/6}\, Q_1^{5/6}\, Q_2^2 \,Q_3^2\,
   Q_4^{13/6}}
   -\frac{1}{\sqrt[12]{q}\, U^{4/3}\, \sqrt[6]{Q_1}\, Q_2^{3/2}\, Q_3\, Q_4^{7/3}} \cr && +\, \frac{6 q^{19/12}}{U^{5/3}\, Q_1^2\,
   Q_2^3\, Q_3^3\, Q_4^{7/3}}-\frac{\sqrt{q}\, Q_1^{2/3}}{U^{11/6}\, \sqrt{Q_2}\, Q_3^{2/3}\, Q_4^{17/6}}
   -\frac{\sqrt{q}}{U^{11/6}\,
   \sqrt[3]{Q_1}\, \sqrt{Q_2}\, Q_3^{5/3}\, Q_4^{17/6}} \cr && -\, \frac{q^{11/6}\, \sqrt[3]{Q_1}}{Q_2\, Q_3^{2/3}\, Q_4^3}+\frac{q
   \sqrt{Q_1}}{U^{4/3}\, Q_2^{3/2}\, Q_3^{2/3}\, Q_4^3}-\frac{5 q}{U^{4/3}\, \sqrt{Q_1}\, Q_2^{3/2}\, Q_3^{5/3}\, Q_4^3}
   \cr && +\,\frac{3
   q^{5/6}}{U^3\, Q_1^{13/6}\, Q_2^{7/2}\, Q_3^{11/3}\, Q_4^3}-\frac{q^{3/2}}{U^{5/6}\, Q_1^{2/3}\, Q_2^{5/2} \,Q_3^{5/3}\,
   Q_4^{19/6}}-\frac{q^{7/12}\, \sqrt[3]{Q_1}}{U^{11/6} \,\sqrt{Q_2}\,
   Q_3^{4/3}\, Q_4^{7/2}} \cr && -\,\frac{q^{13/12}\, \sqrt[6]{Q_1}}{U^{4/3}\,
   Q_2^{3/2}\, Q_3^{4/3}\, Q_4^{11/3}} 
    +\frac{2 \sqrt[4]{q}}{U^{8/3}\, Q_1^{2/3}\, Q_2^2\, Q_3^{7/3}\, Q_4^{11/3}}-\frac{q^{11/12}\,
   Q_1^{5/6}}{\sqrt{U}\, Q_2 \,\sqrt[3]{Q_3}\, Q_4^{23/6}}\cr && +\, \frac{q^{13/6}}{U^{4/3}\, \sqrt[6]{Q_1}\, Q_2^{3/2}\, Q_3^2\,
   Q_4^{13/3}}+\frac{2 q^{17/12}}{U^{8/3}\, \sqrt[3]{Q_1}\, Q_2^2\, Q_3^{8/3}\, Q_4^5}
   -\frac{q^{25/12}\, Q_1^{7/6}}{\sqrt{U}\, Q_2\,
   Q_3^{2/3}\, Q_4^{31/6}}\cr && -\, \frac{q^{13/12}}{U^{7/2}\, Q_1^{4/3}\, Q_2^{7/2}\, Q_3^{11/3}\, Q_4^{31/6}}+\frac{q^3}{U^{13/6}\,
   Q_1^{5/6}\, Q_2^3\, Q_3^{10/3}\, Q_4^{35/6}}+\frac{q^{11/3}\, Q_1^{2/3}}{Q_2^2\, Q_3^{4/3}\, Q_4^6}
   \cr && -\,\frac{q^{29/12}\,
   Q_1^{2/3}}{U^{11/6}\, Q_2^{3/2}\, Q_3^2\, Q_4^{13/2}}-\frac{q^{9/4}}{U^{7/2}\, Q_1\, Q_2^{7/2}\, Q_3^4\,
   Q_4^{13/2}}-\frac{q^{35/12}\, \sqrt{Q_1}}{U^{4/3}\, Q_2^{5/2}\, Q_3^2\, Q_4^{20/3}} \cr && +\, \frac{2 q^{13/4}}{U^{8/3}\, Q_2^3\,
   Q_3^{10/3}\, Q_4^8}+ \cdots \ .
\end{eqnarray}
\end{small}
Finally, the change of variables between the two partition functions reads
\begin{eqnarray}
p &=& q^{1/6} \ , \nonumber\\[4pt]
p_0 &=& q^{1/12} \, Q_1^{8/6} \, Q_2^{3/2} \, Q_3^{5/3} \, Q_4^{5/6} \, U^{5/6} \ , \nonumber\\[4pt]
p_1 &=& q^{1/12}  \, Q_1^{5/6} \, Q_4^{-13/6} \, U^{-1/2} \ , \nonumber\\[4pt]
p_2 &=& q^{2/12} \, Q_1^{-3/6} \, Q_2^{-1/2} \, Q_3^{-4/3} \, Q_4^{-8/6} \, U^{-4/3} \ , \nonumber\\[4pt]
p_3 &=& q^{8/12}  \, Q_1^{-4/6} \, Q_2^{-3/2} \, Q_3^{-4/3} \, Q_4^{-9/6} \, U^{-5/6} \ , \nonumber\\[4pt]
p_4 &=& q^{-19/12}  \, Q_1^{-7/6} \, Q_3^{1/3} \, Q_4^{21/6} \, U^{1/2} \ , \nonumber\\[4pt]
p_5 &=& q^{7/12} \, Q_1^{1/6} \, Q_2^{1/2} \, Q_3^{2/3} \, Q_4^{4/6} \, U^{4/3} \ .
\label{C3Z6varchange}\end{eqnarray}

\subsection{Coulomb branch invariants}

Finally we can present the partition function for the Coulomb branch invariants. In this case the instanton measure (\ref{instmeasure}) has the form
\begin{eqnarray}
\mathcal{K}(\vec\pi;\mbf N) &=& \sum_{l=1}^N \ \sum_{r=0}^5\, |\pi_{l,r}| \ N_{r+b(l)} + \sum_{l,l'=1}^N \ \sum_{r=0}^5\, |\pi_{l,r}|\, \Big( - |\pi_{l',r+b(l)-b(l'\,)-3}| +|\pi_{l',r+b(l)-b(l'\,)-2}| \nonumber \\ && \hspace{6cm} +\, |\pi_{l',r+b(l)-b(l'\,)-1}|  - |\pi_{l',r+b(l)-b(l')}| \Big) \ .
\end{eqnarray}
We can thus write the partition function for noncommutative Donaldson--Thomas invariants of type $\mbf N$ as
\begin{eqnarray}
K^{\rm DT}_{\complex^3 / \zed_6}( \mbf N) &= & \sum_{\vec \pi} \  (-1)^{\mathcal{K}(\vec\pi;\mbf N)} \ p_0^{\sum_{l=1}^N\, |\pi_{l,0-b(l)}|} \ p_1^{\sum_{l=1}^N\, |\pi_{l,1-b(l)}|
} \ p_2^{ \sum_{l=1}^N\, |\pi_{l,2-b(l)}|} \cr& & \qquad \qquad \qquad \times\, p_3^{\sum_{l=1}^N \, |\pi_{l,3-b(l)}|} \ p_4^{\sum_{l=1}^N\, |\pi_{l,4-b(l)}|
} \ p_5^{ \sum_{l=1}^N\, |\pi_{l,5-b(l)}|} \ .
\end{eqnarray}
In physical variables given by the transformation (\ref{C3Z6varchange}), this partition function becomes
\begin{equation}
\cZ_{\complex^3 / \zed_6}(\mbf N) = \sum_{\vec\pi} \ (-1)^{\mathcal{K}(\vec\pi;\mbf N)} \
q^{\cI(\vec\pi;\mbf N)} \ Q_1^{\cB_1(\vec\pi;\mbf N)} \ Q_2^{\cB_2(\vec\pi;\mbf N)} \ Q_3^{\cB_3(\vec\pi;\mbf N)}
\ Q_4^{\cB_4(\vec\pi;\mbf N)} \ U^{\cA(\vec\pi;\mbf N)}
\end{equation}
where
\begin{eqnarray}
\cI(\vec\pi;\mbf N) &=& \frac{1}{12}\,\sum_{l=1}^N \, \Big(3 |\pi_{l,0-b(l)}| +3 |\pi_{l,1-b(l)}| +4 |\pi_{l,2-b(l)}| \cr && \qquad\qquad +\, 10 |\pi_{l,3-b(l)}| -17 |\pi_{l,4-b(l)}| +9 |\pi_{l,5-b(l)}| \Big) \ , \nonumber \\[4pt]
\cB_1(\vec\pi;\mbf N) &=&\frac{1}{6}\,\sum_{l=1}^N \, \Big(8 |\pi_{l,0-b(l)}| +5 |\pi_{l,1-b(l)}| -3 |\pi_{l,2-b(l)}| \cr && \qquad \qquad -\, 4 |\pi_{l,3-b(l)}| -7 |\pi_{l,4-b(l)}| +|\pi_{l,5-b(l)}| \Big) \ , \nonumber\\[4pt]
\cB_2(\vec\pi;\mbf N) &=& \frac{1}{2}\,\sum_{l=1}^N \, \Big(3 |\pi_{l,0-b(l)}| -|\pi_{l,2-b(l)}| -3 |\pi_{l,3-b(l)}| +|\pi_{l,5-b(l)}| \Big) \ , \nonumber \\[4pt]   
\cB_3(\vec\pi;\mbf N) &=& \frac{1}{3}\,\sum_{l=1}^N \, \Big(5 |\pi_{l,0-b(l)}| -4 |\pi_{l,2-b(l)}| -4 |\pi_{l,3-b(l)}| + |\pi_{l,4-b(l)}| +2 |\pi_{l,5-b(l)}| \Big) \ , \nonumber \\[4pt]
\cB_4(\vec\pi;\mbf N) &=& \frac{1}{6}\,\sum_{l=1}^N\, \Big(5  |\pi_{l,0-b(l)}| -13 |\pi_{l,1-b(l)}| -8 |\pi_{l,2-b(l)}| \cr && \qquad \qquad -\,9 |\pi_{l,3-b(l)}| +21 |\pi_{l,4-b(l)}| +4 |\pi_{l,5-b(l)}| \Big) \ , \nonumber \\[4pt]
\cA(\vec\pi;\mbf N) &=& \frac16\,\sum_{l=1}^N\, \Big( 5 |\pi_{l,0-b(l)}|-3|\pi_{l,1-b(l)}|-8 |\pi_{l,2-b(l)}| \cr && \qquad \qquad -\, 5
   |\pi_{l,3-b(l)}|+3 |\pi_{l,4-b(l)}|+8 |\pi_{l,5-b(l)}| \Big) \ .
\end{eqnarray}

\section{Discussion\label{sec:Disc}}

In this paper we have taken a gauge theory approach to the study of
noncommutative Donaldson--Thomas invariants defined on noncommutative
crepant resolutions of orbifold singularities. This gauge theory is
defined on geometries of the form $\complex^3 / \Gamma$, by which we
really mean a gauge theory on $\complex^3$ whose observables are
$\Gamma$-equivariant. In this gauge theory one can study a moduli space of
$\Gamma$-equivariant instantons, and define an enumerative
problem associated with this moduli space. We demonstrated that this
moduli space may also be identified with the moduli space of
representations of a certain quiver. The structure of the quiver is
dictated by the singularity via the McKay correspondence. A certain
topological matrix quantum mechanics based on the quiver data can be used to study the local
structure of the moduli space and hence to compute its virtual numbers.

Our analysis however leaves many open issues. In particular it would
be desirable to develop directly a connection with physical states in
string theory in order to obtain actual partition functions for the invariants,
where the counting parameters are the D~brane charges as
in~\cite{Aganagic:2010qr}. Furthermore our approach seems to be
limited to resolutions of abelian orbifolds and do not include more
general singularities, such as the conifold. While it does not seem impossible that our approach can be generalized to other singularities, major technical problems arise, such as how to impose boundary conditions at infinity.

Another technical point concerns the condition we have to impose in
deriving a compact parametrization of the moduli space, thus excluding
certain classes of sheaves. While it seems that this condition is
not restrictive for the purpose of enumeration of the BPS states
considered in this paper, as it is satisfied for ideal sheaves, it certainly is for more generic physical
states. It would be desirable to have more control over the full
moduli space; we expect this issue to became critical when studying
higher rank invariants beyond the Coulomb phase of the non-abelian gauge theory.

It would also be interesting to study the wall-crossing behaviour of
generalized Donaldson--Thomas invariants from the D~brane perspective
and across different phases. It is natural to expect that the gauge
theory analysis could at least capture the qualitative behaviour. In
the favourable cases where the set of tautological bundles is also a
set of generators for the derived category, the enumerative problem is
already well-posed and what remains to be done is to evaluate the
virtual numbers for different values of the stability parameter, and
use the tilting set to obtain the proper D~brane charges in each
phase. It would be interesting to clarify how variations of the slope
or $\theta$-stability parameters of Section~\ref{subsec:NCDTquivers},
and hence the crossing of chambers, could be understood purely from
the gauge theory perspective as modifications of
the D-term conditions.

It would also be interesting to investigate more deeply the connection
to the crystal melting picture, for example by exploring the crystal
picture in the framework of~\cite{reineke} specifically for orbifold
singularities. The melting rules could be understood as a counting of
coloured plane partitions when an atom is removed. One could then also
explore the high-temperature limit of such a crystal and the algebraic
curve to which it is related; one may further investigate the boundary
conditions at infinity in this way along the lines of~\cite{Aganagic:2010qr}.

Another point which we have left basically untouched is the study of
higher rank invariants of noncommutative crepant resolutions. This is
a formidable problem plagued by technical and conceptual
difficulties. We were able to restrict ourselves to the Coulomb branch
of the non-abelian gauge theory where torus equivariant localization is still a
viable approach. One however could hope that, similarly to what
happens in Seiberg--Witten theory, by combining the Coulomb branch
result with an appropriate modular behaviour of the amplitudes one could derive non-abelian invariants in full generality. Equivalently this problem could be solved via a wall-crossing analysis, where different BPS states corresponding to bound states centred around well-separated D6~branes could coalesce together and form a higher rank stable state.

Finally one cannot help noticing how many concepts and techniques that
have entered the present work seem also to appear in the study of D~branes
probing the singularity from a low-energy perspective, in the approach
pioneered in~\cite{Berenstein:2001jr}. Recent papers have focused on
the role that noncommutative crepant resolutions have in the
properties of low-energy effective gauge theories~\cite{Beil:2008yj,Beil:2008ck,Eager:2010ji}. It would be interesting to understand this correspondence further.

\section*{Acknowledgments}

We thank D.-E.~Diaconescu and B.~Szendr\H{o}i for helpful discussions, and A.~Craw for comments on the manuscript.
Preliminary results concerning the $\complex^3 / \zed_3$ orbifold were presented by M.C. at the ``Fourth
Regional Meeting in String Theory'' in Patras, Greece in June
2007. The work of M.C. was supported in part by the Funda\c{c}\~{a}o
para a Ci\^{e}ncia e Tecnologia (FCT/Portugal). M.C. is grateful to
the Institute des Hautes \'Etudes Scientifiques (IHES) for the warm
hospitality and support during the final stages of this work. The work of R.J.S.
was supported in part by grant ST/G000514/1 ``String Theory Scotland'' from the UK Science and Technology
Facilities Council.

\appendix

\section{Gauge theory on quotient stacks\label{App:Stacks}}

The invariant way to describe orbifolds independently of a particular
presentation is through the language of Deligne--Mumford
stacks. For global orbifolds, obtained as the quotient of a smooth
manifold $M$ by the action of a group $\Gamma$, the relevant stacks
are called quotient stacks. In this appendix we collect some properties
of quotient stacks, focusing mostly on physical perspectives and
streamlining most of the mathematical technicalities. In particular,
we use these notions to present some heuristic justification for our
definition of the gauge theory in Section~\ref{subsec:Colinst} and Section~\ref{sec:Instmodsp}. The point of view adapted here closely follows the approach of~\cite{Sharpe:2001bs}.

Let $M$ be a smooth manifold and $\Gamma $ a finite group acting
properly on $M$; in the main
text we consider the case $M=\mathbb{C}^3$ and $\Gamma\subset
SL(3,\IC)$ abelian. The quotient stack is denoted $\frak{X}=[M/\Gamma]$, as opposed to the quotient space $M/ \Gamma$. One can think of $[M/ \Gamma]$ as
something similar to $M / \Gamma$ away from the orbifold singularities, but with
extra structure at the singularities. The intuition is that any point $x\in\frak{X}$ comes with a finite group $\Gamma_x$, the stabilizer subgroup of $\Gamma$ whose elements are regarded as ``automorphisms of $x$''. 

In contrast to an ordinary
space, the points of a stack do not form a set but objects in a category; the
morphisms in fact are all invertible, hence the category is a
groupoid, in this case the action groupoid $\Gamma\times
M\rightrightarrows M$ whose objects are the points of $M$ and whose
morphisms are given by the actions of elements of $\Gamma$ on $M$. This implies that a single point can have non-trivial automorphisms
(as it happens with the moduli space of sigma-model maps from a worldsheet
into a Calabi--Yau threefold), or two distinct points can be isomorphic. The
category of points of $[M/ \Gamma]$ consists of the orbits of points of
$M$ under the action of $\Gamma$, and there is a one-to-one
correspondence between isomorphism classes of points of $[M/\Gamma]$
and points of $M/ \Gamma$. In other words, the coarse moduli space of the stack $\frak{X}$ is the underlying singular variety $M/\Gamma$. Thus $[M/ \Gamma]$ is similar to $M / \Gamma$ away from
the singularities, but at the fixed points of the $\Gamma$-action on $M$
the stack $[M/
\Gamma]$ has non-trivial automorphisms. 

An important property of stacks is that a sheaf defined on $[M/
\Gamma]$ is the same thing as a $\Gamma$-equivariant sheaf on $M$. This feature extends to all the objects that derive from sheaves, for
example differential forms,
spinors, and functions can all be regarded as sections of sheaves and so
on; in particular a differential form on $[M/ \Gamma]$ is a
$\Gamma$-invariant form on $M$. Moreover, one can also prove that $[M/
\Gamma]$ is smooth; an appropriate definition of smoothness is
provided by the smooth ``orbifold atlas'' $(M,f)$ for the
stack, where $f:M \rightarrow [M/
\Gamma]$ is the canonical surjection (which is a principal $\Gamma$-bundle).

When one constructs the usual string theory sigma-model on a target space $M$,
the spectrum of massless closed string modes is given by the
cohomology of $M$; this is because massless modes are associated
with constant (zero
momentum) maps from the worldsheet to the target space. However when one considers a sigma-model of maps into a
stack $[M/ \Gamma]$ the situation is a bit different because of the
non-trivial automorphisms at the fixed points of the $\Gamma$-action. For generic points of $[M/ \Gamma]$ the only automorphism is the identity
and one basically recovers the quotient space $M/ \Gamma$. But at the fixed
points there are non-trivial automorphisms, and so the appropriate
cohomology is not just the cohomology of the target $[M/ \Gamma]$
but an appropriate generalization that keeps track of the automorphisms of cohomology classes. This is called the orbifold cohomology; it contains a combination of geometric and representation theory data.

These
notions motivate the definition of the inertia stack $I[M/ \Gamma]$
associated to the quotient stack 
$[M/ \Gamma]$, defined as the substack of points $(x,g)\in\frak{X}\times\Gamma$ with $g\in\Gamma_x$. It has the structure of a disjoint union of orbifolds
\begin{equation}
I[M/ \Gamma] \cong \bigsqcup_{[g]}\, \big[M^g\, \big/\, C(g) \big]
\ ,
\end{equation}
where $[g]$ denotes the conjugacy class of an element $g \in \Gamma$, $C(g)$ is the
centralizer of $g$ in $G$, and $M^g$ is the submanifold of points of $M$
which are
invariant under the action of $g$. Twist fields live in the cohomology of this
auxilliary stack, i.e. we define the orbifold cohomology $H_{\rm
  orb}^\bullet([M/\Gamma])$ of $[M/
\Gamma]$ as the ordinary cohomology of $I[M/ \Gamma]$.\footnote{To be
precise there is also a shift in the degree of the cohomology that
is called ``age''; we will ignore the age in order to simplify our
presentation a bit.}

As a particular case, the quotient stack $[{\rm pt}/ \Gamma]$,
with pt some fixed point with trivial $\Gamma$-action, is the
classifying stack of $\Gamma$, denoted $B \Gamma$; it is a moduli space for principal $\Gamma$-bundles.
If $\Gamma$ acts freely on a manifold $M$ then $H^{\bullet}_{\rm orb} ([M / \Gamma]) \cong
H^{\bullet} (M)^{\Gamma}$. In the more general case $H^{\bullet}
(M)^{\Gamma}$ is a subspace of $H^{\bullet}_{\rm orb} ([M/ \Gamma])$ and its
orthogonal complement is the space of twisted sectors.

Assume now that $X$ is a crepant
resolution of the underlying singular space $M/\Gamma$ of a stack $\frak{X}$. The crepant resolution conjecture~\cite{bryan2}
tells us that the Gromov--Witten theories of $\frak{X}$ and of $X$
are equivalent. This means, in particular, that the cohomology
groups $H^{\bullet}_{\rm orb} (\frak{X})$ and $H^{\bullet}(X)$ are
isomorphic, and there is a prescription which takes the Gromov--Witten prepotential of $X$,
$\cF^X$, to the Gromov--Witten prepotential of the orbifold $\frak{X}$,
$\cF^{\frak{X}}$, via a non-trivial transformation.

In the example considered in Section~\ref{sec:C3Z3}, wherein $\frak{X}=[\mathbb{C}^3/ \mathbb{Z}_3]$ and $X=\mathcal{O}_{\PP^2}(-3)$, the construction of this transformation is
precisely the content of the papers~\cite{coates,Aganagic:2006wq}. The orbifold group in this case is $\mathbb{Z}_3=\{1,\zeta,\overline{\zeta}\}$ where
$\zeta= \e^{2 \pi \ii/3}$; it acts on 
$\mathbb{C}^3$ as in (\ref{Z3C3action}). The action of the torus $\torus^3$ lifts
both to $\frak{X}$ and $X$; we work equivariantly with
respect to this toric action.

The inertia stack $I\frak{X}$ has three components
corresponding to the three elements
${1,\zeta,\overline{\zeta}}$. Each component is contractible. A
basis for the equivariant orbifold cohomology $H^{\bullet}_{{\rm orb},\torus^3}(\frak{X})$ is given by the
classes $\{O_1,O_\zeta,O_{\overline{\zeta}}\}$ corresponding to the elements
${1,\zeta,\overline{\zeta}}$, with $O_1 \in H^0_{\rm orb}
(\frak{X})$, $O_\zeta\in H^2_{\rm orb} (\frak{X})$ and $O_{\overline{\zeta}}\in H^4_{\rm orb}(\frak{X})$. Then the
genus zero free energy for the Gromov--Witten series on
$\frak{X}$ is given by
\bea
\cF^{\frak{X}} &=& \sum_{n_1,n_2,n_3 \ge 0}\, \langle O_1^{n_0} \,
O_\zeta^{n_1}\, O_{\overline{\zeta}}^{n_2}\rangle \ \frac{x_1^{n_0}}{n_0 !}\,
\frac{x_\zeta^{n_1}}{n_1 !}\, \frac{x_{\overline{\zeta}}^{n_2}}{n_2 !} \nonumber\\[4pt]
&=& \frac13 \, x_1^3+\frac13\, x_1\,x_\zeta\, x_{\overline{\zeta}} + \sum_{\stackrel{\scriptstyle m,n>0}{m+2n\equiv0\ {\rm mod}\ 3}}\, \langle O_\zeta^m\,O_{\overline{\zeta}}^n\rangle \ \frac{x_\zeta^m}{m!}\, \frac{x_{\overline{\zeta}}^n}{n!} \ .
\eea
The
invariants here are defined via pullback through the evaluation map ev that computes the value of the sigma-model string field embedding the curve in the Calabi--Yau target space. Since we work at genus zero, the unstable terms, with less than three operator insertions, drop out. The counting of ``divisorial classes'' corresponds to the $n=0$ sector of this series and should be compared with the free energy (\ref{calFgorb}) for $g=0$.

The localization calculation is completely determined by what
happens at $[0/\mathbb{Z}_3] \in [\mathbb{C}^3 / \mathbb{Z}_3]$
where $0 \in \mathbb{C}^3$ is fixed by the $\mathbb{Z}_3$-action.
The component of the moduli space of sigma-model fields with target $[\mathbb{C}^3 /
\mathbb{Z}_3]$ that parametrizes constant maps with image
$[0/\mathbb{Z}_3] \in [\mathbb{C}^3 / \mathbb{Z}_3]$ is then
identified with the moduli space of twisted stable maps to the orbifold $B\mathbb{Z}_3 = [0 / \mathbb{Z}_3]$, or equivalently the moduli spaces of admissible $\IZ_3$-covers of genus zero curves. This is analogous
to what happens in the Gromov--Witten theory of a
local Calabi--Yau threefold, where the relevant maps are those which cover
the base of the threefold, i.e. the sigma-model fields factor
through the zero section. See~\cite[Section~4.3]{tseng} and~\cite{jarvis} for a discussion of this point, and~\cite{Bouchard:2007nr} for the extension to higher genus invariants.

Now let us see how this discussion can be used to model our gauge theory formulation. 
The fact that Gromov--Witten theory
is defined through the quotient stack $[\mathbb{C}^3 / \mathbb{Z}_3]$
suggests that one should consider objects on $\mathbb{C}^3$ that
are invariant under the $\mathbb{Z}_3$-action. For example, the $\IZ_3$-invariant part of the character on $\mathbb{C}^3$ is computed in Section~\ref{subsec:InstC3Z3}, see (\ref{calEGammaIt}).
By expanding (\ref{calEGammaIt}) as a power series in $t$ around $t=0$ and
taking the coefficient proportional to $t^3$ one obtains the third
Chern character. If one applies the same procedure to the first and
second Chern characters one finds a non-vanishing result. The
interpretation of this within the context of the localization
calculation is as follows.

To have e.g.~a non-vanishing second Chern
class, one needs a non-trivial toric four-cycle $D$ in the background.
According to our discussion above, the gauge theory
should really be formulated on the quotient stack $[\mathbb{C}^3 / \mathbb{Z}_3]$ and
the localization calculation reduces the target to the classifying stack $[0 / \mathbb{Z}_3] = B\mathbb{Z}_3$. The cohomology of the classifying space $B \mathbb{Z}_3$ can be computed from
\begin{equation}
H^{\rm odd} (B\mathbb{Z}_n , \mathbb{Z}) = 0 \qquad \mbox{and} \qquad H^{\rm even}(B\mathbb{Z}_n , \mathbb{Z}) = \mathbb{Z}_n \ .
\end{equation}
This means that there are effectively four-cycles in the orbifold geometry. As they are purely torsion, they cannot be modelled at the level of differential forms. In what follows we describe a gauge theory realization of the non-trivial orbifold cohomology classes $\{O_1,O_\zeta,O_{\overline{\zeta}}\}$ that arose above in the Gromov--Witten theory.

The computation in Gromov--Witten theory involves insertions of a local operator, which defines the invariants through equivariant localization on the moduli space
of stable maps into
the stack $\frak{X}$. Such insertions correspond to what on a generic threefold $X$ one
would really call (primary) descendent invariants. If one takes them
literally as descendent fields, then the
Gromov--Witten/Donaldson--Thomas correspondence of~\cite{MNOP2}
says that on the gauge theory side one should consider Donaldson--Thomas (primary) descendent invariants. 
On the topological string theory side one takes a basis $\gamma_1, \dots , \gamma_m$ of
$H^{\bullet}(X,\mathbb{Q})$ and defines the primary descendent
fields by integrating products of ${\rm ev}^*(\gamma_{l_i})$. On the gauge theory side one considers
the operator $\mathrm{ch}_{2}
(\gamma): H_{\bullet}
(\frak{M},\mathbb{Q}) \rightarrow H_{\bullet+2-l}
(\frak{M},\mathbb{Q})$ where $\gamma \in H^l (X , \mathbb{Z})$; roughly speaking it is given by
integrating the Chern character of the universal sheaf $\frak{E}$ over
$\gamma$ on $X$ and the virtual fundamental class of the rank one instanton moduli space $\frak{M}$. Then one
defines the descendent invariants of the gauge theory as the integrals
\begin{equation}
(-1)^r\, \int_{\frak{M}}\, \mathrm{ch}_{2} (\gamma_{l_1}) \w \cdots \w
\mathrm{ch}_{2} (\gamma_{l_r}) \ .
\end{equation}
The reduced partition function equals the Gromov--Witten partition function
with the usual change of variables, up to normalization~\cite{MNOP2}.

The operators $\ch_2(\gamma)$ are equivalent to
insertions of an $F_A \wedge F_A$ term integrated over some appropriate
homology class (dual to $\gamma$). To see this, recall that
the toric action on $X$ lifts to the instanton moduli space $\frak{M}$.
Hence one should compute the Chern character of the {\it universal}
sheaf at a fixed point of the toric action on $\frak{M}
\times X$ through localization. But by definition the
universal sheaf $\frak{E}$ restricted to a point of the moduli
space gives precisely the ideal sheaf $\mathcal{I}$ on the
target space $X$, i.e. $\frak{E}|_{\mathcal{I} \times X} \cong
\mathcal{I}$ where $\mathcal{I}$ is now a fixed point of the toric
action on $\frak{M}$. Thus heuristically the descendent invariants are
recovered by D2~charge insertions of $F_A \wedge F_A$ which is
equivalent to the expansion of the term $\exp\big(-\frac12\, \int_D\, F_A
\wedge F_A\big)$ in the gauge theory path integral. This way of interpreting the second Chern character partly justifies the approach to the
gauge theory undertaken in this paper.

\section{Sheaf cohomology\label{App-Cohresults}}

\subsection{Line bundle cohomology of divisors}

In this appendix we collect and prove some results concerning the cohomology
of sheaves on the projective space $\PP^3$. We begin by quoting the
elementary results that will be exploited in the following.
We have
\begin{eqnarray}
\dim_\IC H^0 \big(\PP^3 \,,\, \cO_{\PP^3}(-r)\big) &=& \left( \begin{matrix} 3-r
    \\ -r \end{matrix} \right) \nonumber \\[4pt]
\dim_\IC H^1 \big(\PP^3 \,,\, \cO_{\PP^3}(-r)\big) &=& 0 \ = \ \dim_\IC H^2 \big(\PP^3
\,,\, \cO_{\PP^3}(-r)\big) \ , \nonumber \\[4pt]
\dim_\IC H^3 \big(\PP^3 \,,\, \cO_{\PP^3}(-r)\big) &=& \left( \begin{matrix} r-1 \\ r-4 \end{matrix} \right)\ , 
\end{eqnarray}
and therefore
\begin{eqnarray}
 H^0 \big(\PP^3 \,,\, \cO_{\PP^3}(-r)\big) &=& 0 \qquad \mbox{for}
\quad r>0 \ , \nonumber \\[4pt]
H^1 \big(\PP^3 \,,\, \cO_{\PP^3}(-r)\big) &=& 0  \ = \ H^2 \big(\PP^3
\,,\, \cO_{\PP^3}(-r)\big) \ , \nonumber \\[4pt]
H^3 \big(\PP^3 \,,\, \cO_{\PP^3}(-r)\big) &=& 0 \qquad \mbox{for}
\quad r<4 \ .
\label{HOP30}\end{eqnarray}
These results and others that we use throughout this appendix can be found in~\cite{okonek}.

\subsection{Cohomology of sheaves of differential forms}

The following lemma computes the strongest bounds on the vanishing
cohomology groups for sheaves of differential forms that we were able
to find. Its proof will repeatedly make use of the Euler
sequences for differential forms on $\PP^3$ obtained via truncation of
the Koszul complex (\ref{Koszul}). They are given by
\begin{eqnarray}
\xymatrix{
  0 \ \ar[r] & \ \Omega^1_{\mathbb P^3} \ \ar[r] & \ \mathcal{O}_{\mathbb P^3}
  (-1)^{\oplus 4}\ 
  \ar[r] & \ \mathcal{O}_{\mathbb P^3}\  \ar[r] & \ 0 \ ,
} \label{O1} \\[4pt]
\xymatrix{
  0 \ \ar[r] & \ \Omega^2_{\mathbb P^3} \ \ar[r] & \ \mathcal{O}_{\mathbb P^3}
  (-2)^{\oplus 6}\ 
  \ar[r] & \ \Omega^1_{\mathbb P^3} \ \ar[r] & \ 0 \ ,
} \label{O1O2} \\[4pt]
\xymatrix{
  0 \ \ar[r] & \ \Omega^3_{\mathbb P^3}\ 
  \ar[r] & \ \mathcal{O}_{\mathbb P^3} (-3)^{\oplus 4}\ 
  \ar[r] & \ \Omega^2_{\mathbb P^3} \ \ar[r] & \ 0 \ .
} \label{O2}
\end{eqnarray}

\begin{lemma}
One has the vanishing results
\begin{eqnarray}
H^0 \big(\PP^3 \,,\, \Omega^1_{\PP^3}(-r) \big) &=& 0 \qquad \mbox{for} \quad r>-2 \ , \nonumber \\[4pt]
H^0 \big(\PP^3 \,,\, \Omega^2_{\PP^3}(-r) \big) &=& 0 \qquad \mbox{for} \quad r>-3 \ , \nonumber \\[4pt]
H^1 \big(\PP^3 \,,\, \Omega^1_{\PP^3}(-r) \big) &=& 0 \qquad \mbox{for} \quad r \neq 0 \ , \nonumber \\[4pt]
H^1 \big(\PP^3 \,,\, \Omega^2_{\PP^3}(-r) \big) &=& 0 \ = \
H^2 \big(\PP^3 \,,\, \Omega^1_{\PP^3}(-r) \big) \ , \nonumber \\[4pt]
H^2 \big(\PP^3 \,,\, \Omega^2_{\PP^3}(-r) \big) &=& 0 \qquad \mbox{for} \quad r \neq 0 \ , \nonumber \\[4pt]
H^3 \big(\PP^3 \,,\, \Omega^1_{\PP^3}(-r) \big) &=& 0 \qquad \mbox{for} \quad r<3 \ , \nonumber \\[4pt]
H^3 \big(\PP^3 \,,\, \Omega^2_{\PP^3}(-r) \big) &=& 0 \qquad \mbox{for} \quad r<2 \ .
\label{HOmega0} \end{eqnarray}
\label{HOmega0lemma}\end{lemma}

\Proof{
From the Euler sequence for one-forms (\ref{O1}) we obtain the short exact sequence
\begin{equation}
\xymatrix{
  0 \ \ar[r] & \ \Omega^1_{\mathbb P^3} (-r) \ \ar[r] & \ \mathcal{O}_{\mathbb P^3}
  (-r-1)^{\oplus 4} \ 
  \ar[r] & \ \mathcal{O}_{\mathbb P^3} (-r) \ \ar[r] & \ 0 \ .
} 
\end{equation}
Applying
the snake lemma to write the associated long exact sequence in cohomology, together with
(\ref{HOP30}) we easily conclude
\begin{eqnarray}
H^0 \big(\PP^3 \,,\, \Omega^1_{\PP^3}(-r) \big) &=& 0 \qquad \mbox{for} \quad r>-1 \ , \nonumber \\[4pt]
H^1 \big(\PP^3 \,,\, \Omega^1_{\PP^3}(-r) \big) &=& 0 \qquad \mbox{for} \quad r>0\ , \nonumber \\[4pt]
H^2 \big(\PP^3 \,,\, \Omega^1_{\PP^3}(-r) \big) &=& 0 \ , \nonumber \\[4pt]
H^3 \big(\PP^3 \,,\, \Omega^1_{\PP^3}(-r) \big) &=& 0 \qquad \mbox{for} \quad r<3 \ .
\end{eqnarray}
Since $ \Omega^3_{\mathbb P^3} = \mathcal{O}_{\mathbb P^3} (-4)$, from the Euler sequence for three-forms (\ref{O2}) we obtain the short exact sequence
\begin{equation}
\xymatrix{
  0 \ \ar[r] & \ \cO_{\mathbb P^3} (-4-r) \ 
  \ar[r] & \ \mathcal{O}_{\mathbb P^3} (-3-r)^{\oplus 4} \
  \ar[r] & \ \Omega^2_{\mathbb P^3} (-r) \ar[r] & 0 \ .
}
\label{Euler3O}\end{equation}
Using the associated long exact cohomology sequence and the vanishing
results (\ref{HOP30}) we thus find
\begin{eqnarray}
H^0 \big(\PP^3 \,,\, \Omega^2_{\PP^3}(-r) \big) &=& 0 \qquad \mbox{for} \quad r>-3 \ , \nonumber \\[4pt]
H^1 \big(\PP^3 \,,\, \Omega^2_{\PP^3}(-r) \big) &=& 0  \ , \nonumber\\[4pt]
H^2 \big(\PP^3 \,,\, \Omega^2_{\PP^3}(-r) \big) &=& 0 \qquad \mbox{for} \quad r<0\ , \nonumber \\[4pt]
H^3 \big(\PP^3 \,,\, \Omega^2_{\PP^3}(-r) \big) &=& 0 \qquad \mbox{for} \quad r<1 \ .
\end{eqnarray}
From the Euler sequence for two-forms (\ref{O1O2}) we obtain the final short exact sequence
\begin{equation}
\xymatrix{
  0 \ \ar[r] & \ \Omega^2_{\mathbb P^3}(-r) \ \ar[r] & \ \mathcal{O}_{\mathbb P^3}
  (-r-2)^{\oplus 6} \
  \ar[r] & \ \Omega^1_{\mathbb P^3} (-r) \ \ar[r] & \ 0 \ .
} 
\end{equation}
From the corresponding long exact sequence in cohomology and
(\ref{HOP30}), this in particular implies
\begin{equation}
H^1 \left( \PP^3 \,,\,   \Omega^1_{\mathbb P^3} (-r)   \right) =
H^2 \left( \PP^3 \,,\, \Omega^2_{\mathbb P^3}(-r) \right) \ ,
\end{equation}
but since the left-hand side vanishes for $r > 0$ and the right-hand
side vanishes for $r<0$ the only non-vanishing cohomology group occurs
for $r=0$. In this case we thus conclude
\begin{eqnarray}
H^0 \big(\PP^3 \,,\, \Omega^2_{\PP^3}(-r) \big) &=& 0 \qquad \mbox{for} \quad r>-2 \ , \nonumber \\[4pt]
H^1 \big(\PP^3 \,,\, \Omega^1_{\PP^3}(-r) \big) &=& 0 \qquad \mbox{for} \quad r \neq 0 \ , \nonumber \\[4pt]
H^2 \big(\PP^3 \,,\, \Omega^2_{\PP^3}(-r) \big) &=& 0 \qquad \mbox{for} \quad r \neq 0 \ , \nonumber \\[4pt]
H^3 \big(\PP^3 \,,\, \Omega^1_{\PP^3}(-r) \big) &=& 0 \qquad \mbox{for} \quad r<2 \ .
\end{eqnarray}
Putting everything together we arrive finally at (\ref{HOmega0}).
}

\subsection{Cohomology of ideal sheaves}

We will now study the cohomology of the ideal sheaves of a point. Consider the short exact sequence of sheaves
\begin{equation}
\xymatrix{
  0 \ \ar[r] & \ \cI \
  \ar[r] & \ \mathcal{O}_{\mathbb P^3} \ 
  \ar[r] & \ \cO_z \ \ar[r] & \ 0 \ ,
}
\label{idealseq}\end{equation}
where $\cO_z$ is the skyscraper sheaf of a point $z\in\IC^3$ which is not torsion free. 

\begin{lemma}
The cohomology of the ideal sheaf $\cI$ is given by
\begin{eqnarray}
H^0 \left( \PP^3 \,,\, \cI (-r) \right) &=& 0 \qquad \mbox{for} \quad r > 0 \ , \nonumber \\[4pt]
H^1 \left( \PP^3 \,,\, \cI (-r) \right) &=& \complex \qquad \mbox{for} \quad r > 0 \ , \nonumber \\[4pt]
H^2 \left( \PP^3 \,,\, \cI (-r) \right) &=& 0 \ , \nonumber \\[4pt]
H^3 \left( \PP^3 \,,\, \cI (-r) \right) &=& 0 \qquad \mbox{for} \quad r < 4 \ .
\label{HI0} \end{eqnarray}
\end{lemma}

\Proof{
Take the tensor product of the exact sequence (\ref{idealseq}) with $\cO_{\PP^3} (-r)$ to get
\begin{equation}
\xymatrix{
  0 \ \ar[r] & \ \cI (-r) \
  \ar[r] & \ \mathcal{O}_{\mathbb P^3} (-r) \
  \ar[r] & \ \cO_{\PP^3}(-r) \otimes \cO_z \ \ar[r] & \ 0 \ ,
}
\end{equation}
since $\cO_{\PP^3} (-r)$ is locally free and hence $\underline{\rm Tor}\,_1^{\cO_{\PP^3}} (\cO_{\PP^3} (-r) , \cO_z) = 0$. The skyscraper sheaf is acyclic,
\begin{eqnarray}
H^0 \left( \PP^3 \,,\,  \cO_z  \right) = \complex \qquad \mbox{and} \qquad
H^n \left( \PP^3 \,,\,  \cO_z  \right) = 0 \quad \mbox{for} \ n \neq 0 \ ,
\end{eqnarray}
and this property is unaltered by twisting. Therefore writing
the associated long exact sequence in cohomology and applying (\ref{HOP30}) yields (\ref{HI0}).
}

The next lemma is needed in the proof of Lemma~\ref{HOmegaI0lemma} below, but we defer its proof to Section~\ref{app:Hyper}.

\begin{lemma}
The sheaves $ \Omega_{\PP^3}^m (k) \otimes \cO_z $ are acyclic. 
\label{acycliclemma}\end{lemma}

\begin{lemma}
The groups $H^n (\PP^3 , \Omega^m_{\PP^3} (m) \otimes \cI(-1))$ for $m=1,2$ and $H^n (\PP^3 , \Omega^1_{\PP^3} (1) \otimes \cI(-2))$ are non-zero only for $n=1$, while $H^n (\PP^3 , \Omega^2_{\PP^3} (2) \otimes \cI(-2))$ are non-zero for both $n=1,2$.
\label{HOmegaI0lemma}\end{lemma}

\Proof{
From (\ref{idealseq}) we get the exact sequence of sheaves
\begin{equation}
\xymatrix{
  0 \ \ar[r] & \ \Omega_{\PP^3}^m  (r) \otimes  \cI \
  \ar[r] & \ \Omega_{\PP^3}^m (r) \
  \ar[r] & \ \Omega_{\PP^3}^m (r) \otimes \cO_z \ \ar[r] & \ 0 \ ,
}
\end{equation}
where again $\underline{\rm Tor}\,_1^{\cO_{\PP^3}} (\Omega_{\PP^3}^m (r) , \cO_z) = 0$ since $\Omega_{\PP^3}^m (r) $ is locally free and hence flat. We are interested in computing the sheaf cohomology groups
\begin{eqnarray}
H^{\bullet} \big(\PP^3 \,,\,\cI (-1) \otimes \Omega_{\PP^3}^1 (1)\big) &=& H^{\bullet} \big(\PP^3 \,,\,\cI \otimes \Omega^1_{\PP^3} \big) \ , \nonumber \\[4pt]
H^{\bullet} \big(\PP^3 \,,\,\cI (-2) \otimes \Omega_{\PP^3}^1 (1)\big) &=& H^{\bullet} \big(\PP^3 \,,\,\cI \otimes \Omega^1_{\PP^3} (-1) \big) \ ,
\end{eqnarray}
which come from taking $r=0$ and $r=-1$ respectively. For $r=0,-1$, we
thus consider the long exact cohomology sequence which, by Lemma~\ref{acycliclemma} and Lemma~\ref{HOmega0lemma}, yields
\begin{eqnarray}
H^0 \left( \PP^3 \,,\, \Omega_{\PP^3}^1 (1) \otimes \cI (r-1) \right) &=& 0 \ , \nonumber \\[4pt]
H^1 \left( \PP^3 \,,\, \Omega_{\PP^3}^1 (1) \otimes \cI (r-1) \right) &\neq& 0 \ , \nonumber \\[4pt]
H^2 \left( \PP^3 \,,\, \Omega_{\PP^3}^1 (1) \otimes \cI (r-1) \right) &=& 0 \ , \nonumber \\[4pt]
H^3 \left( \PP^3 \,,\, \Omega_{\PP^3}^1 (1) \otimes \cI (r-1) \right) &=& 0 \ .
\end{eqnarray}
Next we compute
\begin{eqnarray}
H^{\bullet} \big(\PP^3 \,,\,\cI (-1) \otimes \Omega_{\PP^3}^2 (2)\big) &=& H^{\bullet} \big(\PP^3 \,,\,\cI \otimes \Omega^2_{\PP^3} (1) \big) \ , \nonumber \\[4pt]
H^{\bullet} \big(\PP^3 \,,\,\cI (-2) \otimes \Omega_{\PP^3}^2 (2) \big) &=& H^{\bullet} \big(\PP^3 \,,\,\cI \otimes \Omega^2_{\PP^3} \big) \ ,
\end{eqnarray}
which correspond respectively to $r=1$ and $r=0$. For $r=0,1$, we
therefore consider the long exact cohomology sequence, which as above gives
\begin{eqnarray}
H^0 \left( \PP^3 \,,\, \Omega_{\PP^3}^2 (2) \otimes \cI (-2) \right) &=& 0 \ , \nonumber \\[4pt]
H^1 \left( \PP^3 \,,\, \Omega_{\PP^3}^2 (2) \otimes \cI (-2) \right) &\neq& 0 \ , \nonumber \\[4pt]
H^2 \left( \PP^3 \,,\, \Omega_{\PP^3}^2 (2) \otimes \cI (-2) \right) &\neq& 0 \ , \nonumber \\[4pt]
H^3 \left( \PP^3 \,,\, \Omega_{\PP^3}^2 (2) \otimes \cI (-2) \right) &=& 0
\end{eqnarray}
for $r=0$, while
\begin{eqnarray}
H^0 \left( \PP^3 \,,\, \Omega_{\PP^3}^2 (2) \otimes \cI (-1) \right) &=& 0 \ , \nonumber \\[4pt]
H^1 \left( \PP^3 \,,\, \Omega_{\PP^3}^2 (2) \otimes \cI (-1) \right) &\neq& 0 \ , \nonumber \\[4pt]
H^2 \left( \PP^3 \,,\, \Omega_{\PP^3}^2 (2) \otimes \cI (-1) \right) &=& 0 \ , \nonumber \\[4pt]
H^3 \left( \PP^3 \,,\, \Omega_{\PP^3}^2 (2) \otimes \cI (-1) \right) &=& 0
\end{eqnarray}
for $r=1$.
}

\subsection{Hypercohomology of torsion sheaves\label{app:Hyper}}

We now prove Lemma~\ref{acycliclemma}. We want to compute the
cohomology groups $H^{\bullet} (\PP^3 , \Omega_{\PP^3}^m (k) \otimes
\cO_z)$ of the skyscraper sheaf. They are equal to $\Ext_{\cO_{\PP^3}}^{\bullet} (
\Omega_{\PP^3}^m(k)^{\vee} , \cO_z)$. The sheaf
$\Omega_{\PP^3}^m(k)^\vee$ is the holomorphic tangent bundle of $\PP^3$ for $m=1$ and
$k=0$. In particular it is locally free and therefore has a trivial locally free resolution
\begin{equation}
\xymatrix{
0 \ \ar[r] & \ \Omega_{\PP^3}^m(k)^\vee \ \ar[r]^{=} & \
\Omega_{\PP^3}^m(k)^\vee \ \ar[r] & \ 0 \ .
}
\end{equation}
The strategy is to compute first local $\underline{\Ext}$ sheaves and then
global $\Ext$ groups using the local-to-global spectral sequence.

Given this trivial locally free resolution, local $\underline{\Ext}$
sheaves are defined as the cohomology sheaves of the complex
\begin{equation}
\xymatrix{
0 \ \ar[r] & \ \underline{\Hom} \big(\Omega_{\PP^3}^m(k)^\vee \,,\, \cO_z \big) \
\ar[r] & \ 0
}
\end{equation}
and therefore
\begin{equation}
\underline{\Ext}\,^n_{\cO_{\PP^3}} \big(\Omega_{\PP^3}^m(k)^\vee \,,\,
\cO_z \big) = \left\{\  \begin{matrix} \underline{\Hom}
    \big(\Omega_{\PP^3}^m(k)^\vee \,,\, \cO_z \big) =
    \Omega_{\PP^3}^m(k) \otimes \cO_z \ , \quad n=0 \ , \\ 0 \ , \quad
    n=1,2,3 \ . \end{matrix} \right.
\end{equation}
Global $\Ext$ groups are now computed via the local-to-global spectral
sequence
\begin{equation}
E_2^{p,q} = H^p \big(\PP^3 \,,\, \underline{\Ext}\,_{\cO_{\PP^3}}^q
(\cE , \cF) \big) \quad \Longrightarrow \quad \Ext_{\cO_{\PP^3}}^{p+q} (\cE , \cF)
\ .
\end{equation}
However, the spectral sequence in the present case is trivial since
the local $\underline{\Ext}$ sheaves have support only on points; in particular $E^{p,q}_2=0$ for $p >0$. Therefore
\begin{equation}
\Ext^n_{\cO_{\PP^3}} \big(\Omega_{\PP^3}^m(k)^\vee \,,\, \cO_z \big) =
\left\{\  \begin{matrix}  H^0 \big( \PP^3
    \,,\,\Omega_{\PP^3}^m(k)\otimes \cO_z \big) \ , \quad n=0 \ , \\ 0
    \ , \quad n=1,2,3 \ . \end{matrix} \right.
\end{equation}
The result for $n=0$ is tautological, but the conclusion we are
interested in is that all higher cohomology groups vanish,
\begin{equation}
H^{n>0} \big(\PP^3 \,,\,\Omega_{\PP^3}^m (k) \otimes \cO_z \big) = 0 \ ,
\end{equation}
as is required by Lemma~\ref{acycliclemma}. Note that these $\Ext$
groups compute B-model open string spectra between the D0~branes $\cO_z$
and other branes.

We can check this result also directly with the Serre dual. Consider
the cohomology group $\Ext^n_{\cO_{\PP^3}} (\cO_z ,\Omega^m_{\PP^3} (k)^{\vee})$. 
On $\PP^3$ we have the locally free resolution of the skyscraper sheaf
$\cO_z$ given by
\begin{equation}
\xymatrix{
   0 \ \ar[r] & \ \mathcal{O}_{{\mathbb P}^3} (-3) \ \ar[r] & \
   \mathcal{O}_{{\mathbb P}^3} (-2)^{\oplus 3} \ \ar[r] & \
   \cO_{\PP^3} (-1)^{\oplus 3} \ \ar[r] & \ \cO_{\PP^3} \ \ar[r] & \
   \cO_z \
   \ar[r] & \ 0 \ .
}
\end{equation}
Local $\underline{\Ext}$ sheaves are now given by the cohomology sheaves of the complex
\begin{equation}
\xymatrix{
   \underline{\Hom} \big( \cO_{\PP^3} \,,\, \Omega^m_{\PP^3}
   (k)^{\vee} \big) \ \ar[r] & \ \underline{\Hom} \big(
   \cO_{\PP^3}(-1)^{\oplus 3} \,,\, \Omega^m_{\PP^3} (k)^{\vee} \big)
   & 
   \\ \ar[r] & \ \underline{\Hom} \big( \cO_{\PP^3}(-2)^{\oplus 3} \,,\,
   \Omega^m_{\PP^3} (k)^{\vee} \big) \ \ar[r] & \ \underline{\Hom} \big( \cO_{\PP^3}(-3) \,,\, \Omega^m_{\PP^3} (k)^{\vee} \big)  \ .
}
\end{equation}
Arguing as in~\cite[Section~2.4]{Sharpe:2003dr}, the cohomology sheaf of this complex is
\begin{equation}
\underline{\Ext}\,^n_{\cO_{\PP^3}} \big( \cO_z \,,\,
\Omega_{\PP^3}^m(k)^\vee \big) = \left\{\  \begin{matrix} 0 \ , \quad
    n=0,1,2 \ , \\ \cF \ , \quad n=3 \end{matrix} \right.
\end{equation}
where $\cF$ is a certain sheaf that arises as the cohomology sheaf at
the right-most position in the complex (it is essentially
$\cO_z\otimes\Omega_{\PP^3}^m(k)^\vee $); in particular it has zero-dimensional support. It follows that the local-to-global spectral sequence is trivial and
\begin{equation}
\Ext^n_{\cO_{\PP^3}} \big( \cO_z \,,\, \Omega_{\PP^3}^m(k)^\vee \big)
= \left\{\ \begin{matrix} 0 \ , \quad n=0,1,2 \ , \\ H^0 (\PP^3
    , \cF ) \ , \quad n=3 \ . \end{matrix} \right.
\end{equation}
This result is consistent with Serre duality between coherent sheaves
which in the present case implies~\cite{okonek}
\begin{equation}
\Ext^n_{\cO_{\PP^3}} (\cE , \cF) = \Ext^{3-n}_{\cO_{\PP^3}} \big(\cF
\,,\, \cE \otimes \cO_{\PP^3}(-4) \big)
\end{equation}
where $\cO_{\PP^3} (-4)\cong\Omega_{\PP^3}^3$ is the canonical line bundle on $\PP^3$.

\section{Beilinson monad construction\label{App-Beilinson}}

\subsection{Beilinson spectral sequence\label{appsubsec:Beilinson}}

In this appendix we derive the monad parametrization
(\ref{finalcomplexmain})--(\ref{C3sheafcoh}) of sheaves in the moduli
space (\ref{sheavesonP3}). 
The Beilinson theorem implies that for any coherent sheaf
$\mathcal{E}$ on $\mathbb P^3$ there is a spectral sequence
$E_s^{p,q}$ with $E_1$-term (\ref{spectralfirst}) which converges to
(\ref{spectralconv}) for each fixed integer $r\geq0$, where
$\mathcal{E}(-r)=\mathcal{E}\otimes_{\mathcal{O}_{\PP^3}}
\mathcal{O}_{\PP^3}(-r)$. Explicitly, the first term is given by
\begin{equation} E_1^{p,q} = H^q
\big( {\mathbb P}^3 \,,\, \mathcal{E}(-r) \otimes
  \Omega^{-p}_{\PP^3}(-p) \big) \otimes \mathcal{O}_{\PP^3}(p)
\end{equation}
for $p\leq0$. The $E_1$-complexes of the spectral sequence can be
summarized in the diagram
\begin{equation}
\begin{xy}
\xymatrix@C=20mm{
  E_1^{-3,3} \ \ar[r]^{\dd_1} & \ E_1^{-2,3}\  \ar[r]^{\dd_1} & \ E_1^{-1,3}\ 
  \ar[r]^{\dd_1} & \ E_1^{0,3}\  \\
  E_1^{-3,2}\  \ar[r]^{\dd_1} &\  E_1^{-2,2}\  \ar[r]^{\dd_1} &\  E_1^{-1,2}\ 
  \ar[r]^{\dd_1} & \ E_1^{0,2}\ \\
  E_1^{-3,1}\  \ar[r]^{\dd_1} & \ E_1^{-3,1}\  \ar[r]^{\dd_1} & \ E_1^{-1,1}
  \ar[r]^{\dd_1} &\  E_1^{0,1}\ \\
  E_1^{-3,0} \ \ar[r]^{\dd_1} & \ E_1^{-2,0}\  \ar[r]^{\dd_1} &\  E_1^{-1,0}
  \ar[r]^{\dd_1} & \ E_1^{0,0}\ \\
}
\save="x"!LD+<-3mm,0pt>;"x"!RD+<5mm,0pt>**\dir{-}?>*\dir{>}\restore
\save="x"!RD+<0pt,-3mm>;"x"!RU+<0pt,2mm>**\dir{-}?>*\dir{>}\restore
\save!CD+<0mm,-4mm>*{p}\restore \save!CR+<+3mm,0mm>*{q}\restore
\end{xy}  \label{spectralsequence}
\end{equation}
where all other entries are zero for dimensional reasons and
the only nonvanishing differential
\begin{equation}
\dd_1 \,:\, E_1^{p,q} ~\longrightarrow~ E_1^{p+1,q}
\end{equation}
is determined by the morphisms in the Koszul complex
(\ref{Koszul}).

For explicit computations we will again exploit the Euler sequences (\ref{O1})--(\ref{O2}) together with
\begin{eqnarray}
\xymatrix{
  0 \ \ar[r] & \ \mathcal{O}_{\mathbb P^3} (-1)\  \ar[r]^{~~\times z_0} &
  \ \mathcal{O}_{\mathbb P^3}\ 
  \ar[r]^{z_0=0} & \ \mathcal{O}_{\bdiv}\  \ar[r] &\  0 \ .
} \label{boundary} \end{eqnarray}
This sequence defines the plane at infinity $\bdiv=
{[0:z_1:z_2:z_3]} \cong \mathbb P^2 $. We take the tensor product of
the sequence (\ref{boundary})
with $\mathcal{E}(-r)$ to get
\begin{equation}
\xymatrix{
  0 \ \ar[r] & \ \mathcal{E}(-r-1)\  \ar[r] & \ \mathcal{E}(-r)\ 
  \ar[r] & \ \mathcal{E}(-r)\big|_{\bdiv}\  \ar[r] & \ 0 \ ,
}
\label{boundaryE}\end{equation}
where we have used the fact that $\mathcal{O}_{\PP^2}$ is a locally
free sheaf to set ${\rm
  Tor}_{\mathcal{O}_{\PP^2}}^1(\mathcal{E}(-r)
|_{\bdiv},\mathcal{O}_{\bdiv})=0$. The following result is proven
in~\cite{Cirafici:2008sn}.

\begin{lemma}
For $[\cE] \in{\frak M}_{N, k} \big(\mathbb{P}^3\big)$ one has the vanishing results
\bea
H^0 \big( \PP^2 \,,\, \mathcal{E}(-r)\big|_{\bdiv}\, \big) &=& 0
\qquad \mbox{for} \quad r \ge 1 \ , \nonumber \\[4pt]
H^0 \big( \PP^2 \,,\, ( \Omega^1_{\PP^3}(1) \otimes
\mathcal{E}(-r) )\big|_{\bdiv}\, \big) & = & 0 \qquad
\mbox{for} \quad r \ge 1  \ , \nonumber\\[4pt] 
H^0 \left( \PP^3 \,,\, \mathcal{E}(-r) \right) & =& 0 \qquad
\mbox{for} \quad r \ge 1 \ , \nonumber \\[4pt]
H^0 \left( \PP^3 \,,\,
 \Omega^2_{\PP^3}(2) \otimes \mathcal{E}(-r) \right) &=& 0 \qquad
\mbox{for} \quad r \ge 1 \ , \nonumber\\[4pt] 
H^2 \big( \PP^2 \,,\, \mathcal{E}(-r)\big|_{\bdiv}\, \big) & = & 0
\qquad \mbox{for} \quad r \le 2 \ , \nonumber \\[4pt]
H^2 \big( \PP^2
  \,,\,  (\Omega^1_{\PP^3}(1) \otimes \mathcal{E}(-r) )\big|_{\bdiv}
\,\big)&  =&  0 \qquad \mbox{for} \quad r \le 1 \ , \nonumber\\[4pt]
H^3 \left( \PP^3 \,,\, \mathcal{E}(-r) \right) &=& 0 \qquad
\mbox{for} \quad r \le 3 \ , \nonumber \\[4pt]
H^3 \left( \PP^3 \,,\, \Omega^1_{\PP^3}(1) \otimes \mathcal{E}(-r)
\right) &=& 0 \qquad \mbox{for} \quad r \le 2 \ , \nonumber\\[4pt]
H^3 \left( \PP^3 \,,\,
\Omega^2_{\PP^3}(2) \otimes \mathcal{E}(-r) \right) &=& 0 \qquad
\mbox{for} \quad r \le 2 \ .
\eea
\label{HEr0lemma}\end{lemma}

Lemma~\ref{HEr0lemma} imples that for $r=1,2$ all the cohomology groups
$H^0$ and $H^3$ vanish, and hence the first and last rows of the
Beilinson spectral
sequence (\ref{spectralsequence}) are $0$. We will now impose the additional condition
\begin{equation} \label{condition}
H^2 \big(\PP^3 \,,\,\cE (-2)\big) = 0 \ .
\end{equation}
Then the following result implies the vanishing of the second row of
the spectral sequence (\ref{spectralsequence}).

\begin{lemma}
For $[\cE] \in{\frak M}_{N, k} \big(\mathbb{P}^3\big)$ satisfying
$H^2 \big(\PP^3 \,,\,\cE (-2)\big) = 0$ one has
the vanishing results
\bea
H^2 \left( \PP^3 \,,\, \mathcal{E}(-1) \right)=H^2 \left( \PP^3 \,,\, \mathcal{E}(-1) \otimes
    \Omega^1_{\PP^3} (1) \right) =H^2 \big( \PP^3 \,,\, \mathcal{E} (-1)
  \otimes \Omega^2_{\mathbb P^3} (2) \big) = 0 \ .
\eea
\label{H2E0lemma}\end{lemma}

\Proof{
The long exact sequence in cohomology induced by the short exact
sequence of sheaves (\ref{boundaryE}) for $r=1$ contains
\begin{equation}
\xymatrix{
 H^2 \big( \PP^3 \,,\, \mathcal{E}(-2) \big) \ 
  \ar[r] & \ H^2 \left( \PP^3 \,,\, \mathcal{E}(-1) \right) \
  \ar[r] & \ H^2 \big( \PP^2 \,,\, \mathcal{E}(-1)\big|_{\bdiv}\,
  \big) \ ,
}
\end{equation}
and therefore $H^2 \left( \PP^3 \,,\, \mathcal{E}(-1) \right) = 0$ by
(\ref{condition}) and Lemma~\ref{HEr0lemma}. Using (\ref{O1O2}) we consider now the short
exact sequence
\begin{equation}
\xymatrix{
  0 \ \ar[r] & \ \mathcal{E}(-r) \otimes \Omega^2_{\PP ^3} \ \ar[r] &
  \ 
  \mathcal{E}(-r-2)^{\oplus 6} \
  \ar[r] & \ 
  \mathcal{E}(-r) \otimes \Omega^1_{\PP^3} \ \ar[r] & \ 0  \ .
}
\end{equation}
For $r=0$ the corresponding long exact cohomology sequence contains
\begin{equation}
\xymatrix{
  H^2 \left( \PP^3 \,,\, \mathcal{E}(-2)
  \right)^{\oplus 6} \ 
  \ar[r] & \ H^2 \left( \PP^3 \,,\, \mathcal{E}(-1) \otimes
    \Omega^1_{\PP^3} (1) \right) \ \ar[r]
 & \ H^3 \left( \PP^3 \,,\, \mathcal{E}(-2) \otimes \Omega^2_{\PP
^3} (2) \right) \ ,
}
\end{equation}
and therefore $H^2 \left( \PP^3 \,,\, \mathcal{E}(-1) \otimes
    \Omega^1_{\PP^3} (1) \right)  = 0$ by (\ref{condition}) and
  Lemma~\ref{HEr0lemma}. Taking the tensor product of
  (\ref{Euler3O}) for $r=-2$ with $\mathcal{E} (-s)$ gives
\begin{equation} \label{seqEO2}
\xymatrix{
  0 \ \ar[r] & \ \mathcal{E} (-s-2) \ 
  \ar[r] & \ \mathcal{E} (-s-1)^{\oplus 4} \ 
  \ar[r] & \ \mathcal{E} (-s) \otimes \Omega^2_{\mathbb P^3} (2) \
  \ar[r] & \ 0 \ .
}
\end{equation}
The associated long exact sequence in cohomology for $s=1$ contains
\begin{equation} 
\xymatrix{
H^2 \big( \PP^3 \,,\, \mathcal{E}(-2)
  \big)^{\oplus 4} \ 
\ar[r] & \ H^2 \big( \PP^3 \,,\, \mathcal{E} (-1) \otimes
\Omega^2_{\mathbb P^3} (2) \big) \ 
  \ar[r] & \ H^3 \left( \PP^3 \,,\,  \cE (-3) \right) \ ,
}
\end{equation}
and finally we conclude $H^2 \big( \PP^3 \,,\, \mathcal{E} (-1)
\otimes \Omega^2_{\mathbb P^3} (2) \big) = 0$ by (\ref{condition}) and
Lemma~\ref{HEr0lemma}.
}

By Lemma~\ref{HEr0lemma}, Lemma~\ref{H2E0lemma} and Beilinson's
theorem it follows that the cohomology of the differential complex (\ref{spectralsequence})
reduces to
\begin{equation}
\begin{xy}
\xymatrix@C=20mm{
  0  & 0  & 0  & 0 \\
    0  & 0  & 0  & 0 \\
  E_{\infty}^{-3,1} =0  & E_{\infty}^{-2,1} =0&
  E_{\infty}^{-1,1}= \mathcal{E}(-1) & E_{\infty}^{0,1}=0 \ \\
  0  & 0  & 0  & 0 \\
}
\save="x"!LD+<-3mm,0pt>;"x"!RD+<5mm,0pt>**\dir{-}?>*\dir{>}\restore
\save="x"!RD+<0pt,-3mm>;"x"!RU+<0pt,2mm>**\dir{-}?>*\dir{>}\restore
\save!CD+<0mm,-4mm>*{p}\restore \save!CR+<+3mm,0mm>*{q}\restore
\end{xy}
\label{specseqfinal}\end{equation}

\subsection{Generalized ADHM complex}

The following result is proven
in~\cite{Cirafici:2008sn}.

\begin{lemma}
For $[\cE] \in{\frak M}_{N, k} \big(\mathbb{P}^3\big)$ one has
the Euler characters
\begin{eqnarray} 
\chi\big(\mathcal{E}(-2)\big) &=& - k \ , \nonumber\\[4pt]
\chi\left(\mathcal{E}(-1)
\otimes \Omega^2_{\PP^3}(2)\right) &=&- 3 k \ , \nonumber\\[4pt]
\chi\left(\mathcal{E}(-1) \otimes \Omega^1_{\PP^3}(1)\right) &=& - 3 k - N \ ,
\nonumber\\[4pt] \chi\big(\mathcal{E}(-1)\big) &=& - k \ .
\label{Eulerk=2}\end{eqnarray}
\label{Eulerlemma}\end{lemma}

The reduction of the Beilinson spectral sequence from
Section~\ref{appsubsec:Beilinson} is equivalent to the complex
\begin{equation}
\xymatrix{
  0 \ \ar[r] & \ V \otimes \mathcal{O}_{\PP^3} (-3) \ \ar[r]^a &
  \ {B} \otimes \mathcal{O}_{\PP^3} (-2)\ 
  \ar[r]^b &
  \ {C} \otimes \mathcal{O}_{\PP^3} (-1)\ \ar[r]^c &
  \ {D} \otimes \mathcal{O}_{\PP^3} \ \ar[r] & \ 0 \ ,
} \label{finalcomplexforP3}
\end{equation}
where we have defined the complex vector spaces
\bea
V&=& H^1\big(
\PP^3 \,,\, \mathcal{E}(-2)\big) \ \cong \ \IC^k \ , \nonumber\\[4pt] B&=&H^1 \big(
\PP^3 \,,\, \mathcal{E}(-1)\otimes\Omega_{\PP^3}^2(2)\big) \ \cong \ \IC^{3k} \ ,
\nonumber\\[4pt]
C&=&H^1 \big(
\PP^3 \,,\, \mathcal{E}(-1)\otimes\Omega_{\PP^3}^1(1)\big) \ \cong \
\IC^{3k+N} \ ,
\nonumber\\[4pt] D&=&H^1 \big(
\PP^3 \,,\, \mathcal{E}(-1)\big) \ \cong \ \IC^k \ ,
\label{complexvecsp}\eea
whose (stable) dimensions are computed by Lemma~\ref{HEr0lemma}, Lemma~\ref{H2E0lemma} and
Lemma~\ref{Eulerlemma}.
There are some natural identifications. From the short exact sequence
of sheaves (\ref{boundary}) we have
\begin{equation}
\xymatrix{
   H^0 \big( \PP^2 \,,\, \mathcal{E}(-r)\big|_{\bdiv}\,
  \big) \ 
\ar[r] & \ H^1 \left( \PP^3 \,,\, \mathcal{E}(-r-1) \right) & \\
  \ar[r] & \ H^1 \left( \PP^3 \,,\, \mathcal{E}(-r) \right) \ 
  \ar[r] & \ H^1 \big( \PP^2 \,,\, \mathcal{E}(-r)\big|_{\bdiv}\,
  \big) \ .
  }
\label{boundarycohseq}\end{equation}
For $r=1$, the first term vanishes by Lemma~\ref{HEr0lemma}, while $\dim_\IC H^1 \big( \PP^2 \,,\, \mathcal{E}(-1)\big|_{\bdiv}\,
  \big) $ is the instanton charge at infinity which we assume
  to vanish. It follows that $D \cong V$. 

Consider now the vector space $B$ in (\ref{complexvecsp}). The exact
sequence (\ref{seqEO2}) gives
\begin{equation} \label{intermediateB}
\xymatrix{
H^0 \big(\PP^3 \,,\,\cE(-1) \otimes \Omega^2_{\mathbb P^3} (2) \big) \
\ar[r] & \ 
H^1 \big( \PP^3 \,,\, \mathcal{E}(-3)
  \big) \ \ar[r] & \ 
H^1 \big( \PP^3 \,,\, \mathcal{E}(-2)
  \big)^{\oplus 4} \\
\ar[r] & \ H^1 \big( \PP^3 \,,\, \mathcal{E} (-1) \otimes
\Omega^2_{\mathbb P^3} (2) \big) \ 
  \ar[r] & H^2 \left( \PP^3 \,,\,  \cE (-3) \right) \ .
}
\end{equation}
The first term vanishes by Lemma~\ref{HEr0lemma}. Consider the
cohomology sequence (\ref{boundarycohseq}) for $r=2$; the first term
vanishes by Lemma~\ref{HEr0lemma}, while from~\cite[p.~16]{nakajima}
we have the vector space isomorphism
\begin{equation}
H^1 \big( \PP^2 \,,\, \mathcal{E}(-2)\big|_{\bdiv}\,
  \big) = H^1 \big( \PP^2 \,,\, \mathcal{E}(-1)\big|_{\bdiv}\,
  \big)
\end{equation}
whose dimension is again the instanton charge at infinity which we assume to vanish. Therefore $H^1 \left( \PP^3 \,,\, \mathcal{E}(-3) \right) \cong H^1 \left( \PP^3 \,,\, \mathcal{E}(-2) \right) =V$. Finally, the group $ H^2 \left( \PP^3 ,  \cE (-3) \right) $ fits in the exact sequence
\begin{equation}
\xymatrix{
   H^1 \big( \PP^2 \,,\, \mathcal{E}(-2)\big|_{\bdiv}\,
  \big) \ 
\ar[r] & \ H^2 \left( \PP^3 \,,\, \mathcal{E}(-3) \right) \ 
  \ar[r] & \ H^2 \left( \PP^3 \,,\, \mathcal{E}(-2) \right)
  }
\end{equation}
obtained from (\ref{seqEO2}). However, the first term again vanishes
due to the vacuum configuration at infinity, while the third term is
zero due to (\ref{condition}), and therefore 
\begin{equation}
H^2 \left( \PP^3 \,,\, \mathcal{E}(-3) \right) = 0 \ .
\end{equation}
We have just shown that the long exact sequence (\ref{intermediateB}) reduces to the short exact sequence
\begin{equation} 
\xymatrix{
0 \ \ar[r] & \ V \ \ar[r] & \ V^{\oplus 4} \ \ar[r] & \ B \ \ar[r] & \
0 \ .
}
\end{equation}
Since a short exact sequence of vector spaces is always split we
conclude that $B = V\oplus V\oplus V$.

Finally, we determine the vector space $C$ in
(\ref{complexvecsp}). Since we really want the sheaf $\cE$, we twist the complex by $\cO_{\PP^3}(1)$ to get
\begin{equation}
\xymatrix{
  0 \ \ar[r] & \ V \otimes \mathcal{O}_{\PP^3} (-2) \ \ar[r]^a & \ 
  {V}^{\oplus 3} \otimes \mathcal{O}_{\PP^3} (-1) \ 
  \ar[r]^b & \ 
  {C} \otimes \mathcal{O}_{\PP^3} \ \ar[r]^c & \ 
  {V} \otimes \mathcal{O}_{\PP^3} (1) \ \ar[r] & \ 0 \ ,
} \label{finalcomplexforP3a}
\end{equation}
where
\begin{equation}
\underline{\rm im}(a) = \underline{\ker}(b) \qquad \mbox{and} \qquad \cE =
\underline{\ker}(c) \, \big/ \, \underline{\rm im}(b) \ .
\end{equation}
The restriction of this complex to a line $\ell_\infty \cong \PP^1$ at
infinity reads
\begin{equation}
\xymatrix{
  0 \ \ar[r] & \ V \otimes \mathcal{O}_{\PP^3} (-2) \big\vert_{\ell_\infty}
  \ \ar[r]^{a_{\infty}} & \ 
  {V}^{\oplus 3} \otimes \mathcal{O}_{\PP^3} (-1) \big\vert_{\ell_\infty}
  & & \\ & 
  \ar[r]^{b_{\infty}} & \ 
  {C} \otimes \mathcal{O}_{\PP^3} \big\vert_{\ell_\infty} \
  \ar[r]^{c_{\infty}} & \ 
  {V} \otimes \mathcal{O}_{\PP^3} (1) \big\vert_{\ell_\infty} \ \ar[r] & \ 0 \ .
}
\end{equation}
To this complex we can associate the short exact sequences
\begin{eqnarray}
\xymatrix{
  0 \ \ar[r] & \ \underline{\text{ker}}(a_{\infty}) \ \ar[r] & \ V \otimes \mathcal{O}_{\PP^3} (-2) \big\vert_{\ell_\infty} \ \ar[r] & \
 \underline{\text{im}}(a_{\infty}) \
 \ar[r] & \ 0 \ ,
} \label{SESA} \\[4pt]
\xymatrix{
  0 \ \ar[r] & \ \underline{\rm im}(a_{\infty}) \ 
\ar[r] & \ 
  {V}^{\oplus 3} \otimes \mathcal{O}_{\PP^3} (-1) \big\vert_{\ell_\infty} \ 
  \ar[r] & \ \underline{\rm im}(b_{\infty}) \ \ar[r] & \ 0 \ ,
} \label{SESB} \\[4pt]
\xymatrix{
  0 \ \ar[r] & \ \underline{\text{ker}}(c_{\infty}) \ \ar[r] & \
  {C} \otimes \mathcal{O}_{\PP^3} \big\vert_{\ell_\infty} \ \ar[r]^{c_{\infty}} &
  \ {V} \otimes \mathcal{O}_{\PP^3} (1) \big\vert_{\ell_\infty} \ \ar[r] & \ 0 \ ,
} \label{SESC} \\[4pt]
\xymatrix{
  0 \ \ar[r] & \ \underline{\rm im}(b_{\infty}) \ \ar[r] & \ 
   \underline{\ker}(c_{\infty}) \
  \ar[r] & \ \cE \big\vert_{\ell_\infty} \ \ar[r] & \ 0 \ ,
}  \label{SESE}
\end{eqnarray}
where we have used the isomorphisms $\underline{\ker}(b_{\infty})\cong\underline{\text{im}}(a_{\infty})$ and $\underline{\ker}(c_{\infty})\,\big/\, \underline{\rm im}(b_{\infty}) \cong \cE\big|_{\ell_\infty}$.

Since the morphism $a_\infty$ is injective, from (\ref{SESA}) it follows that
\begin{eqnarray}
 H^0 \big( \PP^1 \,,\,
 \underline{\text{im}}(a_{\infty}) \big) &=& 0 \ , \nonumber \\[4pt] \label{resSESA}
V \otimes H^1 \big( \PP^1 \,,\, \mathcal{O}_{\PP^3} (-2) \big\vert_{\ell_\infty} \big) &\cong& V \ \cong \ H^1 \big( \PP^1 \,,\,
 \underline{\text{im}}(a_{\infty}) \big) \ .
\end{eqnarray}
The long exact sequence in cohomology associated with (\ref{SESB}) is
\begin{equation}
\xymatrix{
  0 \ \ar[r] & \ H^0 \big( \PP^1 \,,\, \underline{\rm im}(a_{\infty}) \big) \ \ar[r] & \ 
  {V}^{\oplus 3} \otimes H^0 \big( \PP^1 \,,\, \mathcal{O}_{\PP^3} (-1) \big\vert_{\ell_\infty} \big) \ 
  \ar[r] & \ H^0 \big( \PP^1 \,,\,  \underline{\rm im}(b_{\infty}) \big) \\
  \ar[r] & \ H^1 \big( \PP^1 \,,\, \underline{\rm im}(a_{\infty}) \big) \ \ar[r] & \ 
  {V}^{\oplus 3} \otimes H^1 \big( \PP^1 \,,\, \mathcal{O}_{\PP^3} (-1) \big\vert_{\ell_\infty} \big) \ 
  \ar[r] & \ H^1 \big( \PP^1 \,,\, \underline{\rm im}(b_{\infty}) \big) \ \ar[r] & \ 0 \ ,
}
\end{equation}
which using $H^0 \big( \PP^1 \,,\, \mathcal{O}_{\PP^3} (-1) \big\vert_{\ell_\infty} \big)\cong 0 \cong H^1 \big( \PP^1 \,,\, \mathcal{O}_{\PP^3} (-1) \big\vert_{\ell_\infty} \big)$ implies
\begin{eqnarray} 
H^0 \big( \PP^1 \,,\, \underline{\rm im}(a_{\infty}) \big) = 0 =H^1 \big( \PP^1 \,,\, \underline{\rm im}(b_{\infty}) \big)  \qquad \mbox{and} \qquad
H^0 \big( \PP^1 \,,\,  \underline{\rm im}(b_{\infty}) \big) \cong H^1 \big( \PP^1 \,,\, \underline{\rm im}(a_{\infty}) \big) \ . \nonumber \\
\label{resSESB}
\end{eqnarray}
From (\ref{SESE}) we have
\begin{equation}
\xymatrix{
  0 \ \ar[r] & \ H^0 \big( \PP^1 \,,\,  \underline{\rm im}(b_{\infty}) \big) \ \ar[r] & \ 
   H^0 \big( \PP^1 \,,\, \underline{\ker}(c_{\infty}) \big) \
  \ar[r] & \ H^0 \big( \PP^1 \,,\, \cE \big\vert_{\ell_\infty} \big) & \\
   \ar[r] & \ H^1 \big( \PP^1 \,,\, \underline{\rm im}(b_{\infty}) \big) \ \ar[r] & \ H^1 \big( \PP^1 \,,\, 
   \underline{\ker}(c_{\infty}) \big) \
  \ar[r] & \ H^1 \big( \PP^1 \,,\, \cE \big\vert_{\ell_\infty} \big) \ \ar[r] & \ 0 \ ,
} 
\label{LSEres}\end{equation}
where due to our boundary condition we have $H^1 ( \PP^1 ,  \cE \vert_{\ell_\infty} ) = 0$. 

As in (\ref{Wdecomp}), we define the framing vector space
\begin{equation}
W = H^0 \big( \PP^1 \,,\, \cE \big\vert_{\ell_\infty} \big) \ . 
\end{equation}
If we put together (\ref{resSESA}), (\ref{resSESB}) and (\ref{LSEres}), then we find that
\begin{equation}
 H^1 \big( \PP^1 \,,\, 
   \underline{\ker}(c_{\infty}) \big) = 0
\end{equation}
and that the sequence
\begin{equation}
\xymatrix{
  0 \ \ar[r] & \ V \ \ar[r] & \ 
   H^0 \big( \PP^1 \,,\, \underline{\ker}(c_{\infty}) \big) \
  \ar[r] & \ W \ \ar[r] & \ 0 
} 
\end{equation}
is exact. Since a short exact sequence of vector spaces is always split we conclude that
\begin{equation}
 H^0 \big( \PP^1 \,,\, \underline{\ker}(c_{\infty}) \big) \cong V \oplus W \ .
\end{equation}
Finally, from (\ref{SESC}) we have
\begin{equation}
\xymatrix{
  0 \ \ar[r] & \ H^0 \big( \PP^1 \,,\, \underline{\text{ker}}(c_{\infty}) \big) \ar[r] & \ 
  {C} \otimes H^0 \big( \PP^1 \,,\, \mathcal{O}_{\PP^3} \big\vert_{\ell_\infty}\big) \ \ar[r]^{c_{\infty}} & \ 
  {V} \otimes H^0 \big( \PP^1 \,,\, \mathcal{O}_{\PP^3} (1) \big\vert_{\ell_\infty} \big) & \\
  \ar[r] & \ H^1 \big( \PP^1 \,,\,  \underline{\text{ker}}(c_{\infty}) \big) \ \ar[r] & \ 
  {C} \otimes H^1 \big( \PP^1 \,,\, \mathcal{O}_{\PP^3} \big\vert_{\ell_\infty} \big) \ \ar[r]^{c_{\infty}} & \ 
  {V} \otimes  H^1 \big( \PP^1 \,,\, \mathcal{O}_{\PP^3} (1) \big\vert_{\bdiv} \big) \ \ar[r] & \ 0
} 
\end{equation}
which reduces to the short exact sequence of vector spaces
\begin{equation}
\xymatrix{
  0 \ \ar[r] & \ H^0 \big( \PP^1 \,,\, \underline{\text{ker}}(c_{\infty})\big) \ \ar[r] & \ 
  {C} \ \ar[r]^{c_{\infty}} & \ 
  V\oplus V \ \ar[r] & \ 0 \ ,
} 
\end{equation}
and therefore
\begin{equation}
C \cong  H^0 \big( \PP^1 \,,\, \underline{\text{ker}}(c_{\infty}) \big)  \oplus V\oplus V \cong V\oplus V\oplus V\oplus W \ .
\end{equation}
Putting everything together, we have reduced the complex (\ref{finalcomplexforP3}) to (\ref{finalcomplexmain})--(\ref{C3sheafcoh}).

\end{document}